%% file: 3D-HS_lecture-notes_V2.tex
\documentclass[graybox]{svmult}

\newcommand{\setflag}{\newif\ifwhole \wholetrue}

\include{commands}

\setflag \ifwhole

\begin{document}


\title*{Higher-spin gauge theories in three spacetime dimensions}
\titlerunning{ }
\author{Andrea Campoleoni and Stefan Fredenhagen}
\authorrunning{ }

\institute{A.\ Campoleoni, \at Universit{\'e} de Mons, Service de Physique de l'Univers, Champs et Gravitation, 
Place du Parc 20, 7000 Mons, Belgium,
\email{andrea.campoleoni@umons.ac.be} \and 
S.\ Fredenhagen, \at
University of Vienna, Faculty of Physics, Boltzmanngasse 5, 1090 Vienna, Austria,
\email{stefan.fredenhagen@univie.ac.at} 
}

\maketitle

\vspace{-60pt}

\abstract{These lecture notes provide an introduction to higher-spin gauge theories in three spacetime dimensions, with a focus on their asymptotic symmetries, their holographic description in terms of conformal field theories with \mbox{$\cW$-symmetries} as well as on their couplings to scalar matter.}

\vspace{30pt}

\numberwithin{equation}{section}
\setcounter{minitocdepth}{2}
\dominitoc

\section{Introduction}\label{sec:intro}

Elementary particles are characterised by their mass and spin. For free particles, any value of the spin is allowed, while the structure of interactions strongly depends on the spin.
The interactions of massless particles with spin greater than two, also known as higher-spin particles, are severely constrained by various no-go results; see, e.g., \cite{Bekaert:2010hw, Ponomarev:2022vjb} for a review. 
In spite of this, gauge theories involving higher-spin fields have been investigated over the years,\footnote{See, e.g., \cite{bengtsson_book} for an historical overview.} mainly in view of their possible impact in the study of quantum gravity. 
Indeed, on the one hand, building consistent interactions for massless higher-spin fields typically also requires the presence of a massless spin-two field in the spectrum. This leads to field theories including a putative graviton and displaying a huge gauge symmetry, that is expected to improve the UV behaviour of Einstein's gravity \cite{Fradkin:1990kr}. 
On the other hand, string theory does involve massive excitations of arbitrary spin, and it has been conjectured that it might actually describe a symmetry-broken phase of a higher-spin gauge theory \cite{Gross:1988ue} (see also \cite{Sagnotti:2011jdy, Rahman:2015pzl} for a review of the ensuing developments).
More recently, these suggestions have been reinforced by the natural emergence of higher-spin theories in the AdS/CFT correspondence, as gravity duals of certain weakly coupled conformal field theories; see, e.g., \cite{Giombi:2016ejx} for a review.

In spacetimes of dimension four or higher, there are strong indications that higher-spin gauge theories must involve an infinite number of fields. 
An explicit example of a complete interacting theory of this sort has been provided by Vasiliev and collaborators on constant curvature backgrounds \cite{Vasiliev:1990en, Vasiliev:2003ev} (see also \cite{Bekaert:2004qos, Didenko:2014dwa} for a review). 
Higher-spin generalisations of self-dual Yang-Mills and gravity have also been obtained on flat four-dimensional manifolds with Euclidean or split signature \cite{Ponomarev:2016lrm, Krasnov:2021nsq}, and the status of higher-spin theories on Minkowski backgrounds was recently reconsidered in \cite{Boulanger:2023prx}. 
Conformal gravity admits a higher-spin extension too \cite{Segal:2002gd}, and we refer to \cite{Bekaert:2022poo} for an overview of higher-spin theories and their applications. 
All previous examples entail various peculiarities --~allowing them to evade the plethora of no-go results~-- together with technical difficulties brought by the unbounded spectra of fields.
Studying higher-spin theories on three-dimensional spacetimes gives the unique opportunity to explore some of these peculiarities in a highly simplified context, where, for instance, interacting theories involving only a finite number of higher-spin fields can be built.

Strictly speaking, there are no massless higher-spin particles in three dimensions: the little group does not admit massless representations of arbitrary helicity, so that only scalar and spin-$1/2$ degrees of freedom can propagate \cite{Binegar:1981gv}.
On the other hand, one can still consider field theories similar to those describing higher-spin degrees of freedom in four dimensions and use them as toy models for higher-spin theories, in analogy with the role played by three-dimensional gravity.
One can indeed consider vacuum Einstein's equations also in three dimensions, although they do not allow any local degrees of freedom \cite{Deser:1983tn, Deser:1983nh}. 
Equivalently, no gravitational, spin-two waves propagate in three dimensions. 
Still, black hole solutions do exist \cite{Banados:1992wn, Banados:1992gq} and three-dimensional gravity has been a powerful testing ground for various ideas about their quantum properties as well as a source of insights that found applications in higher dimensions too; see, e.g., \cite{Carlip:1998uc, Kraus:2006wn, Compere:2012jk}.

Analogous considerations extend to arbitrary spin, since handy non-linear completions of the free equations of motion can be defined in three dimensions for any value of the spin, with or without a cosmological constant \cite{Blencowe:1988gj, Bergshoeff:1989ns}. 
They profit from the same simplifications that allow the rewriting of the Einstein--Hilbert action as a Chern--Simons action \cite{Achucarro:1986uwr, Witten:1988hc} and, as anticipated, they do not require to consider an infinite number of fields as in four dimensions.\footnote{The option to define interacting higher-spin theories with field spectra that are not consistent in four dimensions was already noticed in \cite{Aragone:1983sz}.} 
The resulting interacting higher-spin gauge theories have been explored in various directions in recent years and, since they typically involve also spin-two fields, they are often collectively denoted as higher-spin gravity.
A notable research avenue revolves around a holographic duality, dubbed minimal-model holography, that links higher-spin gauge theories coupled to matter on anti-de Sitter (AdS) spacetime to a specific large-$N$ limit of $\cW_N$-minimal models \cite{Gaberdiel:2010pz} (see also \cite{Gaberdiel:2012uj} for a review). 
Such relation elaborates upon the observation that the asymptotic symmetries of three-dimensional higher-spin gauge theories are given by non-linear $\cW$-algebras, rather than by Lie algebras \cite{Henneaux:2010xg, Campoleoni:2010zq}. 
As a result, their putative boundary duals must admit global symmetries of the same type. 
In spite of this peculiarity, minimal-model holography can be considered as a three-dimensional counterpart of the holographic duality between Vasiliev's higher-spin theories in four dimensions and the large-$N$ limit of the three-dimensional $O(N)$ vector model \cite{Klebanov:2002ja} (see also \cite{Sundborg:2000wp, Mikhailov:2002bp, Sezgin:2002rt, Sezgin:2003pt, Giombi:2012ms, Giombi:2016ejx}).
It provides a privileged setup where to test quantitatively the AdS$_3$/CFT$_2$ correspondence, thanks to its virtue of combining the good analytic control one has over $\cW_N$-minimal models with the option to access the semiclassical regime via a large-$N$ limit.   
It also triggered interesting complementary developments as, e.g., the study of black holes in theories in which the higher-spin gauge symmetry makes their usual geometric characterisations meaningless \cite{Gutperle:2011kf, deBoer:2013gz, Bunster:2014mua}, as it might be expected in quantum gravity (see also \cite{Ammon:2012wc, Perez:2014pya, Castro:2016tlm} for a review).

These lecture notes aim at providing an introduction to three-dimensional higher-spin theories allowing to access research papers or specialised reviews on minimal-model holography, like \cite{Gaberdiel:2012uj}.
To this end, three main topics are discussed: sections~\ref{sec:CS} and \ref{sec:symmetries} set the stage by presenting the Chern--Simons formulation of higher-spin gauge theories in three dimensions and their asymptotic symmetries. 
The focus is on higher-spin theories on anti-de Sitter space, but we stress that the same techniques apply to any value of the cosmological constant. 
Section~\ref{sec:quantum} begins with an introduction to quantum $\cW$-algebras and then reviews the bases of minimal-model holography, mainly from the viewpoint of the boundary conformal field theory. Section~\ref{sec:scalar} discusses the subtleties encountered when coupling higher-spin fields to matter, and reviews the Prokushkin--Vasiliev model \cite{Prokushkin:1998bq}. 
This provides an example of such a coupling and has been considered as a natural candidate bulk dual of the conformal field theories introduced in the previous section. 
Additional research directions on three-dimensional higher-spin theories and their links with the material here reviewed are sketched in section~\ref{sec:extra}. We also refer to this section for a more detailed overview of the material presented in these lecture notes. 

Each topic is approached assuming only a textbook knowledge of gravity and conformal field theory, but more advanced developments and applications are discussed towards the end of each section. 
This allows for different reading paths: a reader with experience in three-dimensional higher-spin theories might still profit, e.g., from section~\ref{sec:beyondsl3}, reviewing more sophisticated techniques to characterise the asymptotic symmetries of Chern--Simons theories, or from sections \ref{sec:quantumMiura}--\ref{sec:holography} that delve into the general properties of quantum $\cW$-algebras and their applications in minimal-model holography. 
Section~\ref{sec:scalar} is meant to provide an introduction to the Prokushkin--Vasiliev model, accessible also to readers without any previous exposure to Vasiliev's unfolded formulation of higher-spin gauge theories. 
On the other hand, it introduces a number of techniques that might be useful also in other contexts, like the oscillator realisations of higher-spin algebras discussed in section~\ref{sec:oscillators}. Several exercises complement the text. 
They suggest some computations that might be useful to consolidate the understanding of the material here reviewed, and we collected a number of solutions in section~\ref{sec:solutions}. Exercises with a solution are marked with an asterisk.

\section{Higher spins and Chern--Simons theory}\label{sec:CS}

In this section, we present non-linear actions in three spacetime dimensions that, when linearised around a constant curvature background, reduce to a sum of free actions for massless fields with various spin. The simplest example is the Einstein--Hilbert action in three dimensions: its linearised equations of motion do not admit wave solutions, but they take the same form as in a spacetime of dimension four or higher, where local degrees of freedom are present. When the spin is greater than two, one can reverse the logic: linear equations with an analogous structure as those describing massless spin-$s$ particles in $D \geq 4$ can be defined in $D = 3$ too. They do not admit wave solutions as for $s=2$, but we consider any non-linear completion thereof that preserves the same amount of gauge symmetries as in the free theory as a higher-spin gauge theory.
Remarkably, if one ignores matter couplings, in three dimensions higher-spin gauge theories have essentially the same structure as Einstein's gravity. For this reason, we first briefly recall the main features of the Einstein--Hilbert action in $D = 2+1$ and then we show how to include higher spins.

\subsection{Three-dimensional gravity as a Chern--Simons theory}\label{se:3dGrav}

To generalise the construction of three-dimensional gravity to higher spins, it is convenient to work in a first-order formalism.\footnote{An extensive introduction to three-dimensional gravity, covering both its first and second order formulations, is given in \cite{Carlip:1998uc}.}
In terms of the dreibein $e^a = e_\mu{}^a \intd x^\mu$ and the spin connection $\omega^{ab} = \omega_\mu{}^{ab} \intd x^\mu$, the Einstein--Hilbert action in $D=2+1$ dimensions reads 
\begin{equation} \label{eq:EH1}
S_{\textrm{EH}} = \frac{1}{16\p G} \int \e_{abc} \left( e^a \ww R^{bc} + \frac{1}{3\adsR^2}\, e^a \ww e^b \ww e^c \right) ,
\end{equation}
where $R^{ab} = \intd \omega^{ab} + \omega^{ac} \ww \omega_c{}^b$ is the Riemann curvature, $\adsR$ is related to the cosmological constant by $\L = - \adsR^{-2}$ and $G$ is Newton's constant, which has dimension of length in three dimensions. The equation of motion for the dreibein imposes that the curvature must be constant on shell, thus confirming the absence of local degrees of freedom.

Here and in the following,  $a,b,\ldots$ denote  Lorentz frame indices, while $\mu, \nu, \ldots$ denote base-manifold indices. The latter will often be omitted. We use the following conventions for the Minkowski metric and the Levi--Civita tensor: 
\begin{equation}
\h_{ab} = (-,+,+)
\quad \textrm{and} \quad
\e^{012} = 1\,.
\end{equation}
As usual, $\h_{ab}$ is used to raise and lower Lorentz frame indices.

In three dimensions, one can dualise the spin connection and define
\begin{equation} \label{eq:dual-spin}
\omega^{a} = \frac{1}{2}\, \e^{abc} \omega_{bc} \, , \qquad
R^a = \intd \omega^a + \frac{1}{2}\, \e^{abc} \omega_b \ww \omega_c \, .
\end{equation}
The action \eqref{eq:EH1} can then be rewritten as
\begin{equation} \label{eq:EH2}
S_{\textrm{EH}} = \frac{1}{8\p G} \int \left( e_a \ww R^a + \frac{1}{6\adsR^2}\, \e_{abc}\, e^a \ww e^b \ww e^c \right) ,
\end{equation}
and its equations of motions read
\begin{subequations}
\begin{align}
\cT^a & := \intd e^a + \e^{abc} \omega_b \ww e_c = 0 \, , \label{eq:torsion-spin2} \\
\cR^a & := R^a + \frac{1}{2\adsR^2}\, \e^{abc} e_b \ww e_c = 0 \, . \label{eq:curvature-spin2}
\end{align}
\end{subequations}
The torsion constraint \eqref{eq:torsion-spin2} allows one to rewrite the dualised spin connection in terms of the dreibein, 
while \eqref{eq:curvature-spin2} states that the curvature is constant.

\begin{exercise}\hspace{-2pt}\hyperlink{sol:torsion}{*}
\label{ex:torsion}
Show that the torsion constraint $\cT^a = 0$ implies
\begin{equation}\label{eq:spinsolved}
\omega_\m{}^a = \e^{abc} e^\n{}_{\!b} \left( \pr_{\m} e_{\n\, c} - \pr_{\n} e_{\m\,c} \right) - \frac{1}{2}\, \e^{bcd} \left( e^\n{}_{\!b\,} e^{\,\r}{}_{\!c} \, \pr_\n e_{\r \,d} \right) e_\m{}^a \, , 
\end{equation}
where $e^\m{}_{\!a}$ denotes the inverse dreibein, satisfying $e^\m{}_{\!a} e_\m{}^b = \delta_a{}^b$ and $e^\m{}_{\!a} e_\n{}^a = \delta^\m{}_\n$.
\end{exercise}

As in spacetimes of dimension $D\geq 4$, the Einstein--Hilbert action is invariant under diffeomorphisms and local Lorentz transformations. When $D=3$, the action is also invariant under local translations, so that its symmetries are given by
\begin{equation}\label{eq:localsymmEH}
\delta e^a = \nabla \xi^a + \e^{abc} e_b \L_c  \, , \qquad
\delta \omega^a = \nabla \L^a + \frac{1}{\adsR^2}\, \e^{abc} e_b \x_c \, ,
\end{equation}
where we introduced the Lorentz-covariant derivative $\nabla$, acting on tangent-space vectors as
\begin{equation}
\nabla f^a = \intd f^a + \e^{abc} \omega_b f_c \, .
\end{equation}

\begin{exercise}\hspace{-2pt}\hyperlink{sol:diffeos}{*}
\label{ex:diffeos}
Show that a local translation with parameter $\xi^a = e_\mu{}^a v^\mu$ is equivalent on shell to the diffeomorphism generated by the vector field $v^\mu$,
\begin{equation}
\delta e_\mu{}^a = v^\nu \pr_\nu e_\mu{}^a + e_\n{}^a \pr_\mu v^\nu \, ,
\end{equation}
up to a local Lorentz transformation with parameter $\Lambda^a =\omega_\mu{}^a v^\mu$.
This implies that the invariance under diffeomorphisms does not constitute an additional symmetry of the action.
\end{exercise}

One can further simplify the form \eqref{eq:EH2} of the action by introducing the fields $E = \frac{1}{G}\, e^a P_a$ and $\O = \omega^a L_a$, where $P_a$ and $L_a$ are the generators of the isometry algebra of the vacuum: 
\begin{equation}\label{eq:so22CommutatationRelations}
[ P_a , P_b ] = \left(\frac{G}{\adsR}\right)^{\!\!2}\, \e_{abc} L^c \, , \qquad 
[ L_a , P_b ] = \e_{abc} P^c \, , \qquad 
[ L_a , L_b ] = \e_{abc} L^c \, .
\end{equation}
For $\adsR \to \infty$ one gets the three-dimensional Poincar\'e algebra $\mathfrak{iso}(1,2)$, for $\adsR^{2} < 0$ one gets the de Sitter algebra $\mathfrak{so}(1,3)$, and for $\adsR^{2} > 0$ one gets the anti-de Sitter algebra $\mathfrak{so}(2,2)$. The customary presentation of the vacuum isometry algebras, applying to spacetimes of any dimension, is recovered by dualising the Lorentz generators as $L_a = \frac{1}{2} \e_{abc} M^{bc}$.
One can then rewrite the Einstein--Hilbert action \eqref{eq:EH2} as 
\begin{equation}\label{eq:EH3}
S_{\textrm{EH}} = \frac{1}{8\pi} \int \textrm{Tr} \left( E \ww R + \frac13 \, E \ww E \ww E \right) ,
\end{equation}
where $R =\intd\Omega+\Omega\wedge \Omega= R^a L_a$ and where we introduced the bilinear form\footnote{For the Lie algebras $\mathfrak{iso}(1,2)$ and $\mathfrak{so}(2,2)$, which are not simple, there is also another independent invariant bilinear form that, however, does not lead to the Einstein--Hilbert action \cite{Witten:1988hc}.}
\begin{equation}\label{eq:bilinearForm1}
\textrm{Tr}(P_aP_b) = 0 \, , \qquad
\textrm{Tr}(P_aL_b) = \h_{ab} \, , \qquad
\textrm{Tr}(L_aL_b) = 0 \, .
\end{equation}
Up to boundary terms, the action \eqref{eq:EH3} is equivalent to a Chern--Simons action for the field $\cA = E + \O$ with gauge algebra $\mathfrak{iso}(1,2)$, $\mathfrak{so}(1,3)$ or $\mathfrak{so}(2,2)$, depending on the value of $\adsR$ \cite{Witten:1988hc}:\footnote{In our conventions, the field $\cA$ is dimensionless,
and this explains the overall dimensionless parameter. Newton's constant has been absorbed in the definition of $E$.}
\begin{equation} \label{eq:CS1}
S_{\textrm{EH}} = \frac1{16\pi} \int \textrm{Tr} \left( \cA \ww \intd \cA + \frac23 \cA \ww \cA \ww \cA \right) .
\end{equation}
All solutions of the Chern--Simons field equations,
\begin{equation}\label{CSeom}
    \intd\cA + \cA\wedge \cA = 0\,,
\end{equation}
are flat connections, so that one can see once again that three-dimensional gravity has no local degrees of freedom.

For a negative cosmological constant, one can also profit from the isomorphism $\mathfrak{so}(2,2) \cong \mathfrak{sl}(2,\mathbb{R}) \oplus \mathfrak{sl}(2,\mathbb{R})$. The two copies of $\mathfrak{sl}(2,\mathbb{R})$ are generated by
\begin{equation}
J_a^\pm = \frac{1}{2} \left( L_a \pm \frac{\adsR}{G}\, P_a \right) ,
\end{equation}
that, indeed, satisfy
\begin{equation}
[J^\pm_a , J^\pm_b ] = \e_{ab}{}^c J^\pm_c \,, \qquad [J^\pm_a, J^\mp_b] = 0 \,.
\end{equation}
The bilinear form \eqref{eq:bilinearForm1} then induces
\begin{equation}\label{eq:sl2BilinearForms}
\textrm{Tr} ( J^\pm_a J^\pm_b ) = \pm \frac{1}{2}\, \frac{\adsR}{G} \, \eta_{ab} \,, \qquad
\textrm{Tr} (J^\pm_a J^\mp_b) = 0 \,. 
\end{equation}
Splitting the $\mathfrak{so}(2,2)$ connection as $\cA = A^a J^+_a + \tilde{A}^a J^-_a$ gives
\begin{equation}
A^a = \omega^a + \frac{1}{\adsR}\, e^a\,, \qquad
\tilde{A}^a = \omega^a - \frac{1}{\adsR}\, e^a \, ,
\end{equation}
and substituting this splitting of $\cA$ in the action \eqref{eq:CS1} one can rewrite it as the difference of two Chern--Simons actions \cite{Achucarro:1986uwr, Witten:1988hc}. The latter form of the three-dimensional Einstein--Hilbert action with negative cosmological constant is the most common one in the literature:
\begin{equation}\label{eq:differenceCS}
S_{\textrm{EH}} = S_{\textrm{CS}}[A] - S_{\textrm{CS}}[\tilde{A}] \, ,
\end{equation}
with  
\begin{equation}\label{eq:defCS}
S_{\textrm{CS}}[A] = \frac{k}{4 \pi} \int \tr \left( A \ww \intd A + \frac{2}{3}\, A \ww A \ww A \right) . 
\end{equation}
We introduced here the fields $A = A^a J_a$ and $\tilde{A} = \tilde{A}^a J_a$, taking values in a single $\mathfrak{sl}(2,\mathbb{R})$ algebra with bilinear form
\begin{align}
\label{eq:bilinearFormSl2}
\tr(J_a J_b) = \frac{1}{2}\, \eta_{ab} \,.
\end{align}
The opposite sign in the scalar products \eqref{eq:sl2BilinearForms} of the generators $J^\pm_a$ is taken into account by the relative sign between the two Chern--Simons actions in \eqref{eq:differenceCS}. Furthermore, we absorbed the dimensionless factor $\frac{\adsR}{G}$ in the definition of the level $k$:
\begin{equation}\label{CS:levelandNewtonconstant}
k=\frac{\adsR}{4G} \, .
\end{equation}
Let us stress again that the Einstein--Hilbert action can only be rewritten as the difference of two $\mathfrak{sl}(2,\mathbb{R})$ Chern--Simons actions  when the cosmological constant is negative, i.e.\ for $\adsR^2>0$. Nevertheless, it can always be rewritten as a Chern--Simons action with a $\mathfrak{iso}(1,2)$, $\mathfrak{so}(1,3)$ or $\mathfrak{so}(2,2)$ gauge algebra, since \eqref{eq:CS1} holds for arbitrary values of $\adsR^2$.

Yet another useful rewriting of the Einstein--Hilbert action results from considering the combinations
\begin{equation} \label{A-Atilde}
e =  \frac{\adsR}{2} \left( A - \tilde{A} \right) ,  \qquad
\omega = \frac{1}{2} \left( A + \tilde{A} \right) ,
\end{equation}
where all fields are again assumed to take values in the single $\mathfrak{sl}(2,\mathbb{R})$ algebra with Killing metric \eqref{eq:bilinearFormSl2}.

\begin{exercise}\hspace{-2pt}\hyperlink{sol:sl2EHaction}{*}
\label{ex:sl2EHaction} 
Show that one can rewrite the action \eqref{eq:differenceCS} as
\begin{equation} 
\label{eq:sl2EHaction}
S_{EH} = \frac{1}{4\pi G} \int \tr \left( e \ww R + \frac1{3\adsR^2} \, e \ww e \ww e \right) ,
\end{equation}
and argue that this form of the action actually holds for arbitrary values of the cosmological constant.
\end{exercise}

\begin{exercise}\label{ex:rewritingGaugeTransformation}
Check that the infinitesimal gauge transformations of the Chern--Simons fields,
\begin{equation}
\delta A = \intd \l + [A , \l] \, ,
\qquad
\delta \tilde{A} = \intd \tilde{\l} + [ \tilde{A} , \tilde{\l} ] \, ,
\end{equation}
can be rewritten as
\begin{equation}\label{symm-EH}
\delta e = \intd \x + [\omega , \x] + [e, \L] \, ,
\qquad
\delta \omega = \intd \L + [\omega , \L] + \frac{1}{\adsR^2}\, [e, \x] \, ,
\end{equation}
with
\begin{equation}
\x = \frac{\adsR}{2} \left( \l - \tilde{\l} \right) , 
\qquad
\L = \frac{1}{2} \left( \l + \tilde{\l} \right) ,
\end{equation}
and check that the transformations \eqref{symm-EH} correspond to the symmetries \eqref{eq:localsymmEH} of the Einstein--Hilbert action.
\end{exercise}

\subsection{Including higher-spin fields}\label{sec:HS}

To include higher-spin gauge fields, we begin by presenting their free equations of motion in two approaches, first generalising the ``metric'', second-order formulation of Einstein's gravity, and then generalising the ``frame'', first-order formulation that we reviewed in the previous section. We then show that Chern--Simons actions give interacting theories that decompose into a sum of free higher-spin frame-like actions upon linearisation.

\subsubsection{Metric-like formulation}\label{sec:FronsdalFormulation}

A massless spin-$s$ particle in Minkowski space can be described by a symmetric tensor field $\phi_{\mu_1 \cdots \mu_s}$ fulfilling the Fronsdal equation \cite{Fronsdal:1978rb}
\begin{align}
\label{eq:fronsdalEqFlat}
\Box \phi_{\mu_1 \cdots \mu_s} - s \, \partial^\sigma \partial_{(\mu_1} \phi_{\mu_2 \cdots \mu_s)\sigma} + \frac{s(s-1)}2\, \partial_{(\mu_1} \partial_{\mu_2} \phi_{\mu_3 \cdots \mu_s)}{}^{\sigma}{}_{\sigma} = 0 \,.
\end{align}
Here and below, indices enclosed between parentheses are symmetrised, and dividing by the number of terms in the sum is understood (weight-one convention). For instance, $A_{(\mu} B_{\nu)} := \frac{1}{2} \left( A_{\mu} B_{\nu} + A_{\nu} B_{\mu} \right)$. This equation is left invariant by gauge transformations
\begin{equation}\label{eq:gaugeFronsdal}
\delta \phi_{\mu_1 \cdots \mu_s} = s \, \partial_{(\mu_1} \epsilon_{\mu_2 \cdots \mu_s)}
\end{equation}
with a traceless gauge parameter,
\begin{equation}
\epsilon_{\mu_1 \cdots \mu_{s-3}}{}^\sigma{}_\sigma=0 \,.
\end{equation}
For $s=1$ and $s=2$, Fronsdal's equations reduce, respectively, to Maxwell's equations and to linearised Einstein's equations.

The equations of motion \eqref{eq:fronsdalEqFlat} can be defined on a Minkowski space of any dimension $D$. Together with the gauge symmetry \eqref{eq:gaugeFronsdal}, they imply that, out of the $\binom{D+s-1}{s}$ independent components of the field $\phi_{\mu_1 \cdots \mu_s}$, only the \mbox{$\binom{D+s-4}{s} \frac{D+2s-4}{D+s-4}$} degrees of freedom of a massless spin-$s$ particle propagate on shell (see, e.g., \cite{Rahman:2015pzl}). When \mbox{$D=3$}, Fronsdal's equations can be defined as well but, consistently with the previous counting, they do not admit wave solutions like the spin-two Fierz--Pauli equations.

In the following, we shall be mainly interested in massless higher-spin fields propagating on an AdS background with metric $\bar{g}_{\mu \nu}$. Their equations of motion can be obtained by replacing partial derivatives with Levi--Civita covariant derivatives $\bar{\nabla}$ for the AdS metric in \eqref{eq:fronsdalEqFlat} and by adding suitable mass terms to restore the gauge symmetry
\begin{equation}\label{eq:gaugeFronsdalAdS}
\delta \phi_{\mu_1 \dots \mu_s} = s \, \bar{\nabla}_{\!(\mu_1} \epsilon_{\mu_2 \cdots \mu_s)} \quad \text{with} \quad \bar{g}^{\rho \sigma} \epsilon_{\mu_1 \cdots \mu_{s-3}\rho \sigma} = 0 .
\end{equation}
The additional mass terms allow one to compensate the contributions to the gauge variation produced by the commutator of covariant derivatives,
\begin{equation}
[\bar{\nabla}_{\!\mu}, \bar{\nabla}_{\!\nu}] v_\rho = \frac{1}{\adsR^2} \left( \bar{g}_{\nu \rho} v_\mu - \bar{g}_{\mu \rho} v_\nu \right) .
\end{equation}
The resulting equations of motion for a massless spin-$s$ field in AdS \cite{Fronsdal:1978vb} are then given by
\begin{equation}
\left( \Box - m^2_s\right) \phi_{\mu_1 \cdots \mu_s} - s \, \bar{\nabla}^\sigma \bar{\nabla}_{\!(\mu_1} \phi_{\mu_2 \cdots \mu_s)\sigma} + \frac{s(s-1)}2 \, \bar{\nabla}_{\!(\mu_1}\!\! \bar{\nabla}_{\!\mu_2} \phi_{\mu_3 \cdots \mu_s)}{}^{\sigma}{}_{\sigma} = 0 \, ,
\label{eq:fronsdalEqAds}
\end{equation}
with the mass coefficient\footnote{In AdS the order of derivatives does matter: often in the literature the derivatives in the second term of \eqref{eq:fronsdalEqAds} are exchanged and this gives a different mass term involving also the trace of the field. Notice also that, for $s=2$, \eqref{eq:fronsdalEqAds} agrees with the linearisation of the vacuum Einstein equations $R_{\m\n} - \frac{2\Lambda}{D-2}\,g_{\m\n} = 0$.}
\begin{equation}
m_s^2 = \frac{2(s-1)(D+s-3)}{\adsR^2} \, .
\end{equation}
In AdS$_3$ one thus obtains $m_s^2 = 2s(s-1)\adsR^{-2}$.

\begin{exercise}\label{ex:gaugeInvarianceFronsdalAdS}
Check the invariance of the equations \eqref{eq:fronsdalEqAds} under the gauge transformations \eqref{eq:gaugeFronsdalAdS}.
\end{exercise}

\subsubsection{Frame-like formulation}\label{sec:frame-like}

In three dimensions, the equations of motion \eqref{eq:fronsdalEqAds} can be reformulated in a first-order form by introducing the higher-spin vielbeins and spin connections
\begin{equation}\label{eq:def-fields}
e^{a_1 \cdots a_{s-1}} = e_\mu{}^{a_1 \cdots a_{s-1}} \intd x^\mu \,, \qquad
\omega^{a_1 \cdots a_{s-1}} = \omega_\mu{}^{a_1 \cdots a_{s-1}} \intd x^\mu \,,
\end{equation}
which are symmetric and traceless in their Lorentz indices $a_i$ and admit the gauge symmetries
\begin{subequations} \label{eq:gauge-s}
\begin{align}
\delta e^{a_1 \cdots a_{s-1}} & = \nabla \xi^{a_1 \cdots a_{s-1}} + (s-1)\, \bvb_{b}\,\epsilon^{bc(a_1} \Lambda^{a_{2} \cdots a_{s-1})}{}_c \, , \\
\delta \omega^{a_1 \cdots a_{s-1}} & = \nabla \Lambda^{a_1 \cdots a_{s-1}} + \frac{s-1}{\adsR^2}\, \bvb_{b}\,\epsilon^{bc(a_1} \xi^{a_{2} \cdots a_{s-1})}{}_c \, , 
\end{align}
\end{subequations}
where $\bvb$ and $\bsc$ denote the dreibein and spin connection of the AdS background, and the gauge parameters $\x^{a_1 \cdots a_{s-1}}$ and $\L^{a_1 \cdots a_{s-1}}$ are symmetric and traceless like the fields. Moreover, $\nabla$ denotes the Lorentz covariant derivative
\begin{equation}\label{eq:lorentzDer}
\nabla f^{a_1 \cdots a_{n}} = \intd f^{a_1 \cdots a_{n}} + n\, \epsilon^{bc (a_1} \, \bsc_b \ww f^{a_2 \cdots a_{n})}{}_{c} \, ,
\end{equation}
which satisfies
\begin{equation}\label{eq:lorentzsquared}
\nabla^2 f^{a_1 \cdots a_n} = \frac{n}{\adsR^2}\, \bvb^b \ww \bvb^{(a_1} \ww f^{a_2 \cdots a_n)}{}_b \, .
\end{equation}
For \mbox{$s=2$} one recovers the dreibein and the dualised spin connection introduced in \eqref{eq:dual-spin}, together with the linearisation of the transformations \eqref{eq:localsymmEH}. When $D>3$, one must actually introduce spin connections of the form $\omega^{b,a_1\cdots a_{s-1}}$ satisfying $\omega^{(b,a_1\cdots a_{s-1})} = 0$ (see, e.g., \cite{Didenko:2014dwa,Rahman:2015pzl} and appendix~\ref{app:arbitraryD} for more details). When $D=3$, the latter irreducibility condition allows one to dualise them to recover the fields introduced in \eqref{eq:def-fields}. 

For any value of the spin, one can then consider the action
\begin{equation} \label{eq:frameAction}
\begin{split}
S = \frac{1}{8\p G} \int & \Big( e_{a_1 \cdots a_{s-1}} \ww \nabla \omega^{a_1 \cdots a_{s-1}} + \frac{s-1}{2}\, \e_{bcd} \bvb^b \ww \omega^c{}_{a_1 \cdots a_{s-1}} \ww \omega^{da_1 \cdots a_{s-1}} \\
& + \frac{s-1}{2\adsR^2}\, \e_{bcd} \bvb^b \ww e^c{}_{a_1 \cdots a_{s-1}} \ww e^{da_1 \cdots a_{s-1}} \Big) \, ,
\end{split}
\end{equation}
which is left invariant by the gauge transformations \eqref{eq:gauge-s} and, for $s=2$, corresponds to the linearisation of the Einstein--Hilbert action \eqref{eq:EH2}. It also corresponds to the three-dimensional version of the linearised action for a massless spin-$s$ particle in arbitrary dimension that we discuss in section~\ref{app:arbitraryD}. When $D = 3$, one could also consider other gauge invariant quadratic actions in the fields \eqref{eq:def-fields}, while for generic values of $D$ the action is uniquely fixed. 

One can also introduce the following curvatures for the fields $e$ and $\omega$:
\begin{subequations}
\label{eq:frameLikeCurvatures}
\begin{align}
\cT^{a_1 \cdots a_{s-1}} &= \nabla e^{a_1 \cdots a_{s-1}} + (s-1)\, \epsilon^{bc (a_1} \, \bvb_b \ww \omega^{a_2 \cdots a_{s-1})}{}_{c} \,, \label{torsion_3D} \\
\cR^{a_1 \cdots a_{s-1}} &= \nabla \omega^{a_1 \cdots a_{s-1}} + \frac{s-1}{\adsR^2} \, \epsilon^{bc (a_1} \, \bvb_{b} \ww e^{a_2 \cdots a_{s-1})}{}_{c} \,. \label{curvature_3D}
\end{align}
\end{subequations}
These two-forms are invariant under the transformations \eqref{eq:gauge-s} and satisfy the Bianchi identities
\begin{subequations}\label{eq:Bianchi-lin}
\begin{align} 
\nabla \cT^{a_1 \cdots a_{s-1}} + (s-1)\, \e^{bc(a_1} \bvb_b \ww \cR^{a_2 \cdots a_{s-1})}{}_c & = 0 \, , \\
\nabla \cR^{a_1 \cdots a_{s-1}} + \frac{s-1}{\adsR^2}\, \e^{bc(a_1} \bvb_b \ww \cT^{a_2 \cdots a_{s-1})}{}_c & = 0 \, .
\end{align}
\end{subequations}
In analogy with the spin two example, the equations of motion of the action \eqref{eq:frameAction} are the zero-curvature conditions 
\begin{subequations}
\label{eq:frameLikeFreeEOM}
\begin{align}
\cT^{a_1\cdots a_{s-1}} &= 0 \,, \label{eq:curvatureConstraint1}\\
\cR^{a_1\cdots a_{s-1}} &= 0 \,. \label{eq:curvatureConstraint2}
\end{align}
\end{subequations}
This is again a peculiarity of the three-dimensional setup: when $D > 3$, in analogy with gravity, the equations of motion only imply the vanishing of a certain projection of the analogue of the curvature \eqref{eq:curvatureConstraint2} --~see again section~\ref{app:arbitraryD} for more details. 

\begin{exercise}\label{ex:gaugeInvarianceFrame}
Prove the gauge invariance of the action \eqref{eq:frameAction} under the transformations \eqref{eq:gauge-s} and check that its equations of motion are those in \eqref{eq:frameLikeFreeEOM}.
\end{exercise}

Equations \eqref{eq:frameLikeFreeEOM} are equivalent to the Fronsdal equation \eqref{eq:fronsdalEqAds}. This can be shown by following the same steps as in exercise~\ref{ex:torsion}, i.e., by first expressing the generalised spin connection in terms of the higher-spin vielbein and its first derivative using the torsion constraint \eqref{eq:curvatureConstraint1}, and then by substituting the resulting
$
\omega^{a_1\cdots a_{s-1}} = \omega^{a_1\cdots a_{s-1}}(e,\nabla e)
$ 
in \eqref{eq:curvatureConstraint2} (see \cite{Campoleoni:2010zq} for more details or section~\ref{app:arbitraryD} for the analogous, and somehow simpler, solution~\eqref{solving_torsion} of the torsion constraint in any $D$).
The Fronsdal equation can eventually be recovered by rewriting \eqref{eq:curvatureConstraint2} in terms of
\begin{align}
\label{eq:frondsdalFromHSvielbein}
\phi_{\mu_1 \cdots \mu_s} = \bvb^{a_1}{}_{(\mu_1} \cdots\, \bvb^{a_{s-1}}{}_{\mu_{s-1}} \, e_{\mu_s) \, a_1 \cdots a_{s-1}} \, .
\end{align}
%

\subsubsection{Chern--Simons formulation}\label{sec:CSformulation}

As for gravity, one can combine the spin connections $\omega^{a_1\cdots a_{s-1}}$ and the vielbeins $e^{a_1\cdots a_{s-1}}$ into the linear combinations
\begin{equation}
A^{a_1 \cdots a_{s-1}}=\Big(\omega+\frac1\adsR \, e\Big)^{a_1 \cdots a_{s-1}} \,, \qquad 
\tilde{A}^{a_1 \cdots a_{s-1}}=\Big(\omega-\frac1\adsR \, e\Big)^{a_1 \cdots a_{s-1}} \,.
\end{equation}
We now wish to introduce Chern--Simons fields $A$ and $\tilde{A}$ including the gravitational connections discussed in section~\ref{se:3dGrav}, together with the fields above contracted with the generators $J_{a_1 \cdots a_{s-1}}$ of a higher-spin Lie algebra that we still need to identify.

These generators, besides being symmetric under permutations of their indices, cannot be all independent as they must satisfy the constraints $\h^{ab} J_{abc_1 \cdots c_{s-3}} = 0$ induced by the contraction with traceless fields.
Moreover, we can fix the commutation relations between $J_a$ and $J_{b_1 \cdots b_{s-1}}$ by imposing that the linearisation of the Chern--Simons equations of motion reproduces the free equations of motion for  higher-spin fields on the AdS background. The vanishing of the curvatures for $A$ and $\tilde{A}$, when expressed in terms of $e$ and $\omega$, imposes
\begin{equation} \label{curv_gen}
\intd\omega + \omega \ww \omega + \frac{1}{\adsR^2}\, e \ww e =0 \, , \quad 
\intd e + e\ww \omega + \omega \ww e =0
\end{equation}
(see also exercise~\ref{ex:sl2EHaction}, where the precise form of the gauge algebra did not play any role).
Expanding the fields in terms of the generators $t_A$ of the Lie algebra satisfying $[t_A,t_B]=f_{AB}{}^{C}t_C$, we obtain
\begin{equation} \label{eom_before_linearisation}
\intd\omega^A + \frac{1}{2} f_{BC}{}^{A}\,\omega^B \ww \omega^C + \frac{1}{2\adsR^2} f_{BC}{}^A\,e^B \ww e^C=0 \, ,\quad
\intd e^A + f_{BC}{}^A \, e^B \ww \omega^C =0 \, .
\end{equation}
Linearising around a background $\bvb=\bvb^a J_a$, $\bsc=\bsc^a J_a$ one obtains
\begin{equation}\label{eom_after_linearisation}
\intd\omega^A + f_{bC}{}^{A}\,\bsc^b \ww \omega^C + \frac{1}{\adsR^2} f_{bC}{}^A\,\bvb^b \ww e^C=0 \, ,\quad
\intd e^A + f_{bC}{}^A \, \bsc^b \ww e^C + f_{bC}{}^A \, \bvb^b \ww e^C =0 \, ,
\end{equation}
where capital indices can be traded for the multi-indices $(a_1 \cdots a_{s-1})$.
Comparing to \eqref{eq:frameLikeFreeEOM} we can read off the Lie bracket
\begin{align}
\label{eq:hscommutatorsl2}
[J_a,J_{b_1 \cdots b_{s-1}}] = (s-1)\, \epsilon^d{}_{a(b_1} J_{b_2 \cdots b_{s-1})d} \,.
\end{align}

The zero-curvature equations of motion \eqref{curv_gen} may be obtained starting from several actions, as noticed in the gravity case in \cite{Witten:1988hc}. On the other hand, we also wish to recover upon linearisation the action \eqref{eq:frameAction}, that directly generalises the Einstein--Hilbert action and corresponds to the $D=3$ instance of the free higher-spin action \eqref{action_anyD}. To this end, one can start from an action of the form \eqref{eq:sl2EHaction} that, as shown in exercise~\ref{ex:sl2EHaction}, can also be written as the difference of two Chern--Simons actions when the cosmological constant is negative (the rewriting does not rely on the form of the gauge algebra). 
A direct comparison with the linearised action \eqref{eq:frameAction} then fixes the invariant bilinear form in \eqref{eq:sl2EHaction} to
\begin{subequations} \label{killing_J}
\begin{align}
& \tr\left( J_a J_b \right) = \frac{1}{2}\, \h_{ab} \, , \quad
\tr\left( J_a J_{b_1 \cdots b_{s-1}} \right) = 0 \, , \\
& \tr\left( J_{a_1 \cdots a_{s-1}} J_{b_1 \cdots b_{s-1}} \right) = \frac{1}{2}\, \h_{a_1(b_1|} \h_{a_2|b_2} \cdots \h_{b_{s-1})a_{s-1}} + \cdots \, , \label{killing-s}
\end{align}
\end{subequations}
where in the last line we omitted the terms needed to separately enforce a traceless projection in the indices $a_i$ and $b_i$ (when writing the action the projection is already automatically taken into account by the contraction with the traceless fields).\footnote{Notice that we fixed conventionally the overall sign in \eqref{killing-s} independently of the value of $s$. While keeping the same relative sign between the kinetic terms of different fields is a reasonable requirement, we stress that this is not mandatory in the current context in which there are no propagating degrees of freedom and, therefore, no risks to introduce ghosts. At the level of the gauge algebra, introducing relative signs in \eqref{killing-s} would correspond to selecting different real forms of the same complex Lie algebra.}

\begin{exercise}\hspace{-2pt}\hyperlink{sol:linearizationHS}{*}
\label{ex:linearizationHS}
Check \eqref{eq:hscommutatorsl2} and \eqref{killing_J} starting, respectively, from \eqref{eom_after_linearisation}, \eqref{eq:frameLikeFreeEOM} and \eqref{eq:sl2EHaction}, \eqref{eq:frameAction}.
\end{exercise}

A Chern--Simons theory with gauge algebra $\mathfrak{g}\oplus\mathfrak{g}$ and action \eqref{eq:differenceCS} thus reproduces Fronsdal's equations in AdS$_3$ upon linearisation, provided that $\mathfrak{g}$ admits the non-degenerate invariant bilinear form \eqref{killing_J} and its generators can be collected in traceless symmetric tensors $J_{a_1 \cdots a_{s-1}}$ satisfying the Lie bracket \eqref{eq:hscommutatorsl2}.
At this point, one should classify all algebras fulfilling the previous conditions to classify all possible interacting higher-spin gauge theories.\footnote{A priori this classification would only cover interacting theories that can be cast in the Chern--Simons form, but one can prove that the latter exhaust the space of all possible interactions in the absence of matter couplings \cite{Fredenhagen:2019hvb,Grigoriev:2020lzu}.} In principle, this task can be achieved by solving the Jacobi identities with the initial data \eqref{eq:hscommutatorsl2}; still, we stress that any non-compact Lie algebra with a non-degenerate bilinear form fits into this scheme upon selecting a distinguished $\mathfrak{sl}(2,\mathbb{R})$ subalgebra to be interpreted as the gravitational one, possibly modulo the flip in the overall sign of \eqref{killing-s} for some values of $s$. We shall comment more on this general setup at the end of this section, while we now illustrate the key features of three-dimensional higher-spin algebras in a prototypical example.

To begin with, we observe that a symmetric and traceless product of $\mathfrak{sl}(2,\mathbb{R})$ generators satisfies the commutator \eqref{eq:hscommutatorsl2}, so that we can set
\begin{align}
\label{eq:symProd}
J_{a_1 \cdots a_{s-1}} \sim J_{\{a_1} \cdots J_{a_{s-1}\}} \, ,
\end{align}
where braces denote a symmetrisation and a traceless projection. 

Defining \eqref{eq:symProd} requires introducing an associative product, out of which one can realise the Lie bracket as a commutator. The simplest way of doing so is to consider a finite-dimensional representation of $\mathfrak{sl}(2,\mathbb{R})$ and use the matrix product. For concreteness, let us first focus on the three-dimensional representation. We can then consider the combinations
\begin{align}
\label{eq:matrices}
J_a\,, \quad 
J_{\{a}J_{b\}}= J_{(a}J_{b)}-\frac{1}{3} \, \eta_{ab} \, C_2\,,
\end{align}
where 
\begin{equation}\label{Casimir}
C_2 = \h^{ab} J_a J_b  
\end{equation}
is the $\mathfrak{sl}(2,\mathbb{R})$ Casimir operator.
As you are invited to check in the following exercise, these eight matrices are all traceless and linearly independent: as a result, they span all traceless 3$\times$3-matrices and generate $\mathfrak{sl}(3,\mathbb{R})$. The resulting Chern--Simons theory contains fields of spin two and three corresponding, respectively, to the generators $J_a$ and $J_{\{a}J_{b\}}$. 

\begin{exercise}\label{ex:sl3generators}
Use the following three-dimensional representation of $\mathfrak{sl}(2,\mathbb{R})$,
\begin{equation}\label{3drepofsl2}
J_0 = \frac{1}{\sqrt{2}} \begin{pmatrix}
    0 & -1 & 0 \\
    1 & 0 & -1  \\
    0 & 1 & 0  \\
    \end{pmatrix} , \quad
J_1 = \frac{1}{\sqrt{2}} \begin{pmatrix}
    0 & 1 & 0 \\
    1 & 0 & 1  \\
    0 & 1 & 0  \\
    \end{pmatrix} , \quad
J_2 = \begin{pmatrix}
    1 & 0 & 0 \\
    0 & 0 & 0  \\
    0 & 0 & -1  \\
    \end{pmatrix} ,
\end{equation}
to show that the matrices in \eqref{eq:matrices} are linearly independent and traceless.
\end{exercise}

Starting from a $N$-dimensional representation of $\mathfrak{sl}(2,\mathbb{R})$ leads to an analogous construction: symmetrised and traceless products of $s-1 \leq N$ generators are independent and form the fundamental representation of $\mathfrak{sl}(N,\mathbb{R})$ \cite{PhD_Hoppe}. As a result, a Chern--Simons theory with gauge algebra $\mathfrak{sl}(N,\mathbb{R}) \oplus \mathfrak{sl}(N,\mathbb{R})$ describes gauge fields of spin $s=2,3,\ldots,N$. Each of them corresponds, respectively, to the components proportional to
\begin{align}
\label{eq:slNrep} 
J_{a_1 \cdots a_{s-1}} = \sqrt{\frac{(2s-1)!}{6(s-1)!^2}\prod^{s-1}_{i=2} \frac{2}{N^2-i^2}} \ J_{\{a_1} \cdots J_{a_{s-1}\}} \, , \quad 2 \leq s \leq N\, ,
\end{align}
where we fixed the normalisation such that it agrees with the normalisation of the bilinear invariant form introduced in \eqref{killing_J}.

This construction can be generalised to include infinite-dimensional representations of $\mathfrak{sl}(2,\mathbb{R})$ so as to obtain higher-spin theories with an unbounded spectrum of fields. To this end, let us recall that in a $N$-dimensional irreducible representation of $\mathfrak{sl}(2,\mathbb{R})$ the quadratic Casimir \eqref{Casimir} takes the value 
\begin{equation}\label{Casimir_finite}
C_2 = \frac{1}{4} \left(N^2 - 1\right) \mathbb{I}\, .
\end{equation}
The previous construction can thus be recovered by defining the combinations \eqref{eq:slNrep} in the universal enveloping algebra\footnote{For any Lie algebra $\mathfrak{g}$, its universal enveloping algebra $\mathcal{U}(\mathfrak{g})$ is built as follows. One first introduces the space of linear combinations of all possible products of generators, defined by concatenation, and then quotients out the two-sided ideal generated by $J_a J_b - J_b J_a - f_{ab}{}^c J_c$ to ensure that the Lie bracket is realised as the commutator. One can show that $\mathcal{U}(\mathfrak{g})$ is spanned by symmetrised products of the generators since any antisymmetrised combination can be replaced by the commutator. For more details, see, e.g., \cite[chapter 14]{Fuchs:1997jv}.} of $\mathfrak{sl}(2,\mathbb{R})$ and then imposing \eqref{Casimir_finite} by factoring out the corresponding relation. One can then generalise the procedure to include other values of $C_2$ so as to define the associative algebras
\begin{align}\label{CS:Blambda}
B[\lambda] = \frac{\mathcal{U}(\mathfrak{sl}(2,\mathbb{R}))}{\langle C_2 - \frac{1}{4}\! \left(\lambda^2-1\right) \mathbb{I} \rangle}\, .
\end{align}
We can turn this associative algebra into a Lie algebra with Lie bracket given by the commutator. As a Lie algebra, $B[\lambda]$ decomposes as
\begin{equation} \label{fromB-to-hs}
B[\lambda] = \mathbb{R} \oplus \mathfrak{hs}[\lambda] \,,
\end{equation}
since the identity $\mathbb{I}$ commutes with all generators. For all values of $\lambda$, the Lie algebra $\mathfrak{hs}[\lambda]$ is a higher-spin algebra spanned by generators $J_a$, $J_{\{a}J_{b\}}$,\ldots, because any antisymmetric combination can be expressed in terms of the Lie bracket, and any trace part can be expressed in terms of the Casimir operator.\footnote{This infinite-dimensional algebra has been introduced from different perspectives in \cite{Feigin_1988, Bergshoeff:1989ns, Pope:1989sr, Fradkin:1990ir, Bordemann:1989zi, Khesin:1994ey}. In section~\ref{sec:oscillators}, we shall also discuss in detail an alternative construction of $\mathfrak{hs}[\lambda]$ in terms of oscillator variables.} A Chern--Simons theory with gauge algebra $\mathfrak{hs}[\lambda] \oplus \mathfrak{hs}[\lambda]$ thus describes an interacting higher-spin theory for gauge fields of spin $s=2,3,4,\dots$, while different values of $\lambda$ give inequivalent theories since the gauge algebras are not isomorphic (see, e.g., \cite{Bergshoeff:1989ns, Bordemann:1989zi}).

To build a Chern--Simons action, one also needs to define an invariant, non-degenerate bilinear form on $\mathfrak{hs}[\lambda]$. This can be done introducing a trace in $B[\lambda]$ defined by
\begin{equation}\label{traceBlambda}
\tr\, \mathbb{I} = \textrm{const} \,, \quad \tr J_{\{a_1}\cdots J_{a_{s-1}\}} = 0 \, .
\end{equation} 
The corresponding invariant bilinear form on $\mathfrak{hs}[\lambda]$ is non-degenerate for non-integer $\lambda$ \cite{Vasiliev:1989re}. If $\lambda=N>1$ is an integer, the bilinear form degenerates, and there is a non-trivial ideal 
\begin{equation}
I_N = \{ V \in  \mathfrak{hs}[N] |\, \text{for all}\ W\in \mathfrak{hs}[N]:\, \tr (VW)=0\}\, .
\end{equation} 
Factoring out this ideal leads to $\mathfrak{sl}(N,\mathbb{R})$,
\begin{equation}
\frac{\mathfrak{hs}[\lambda=N]}{I_N} \cong \mathfrak{sl}(N,\mathbb{R}) \,,
\end{equation}
and thus reproduces the previous construction in terms of products of matrices of finite-dimensional representations of the $\mathfrak{sl}(2,\mathbb{R})$ algebra.

We now introduce a convenient basis for $\mathfrak{hs}[\lambda]$ that we shall use extensively later.
To begin with, we define the linear combinations
\begin{equation} 
\label{LmintermsofJa}
L_{\pm 1} = J_0 \pm J_{1} \, , \quad L_0 = J_2
\end{equation}
that form a basis for the $\mathfrak{sl}(2,\mathbb{R})$ algebra with commutation relations
\begin{equation}\label{L-basis}
[L_m , L_n] = (m-n) L_{m+n} \, .
\end{equation}
The 
invariant bilinear form 
\eqref{eq:bilinearFormSl2} 
in this basis takes the form
\begin{equation}\label{sl2trace}
    \tr L_0L_0 = \frac{1}{2}\,, \quad \tr L_1 L_{-1} = -1\,, \quad \tr L_0 L_{\pm 1}=0\, .
\end{equation}
For any fixed $s$ the generators $J_{\{a_1}\cdots J_{a_{s-1}\}}$ form a $(2s-1)$-dimensional irreducible representation of $\mathfrak{sl}(2,\mathbb{R})$ with respect to the adjoint action. This directly follows from standard consideration on the tensoring of $\mathfrak{sl}(2,\mathbb{R})$ representations, and can also be made more manifest by using as a basis the generators $W^{s}_m$ (with $-s+1\leq m \leq s-1$) defined as follows. One first introduces the generator
\begin{equation}\label{Wsrecursionstart}
W^{s}_{s-1} = (L_1)^{s-1}
\end{equation}
and then defines recursively further $2s-2$ generators using
\begin{equation}\label{Wsrecursion}
W^s_{m-1} = -\frac{1}{m+s-1} [L_{-1} , W^s_m]
\end{equation}
until one gets $W^s_{1-s} = (L_{-1})^{s-1}$. These generators are in one-to-one correspondence with the independent components of the symmetric and traceless tensor $J_{a_1 \cdots a_{s-1}}$ of rank $s-1$: their number is $2s-1$ and they are all independent because they are eigenvectors of $L_0$ with different eigenvalues (see the following exercise).

\begin{exercise}\hspace{-2pt}\hyperlink{sol:LWcommrel}{*}
\label{ex:LWcommrel}
Check that \eqref{Wsrecursionstart} and \eqref{Wsrecursion} imply
\begin{equation} \label{LWcommrel}
[L_{m},W^s_{n}] = ((s-1)m-n) W^s_{m+n}\, .
\end{equation}
The definition we gave thus uniquely fixes the commutation relations of the basis elements with the  $\mathfrak{sl}(2,\mathbb{R})$ generators. In particular, the ``mode number'' $n$ corresponds to minus the eigenvalue of $L_0$ under the adjoint action.
One can then check that \eqref{LWcommrel} implies
\begin{equation}
[L_0 , [W^{s}_{m},W^{t}_{n}] ] = - (m+n) [W^{s}_{m},W^{t}_{n}] 
\end{equation}
and therefore that the mode number is additive under the Lie bracket.
\end{exercise}

In general, all commutation relations of $\mathfrak{hs}[\lambda]$ can be cast in the form\footnote{The structure of the commutators shows that one can rescale $W^{2n+1}_m \to i
\,W^{2n+1}_m$ to obtain another real form of $\mathfrak{hs}[\lambda]$, which allows for unitary representations for $0<\lambda<1$ \cite{Monnier:2014tfa}. For $\lambda = N$, it corresponds to the algebra $\mathfrak{su}(\frac{N-1}{2},\frac{N+1}{2})$ for $N$ odd and $\mathfrak{su}(\frac{N}{2},\frac{N}{2})$ for $N$ even. One can also see from~\eqref{generalcommutatorhs} that it is possible to consistently truncate the algebra to only contain even-spin generators, which for integer $\lambda$ results in a real form of $\mathfrak{so}(2N)$.
We refer, e.g., to \cite{Ahn:2011pv, Gaberdiel:2011nt, Campoleoni:2011hg, Vollenweider:2015, Kelm:2016pad} for more details and for applications of these observations.}
\begin{equation} \label{generalcommutatorhs}
[W^{s_1}_{m},W^{s_2}_{n}] = \sum_{k=0}^{\min(s_1,s_2)-2} c_\lambda[m,n]\, W^{s_1+s_2-2k-2}_{m+n} \, .
\end{equation} 
We refer to, e.g., \cite{Pope:1989sr} for more details about this statement, while here we propose to analyse its simplest manifestation in the following exercises.

\begin{exercise}\hspace{-2pt}\hyperlink{sol:WWcommrel}{*}\label{ex:WWcommrel} 
Prove that the commutator $[W^3_{m},W^3_{n}]$ must be a combination of $W^4_{m+n}$ and $L_{m+n}$, and then check that the $m,n$-dependence of the structure constants is completely fixed by \eqref{LWcommrel} via the Jacobi identities. That is, prove that
\begin{equation}\label{W3W3commutator}
[W^3_{m},W^3_{n}] = \alpha[\lambda] (m-n) W^4_{m+n} + \frac{\beta[\lambda]}{12} (m-n) (2m^{2}+2n^{2}-mn-8)L_{m+n}\, .
\end{equation} 
\end{exercise}

\begin{exercise}\hspace{-2pt}\hyperlink{sol:W3mexplicit}{*}
\label{ex:W3mexplicit}
Determine $\alpha[\lambda]$ and $\beta[\lambda]$ by computing the commutator \eqref{W3W3commutator} using concrete expressions for $W^3_m$ and $W^4_n$ obtained starting from~\eqref{Wsrecursionstart} and the recursion relation \eqref{Wsrecursion}. Notice that the result,
\begin{equation}
\alpha[\lambda]=2\,, \quad \beta[\lambda]=\frac{4-\lambda^2}{5}\,,
\end{equation}
can be obtained using only a selected number of $m$ and $n$ components; one also has to use that
\begin{equation}
C_2 = L_0^2 - \frac{1}{2}\left(L_1 L_{-1} + L_{-1} L_1 \right) = \frac{1}{4}\left( \lambda^2-1 \right)\mathbb{I}\, .
\end{equation}
\end{exercise}

General expressions for the structure constants of $\mathfrak{hs}[\lambda]$ were postulated in~\cite{Pope:1989sr} and later proven in~\cite{Basile:2016goq}. The invariant bilinear form can be defined via the trace in~\eqref{traceBlambda} \cite{Vasiliev:1989re}. If we normalize the trace such that it coincides with the standard form~\eqref{eq:bilinearFormSl2} on the $\mathfrak{sl}(2)$ subalgebra, it is explicitly given by (see for example~\cite{Campoleoni:2011hg})
\begin{equation}\label{trace}
\tr \big(W^s_m \,W^t_n\big) = (-1)^{m}   \frac{6(s-1)!^2}{(2s-1)!}\frac{(s+m-1)!(s-m-1)!}{(2s-2)!}   \prod_{i=2}^{s-1}(\lambda^2-i^2)\,\delta^{s,t}\delta_{m+n,0}\, .
\end{equation}
From this expression one can directly read off that the bilinear form is degenerate for any integer $\lambda=N>1$.

Taking symmetrised products of $\mathfrak{sl}(2)$ generators 
is a natural way to build higher-spin algebras in three dimensions and mimics the construction of higher-spin algebras in any spacetime dimensions \cite{Eastwood:2002su, Iazeolla:2008ix, Boulanger:2011se, Joung:2014qya}.\footnote{In a spacetime of dimension $D$, higher-spin algebras are defined as quotients of the universal enveloping algebra of $\mathfrak{so}(2,D-1)$ and are used to organise the non-linear deformations of the free field equations,
see, e.g., \cite{Didenko:2014dwa}. See also \cite{Campoleoni:2021blr} for an explicit comparison between the previous construction and the one that applies in generic spacetime dimensions.} On the other hand, we can start with any Lie algebra $\mathfrak{g}$ admitting a non-degenerate bilinear form and that contains a $\mathfrak{sl}(2,\mathbb{R})$ subalgebra. The decomposition of $\mathfrak{g}$ into irreducible representations of $\mathfrak{sl}(2,\mathbb{R})$ with respect to the adjoint action then determines the field content: the $2s-1$ generators belonging to each irreducible representation can indeed be grouped into a tensor $J_{a_1\cdots a_{s-1}}$ satisfying \eqref{eq:hscommutatorsl2} or, equivalently, into a tensor $W^{s}_m$ satisfying \eqref{LWcommrel}. Let us stress that different embeddings of $\mathfrak{sl}(2,\mathbb{R})$ into $\mathfrak{g}$ lead to different spectra (see, e.g., \cite{Campoleoni:2011hg, Castro:2012bc}). For example, under the regular embedding of $\mathfrak{sl}(2,\mathbb{R})\subset \mathfrak{sl}(3,\mathbb{R})$ as $2\times 2$-matrix blocks,\footnote{For $N=3$, the construction that we discussed before, in which $\mathfrak{sl}(2,\mathbb{R})$ is embedded into $\mathfrak{sl}(3,\mathbb{R})$ as the $3 \times 3$ matrices \eqref{3drepofsl2}, corresponds to another embedding called the principal one.} we have the decomposition
\begin{equation}
    \mathfrak{sl}(3,\mathbb{R}) \simeq \mathbf{3} + 2\cdot \mathbf{2} + \mathbf{1}\, ,
\end{equation}
where we labelled $\mathfrak{sl}(2)$ representations by their dimension in boldface.
We can interpret the corresponding Chern--Simons theory as a gauge theory containing gravity coupled to two bosonic spin-$\frac{3}{2}$ fields and one spin-$1$ field. 

In the following, we shall anyway focus on higher-spin theories with $\mathfrak{hs}[\lambda]$ or $\mathfrak{sl}(N,\mathbb{R})$ gauge algebras, often resorting to their simplest $N = 3$ instance to introduce new concepts. For this reason, we close this section with an exercise aiming at fixing all conventions for the $\mathfrak{sl}(3,\mathbb{R}) \times \mathfrak{sl}(3,\mathbb{R})$ Chern--Simons theory describing the coupling of a spin-three field to gravity.  

\begin{exercise}\label{ex:sl3matrices}
Starting from the three-dimensional representation of $\mathfrak{sl}(2,\mathbb{R})$ given in~\eqref{3drepofsl2}, 
check that the explicit matrix representations of the generators $L_m$ and $W^3_m$ is 
\begin{alignat}{5}
L_1 & = \sqrt{2}
\begin{pmatrix}
    0 & 0 & 0 \\
    1 & 0 & 0  \\
    0 & 1 & 0  \\
\end{pmatrix} , \ &
L_{0} & = 
\begin{pmatrix}
    1 & 0 & 0 \\
    0 & 0 & 0  \\
    0 & 0 & -1  \\
\end{pmatrix} , \ &
L_{-1} & = \sqrt{2}
\begin{pmatrix}
    0 & -1 & 0 \\
    0 & 0 & -1  \\
    0 & 0 & 0  \\
\end{pmatrix} , \nn \\
\label{sl3matrixrealisation}
W^3_{2} & = 
\begin{pmatrix}
    0 & 0 & 0 \\
    0 & 0 & 0  \\
    2 & 0 & 0  \\
\end{pmatrix} , \ &
W^3_{1} & = \frac{1}{\sqrt{2}}
\begin{pmatrix}
    0 & 0 & 0 \\
    1 & 0 & 0  \\
    0 & -1 & 0  \\
\end{pmatrix} , \quad &
W^3_{0} & = \frac{1}{3}
\begin{pmatrix}
    1 & 0 & 0 \\
    0 & -2 & 0  \\
    0 & 0 & 1  \\
\end{pmatrix} , \\
W^3_{-2} & = 
\begin{pmatrix}
    0 & 0 & 2 \\
    0 & 0 & 0  \\
    0 & 0 & 0  \\
\end{pmatrix} , \ &
W^3_{-1} & = \frac{1}{\sqrt{2}}
\begin{pmatrix}
    0 & -1 & 0 \\
    0 & 0 & 1  \\
    0 & 0 & 0  \\
\end{pmatrix} \,. && \nn
\end{alignat}
Using this representation, verify that the normalised trace in~\eqref{trace} is related to the matrix trace $\mathrm{tr}_{3\times 3}$ by a factor $\frac{1}{4}$,
\begin{subequations}\label{sl3trace}
\begin{alignat}{2}
    \tr (L_m L_n) &= \frac{1}{4}\, \mathrm{tr}_{3\times 3} (L_m L_n) &&= (-1)^m \frac{(1+m)!(1-m)!}{2}\,\delta_{m+n,0}\,,\\ 
    \tr (W^3_m W^3_n) &= \frac{1}{4}\, \mathrm{tr}_{3\times 3} (W^3_m W^3_n) &&= (-1)^m \frac{(2+m)!(2-m)!}{4!}\,\delta_{m+n,0}\,.
\end{alignat}
\end{subequations}
By explicit computation, check the commutation relations of $\mathfrak{sl}(3,\mathbb{R})$ in this basis:
\begin{subequations}
\label{sl3commrel}
\begin{align}
[L_{m},L_{n}] & = (m-n) L_{m+n} \, , \\[4pt]
[L_{m},W^3_{n}] & = (2m-n) W^3_{m+n} \, , \\
[W^3_{m},W^3_{n}] &= -\frac{1}{12} (m-n) (2m^{2}+2n^{2}-mn-8)L_{m+n} \, .
\end{align}
\end{subequations}
\end{exercise}

\section{Asymptotic symmetries}\label{sec:symmetries}

In this section, we study the asymptotic symmetries of three-dimensional higher-spin gauge theories on a spacetime which is asymptotically AdS. These are transformations of the fields that --~although they take the same form as a gauge transformation~-- correspond to global symmetries of the system rather than parameterising a redundancy in its description. They are identified among the gauge transformations that preserve the boundary conditions on the fields (that one has to impose due to the non-compact nature of spacetime) while acting on the boundary data. As we shall discuss, transformations of this sort contain a subset that is canonically generated by global charges defined on the boundary of spacetime and that has to be identified with global symmetries.

We work in the Chern--Simons formulation that we introduced in the last section, first explaining the basic concepts using boundary conditions that lead to an affine Lie algebra of asymptotic symmetries. We then include additional constraints that characterise asymptotically AdS field configurations. These conditions lead to a reduction of the affine Lie algebra to a classical $\cW$-algebra known as Drinfeld--Sokolov reduction. In the case of gravity, this reproduces the Virasoro algebra first identified by Brown and Henneaux \cite{Brown:1986nw}, whereas for higher-spin gravity we get asymptotic non-linear $\cW$-algebras as first observed in \cite{Henneaux:2010xg, Campoleoni:2010zq}.

We follow the approach of \cite{Banados:1998gg, Campoleoni:2010zq}, while distinguishing the role of the time and radial coordinates, as later suggested in \cite{Henneaux:2013dra, Bunster:2014mua}. We assume here that the reader is familiar with Dirac's treatment of Hamiltonian systems with constraints, 
while recalling the basic facts we need in section~\ref{sec:hamiltonian}.

\subsection{Chern--Simons theory with boundary}\label{sec:CSbdy}

We wish to describe gravity and higher-spin gravity on three-dimensional geometries which are asymptotically anti-de Sitter.
The conformal completion of AdS$_3$ is diffeomorphic to a solid cylinder $\Sigma = \mathbb{R}\times D_{2}$ with boundary $\partial \Sigma = \mathbb{R}\times S^{1}$: for this reason, in the following we shall consider Chern--Simons theories defined on this manifold.\footnote{A full characterisation of the space of solutions might require to also consider different topologies, but asymptotically the manifolds should take the form of a solid cylinder. For this reason, focussing on Chern--Simons theories on a solid cylinder suffices to discuss asymptotic symmetries.} We also choose coordinates $t$, $\theta$, $\rho$, where $t$ is the time coordinate along $\mathbb{R}$, $\theta$ is the $2\pi$-periodic angular coordinate on the disk $D_{2}$, and $\rho$ is the radial coordinate on $D_{2}$, for which we find it convenient to choose the range from $\rho=0$ (centre of the disk) to $\rho =\infty$ (boundary of the disk).

To begin with, we consider a generic Chern--Simons action,
\begin{equation} \label{covariant-CS}
S_{\text{CS}}=\frac{k}{4\pi} \int_{\Sigma} \tr \left(A\wedge \intd A +\frac{2}{3}\, A\wedge A\wedge A \right) ,
\end{equation}
where $A$ is a Lie-algebra-valued field, for a Lie algebra $\mathfrak{g}$ admitting an invariant non-degenerate bilinear form here denoted by $\tr$. This action is already in Hamiltonian form, with conjugate fields $A_{\theta}$ and $A_{\rho }$, and a Lagrange multiplier $A_{t}$:
\begin{equation} \label{CS_Hamiltonian}
S_{\text{CS}} = \frac{k}{4\pi} \left( \int_{\Sigma} \intd t \intd \rho \intd \theta\ \tr \big(A_{\theta}\dot{A}_{\rho}-A_{\rho}\dot{A}_{\theta} +2\,A_{t}F_{\rho \theta} \big) 
-\int_{\partial \Sigma} \intd t \intd\theta \ \tr \big( A_{t}A_{\theta} \big) \right) ,
\end{equation}
where $\dot{A}_{\rho}$, $\dot{A}_{\theta}$ denote the $t$-derivative of $A_{\rho}$ and $A_{\theta}$, respectively. The bulk Hamiltonian vanishes as it should be the case for any action that is invariant under diffeomorphisms while, as anticipated, $A_{t}$ only enters algebraically and enforces the constraint
\begin{equation}\label{first-class}
F_{\rho \theta}:=\partial_{\rho}A_{\theta}-\partial_{\theta}A_{\rho}+[A_{\rho},A_{\theta}] = 0 \, .
\end{equation}
The variation of the action reads
\begin{multline}\label{CSVariation}
    \delta S_{\text{CS}} = \frac{k}{2\pi}\int_\Sigma \intd t \intd\rho \intd\theta\ \tr \big( (\dot{A}_\rho - \partial_\rho A_t +[A_t,A_\rho])\delta A_\theta - (\dot{A}_\theta -\partial_\theta A_t +[A_t,A_\theta])\delta A_\rho \\+ F_{\rho\theta} \delta A_t \big) - \frac{k}{4\pi}\int_{\partial\Sigma} \intd t\intd\theta\ \tr \big( A_\theta \delta A_t - A_t \delta A_\theta \big) \, .
\end{multline}
The bulk part leads to the equations of motion for $A_\rho$ and $A_\theta$ and to the constraint $F_{\rho\theta}=0$. At this stage, we do not want to put boundary conditions on the field $A_\theta $. To cancel the boundary contribution that contains $\delta A_\theta$, we then add the boundary term
\begin{equation} \label{bdy-term_action}
 S_\mathrm{bdy} = -\frac{k}{4\pi} \int_{\partial \Sigma}\intd t\intd\theta\ \tr \big( A_t A_\theta \big)
\end{equation}
to the action. The total boundary contribution to the variation of the action now only consists of a term that contains $\delta A_t$. We can set the latter to zero by fixing $A_t$ at the boundary, i.e.~by imposing $\delta A_t|_{\partial \Sigma}=0$. Whereas $A_t$ plays the role of a Lagrange multiplier in the bulk, its boundary value can thus be viewed as a parameter of the theory.

Since $A$ is a Lie-algebra-valued field, we can decompose it with respect to a basis of generators $t_{A}$ of $\mathfrak{g}$ as $A=A^{A}t_{A}$. The equal-time Poisson brackets between the fields $A^{A}_{\rho }$ and $A^{B}_{\theta}$ are given by\footnote{We restrict here to a fixed time slice and suppress the time dependence of the fields; it will later be restored when we study the time evolution.}
\begin{equation}\label{sym:elementaryPB}
\big\{A^{A}_{\rho} (\rho ,\theta),A^{B}_{\theta} (\rho ',\theta ') \big\} = \frac{2\pi}{k}\,\gamma^{AB}\,\delta (\rho -\rho ')\,\delta (\theta -\theta ')\, ,
\end{equation}
where $\gamma^{AB}$ is the inverse of the matrix $\gamma_{AB}=\tr(t_{A}t_{B})$. For phase-space functionals $\cF$ and $\cH$ with well-defined functional derivatives with respect to $A_{\rho}$ and $A_{\theta}$, the Poisson bracket is
\begin{equation}\label{sym:PBforfunctionals}
\{\cF,\cH \} = \frac{2\pi}{k} \int_{D_{2}} \intd \rho \intd\theta \ \tr \left(\frac{\delta \cF}{\delta A_{\rho} (\rho,\theta)}\frac{\delta \cH}{\delta A_{\theta} (\rho ,\theta)} - \frac{\delta \cF}{\delta A_{\theta} (\rho ,\theta)}\frac{\delta \cH}{\delta A_{\rho} (\rho ,\theta)} \right) .
\end{equation}

\begin{exercise}
Show that~\eqref{sym:elementaryPB} implies~\eqref{sym:PBforfunctionals}.
\end{exercise}

$F_{\rho \theta}$ is a single constraint, and it is compatible with time evolution. As such, it can only be a first-class constraint, and we can build from it phase-space functionals $G(\Lambda)$ which generate the gauge transformations $A\to A+ d\Lambda +[A,\Lambda ]$ with the help of the Poisson brackets. Let us discuss this in more detail because it becomes subtle in the presence of a boundary. Indeed, in this case one has to write the generators of gauge transformations as \cite{Banados:1994tn}
\begin{equation}\label{defG}
G (\Lambda) = \frac{k}{2\pi} \int_{D_2} \intd\rho  \intd\theta \ \tr (\Lambda \,F_{\rho \theta }) +Q (\Lambda )\, ,
\end{equation}
where we introduced a boundary term to ensure that $G$ has a well-defined functional derivative with respect to $A_{\rho}$ and $A_{\theta}$, i.e.~to ensure that the variation of $G(\Lambda)$,  
\begin{equation} \label{varG}
\begin{split}
\delta G & = \frac{k}{2\pi} \int_{D_2} \intd\rho \intd\theta \ \tr \Big((\partial_{\theta}\Lambda +[A_{\theta} ,\Lambda])\delta A_{\rho} - (\partial_{\rho}\Lambda +[A_{\rho},\Lambda])\delta A_{\theta} \Big) \\
& \phantom{=}\ + \frac{k}{2\pi}\int_{S^{1}}\intd\theta \Big( \tr (\Lambda \delta A_{\theta}) + \delta Q(\Lambda) \Big) \, ,
\end{split}
\end{equation}
does not contain any boundary terms. This condition fixes the functional variation of $Q(\Lambda)$ to be
\begin{equation} \label{sym:var-charge}
\delta Q(\Lambda) = -\frac{k}{2\pi}\int_{S^{1}}\intd\theta \ \tr (\Lambda \delta A_{\theta} ) \, .
\end{equation}
In the current setup, in which we do not impose any constraint on $A_\theta$, the parameter $\L$ is just an arbitrary Lie-algebra-valued function, so that the variation \eqref{sym:var-charge} can be straightforwardly integrated to get\footnote{One could also add an arbitrary function on the boundary, provided that it does not depend on the boundary value of the dynamical field $A_\theta$. We omitted it in agreement with the most common expression for $Q(\Lambda)$ in the literature, but this issue will be relevant when we shall discuss the time dependence of the charge in \eqref{improved-charge} below.}
\begin{equation}\label{sym:boundarycharge}
Q (\Lambda)=-\frac{k}{2\pi}\int_{S^{1}}\intd\theta \ \tr (\Lambda A_{\theta}) \, .
\end{equation}
Notice, however, that the expression \eqref{sym:var-charge} for the variation of the charge also applies to more general contexts that we shall encounter later, in which the gauge parameter may depend on the fields. This is so because any possible contribution in $\delta \Lambda$ in the variation of $G$ would anyway be proportional to the constraint, as it is manifest from~\eqref{defG}. On the other hand, for field-dependent gauge parameters --~as, e.g., those corresponding to diffeomorphisms (see exercise~\ref{ex:diffeos})~-- the integration of $\delta Q(\Lambda)$ may lead to different results compared to \eqref{sym:boundarycharge} or require additional boundary conditions.  

\begin{exercise}\label{ex:functionalderivativeofG}
Verify the variation \eqref{varG} and show that it implies
\begin{subequations}
\begin{alignat}{2}
\delta_\Lambda A_\rho &= \{G (\Lambda),A_{\rho} \} &&= \partial_{\rho}\Lambda + [A_{\rho},\Lambda]\,, \\
\delta_\Lambda A_\theta &= \{G (\Lambda),A_{\theta} \} &&= \partial_{\theta}\Lambda + [A_{\theta},\Lambda]\, .
\end{alignat}
\end{subequations}
\end{exercise}

The equations of motion for $A_\rho$ and $A_\theta$, which follow from the variation of the action $S = S_{\mathrm{CS}} + S_{\mathrm{bdy}}$, show that the time evolution is a gauge transformation with $A_t$ having the interpretation of a gauge parameter per unit time,
\begin{equation} \label{eom-CS}
    \dot{A}_\rho = \{ G(A_t), A_\rho\} \, ,\quad \dot{A}_\theta = \{ G(A_t), A_\theta \} \, .
\end{equation}
This comes without surprise because the Hamiltonian vanishes in the action \eqref{CS_Hamiltonian} and the total bulk Hamiltonian of the system is a pure constraint, $H_{\textrm{tot}} \propto G(A_t)$.
On the other hand, when one applies a possibly time-dependent gauge transformation with parameter $\Lambda(t,\rho,\theta)$, one also changes the precise time evolution of $A_\theta$ and $A_\rho$ along the gauge orbit. To be consistent with the time evolution determined by \eqref{eom-CS}, the gauge transformation has to be accompanied by a variation of $A_t$:
\begin{equation} 
 \delta A_t = \partial_t \Lambda + [A_t,\Lambda]\, .
\end{equation}
The gauge transformations of the fields and of the Lagrange multiplier combine into 
\begin{equation}\label{CS-gauge-transf}
    \delta A_\mu = \partial_\mu \Lambda + [A_\mu,\Lambda]\, ,
\end{equation}
which is the usual covariant infinitesimal gauge symmetry of the Chern--Simons action \eqref{covariant-CS}. 

We have discussed how to recover in the Hamiltonian formalism the gauge symmetry \eqref{CS-gauge-transf}, but the boundary term \eqref{sym:boundarycharge} in the generator of gauge transformations leads to several subtleties in their interpretation. This already becomes apparent in the Poisson bracket of two gauge generators,
\begin{equation}\label{sym:algebraofG}
\{G (\Lambda),G (\Gamma) \} = G ([\Lambda ,\Gamma]) + \frac{k}{2\pi}\int_{S^1} \intd\theta \ \tr \left(\Lambda \,\partial_{\theta}\Gamma\right) \,,
\end{equation}
that develops a central extension which depends on the boundary values of $\Lambda$ and $\Gamma$.
\begin{exercise}\hspace{-2pt}\hyperlink{sol:PoissonbracketofGs}{*}
\label{ex:PoissonbracketofGs}
Prove~\eqref{sym:algebraofG}.
\end{exercise}

The most important consequence of a non-trivial $Q(\Lambda)$ is that the corresponding transformation is actually not a gauge symmetry,  but a global symmetry canonically generated by the boundary charge $Q(\Lambda)$. 
In other words, a variation \eqref{CS-gauge-transf} associated to a non-trivial boundary charge sends solutions of the equations of motion into physically inequivalent solutions even if it takes the same form as a gauge transformation. Symmetries of this kind are known as asymptotic symmetries.

To analyse this statement in detail, it is convenient to add a gauge-fixing condition on top of the first-class constraint \eqref{first-class}. As we shall see shortly, the additional constraint we are going to impose completely fixes the proper gauge symmetry, so that no redundancy in the description of the space of solutions is left. The gauge-fixing, however, still allows for residual  transformations of the form \eqref{CS-gauge-transf}, to be interpreted as global (or asymptotic) symmetries. 

To this end, we choose a specific form for $A_{\rho }$ which, close to the boundary, we take as
\begin{equation} \label{radial-gauge}
A_{\rho} (\rho) = b^{-1} (\rho)\partial_{\rho}b (\rho) 
\end{equation}
for a chosen group-valued function $b (\rho)$. Close to the boundary one can always reach this gauge off-shell: to bring a gauge field $A'$ into the form \eqref{radial-gauge}, one has to perform a gauge transformation satisfying 
\begin{equation}
U^{-1} A'_\rho U + U^{-1} \pr_\rho U = b^{-1} \pr_\rho b \, .
\end{equation}
Writing $U = U'b$, this condition holds if
\begin{equation}
\pr_\rho U' = - A'_\rho U' \, ,
\end{equation}
which can be solved by a path-ordered exponential,
\begin{equation}
U = \cP e^{-\int_{\rho_0}^\rho A'_{r} \intd r} U_0 b \, .
\end{equation}
$U_0$ is a constant of integration that must be chosen so as to preserve the boundary condition $\delta A_t|_{\partial \Sigma} = 0$, and one can always do this on the solid cylinder.\footnote{If other boundaries are present, it might not be possible to select a $U_0$ compatible with all boundary conditions. In other words, in some cases the gauge \eqref{radial-gauge} might not be reachable, or it may be necessary to select specific group-valued functions in order to reconstruct globally certain classes of solutions. We refer, e.g., to \cite{Castro:2016ehj, Banados:2016nkb} for more details.}

On the constraint surface, where we implement $F_{\rho \theta}=0$, the $\rho$-dependence of $A_{\theta}$ is also fixed,
\begin{equation}
\partial_{\rho}A_{\theta} +[A_{\rho},A_{\theta}]=0 
\quad \Longrightarrow \quad
\partial_{\rho} (b\,A_{\theta}\,b^{-1}) = 0\, .
\end{equation}
$A_{\theta}$ is then of the form
\begin{equation}
A_{\theta} (t,\rho ,\theta) = b^{-1} (\rho)\, a_\theta(t,\theta)\,b (\rho)
\end{equation}
with some $\rho$-independent term $a_\theta(t,\theta)$, where we restored the time dependence to stress that this condition is respected along time evolution.

The residual transformations of the form \eqref{CS-gauge-transf} that leave $A_\rho$ invariant satisfy
\begin{equation}
    \partial_\rho \Lambda + [A_\rho , \Lambda ] =0\, ,
\end{equation}
hence $\Lambda$ has to be of the form
\begin{equation}\label{sym:residualgaugeparameter}
\Lambda (t,\rho ,\theta) = b^{-1} (\rho)\lambda (t,\theta) b (\rho) \, .
\end{equation}
Similarly, the Lagrange multiplier $A_t$ needs to satisfy
\begin{equation}\label{multiplier_radialgauge}
A_{t} (t,\rho ,\theta) = b^{-1} (\rho)\, a_t (t,\theta)\,b (\rho)\,, 
\end{equation}
such that the gauge choice for $A_\rho$ is preserved along time evolution. The function $a_t$ parameterises the freedom to choose the boundary value of $A_t$.

Now we want to argue that the residual transformations generated by gauge parameters of the form \eqref{sym:residualgaugeparameter} cease to be gauge symmetries, and they parameterise global symmetries instead. One way to see this is by looking at the evolution in time. If we have a proper gauge symmetry, we can choose an arbitrary time dependence for the gauge parameter because, on each time slice, configurations related by gauge transformations should be completely equivalent. The transformation of $a_t$ induced by a $\Lambda$ of the form~\eqref{sym:residualgaugeparameter} is however
\begin{equation}\label{sym:gaugetransf}
a_t (t,\theta) \to a_t (t,\theta ) + \partial_{t}\lambda (t,\theta) +[a_t (t,\theta ),\lambda (t,\theta)]
\, .
\end{equation}
As $a_t$ is a fixed parameter of the theory in the current setup, it should be invariant under this transformation. This fixes the time dependence of $\lambda$, which therefore cannot correspond to a proper gauge transformation. 

Moreover, adding the constraint \eqref{radial-gauge} to the first-class constraint \eqref{first-class} one obtains a set of second-class constraints, as it should be the case for any complete gauge-fixing. Indeed, the Poisson bracket of the constraints can be computed as 
\begin{equation} \label{bracket-constraints}
\{ G(\Xi) , A_\rho \} = \pr_\rho \Xi + [A_\rho, \Xi] =: D_\rho\Xi \, ,
\end{equation}
where we introduced a smearing function $\Xi$ for the constraint $F_{\rho\theta} = 0$, while we omitted the term $b^{-1}\partial_\rho b$ in the constraint \eqref{radial-gauge} since it does not depend on phase-space variables. Notice that here $G(\Xi)$ has to depend only on the first-class constraint \eqref{first-class}, so that the smearing function $\Xi$ must vanish at the boundary. The unique solution of the first-order differential equation $D_\rho \Xi = 0$ with this boundary condition is $\Xi = 0$ everywhere, thus showing that the bracket \eqref{bracket-constraints} is not degenerate.

Conversely, non-trivial transformations preserving the gauge-fixing \eqref{radial-gauge} have a non-vanishing boundary charge~\eqref{sym:boundarycharge}, so that they are not generated by a first-class constraint:
\begin{equation}\label{sym:bcharge}
Q (\Lambda) = -\frac{k}{2\pi} \int_{S^{1}}\intd\theta \ \tr\! \left(b^{-1}\lambda b\,b^{-1}a_\theta b\right) = -\frac{k}{2\pi} \int_{S^1} \intd\theta \ \tr\! \left(\lambda \,a_\theta \right) =: Q(\lambda)\,.
\end{equation}
This is yet another way of concluding that they cannot be proper gauge transformations (see, e.g., \cite{Benguria:1976in, Banados:1994tn, Blagojevic:2002du} for an ampler discussion of this point in the Hamiltonian formalism). 

Notice that the charge given in \eqref{sym:bcharge} is conserved on-shell if one chooses the function $a_t$ to be constant: 
\begin{equation}
\frac{d}{dt}\, Q(\lambda) = - \frac{k}{2\pi} \int_{S^1} \intd\theta \ \tr \left( \l\,  \partial_\theta a_t \right) ,
\end{equation}
where we stress that one has to impose the equations of motion \eqref{eom-CS} to get this result.
For generic values of $a_t$ the charge is not conserved, but its time variation only depends on the latter and not on the boundary value of the dynamical field $a_\theta$ (see also \cite{Grumiller:2016pqb}). As such, since $a_t$ is just a parameter of the theory, one can still define a quantity that is conserved on shell along time evolution:\footnote{Boundary conditions leading to truly non-conserved surface charges in three dimensions have been considered, e.g., in \cite{Alessio:2020ioh, Ruzziconi:2020wrb, Geiller:2021vpg, Campoleoni:2022wmf}.}
\begin{equation} \label{improved-charge}
\hat{Q}(\lambda) = Q(\lambda) + \frac{k}{2\pi} \int_{t_0}^t \intd t' \int_{S^1} \intd\theta \ \tr \left( \lambda\,  \partial_\theta a_{t'} \right) .
\end{equation} 

Since the constraints \eqref{first-class} and \eqref{radial-gauge} form a set of second-class constraints, one can define a Poisson structure on the reduced phase space, here parameterised by the functions $a_\theta(t,\theta)$, via the Dirac bracket. The latter can be efficiently computed by realising that the Dirac bracket of the charges $Q(\Lambda)$ with any phase-space functional corresponds to the Poisson bracket. Indeed, denoting collectively the two constraints \eqref{first-class} and \eqref{radial-gauge} by $\phi_i$, the Dirac bracket $\{\cdot ,\cdot \}_{*}$ is defined as (see also~\eqref{DefDiracbracket})
\begin{equation} \label{dirac-bracket}
\{ Q(\L) , \cF \}_* = \{ G(\Lambda) , \cF \}_* = 
\{ G(\Lambda) , \cF \} - \{ G(\Lambda) , \phi_i \}\, C^{ij} \{ \phi_j , \cF \} \, ,
\end{equation}
where $C^{ij}$ is the inverse of the matrix $C_{ij} = \{ \phi_i , \phi_j \}$. On the other hand, the constraints are invariant under the residual symmetries generated by gauge parameters of the form \eqref{sym:residualgaugeparameter},
\begin{equation}
\{ G(\Lambda) , \phi_i \} = \delta_\Lambda \phi_i \approx 0 \, ,
\end{equation}
where we used the symbol $\approx$ to stress that the identity holds on the constraint surface. Eq.~\eqref{sym:residualgaugeparameter} was indeed obtained by demanding the invariance of the constraint \eqref{radial-gauge}, while the constraint $F_{\rho\theta} = 0$ is also invariant on the constraint surface since 
\begin{equation}
\delta_\Lambda F_{\m\n} = [F_{\mu\nu}, \Lambda] \, .
\end{equation}
As a result, the boundary charges canonically generate the global symmetries induced by non-trivial $\l(t,\theta)$, that is
\begin{equation}\label{sym:transformofFbyboundarycharge}
\delta_\lambda \cF = \{ G(\Lambda) , \cF \} \approx \{ Q(\lambda) , \cF \}_* \, , 
\end{equation}
and the algebra of boundary charges on the constrained phase space is the same as that of the generators $G(\Lambda)$ (given in~\eqref{sym:algebraofG}),
\begin{equation}\label{sym:algebraofQ}
\{Q(\lambda_1),Q(\lambda_2) \}_{*} = Q ([\lambda_1 ,\lambda_2]) + \frac{k}{2\pi}\int_{S^1} \intd\theta \ \tr (\lambda_1 \,\partial_{\theta}\lambda_2) \, .
\end{equation}
From this result we can infer the Poisson structure on the reduced phase space parameterised by $a_\theta$, and a similar strategy can be employed also when some constraints on $a_\theta$ are imposed as part of the boundary conditions of the theory (as in the next section).

In the current example, it is convenient to expand $a_\theta (t,\theta)$ in Fourier modes and in a basis $t_{A}$ of generators of $\mathfrak{g}$, 
\begin{equation}\label{sym:modedecomposition}
a_\theta (t,\theta) = \frac{1}{k} \sum_{p\in\mathbb{Z}}a^{A}_{p}(t) e^{-ip\theta}\,t_{A}\, .
\end{equation}
The modes $a^{A}_{p}$ can be expressed in terms of the boundary charges as
\begin{equation}
a^{A}_{p} = -\, \g^{AB} Q (t_B\,e^{ip\theta}) \, ,
\end{equation}
where we recall that $\g^{AB}$ is the inverse of the Killing metric $\g_{AB} = \tr(t_At_B)$.
We can then determine the brackets between the modes from~\eqref{sym:algebraofQ}:
\begin{equation}\label{sym:affinealgebra}
\begin{split}
\{a^{A}_{p},a^{B}_{q} \}_{*} &= Q ([t^{A},t^{B}]e^{i (p+q)\theta}) + \frac{k}{2\pi} \int \intd\theta \ \tr (t^{A}t^{B})\, iq\,e^{i (p+q)\theta}\\
&= -f^{AB}{}_{C}\,a^{C}_{p+q} +iqk\,\gamma^{AB}\,\delta_{p+q,0}\, .
\end{split}
\end{equation}
This gives the central extension of the loop algebra of $\mathfrak{g}$ (which one calls the affine Lie algebra $\hat{\mathfrak{g}}$ , see also~\eqref{quantum:affineLie}), or more precisely, the modes $a^{A}_{p}$ provide a representation (via the Dirac bracket) of the central extension of the loop algebra at a fixed level $k$.
To summarise, on the reduced phase space of a Chern--Simons theory with gauge algebra $\mathfrak{g}$ and with fixed boundary values of $A_{t}$ we have found a realisation of the affine Lie algebra $\hat{\mathfrak{g}}_{k}$.

\subsection{Asymptotically AdS solutions: the Drinfeld--Sokolov reduction}

In the previous section, we discussed how some of the residual gauge transformations compatible with the boundary conditions might become asymptotic symmetries. To illustrate this point, we used boundary conditions in which we completely fixed $A_\rho$ while leaving the boundary value of $A_\theta$ arbitrary. These boundary conditions are useful to discuss the subtleties that are brought by the presence of a boundary, but they are not appropriate to describe asymptotically AdS$_3$ spaces, at least according to the definition given in the seminal paper by Brown and Henneaux \cite{Brown:1986nw}.\footnote{See, e.g., \cite{Troessaert:2013fma, Grumiller:2016pqb, Alessio:2020ioh, Ruzziconi:2020wrb, Geiller:2021vpg, Campoleoni:2022wmf} and references therein for generalisations of the Brown--Henneaux boundary conditions.} In the following, we recall how asymptotically AdS$_3$ spaces can be described in the Chern--Simons formulation of gravity, and we then generalise their definition to the higher-spin gauge theories that we introduced in section~\ref{sec:CSformulation}.

To begin with, we recall that a global parameterisation of AdS$_{3}$ is given by
\begin{equation}
\intd s^2 = - \cosh^2\rho \, \intd t^2 + \adsR^2 \left( \intd \rho^2 + \sinh^2\rho \, \intd\theta^2\right) , 
\end{equation}
where the radial coordinate $\rho$ is taken to be dimensionless, but it still takes the same range as in the previous subsection.
Rewriting the hyperbolic functions in terms of exponentials and performing for later convenience the shift $\rho \to \rho + \log(2)$ we obtain
\begin{equation}\label{sym:metricAdS}
\intd s^{2} = \adsR^{2} \intd\rho^{2} + \left( e^{2\rho} + \frac{1}{16}\, e^{-2\rho} \right) \left(-\intd t^2 + \adsR^2 \intd\theta^2 \right) - \frac{1}{2} \left(\intd t^2 + \adsR^2 \intd\theta^2 \right) .
\end{equation}
In the Chern--Simons formulation, this metric can be reproduced starting from the $\mathfrak{sl}(2)$-valued connections 
\begin{subequations} \label{AdS-connections}
\begin{align} 
A^{\text{AdS}} & =  L_{0}\,\intd\rho 
+ \left(e^{\rho}L_{1}+\tfrac{1}{4}e^{-\rho}L_{-1}\right)\left(\frac{\intd t}{\adsR} + \intd\theta\right) , \\
\tilde{A}^{\text{AdS}} & = - L_{0}\,\intd\rho 
 -\left(\tfrac{1}{4}e^{-\rho}L_{1}+e^{\rho}L_{-1} \right) \left( \frac{\intd t}{\adsR} - \intd\theta \right) ,
\end{align}
\end{subequations}
that are related to the vielbein and the spin connection as in \eqref{A-Atilde}. We used here the basis \eqref{L-basis} for $\mathfrak{sl}(2)$, that we recall for the reader's convenience:
\begin{equation} \label{L-basis_2}
[L_m , L_n] = (m-n) L_{m+n} \, .
\end{equation}

\begin{exercise}\label{ex:AdSmetricfromvielbein}
Check eqs.~\eqref{AdS-connections} by first verifying that they satisfy the zero-curvature condition. Then verify that the corresponding vielbein reproduces the metric \eqref{sym:metricAdS} with the conventions
\begin{equation}
\intd s^2 = \eta_{ab} e^a e^b = 
2\, \tr (e^2) \,,
\end{equation}
where the vielbein is defined as in \eqref{A-Atilde} while the trace is defined as in \eqref{sl2trace} (see also \cite{Campoleoni:2010zq, Campoleoni:2014tfa} for a more general discussion on how to rewrite metric-like fields as traces over Lie-algebra-valued fields).
\end{exercise}

The connections $A^{\text{AdS}}$ and $\tilde{A}^{\text{AdS}}$ satisfy the gauge-fixing condition \eqref{radial-gauge} that we imposed in the previous section with, respectively, the group-valued elements
\begin{align} \label{b(rho)}
b(\rho) &= e^{\rho L_0}\,, & \tilde{b}(\rho) &= b^{-1}(\rho)\,.
\end{align}
These connections thus fit in our previous discussion with the following expressions for the canonical variables:
\begin{subequations} \label{gauge-fixed-A}
\begin{alignat}{5}
A_{\rho} & = b^{-1} (\rho) \partial_{\rho} b(\rho) \,, \qquad
& \tilde{A}_{\rho} & = b(\rho) \partial_{\rho} b^{-1}(\rho) \,, \label{gauge-fixed-A_1} \\[5pt] 
A_{\theta} & = b^{-1}(\rho )a(t,\theta) b(\rho ) \,, \qquad
& \tilde{A}_{\theta} & = b(\rho )\tilde{a}(t,\theta) b^{-1}(\rho )\,, \label{gauge-fixed-A_2}
\end{alignat}
\end{subequations}
with
\begin{equation}
a^{\text{AdS}} = L_{1} + \tfrac{1}{4}L_{-1} \, , \qquad
\tilde{a}^{\text{AdS}} = \tfrac{1}{4} L_{1} + L_{-1} \, .
\end{equation}

The asymptotically AdS$_3$ spacetimes of \cite{Brown:1986nw} can be recovered by imposing the natural conditions that the corresponding connections take the form \eqref{gauge-fixed-A} and approach those of global anti-de Sitter space close to the boundary, that is
\begin{equation} \label{sym:AdSconditions}
A_{\theta}-A^{\text{AdS}}_{\theta} = \mathcal{O}(1)
\quad \textrm{and} \quad
\tilde{A}_{\theta}-\tilde{A}^{\text{AdS}}_{\theta} = \mathcal{O}(1)
\quad \text{for}\ \rho \to \infty
\end{equation}
at any time $t$.
The conditions \eqref{b(rho)}, \eqref{gauge-fixed-A} and \eqref{sym:AdSconditions} can be extended verbatim to higher-spin theories in the Chern--Simons formulation \cite{Campoleoni:2010zq} and this will be our definition of asymptotically AdS$_3$ field configurations.\footnote{The Brown--Henneaux boundary conditions were first translated in the Chern--Simons formulation in slightly different terms in \cite{Coussaert:1995zp}. This reformulation later allowed to characterise asymptotically AdS$_3$ configurations in supergravity \cite{Banados:1998pi, Henneaux:1999ib} and higher-spin theories \cite{Henneaux:2010xg, Campoleoni:2010zq}.}

Since the $\rho$-dependence is completely encoded in the group-valued $b(\rho)$ defined in \eqref{b(rho)}, the constraints \eqref{sym:AdSconditions} translate into constraints on the form of $a(t,\theta)$ and $\tilde{a}(t,\theta)$.  For simplicity, we now restrict the discussion to $\mathfrak{sl}(N)$ or $\mathfrak{hs}[\lambda]$ valued connections. These can be conveniently expanded in the basis $W^s_m$ that we introduced in \eqref{Wsrecursion}, whose generators satisfy
\begin{equation}\label{LWcommrel_2}
[L_{m},W^s_{n}] = ((s-1)m-n) W^s_{m+n}\, .
\end{equation}
In this basis, a generic $\mathfrak{sl}(N)$-valued connection takes the form
\begin{equation}\label{sym:Wgenerators}
a (t,\theta) = \sum_{m=-1}^{1}\ell^{m} (t,\theta)L_{m} + \sum_{l=2}^{N-1} \sum_{m=-l}^{l} w_{l+1}^{m} (t,\theta)W^{l+1}_{m}\,,
\end{equation}
while the AdS conditions~\eqref{sym:AdSconditions} constrain $a$ and $\tilde{a}$ to be of the form
\begin{subequations}\label{sym:DScondition}
\begin{align}
a(t,\theta) & = L_{1} + \sum_{m=-1}^{0}\ell^{m} (t,\theta)L_{m} +  \sum_{l=2}^{N-1}\sum_{m=-l}^{0}w_{l+1}^{m}(t,\theta)W^{l+1}_{m} \, , \label{DScond_1} \\
\tilde{a}(t,\theta) & = L_{-1} + \sum_{m=0}^{1}\tilde{\ell}^{m} (t,\theta)L_{m} +  \sum_{l=2}^{N-1}\sum_{m=0}^{l}\tilde{w}_{l+1}^{m}(t,\theta)W^{l+1}_{m} \, . \label{DScond_2}
\end{align}
\end{subequations}
The second expansion can be recovered by sending $m \to - m$ in the first, so that in the following we shall focus only on the connection $a(t,\theta)$. The corresponding results for $\tilde{a}(t,\theta)$ can be easily recovered. In the $\mathfrak{hs}[\lambda]$ case, the connections take the same form, but one has to consider an unbounded sum over $l$.

\begin{exercise}\hspace{-2pt}\hyperlink{sol:DScondition}{*}
\label{ex:DScondition}
Check that the AdS conditions \eqref{sym:AdSconditions} imply the expansion \eqref{sym:DScondition}.
\end{exercise}

We gave a definition of the space of asymptotically AdS$_3$ configurations that fits within the discussion of section \ref{sec:CSbdy}, but where $A_\theta$ is bounded to satisfy the constraint \eqref{sym:AdSconditions} on any time slice. The boundary term \eqref{bdy-term_action} continues to guarantee a well-posed action principle provided that the Lagrange multiplier $A_t$ is fixed at the boundary, but the latter is not any more arbitrary. Preserving the form \eqref{sym:AdSconditions} of $A_\theta$ under time evolution imposes some constraints on the Lagrange multiplier $A_t$ that we discuss in specific examples in the following. For what concerns the asymptotic symmetries, the additional constraints lead to a reduction of the affine Lie algebra $\hat{\mathfrak{g}}_{k}$ that we identified in the previous section, which in mathematics is known as the Drinfeld--Sokolov reduction (see, e.g., \cite{Balog:1990mu, Bouwknegt:1992wg, khesin2009:book} and references therein). 
 
We discuss it in detail in the following, starting from the illuminating examples of gravity and of the $\mathfrak{sl}(3) \oplus \mathfrak{sl}(3)$ Chern--Simons theory describing the coupling of a spin-three field to gravity. We then conclude with a discussion of the $\mathfrak{sl}(N)$ case and some comments on the $\mathfrak{hs}[\lambda]$ case.

\subsection{The $\mathfrak{sl}(2)$ example: asymptotic symmetries of gravity}

We now derive the asymptotic symmetries of gravity with the Brown--Henneaux boundary conditions \eqref{sym:AdSconditions} by focussing on the connection $A_\mu$. As already mentioned, the contribution of the connection $\tilde{A}_\mu$ can be recovered following similar steps. The following analysis amounts to implement the Drinfeld--Sokolov reduction of the $\widehat{\mathfrak{sl}}(2)_k$ affine algebra \eqref{sym:affinealgebra} to the Virasoro algebra.

In this example, on any fixed time slice, the generic expansion \eqref{sym:Wgenerators} of the \mbox{$\rho$-independent} part of $A_\theta = b^{-1}(\rho) a(\theta) b(\rho)$ reads
\begin{equation}\label{generic_sl2}
a (\theta) = \ell^{1} (\theta)L_{1} + \ell^{0} (\theta )L_{0} + \ell^{-1} (\theta)L_{-1}\, . 
\end{equation}
Before imposing the Drinfeld--Sokolov condition \eqref{DScond_1}, the phase space is parameterised by this $a(\theta)$. The corresponding modes, defined by the expansion
\begin{equation} \label{sl2-modes}
\ell^{m}(\theta) = \frac{1}{k} \sum_{p \in \mathbb{Z}} \ell^m_p e^{-ip\theta}  \,,
\end{equation}
satisfy the affine Lie algebra \eqref{sym:affinealgebra}, that in this case reads
\begin{subequations}\label{sym:sl2affine}
\begin{align}
\{\ell^{1}_{p},\ell^{0}_{q} \} &= 2\,\ell^{1}_{p+q} \,, &  
\{\ell^{1}_{p},\ell^{-1}_{q} \} &= \ell^{0}_{p+q} +ipk\,\delta_{p+q,0} \,, \\[5pt]
\{\ell^{0}_{p},\ell^{-1}_{q} \} &= 2\,\ell^{-1}_{p+q} \,, & 
\{\ell^{0}_{p},\ell^{0}_{q} \} &= -2ipk\,\delta_{p+q,0} \,,
\end{align}
\end{subequations}
where here and in the following we omit the $*$ subscript denoting the Dirac bracket. Indeed, the origin of this phase space as a constrained surface in an ampler phase space will be immaterial in the following.

The Drinfeld--Sokolov condition~\eqref{sym:DScondition} implies
\begin{equation} \label{DSconstraint}
\ell^{1} (\theta)-1 \approx 0 
\quad \Longleftrightarrow \quad 
\ell^{1}_{p}-k\,\delta_{p,0} \approx 0 \, ,
\end{equation}
where we used the symbol $\approx$ to stress that we are imposing a constraint on the phase space parameterised by the modes 
\eqref{sl2-modes} with the Poisson structure \eqref{sym:sl2affine}.
This is a first-class constraint because
\begin{equation}
\{\ell^{1}_{p} , \ell^{1}_{q} \} = 0 \, ,
\end{equation}
so that it generates a gauge symmetry on the phase space. From the viewpoint of the connection $a(\theta)$, this corresponds to all transformations generated by parameters of the form \eqref{sym:residualgaugeparameter} that preserve the constraint \eqref{DSconstraint} \cite{Balog:1990mu}.

A convenient way to proceed is to impose an additional gauge-fixing condition to work with a reduced phase space on which one can compute the Dirac bracket. For instance, the bracket of the mode $\ell^{1}_{p}-k\delta_{p,0}$ of the constraint with $\ell^0_q$ reads
\begin{equation} \label{constr-bracket}
\{\ell^{1}_{p},\ell^{0}_{q} \} = 2\, \ell^{1}_{p+q} \approx 2\,k\,\delta_{p+q,0} \, ,
\end{equation}
thus implying that $\ell^{0}_{-p}$ can be set to any value by using the gauge transformation generated by the mode $\ell^{1}_{p}$ of the constraint, that is $\delta \ell^{0}_{-p} = 2\,k\,\epsilon_{-p} $, with a constant $\epsilon_{-p}$. A possible complete gauge-fixing is therefore\footnote{In the $\mathfrak{sl}(2)$ case, a complete gauge-fixing can be achieved only by fixing $\ell^0$. On the other hand, it is also possible to analyse the Drinfeld--Sokolov reduction by performing the partial gauge-fixing $\ell^{-1} = 0$. This leads to a simpler presentation of the reduced algebra, although one still has to handle the residual gauge symmetry; see, e.g.,~\cite{Campoleoni:2017xyl}.}
\begin{equation}\label{sym:DSgauge}
\ell^{0} = 0\, .
\end{equation}
This gauge choice corresponds to the stronger condition $A_{\theta}-A^{\text{AdS}}_{\theta}\xrightarrow{\rho \to\infty}0$, and the once more reduced phase space is parameterised only by $\cL(\theta):=\ell^{-1}(\theta)$. 

We now want to understand the structure of the Dirac bracket on this reduced space. We recall that given a family $\{\phi_i \}$ of second-class constraints, the Dirac bracket of two phase-space functionals $f$ and $g$ reads (see~\eqref{DefDiracbracket})
\begin{equation}\label{sym:defDirac}
\{f,g \}_{*} = \{f,g \} - \{f,\phi_i \} C^{ij}\{\phi_j , g \}\, ,
\end{equation}
where $C^{ij}$ is the inverse of the matrix defined by $C_{ij}=\{\phi_i,\phi_j \}$. In this case, it is convenient to label the constraints using the mode labels, introducing 
\mbox{$\phi_{(1,p)}=\ell^{1}_{p}-k\,\delta_{p,0}$} and $\phi_{(0,p)}=\ell^{0}_{p}$. Then
\begin{equation} \label{matrix-constraints}
C_{(\,\cdot\,,p) (\,\cdot\,,q)} \approx 2k\,\delta_{p+q,0} \begin{pmatrix}
0 & 1\\ -1 & -ip \end{pmatrix}
\end{equation}
and 
\begin{equation}\label{sym:inverseCsl2}
C^{(\,\cdot\,,p) (\,\cdot\,,q)} \approx \frac{1}{2k}\,\delta_{p+q,0}\begin{pmatrix}
ip & -1\\ 1 & 0 \end{pmatrix}\, .
\end{equation}
The Dirac bracket for the modes $\cL_{p} = \ell_p^{-1}$ is then obtained as
\begin{equation}\label{sym:finalDiracsl2}
\{\cL_{p},\cL_{q} \}_{*} \approx i (p-q)\,\cL_{p+q} -i
\,\frac{k}{2}p^{3}\,\delta_{p+q,0}\, .
\end{equation}

\begin{exercise}
Confirm the result for the Dirac bracket~\eqref{sym:finalDiracsl2} using~\eqref{sym:sl2affine} and~\eqref{sym:inverseCsl2}.
\end{exercise}

By defining $\hat{\mathcal{L}}_{p}=-\cL_{p}+\frac{k}{4}\,\delta_{p,0}$ we find the standard form of the Virasoro algebra,
\begin{equation}\label{sym:standardVirasoro}
i\{\hat{\mathcal{L}}_{p},\hat{\mathcal{L}}_{q} \}_{*} = (p-q)\hat{\mathcal{L}}_{p+q} + \frac{c}{12}\big( p^{3}-p\big)\,\delta_{p+q,0}\, ,
\end{equation}
with central charge $c=6k$.

We can perform a similar analysis for the modes of the field $\tilde{A}$ leading to another copy of the Virasoro algebra. This reproduces the classical result by Brown and Henneaux \cite{Brown:1986nw}: the asymptotic symmetry algebra of three-dimensional gravity with a negative cosmological constant is given by two copies of the Virasoro algebra with central charge $c=\frac{3\adsR}{2G}$ (see~\eqref{CS:levelandNewtonconstant} for the relation between $k$, $\adsR$ and $G$).

Before we move on to higher-spin theories, we now rederive this result following a different approach that is easier to generalise to more involved gauge connections. Along the way, we also exhibit the constraints on the Lagrange multiplier $A_t$ induced by the additional constraints that we are imposing on $A_\theta$. 
The starting point is similar: we still use the gauge transformations generated by the Drinfeld--Sokolov constraint to fix the $\rho$-independent part of $A_\theta$ as 
\begin{equation} \label{sym:sl2formofa}
a (t,\theta) = L_{1} +  \cL (t,\theta)L_{-1}\, , 
\end{equation}
where we reinstated the time dependence to stress that we are imposing this condition all along time evolution. 
As we have already discussed, these transformations are generated by a first-class constraint and, therefore, they have to be interpreted as proper gauge transformations rather than asymptotic symmetries. In other words, fixing the form of $a$ as in \eqref{sym:sl2formofa} is not affecting the global symmetries of the system. The same conclusion can also be reached by checking explicitly that the transformation setting $\ell^0 = 0$ does not have an associated boundary charge.
Generic infinitesimal gauge transformations preserving the radial gauge-fixing \eqref{radial-gauge} indeed act on $a$ as
\begin{equation} \label{HW_spin2}
\delta_\lambda a (t,\theta) = \partial_{\theta}\lambda (t,\theta) + [a (t,\theta),\lambda (t,\theta)]\, , 
\end{equation}
where we recall that $\lambda$ is the $\rho$-independent part of the gauge parameter, see \eqref{sym:residualgaugeparameter}. Choosing
\begin{equation} \label{var-l0}
\lambda = \epsilon\, L_{-1} \quad \Longrightarrow \quad \delta_\lambda a = 2\epsilon\, L_0 + \left( \partial_\theta \epsilon + 2\, \ell^0 \epsilon \right) L_{-1}\,,
\end{equation}
one obtains a variation that can be used to set a generic connection \eqref{generic_sl2} in the form \eqref{sym:sl2formofa} while giving a vanishing $\delta Q(\lambda)$ when substituted in \eqref{sym:var-charge}. The constants $\epsilon_{p}$ that we encounter below \eqref{constr-bracket} can be interpreted as the Fourier modes of the gauge parameter $\epsilon(\theta)$. 

We can now identify the residual gauge transformations preserving the gauge-fixed form of the connection \eqref{sym:sl2formofa}. Setting
\begin{equation} \label{lambda-expansion}
\lambda (t,\theta) = \epsilon (t,\theta)L_{1} + \epsilon^{0} (t,\theta)L_{0} + \epsilon^{-1} (t,\theta)L_{-1}
\end{equation}
and requiring $\delta a$ to only have a $L_{-1}$ component fixes $\epsilon^{0}$ and $\epsilon^{-1}$ in terms of $\epsilon$ as
\begin{subequations} \label{epsilons}
\begin{align}
\epsilon^{0}  &= - \epsilon' \, , \\
\epsilon^{-1} &= \frac{1}{2}\, \epsilon'' + \epsilon \, \cL\, ,  
\end{align}
\end{subequations}
where $f':=\partial_\theta f$. The residual gauge transformations then act on the $L_{-1}$ component of the connection as follows:
\begin{equation} \label{deltaL}
\delta \cL = \epsilon \, \cL' + 2 \,\epsilon' \,\cL +\frac{1}{2}\,\epsilon''' \, .
\end{equation}

\begin{exercise}
Check eqs.~\eqref{epsilons} and \eqref{deltaL}.
\end{exercise}

Eqs.~\eqref{epsilons} characterise the asymptotic symmetries, and they also fix the form of the Lagrange multiplier $A_t$. As we discussed around \eqref{eom-CS}, time evolution corresponds indeed to a gauge transformation generated by $A_t$. Preserving the form \eqref{HW_spin2} of $A_\theta$ for any $t$ therefore implies
\begin{equation}
A_t = b^{-1}(\rho) \left[\, \mu(t,\theta) \Big( L_1 + \cL(t,\theta) L_{-1} \Big) - \mu'(t,\theta) L_0 + \frac{1}{2}\, \mu''(t,\theta) L_{-1} \right] b(\rho) \, ,
\end{equation}
where $\mu(t,\theta)$ is an arbitrary boundary function, often dubbed chemical potential because of the way it enters the variation of the entropy of black hole solutions \cite{Henneaux:2013dra, Bunster:2014mua}. The previous expression can be obtained by replacing \eqref{epsilons} in the general form \eqref{multiplier_radialgauge} taken by the Lagrange multiplier as a result of the radial gauge-fixing. Via \eqref{eom-CS} this result also implies the following time evolution for the function $\cL$ parameterising the connection:
\begin{equation}
\dot{\cL}(t,\theta) = \mu(t,\theta) \, \cL'(t,\theta) + 2 \,\mu'(t,\theta) \,\cL +\frac{1}{2}\,\mu'''(t,\theta) \, .
\end{equation}
Although $\mu(t,\theta)$ can be an arbitrary function, it is customary to set it to one. This choice implies $A_t - A_\theta = 0$ and $\dot{\cL} = \cL'$, so that $\cL$ becomes a chiral function. This is the option chosen, e.g., in \cite{Banados:1994tn, Banados:1998gg} that was later generalised to higher spins in \cite{Henneaux:2010xg, Campoleoni:2010zq}. While this convenient choice does not affect the analysis of asymptotic symmetries, it still imposes a restriction on the space of allowed solutions. This is harmless in gravity, while its analogue in higher spin theories does not allow one to access black holes solutions \cite{Bunster:2014mua} (see also \cite{Gutperle:2011kf, Ammon:2011nk}). 

We now go back to the residual transformations of the form \eqref{HW_spin2} that preserve the Drinfeld--Sokolov boundary conditions on any time slice. Since $\delta a$ only has a component along $L_{-1}$, the variation \eqref{sym:var-charge} of the charge is
\begin{equation}
\delta Q(\lambda) = \frac{k}{2\pi} \int \intd\theta\, \epsilon\, 
\delta \cL 
\end{equation}
thanks to $\tr(L_1 L_{-1}) = -1$ (see \eqref{sl2trace}).
Even if the $\mathfrak{sl}(2)$-valued parameter $\lambda$ depends on the function $\cL(\theta)$, 
the variation of the charge is insensitive to this and can be integrated to
\begin{equation}\label{integratedcharge}
Q(\epsilon) = \frac{k}{2\pi} \int \intd\theta\, \epsilon(\theta)\, \cL(\theta)
\end{equation}
on each time slice. Being associated to a non-vanishing boundary charge, the residual transformations generated by a $\mathfrak{sl}(2)$-valued parameter satisfying \eqref{lambda-expansion} and \eqref{epsilons} are therefore asymptotic symmetries. 

The additional constraint we imposed compared to section~\ref{sec:CSbdy} is, by construction, invariant under these residual symmetries. As a result, the considerations that led to \eqref{sym:transformofFbyboundarycharge} are still valid, and the boundary charge canonically generates the asymptotic symmetries:
\begin{equation}\label{deltaLcanonical}
\delta \cL = \{ Q(\epsilon) , \cL \}_{*} \, . 
\end{equation}
In analogy with the derivation of the $\widehat{\mathfrak{sl}}(2)_k$ algebra in \eqref{sym:affinealgebra}, we can use this result to fix the Poisson structure of the functions parameterising the phase space, obtaining
\begin{equation}\label{PoissonL}
\{ \cL(\theta) , \cL(\theta') \}_{*} = - \frac{2\pi}{k} \left( \delta(\theta-\theta')\, \cL'(\theta) + 2\,\delta'(\theta-\theta')\, \cL(\theta) + \frac{1}{2}\,\delta'''(\theta-\theta')  \right) \, ,
\end{equation}
where $\delta'(\theta-\theta') = \partial_\theta\delta(\theta-\theta')$. One can then recover the Virasoro algebra in the form \eqref{sym:finalDiracsl2} by expanding $\cL$ as 
\begin{equation}\label{sym:expansionofell}
\cL(\theta) = \frac{1}{k}\sum_{p\in \mathbb{Z}} \cL_{p}\,e^{-ip\theta} \, .
\end{equation} 

\begin{exercise}\hspace{-2pt}\hyperlink{sol:checkPoissonL}{*}\label{ex:checkPoissonL}
Check \eqref{PoissonL} by substituting \eqref{deltaL} in \eqref{deltaLcanonical} and introducing an integral over $\theta$ also on the left-hand side.
\end{exercise} 

Alternatively, the same result can be obtained by evaluating the variation \eqref{deltaLcanonical} for $\epsilon (\theta) = \epsilon_{(0)} \, e^{im\theta}$ with a fixed $m\in\mathbb{Z}$ and a constant $\epsilon_{(0)}$, and at the same time expanding $\cL$ again as in \eqref{sym:expansionofell}.

We then find
\begin{equation}\label{sym:deltal}
\delta \cL_{n} = i\epsilon_{(0)} \, (m-n)\, \cL_{m+n} -\frac{i\epsilon_{(0)} k}{2}m^{3}\delta_{m,-n} \, .
\end{equation}
On the other hand, for the chosen $\epsilon$ the boundary charge takes the form
\begin{equation}\label{sym:Qlambda}
Q (\epsilon) = \epsilon_{(0)} \,\cL_{m}  \, .
\end{equation}
By substituting~\eqref{sym:deltal} and~\eqref{sym:Qlambda} in~\eqref{deltaLcanonical} we obtain again \eqref{sym:finalDiracsl2}.

\subsection{The $\mathfrak{sl}(3)$ case}

We now move on to a Chern--Simons theory with a $\mathfrak{sl}(3,\mathbb{R}) \oplus \mathfrak{sl}(3,\mathbb{R})$ gauge algebra, describing the coupling of a spin-three gauge field to gravity. As in the previous example, we focus on a single copy of the Chern--Simons connection and we consider the basis \eqref{sl3commrel} for the eight-dimensional $\mathfrak{sl}(3,\mathbb{R})$ algebra where, for simplicity, we rename as $W_m:=W^3_{m}$ the five generators that complement the gravitational $\mathfrak{sl}(2,\mathbb{R})$ subalgebra (spanned by the generators $L_m$ as in the previous section). The invariant bilinear form is given in~\eqref{sl3trace}.

As before, we expand the $\rho$-independent part of $A_\theta$ as
\begin{equation}
a (\theta) = \sum_{m=-1}^{1}\ell^{m} ( \theta)L_{m} + \sum_{m=-2}^{2}w^{m} (\theta)W_{m}\, .
\end{equation}
The Drinfeld--Sokolov constraint~\eqref{sym:DScondition} takes the form
\begin{equation}\label{DSconditionSL3}
\ell^{1} (\theta) - 1 =0 \,, \qquad 
w^{1} (\theta) = 0 \,, \qquad 
w^{2} (\theta) =0 \, .
\end{equation}
In terms of the modes of $\ell^{m}$ and $w^{m}$ defined as in in~\eqref{sym:modedecomposition}, the constraints read
\begin{equation}
\ell^{1}_{p}-k\delta_{p,0} =0 \qquad w^{1}_{p}=0 \qquad w^{2}_{p}=0 \, .
\end{equation}
As in the $\mathfrak{sl}(2)$ case, these are first-class constraints: the Poisson brackets between them are zero except for the bracket $\{w_{p}^{1},\ell_{q}^{1} \}$ which is proportional to $w^{2}_{p+q}$ and thus vanishes on the constraint surface. 
To proceed, one can impose further gauge-fixing constraints, so as to obtain a set of second-class constraints allowing one to compute the Dirac bracket on the reduced phase space as we did to get \eqref{sym:finalDiracsl2}. 
The main difference is that in this case, one can impose different gauge-fixings, leading to different bases for 
the asymptotic symmetry algebra. We refer to \cite{Campoleoni:2010zq} for a computation of the Dirac bracket along the lines of \eqref{matrix-constraints} with the additional gauge-fixing $\ell^0 = w^0 = 0$. Here we focus on the approach deriving the Poisson bracket on the reduced phase space via the canonical realisation of the residual gauge symmetries, along the lines of \eqref{deltaLcanonical}. Even when employing this strategy, one has to first completely fix the proper gauge symmetry. In the following, we review successively two widely used complete gauge-fixings: the \textsl{highest-weight gauge} and the \textsl{single-row gauge} (also known as \textsl{u-gauge}).\footnote{In the $\mathfrak{sl}(3)$ case these two options exhaust all possible complete gauge-fixings, while for gauge algebras of bigger rank other options are possible, see, e.g., \cite{Balog:1990mu}. As for gravity, one can also characterise the asymptotic symmetries without fixing completely the proper gauge symmetry, see, e.g., \cite{Campoleoni:2017xyl}.}

\subsubsection{The highest-weight gauge}

When $a$ satisfies the Drinfeld--Sokolov constraint~\eqref{DSconditionSL3}, we can write it as
\begin{equation}\label{awithDSSL3}
    a(\theta) = L_1 + \sum_{m=-1}^{0}\ell^{m} ( \theta)L_{m} + \sum_{m=-2}^{0}w^{m} (\theta)W_{m} =:L_1 + u(\theta)\,,
\end{equation}
where $u$ is a linear combination of $\mathfrak{sl}(3)$ basis elements with non-positive mode numbers. A transformation of the form~\eqref{HW_spin2} with $\lambda(\theta)=\epsilon(\theta)L_{-1}$ leads to 
\begin{equation}
    \delta_\lambda a = \partial_\theta \epsilon \,L_{-1} + 2 \epsilon \,L_0 + \epsilon [u,L_{-1}]\, .
\end{equation}
As $[u,L_{-1}]$ is a combination of $\mathfrak{sl}(3)$ basis elements with strictly negative mode numbers, there is only one term in $L_0$, and such transformations can be used to set the $L_0$-component of $a$ to $0$, $\ell^0 = 0$.
Similarly, we can achieve $w^0=0$ by transformations with $\lambda=\epsilon W_{-1}$ because of $[L_1,W_{-1}]=3W_0$. Furthermore, by transformations with $\lambda=\epsilon W_{-2}$ we can set $w^{-1}=0$. Notice that in each step the involved infinitesimal gauge transformations can be integrated to finite ones because the components along, respectively, $L_0$, $W_0$  and $W_{-1}$ of the gauge variations do not depend on the fields. Moreover, these transformations are associated to vanishing boundary charges for the same mechanism that we discussed for $\mathfrak{sl}(2)$ in \eqref{var-l0}. We finally arrive at the \textsl{highest-weight gauge} in which we constrain all components of $a$ except the ``highest modes'' $\cL :=\ell^{-1}$ and $\cW :=w^{-2}$, so that $a$ has the form
\begin{equation}\label{sym:sl3hwgofa}
a (\theta) = L_{1} + \cL (\theta) L_{-1} + \cW (\theta) W_{-2} \, .
\end{equation}
Similarly to the second approach in the last subsection, we can infer the Dirac brackets on the constraint surface from the transformations that leave the constrained form~\eqref{sym:sl3hwgofa} invariant. In order to do this, we start with a general gauge parameter
\begin{equation}\label{sym:sl3gaugeparameter}
\lambda (\theta) = \epsilon (\theta)L_{1} + \sum_{m=-1}^{0}\epsilon^{m} (\theta)L_{m} + \chi (\theta) W_{2}  + \sum_{m=-2}^{1}\chi^{m} (\theta) W_{m} \, .
\end{equation}
The condition that the transformation leaves the form~\eqref{sym:sl3hwgofa} untouched means that $\delta_{\lambda}a$ only has components corresponding to the generators $L_{-1}$ and $W_{-2}$, the coefficients in front of all other generators vanish. This leads to a system of differential equations that determines $\epsilon^{m}(\theta)$ and $\chi^{m}(\theta)$ in terms of the functions $\epsilon(\theta)$ and $\chi(\theta)$ and their derivatives. The transformation of $a$ then takes the form 
\begin{subequations}
\begin{align}
\label{sym:sl3transformofcL}
\delta \cL &= \epsilon\, \cL' + 2\,\epsilon' \,\cL + \frac{1}{2}\,\epsilon''' - 2 \,\chi \,\cW' - 3 \,\chi' \,\cW \, , \\
\delta \cW & =  \epsilon \,\cW' +3 \epsilon' \,\cW  +\frac{1}{12} \left(2\,\chi\,\cL''' + 9\,\chi'\,\cL'' + 15 \,\chi'' \, \cL' + 10\,\chi'''\, \cL \right) \nonumber \\
& \phantom{=} +\frac{1}{24}\,\chi^{(5)} +\frac{8}{3} \left(\chi \,\cL\,\cL' + \chi'\, \cL^2 \right) ,
\label{sym:sl3transformofcW}
\end{align}
\end{subequations}
where $\chi^{(5)}$ denotes the fifth derivative of $\chi$.

\begin{exercise}\hspace{-2pt}\hyperlink{sol:deltaaSL3}{*}
\label{ex:deltaaSL3}
Derive the expressions for $\epsilon^m(\theta)$, $\chi^m(\theta)$ in terms of $\epsilon(\theta)$, $\chi(\theta)$ and verify~\eqref{sym:sl3transformofcW}.
\end{exercise}

Now we can follow the same procedure that led to the Dirac brackets~\eqref{sym:finalDiracsl2} in the last section. To obtain the Dirac bracket involving the modes of $\cL$ we start with a transformation with parameters $\epsilon (\theta)= \epsilon_{(0)}\,e^{im\theta}$ and $\chi=0$. From~\eqref{sym:sl3transformofcL} and~\eqref{sym:sl3transformofcW}  we obtain the transformation of the modes $\cL_{n}$ and $\cW_{n}$ (where the expansion into Fourier modes is done similarly to~\eqref{sym:expansionofell}):
\begin{subequations}
\begin{align}
\delta \cL_{n} &= i\,\epsilon_{(0)} \, (m-n)\, \cL_{m+n} -\frac{i\epsilon_{(0)} k}{2}m^{3}\,\delta_{m,-n} \, , \\
\delta \cW_{n} &= i\,\epsilon_{(0)}\, (2m-n)\,\cW_{m+n} \, .\label{sym:deltaw}
\end{align}
\end{subequations}
The transformation of $\cL_{n}$ is identical to the $\mathfrak{sl}(2)$ case (see~\eqref{sym:deltal}).

\begin{exercise}\hspace{-2pt}\hyperlink{sol:deltaw}{*}\label{ex:deltaw}
Check~\eqref{sym:deltaw}.
\end{exercise}

Whereas the transformation of $\cL$ leads to the same Dirac brackets~\eqref{sym:finalDiracsl2} as in the $\mathfrak{sl} (2)$ example, we can read off the Dirac brackets of $\cL_{m}$ and $\cW_{n}$ from~\eqref{sym:deltaw}:
\begin{equation}\label{sym:bracketellw}
\{\cL_{m} , \cW_{n} \}_{*}  = i\, (2m-n)\, \cW_{m+n} \, .
\end{equation}
Similarly, setting $\epsilon (\theta) =0$ and $\chi (\theta)=\chi_{(0)}\,e^{im\theta}$ we obtain the transformations
\begin{subequations}
\begin{align}
\delta \cL_{n} &= i\,\chi_{(0)}\, (2n-m)\, \cW_{m+n} \, ,\\
\delta \cW_{n} &= i\,\chi_{(0)}\frac{1}{12} \bigg(\frac{k}{2}m^{5}\,\delta_{m+n,0} - (m-n) (2m^{2}+2n^{2}-mn)\cL_{m+n} \nonumber\\
&\qquad \qquad \quad + \frac{16}{k} (m-n)\sum_{q\in \mathbb{Z}}\cL_{m+n+q}\cL_{-q} \bigg) \, .\label{sym:transfofw}
\end{align}
\end{subequations}

\begin{exercise}\hspace{-2pt}\hyperlink{sol:transfofw}{*}
\label{ex:transfofw}
Verify~\eqref{sym:transfofw}.
\end{exercise}

From here we can read off the Dirac brackets of $\cW_{m}$ with $\cL_{n}$ (consistent with~\eqref{sym:bracketellw}) and the Dirac brackets of $\cW_{m}$ and $\cW_{n}$. As before, we define $\hat{\mathcal{L}}_{m}=-\cL_{m}+\frac{k}{4}\,\delta_{m,0}$, and we introduce the quadratic field
\begin{equation}\label{sym:DefofLambda}
    \Lambda_{p}=\sum_{q\in\mathbb{Z}}\hat{\mathcal{L}}_{p+q}\hat{\mathcal{L}}_{-q}\, .
\end{equation}
We finally obtain the classical $\mathcal{W}_{3}$-algebra in standard form:
\begin{subequations}
\label{sym:W3Diracbrackets}
\begin{align}
i\{\hat{\mathcal{L}}_{m},\hat{\mathcal{L}}_{n} \}_{*} &= (m-n)\hat{\mathcal{L}}_{m+n} + \frac{c}{12}\big( m^{3}-m\big)\,\delta_{m+n,0}\,,\\
i\{ \hat{\mathcal{L}_{m}},\mathcal{W}_{n}\}_{*} &= (2m-n)\, \mathcal{W}_{m+n}\,,\\
i\{\mathcal{W}_{m},\mathcal{W}_{n} \}_{*} & =  \frac{1}{12}\bigg( (m-n) (2m^{2}+2n^{2}-mn-8)\hat{\mathcal{L}}_{m+n} + \frac{96}{c} (m-n)\Lambda_{m+n} \nonumber\\
&\qquad \qquad +\frac{c}{12}m (m^{2}-1) (m^{2}-4)\delta_{m+n,0} \bigg) \, .\label{sym:WWbracket}
\end{align}
\end{subequations}

\begin{exercise}\hspace{-2pt}\hyperlink{sol:WWbracket}{*}
\label{ex:WWbracket}
Derive~\eqref{sym:WWbracket} from~\eqref{sym:transfofw}.
\end{exercise}

Notice that this non-linear algebra includes a Virasoro subalgebra. Furthermore, in the $c \to \infty$ limit the non-linear terms vanish, and if one focuses on the modes $m,n \in {-2,\ldots,2}$ in \eqref{sym:WWbracket} the term proportional to the central charge does not contribute. Therefore, in this limit one identifies a $\mathfrak{sl}(3)$ subalgebra, cf.~\eqref{sl3commrel}, which is called the wedge algebra of $\cW_3$.

\subsubsection{The single-row gauge}

The Drinfeld--Sokolov condition restricts $a$ to the form $a(\theta)=L_1+u(\theta)$ as in~\eqref{awithDSSL3} where $u$ is a combination of generators with non-positive mode numbers. By inspecting the explicit matrix realisation of $\mathfrak{sl}(3)$ in~\eqref{sl3matrixrealisation}, we observe that $u$ is a general traceless upper triangular matrix. As in the previous discussion of the highest-weight gauge, we can achieve that $\ell^0=0$ and $w^0=0$, such that $u$ is strictly upper triangular. Concretely, it has the form
\begin{equation}
    u = \begin{pmatrix}
        0 & -\frac{1}{\sqrt{2}} (2\ell^{-1}+w^{-1}) & 2 w^{-2} \\ 
        0 & 0 & -\frac{1}{\sqrt{2}}(2\ell^{-1}-w^{-1})\\ 
        0 & 0 & 0 
    \end{pmatrix} .
\end{equation}
The remaining gauge freedom can be used to fix the coefficient $w^{-1}$ at will, and we can choose it as $w^{-1}=2\ell^{-1}$ such that the only non-vanishing entries of $u$ are in the first row (\textsl{single-row gauge}),
\begin{equation}
    u=\begin{pmatrix}
        0 & u_2 & u_3\\
        0 & 0 & 0\\ 
        0 & 0 & 0 
    \end{pmatrix} .
\end{equation}
Now we determine the most general transformation parameter $\lambda$, 
such that the transformation $\delta_\lambda a$ only has components in the first row. In matrix components, we have
\begin{equation}\label{sym:sl3singlerowdeltaa}
    \delta_\lambda a_{ij} = \partial \lambda_{ij}+\sqrt{2} \lambda_{i-1\,j} -\sqrt{2} \lambda_{i\,j+1} - u_{j}\lambda_{i1}+\delta_{i1}(u_2\lambda_{2j}+u_3\lambda_{3j})\, ,
\end{equation}
where from now on we shall denote with $\partial$ a derivative with respect to $\theta$. We then require that all entries of $\delta a$, except the two last entries in the first row, vanish. One can solve the corresponding system of equations and express the most general allowed matrix $\lambda$ in terms of the coefficients $\lambda_{21}$ and $\lambda_{31}$ in the first column, which are unconstrained. The corresponding transformations of $u_2$ and $u_3$ are then given as 
\begin{subequations}
\begin{align}
    \delta u_2 &= -\frac{1}{2\sqrt{2}}\lambda^{(4)}_{31} + \frac{1}{2}\partial^2(u_2\lambda_{31})+\frac{3}{\sqrt{2}}u_3\lambda'_{31} + \sqrt{2}u'_3 \lambda_{31}\nonumber\\ 
    &\quad -\lambda^{(3)}_{21} +\sqrt{2}u_2\lambda'_{21}+\frac{1}{\sqrt{2}}u'_2 \lambda_{21} \, ,\label{deltalambdau1}\\ 
    \delta u_3 &= \frac{1}{6} \lambda _{31}^{(5)} -\frac{\partial^3(u_2\lambda_{31})}{3\sqrt{2}} -\frac{1}{3\sqrt{2}} u_2 \lambda _{31}^{(3)} +\frac{1}{3} u_2\partial(u_2
   \lambda _{31})-u_3' \lambda _{31}'-\frac{1}{2} \lambda
   _{31} u_3''\nonumber \\ 
   & \quad +\frac{\lambda _{21}^{(4)}}{2 \sqrt{2}}+\frac{u_3'\lambda _{21} }{\sqrt{2}}+\frac{3 u_3 \lambda _{21}'}{\sqrt{2}}-\frac{1}{2} u_2 \lambda
   _{21}''\,.
   \label{deltalambdau2}
\end{align}
\end{subequations}

\begin{exercise}\hspace{-2pt}\hyperlink{sol:sl3ugauge}{*}
\label{ex:sl3ugauge}
Work out the details that lead to~\eqref{deltalambdau1} and~\eqref{deltalambdau2}.
\end{exercise}

From here one can read off the Dirac brackets for $u_2$ and $u_3$. We obtain the same algebraic structure as in the highest-weight gauge if we identify
\begin{subequations}
\begin{align}
    \cL(\theta) &= -\frac{1}{2\sqrt{2}}\,u_2(\theta) \, , \\ 
    \cW(\theta) &= \frac{1}{2} \bigg(u_3(\theta) + \frac{1}{2\sqrt{2}}\, u_2'(\theta)\bigg) \, .
\end{align}
\end{subequations}
Expressed in terms of modes, one recovers the Dirac brackets~\eqref{sym:W3Diracbrackets}.

\subsection{Beyond $\mathfrak{sl}(3)$}\label{sec:beyondsl3}

A similar analysis can be done for $\mathfrak{sl} (N)$ with $N>3$ both in the highest-weight gauge and in the single-row gauge (see also \cite{Balog:1990mu} for a general discussion on the options to completely fix the Drinfeld--Sokolov gauge freedom). In the highest-weight gauge, $a$ is constrained to the form 
\begin{equation}
a (\theta) = L_{1} + \cL(\theta)\,L_{-1} + \sum_{s=3}^{N} 
\cW_{s} (\theta) W^s_{-s+1} \, .
\end{equation}
The resulting asymptotic symmetry algebra is the classical $\mathcal{W}_{N}$-algebra generated by the $N-1$ fields $\cL$ and $\cW_{s}$ with $3 \leq s \leq N$. The Dirac brackets are nonlinear in the fields, with the maximum order of nonlinearity being $N-1$, and concrete expressions for the structure constants can be found in~\cite{Campoleoni:2011hg}.

In the single-row gauge, the maximal nonlinearity in the Dirac brackets is quadratic. This can be seen in a formulation based on pseudo-differential operators that we now review. 
Starting from $a(\theta)=L_1 + u(\theta)$, where the only non-vanishing entries of $u$ are the off-diagonal terms of the first row, we determine as before the most general matrix $\lambda$ such that the corresponding transformation $\delta_\lambda a$ does not change the form of $u$. To simplify the computation, we choose a basis of the $\mathfrak{sl}(2)$-subalgebra in $\mathfrak{sl}(N)$ where $L_1$ is realised as 
\begin{equation}
    L_1 = \begin{pmatrix}
        0\\ 
        -1 & 0 \\ 
        0 & -1 & 0 \\ 
         & \ddots & \ddots & \ddots \\ 
         & & 0 & -1 & 0
    \end{pmatrix} .
\end{equation}
Similarly to~\eqref{sym:sl3singlerowdeltaa}, the transformation of the components of $a$ now reads
\begin{equation}\label{sym:slNsinglerowdeltaa}
    \delta_\lambda a_{ij} =  \lambda_{ij}'- \lambda_{i-1\,j} + \lambda_{i\,j+1} - u_{j}\lambda_{i1}+\delta_{i1}\sum_{k=2}^{N} u_k \lambda_{kj}\, .
\end{equation}
We introduce the following differential and pseudo-differential operators in which also negative powers of the derivative operator $\partial$ appear:
\begin{subequations}
\begin{align}
\label{sym:deflambdai}
    \lambda_i &= \sum_{j=1}^{N}\lambda_{ij}\partial^{N-j} \, , \\ 
    \lambda^1 &= \sum_{i=1}^{N}\partial^{-N+i-1}\lambda_{i1}\, .
\end{align}
\end{subequations}
One can formally compute with such objects if one implements the following rule for commuting differentials with functions,
\begin{equation}
    \partial^k f = \sum_{i=0}^\infty \binom{k}{i} f^{(i)} \partial^{k-i}  \,.
\end{equation}
For a positive integer $k$ the sum truncates, and we get the usual rule. For negative $k$, we formally consider combinations of differential operators with arbitrarily negative powers. We also introduce the operator
\begin{equation}\label{sym:defofL}
    L = \partial^N + u_2 \partial ^{N-2}+ \dots + u_{N}\, .
\end{equation}
The vanishing of $\delta_\lambda a_{ij}$ given in~\eqref{sym:slNsinglerowdeltaa} for $i\geq 2$ can be expressed in terms of differential operators as 
\begin{equation}\label{sym:pdoconstraint}
0=\sum_{j=1}^{N}\delta_\lambda a_{ij}\partial^{N-j}   = \partial \lambda_i -\lambda_{i-1}-\lambda_{i1}L \,.
\end{equation}
\begin{exercise}\hspace{-2pt}\hyperlink{sol:pdoconstraint}{*}\label{ex:pdoconstraint}
Show~\eqref{sym:pdoconstraint}.
\end{exercise}
Composing~\eqref{sym:pdoconstraint} with $\partial^{-N+i-1}$ from the left and summing over $i$ we find  
\begin{align}
    0&=\sum_{i=2}^N \big(\partial^{-N+i} \lambda_i -\partial^{-N+i-1}\lambda_{i-1}\big) -\big(\lambda^1-\partial^{-N} \lambda_{11}\big) L \nonumber\\ 
    &= \lambda_N - \partial^{-N+1}\lambda_1 -\big(\lambda^1-\partial^{-N} \lambda_{11}\big) L  
    \label{sym:pdosummedidentity} \,.
\end{align}
For a pseudo-differential operator $X=\sum_{i=-\infty}^{M}x_i\partial^i$, we denote by $(X)_+=\sum_{i=0}^M x_i \partial^i$ its truncation to the part containing non-negative powers of $\partial$. Similarly, $(X)_{-}=X-(X)_+$ denotes the projection to the part containing only negative powers of $\partial$. From truncating~\eqref{sym:pdosummedidentity} in this way, we obtain
\begin{equation}
    \lambda_N = \big(\lambda^1\,L\big)_+\, ,
\end{equation}
which expresses the last row of $\lambda$ in terms of its first column. We can obtain the other rows by using~\eqref{sym:pdoconstraint} as a recursion relation,
\begin{equation}\label{sym:recursionrelation}
    \lambda_{i-1} = \partial \lambda_i - \lambda_{i1}L\,.
\end{equation}
The solution is
\begin{equation}\label{sym:solutionofrecursion}
    \lambda_{N-i} = \partial^i \big( \lambda^1\,L\big)_+ - \big(\partial^i\lambda^1\big)_+L\,.
\end{equation}
\begin{exercise}\hspace{-2pt}\hyperlink{sol:solutionofrecursion}{*}\label{ex:solutionofrecursion}
Check that~\eqref{sym:solutionofrecursion} solves the recursion relation~\eqref{sym:recursionrelation}.
\end{exercise}
Similarly to~\eqref{sym:pdoconstraint}, the transformation of the $u_j$ can be summarised as
\begin{equation}\label{sym:transformationofLunsimplified}
    \delta_\lambda L = \sum_j \delta_\lambda a_{1j}\partial^{N-j} = \partial \lambda_1 -\lambda_{11}L + \sum_{k=2}^N u_k \lambda_k\,.
\end{equation}
With the help of the expression for $\lambda_i$ in terms of $\lambda^1$, one can show that
\begin{equation}\label{sym:transformationofL}
    \delta_\lambda L = L\big(\lambda^1 L\big)_+ -\big(L\lambda^1\big)_+ L\, .
\end{equation}
\begin{exercise}\hspace{-2pt}\hyperlink{sol:transformationofL}{*}\label{ex:transformationofL}
Show~\eqref{sym:transformationofL}.
\end{exercise}
We have not yet implemented the condition $\delta_\lambda a_{11}=0$. Therefore, the results above are also valid if we consider $\mathfrak{gl}(N)$ instead of $\mathfrak{sl}(N)$ and allow for a non-zero entry $u_1$ in the top-left corner of the matrix $u$. The Poisson brackets (often referred to as second Gelfand--Dickey Poisson structure) that we would then obtain from the transformation~\eqref{sym:transformationofL} lead to a $\mathcal{W}$-algebra with fields $u_1,\dots,u_{N}$.

To complete the reduction for $\mathfrak{g}=\mathfrak{sl}(N)$, it remains to require $\delta_\lambda a_{11}=0$ or $u_1=0$. This translates into the condition that the coefficient of the highest possible power $\partial^{N-1}$ in $\delta_\lambda L$ vanishes. Using $(X)_+=X-(X)_-$ in~\eqref{sym:transformationofL}, we get 
\begin{align}
    \delta_\lambda L &= -L\big(\lambda^1 L\big)_- +\big(L\lambda^1\big)_- L\nonumber\\
    &= -(\partial^N + \dots) \big((\res \lambda^1 L)\partial^{-1}+\dots\big) +  \big((\res L \lambda^1)\partial^{-1}+\dots\big) (\partial^N + \dots)\nonumber\\ 
    &= \res [L,\lambda^1] \partial^{N-1} + \dots 
\end{align}
Here, we introduced the notation 
\begin{equation}\label{sym:defofres}
    \res \Big( \sum_i x_i \partial^i \Big) = x_{-1} 
\end{equation}
for the coefficient of $\partial^{-1}$ of a pseudo-differential operator. When we enforce $\delta_\lambda a_{11}=0$, we obtain
\begin{align}\label{sym:slNreduction}
    0&=\res [L,\lambda^1]\nonumber\\ 
    &=\res [\partial^N,\partial^{-N}\lambda_{11}] + \res [L,\hat{\lambda}^1]\nonumber\\ 
    &=N\lambda_{11}' + \res [L,\hat{\lambda}^1]\, ,
\end{align}
which allows expressing $\lambda_{11}'$ in terms of the remaining coefficients of the first column collected in
\begin{equation}\label{sym:defoflambda1}
    \hat{\lambda}^1 = \sum_{i=2}^N \partial^{-N+i-1}\lambda_{i1}\,.
\end{equation}
On the other hand, we can rewrite the transformation~\eqref{sym:transformationofL} of $L$ as
\begin{align}
    \delta_\lambda L &= L\big(\partial^{-N}\lambda_{11}L\big)_+ - \big(L\,\partial^{-N}\lambda_{11}\big)_+L + L\big(\hat{\lambda}^1 L\big)_+ - \big(L\,\hat{\lambda}^1\big)_+ L\nonumber\\ 
    &= [L,\lambda_{11}] + L\big(\hat{\lambda}^1 L\big)_+ - \big(L\,\hat{\lambda}^1\big)_+ L\,,
\end{align}
where we separated the contribution of $\lambda_{11}$ from the rest. Using 
\begin{equation}
    [\partial^k,f]= \sum_{j=0}^{k-1} \partial^j \,f'\,\partial^{k-1-j}\,,
\end{equation}
we find (defining $u_0=1$)
\begin{align}
    [L,\lambda_{11}] &= \sum_{i=0}^N u_i [\partial^{N-i},\lambda_{11}]\nonumber\\ 
    &= \sum_{i=0}^N u_i \sum_{j=0}^{N-i-1} \partial^j\,\lambda_{11}'\,\partial^{N-i-1-j}\nonumber\\ 
    &= \sum_{k=0}^{N-1} \sum_{i=0}^k u_i \partial^{k-i}\,\lambda_{11}'\,\partial^{N-1-k}\nonumber\\ 
    &= \sum_{k=0}^{N-1} \big(L\,\partial^{k-N}\big)_+ \lambda_{11}'\,\partial^{N-1-k}\,.
\end{align}
In total, we obtain
\begin{equation}\label{sym:finaltransformationofL}
    \delta_\lambda L = \sum_{k=0}^{N-1} \big(L\,\partial^{k-N}\big)_+ \lambda_{11}'\,\partial^{N-1-k} + L\big(\hat{\lambda}^1 L\big)_+ - \big(L\,\hat{\lambda}^1\big)_+ L\, ,
\end{equation}
where
\begin{equation}
    \lambda_{11}' = -\frac{1}{N} \res \,[L,\hat{\lambda}^1]\, .
\end{equation}
The transformation~\eqref{sym:finaltransformationofL} encodes the Dirac brackets. Because $L$ depends linearly on the fields, it is obvious from the formula above that in this basis the Dirac bracket is at most quadratic in the fields $u_i$. 

An analogous analysis can be performed for the higher-spin algebra $\mathfrak{hs} [\lambda]$. One obtains similar expressions for the transformations, where the exponent $N$ is formally replaced by the parameter $\lambda$ and the sums become untruncated (formal) series (thus generalising to pseudo-differential operators involving not necessarily integer powers of $\partial$). The Drinfeld--Sokolov reduction for this case (without employing $u_1=0$) has been performed in~\cite{Khesin:1994ey}. Roughly, one can argue that the coefficients in the Dirac brackets $\{ u_i,u_j \}_*$ are polynomials in the parameter $\lambda$, or equivalently that the coefficients in $\delta_\lambda u_i$ for a transformation involving only finitely many non-zero $\lambda_{j1}$ ($j\not= 1$) are polynomials in $\lambda$. Replacing $N$ by $\lambda$ in the exponents appearing in~\eqref{sym:transformationofL} yields an expression that precisely has this property and by construction coincides with it for positive integer $\lambda=N$. The resulting $\mathcal{W}$-algebra contains infinitely many fields including $u_1$, it is called the $\mathcal{W}_{1+\infty}[\lambda]$-algebra. Its further reduction by implementing $u_1=0$ leads to the $\mathcal{W}_\infty[\lambda]$-algebra~\cite{Figueroa-OFarrill:1992uuf}.

For computational purposes, it is useful to realise the algebras $\mathcal{W}_N$ and $\mathcal{W}_\infty[\lambda]$ by a free-field construction. This is achieved by the Miura transformation~\cite{Dickey:1997wia}, 
\begin{equation}\label{sym:classicalMiura}
    (\partial +v_1)\cdots (\partial +v_N) = \partial^N + \sum_{s=2}^N u_s(\theta)\partial^{N-s}\, .
\end{equation}
Expanding the left-hand side, one can read off expressions for $u_s$ in terms of the fields $v_i$. The Poisson bracket for the $u_s$ is then induced from the simple (free field) Poisson brackets
\begin{equation}\label{sym:freePoisson}
    \{ v_k (\theta), v_l(\theta') \}= \frac{2\pi N(N^2-1)}{6k} \bigg(\delta_{kl}-\frac{1}{N}\bigg) \delta'(\theta-\theta')\, .
\end{equation}
The possibility to express the fields $u_s$ in terms of free fields allows one to construct a quantum version of the classical $\mathcal{W}_N$-algebra by quantising the free fields. This will be discussed in section~\ref{sec:quantumMiura}.

The fields $v_k$ can be understood as the elements of a diagonal matrix $D$ for a gauge choice $a=L_1 + D$ in the Drinfeld--Sokolov reduction. This \textsl{diagonal gauge} however is not a complete gauge-fixing. The algebra $\mathcal{W}_N$ is the subalgebra of the Poisson algebra generated by the $v_k$ that is invariant under the residual gauge symmetries \cite{Campoleoni:2017xyl} (see also~\cite{Balog:1990mu,Bershadsky:1989mf,Lukyanov:1990tf}).

The fact that the Poisson brackets for the fields $u_j$ induced by the Miura transformation coincide with the Poisson brackets obtained by the Drinfeld--Sokolov reduction in the single-row gauge is known as Kupershmidt--Wilson theorem in the context of pseudo-differential operators \cite{Dickey:1997wia}. This is explained in section~\ref{sec:Miura}.

The Miura transformation provides expressions for the fields $u_j$ in terms of polynomials of $v_k$ and their derivatives. The coefficients in the Poisson brackets then depend rationally on $N$ which follows from the explicit formula~\eqref{Miura:Poissonbracket} for the Poisson bracket and the condition~\eqref{Miura:ambiguityFD}. Replacing $N$ in these expressions by $\lambda$, one obtains the classical $\mathcal{W}_\infty[\lambda]$ based on a free-field construction. Similarly to what we discussed after \eqref{sym:W3Diracbrackets}, the wedge algebra of $\mathcal{W}_\infty[\lambda]$ is the algebra $\mathfrak{hs}[\lambda]$.

\section{Quantum $\cW$-algebras and minimal-model holography}\label{sec:quantum}

\subsection{Motivation}

In the previous section, we have seen the appearance of classical
$\mathcal{W}$-algebras as asymptotic symmetries of higher-spin gauge theories on
AdS$_{3}$ backgrounds. These are classical Poisson algebras, and the
obvious question arises whether they would also appear in a quantised
theory. Even without having the quantisation of higher-spin gauge
theories under control, we can ask whether quantum
versions of the classical $\mathcal{W}$-algebras exist.

The simplest case is the Virasoro algebra that we found in the last section (see~\eqref{sym:standardVirasoro}),
\begin{equation}
i\{\hat{\mathcal{L}}_{m},\hat{\mathcal{L}}_{n} \} = (m-n)\hat{\mathcal{L}}_{m+n} 
 + \delta_{m,-n}\frac{c}{12} m (m^{2}-1)\, .
\end{equation}
In this case we can just replace $\hat{\mathcal{L}}_{m}$ by quantum operators $L_{m}$ and
replace $i\{\cdot,\cdot \}$ by the commutator $[\cdot ,\cdot]$, so
that the $L_{m}$ satisfy the commutation relations
\begin{equation}\label{quantum:Virasoro}
[L_{m},L_{n}] = (m-n) L_{m+n} + \delta_{m,-n}\frac{c}{12} m
(m^{2}-1)\, .
\end{equation}
Such a quantum Virasoro algebra (or better two copies
--~left-moving and right-moving~-- of it) appears in two-dimensional conformal quantum field theories. The appearance of this algebra is expected in the spirit of a holographic correspondence between a gravitational theory on asymptotically AdS$_{d+1}$ spacetimes and a conformal field theory on its conformal $d$-dimensional boundary, for $d=2$.\footnote{For a general introduction into the holographic AdS/CFT correspondence, see, e.g., \cite{Ammon:2015wua,Nastase:2015wjb,Penedones:2016voo}.}

Finding a quantum version of the Virasoro algebra was simple because
the corresponding Poisson algebra is linear. This is far less trivial
for non-linear $\mathcal{W}$-algebras. When we replace the classical modes by quantum
operators, we face two problems: on the one hand, there is some freedom
in translating the classical Poisson brackets to quantum commutators
because we have to choose an ordering in the non-linear terms. On
the other hand, after having defined the commutators using some
ordering prescription, the classical Jacobi identity does not guarantee that the quantum Jacobi identity is satisfied by the commutators. As an
example, we look at the $\mathcal{W}_{3}$-algebra. Recall from the previous section the Poisson bracket
of two spin-3 modes, which reads (see~\eqref{sym:WWbracket})
\begin{multline}\label{quantum:W3Poissonbracket}
i\{\mathcal{W}_{m},\mathcal{W}_{n} \} = \frac{1}{12}
\Big((m-n) (2m^{2}+2n^{2}-mn-8)\hat{\mathcal{L}}_{m+n}
+\frac{96}{c} (m-n)\Lambda_{m+n} \\
+ \frac{c}{12}m (m^{2}-1)
(m^{2}-4)\delta_{m,-n} \Big) \, ,
\end{multline}
with
$\Lambda_{p}=\sum_{q\in\mathbb{Z}}\hat{\mathcal{L}}_{p+q}\hat{\mathcal{L}}_{-q}$.
\begin{exercise}\hspace{-2pt}\hyperlink{sol:bracket_L_Lambda}{*}\label{ex:bracket_L_Lambda}
Show that
\begin{equation}\label{quantum:bracket_L_Lambda}
i\{\hat{\mathcal{L}}_{m},\Lambda_{n} \} = (3m-n)\Lambda_{m+n} +
\frac{c}{6}m(m^{2}-1)\hat{\mathcal{L}}_{m+n} \, .
\end{equation}
\end{exercise}

When we replace $\hat{\mathcal{L}}_m$ and $\mathcal{W}_n$ by quantum
operators $L_m$ and $W_n$, respectively, we first have to decide what we mean by $\Lambda$: which
ordering prescription do we use? Notice that there is from the start a
problem to make sense of the infinite series: the naive expressions
$\sum_{p\in\mathbb{Z}}L_{m+p}L_{-p}$ and
$\sum_{p\in\mathbb{Z}}L_{-p}L_{m+p}$ differ by $\sum_{p\in\mathbb{Z}}
(m+2p)L_{m}$ (for $m\not= 0$), which is ill-defined. We shall require
that the operators $\Lambda_{m}$ are well-defined on states that are
annihilated by modes $L_{p}$ for large enough mode number $p$. Then
$\Lambda$ has to be of the form (we do not distinguish the classical $\Lambda$ and the quantum counterpart in notation, and hope
that the meaning will be clear from the context)
\begin{equation}\label{quantum:Lambda_modes}
\Lambda_{m} = \sum_{p\in\mathbb{Z}} :L_{m+p}L_{-p}: + f (m)L_{m}\, ,
\end{equation}
where the colons denote mode normal ordering, i.e.\ operators $L_{q}$
are ordered such that the highest mode numbers appear on the right.
Requiring the Jacobi identity to hold will then, on the one hand, fix $f(m)$ (which will be determined in exercise~\ref{ex:modes_of_Lambda}), and, on the other hand, it leads to a shift in the structure
constants: the coefficient $\frac{96}{c}$ in front of $\Lambda$ in the
Poisson bracket~\eqref{quantum:W3Poissonbracket} of two
$\mathcal{W}^{(3)}$-modes is replaced in the quantum commutator by
$\frac{96}{c+22/5}$.
Working out the restrictions imposed by the
Jacobi identity directly is straightforward but tedious. A more
elegant way of understanding how $\Lambda$ has to be defined, and how
the shift by $22/5$ arises, is by using the language of
two-dimensional conformal field theories, which we shall now introduce.

\subsection{$\mathcal{W}$-algebras and operator product expansion}

A conformal field theory is a quantum field theory which is invariant
under conformal transformations. In two dimensions, in the Euclidean
formulation, we can consider a field theory on the compactified complex plane
$\overline{\mathbb{C}}=\mathbb{C}\cup \{\infty \}$, and the group of
global conformal transformations mapping $\overline{\mathbb{C}}$ to
itself is $PSL(2,\mathbb{C})$. There is a special set of fields, the quasi-primary fields, which transform covariantly under such
conformal transformations $z\mapsto w=f (z)$,
\begin{equation}
\phi_{h,\bar{h}} (z,\bar{z}) \mapsto \tilde{\phi}_{h,\bar{h}}(w,\bar{w}) =
\big(f' (z) \big)^{-h}\big(\overline{f' (z)}
\big)^{-\bar{h}}\phi_{h,\bar{h}} (z,\bar{z}) \, .
\end{equation} 
They are characterised by the conformal weights $h,\bar{h}$ (real and
non-negative in a unitary theory), or equivalently by the scaling
dimension $\Delta=h+\bar{h}$ and the spin $s=h-\bar{h}$. All other
fields in the theory are obtained from quasi-primary fields by taking
derivatives.

In a unitary two-dimensional scale-invariant theory, conservation of a current
implies that its $z$- and its $\bar{z}$-component are separately
conserved. Therefore, the currents can be split into a set of holomorphic
currents $W_{s_{i}}(z)$ (of spin $s_{i}$ and scaling dimension
$\Delta_{i}=s_{i}$), and a set of antiholomorphic currents
$\overline{W}_{s_{i}} (\bar{z})$ (of spin $-s_{i}$ and scaling dimension
$\Delta_{i}=s_{i}$). Let us focus on the holomorphic currents. The set
of such currents is closed under operator product expansion (OPE): if
we take the operator product of two holomorphic fields, only
holomorphic fields appear in the short-distance expansion. These
fields then define an algebraic structure that is called (quantum)
$\mathcal{W}$-algebra. For a basic introduction to conformal field theories, see, e.g., \cite{DiFrancesco:1997nk, Gaberdiel:1999mc, Blumenhagen:2009zz}, while we refer to \cite{Lukyanov:1990tf, Bouwknegt:1992wg, Arakawa2017, Prochazka:2024xyd} for an ampler introduction to quantum $\cW$-algebras.

There are different ways to think about fields in a quantum field
theory, one way is to think of them as objects that can be plugged into correlation functions resulting in some function (or
distribution) depending on the insertion points. In many cases it is
useful to have an operator realisation, where we can view the
fields as operator-valued functions (or distributions) which act on
some Hilbert space, and the correlation functions can be understood as
expectation values of products of these field operators. In
two-dimensional Euclidean conformal field theories, there is a
particularly advantageous operator realisation called radial
quantisation. Here we think of the radial direction as representing
(Euclidean) time, and scaling corresponds to time translation. This
quantisation picture results from the ordinary quantisation picture on
the Euclidean cylinder $\mathbb{R}\times S^{1}$ by a conformal
transformation mapping the infinite Euclidean past on the cylinder to the
origin of the plane. Correlation functions are then given by
expectation values of radially ordered products of field operators.

These field operators can be expanded in modes,
\begin{equation}
W^{(s)} (z) = \sum_{n\in\mathbb{Z}}W^{(s)}_{n}z^{-n-s} \, ,
\end{equation}
with operators $W^{(s)}_{n}$ resulting from contour integrals of
$W^{(s)} (z)$,
\begin{equation}
W^{(s)}_{n} = \frac{1}{2\pi i}\oint \intd z\; z^{n+s-1} W^{(s)} (z) \, .
\end{equation}
These mode operators~$W^{(s)}_{n}$ satisfy commutation relations
determined by the OPEs of the fields (see below). A $\mathcal{W}$-algebra can be equivalently described in terms of commutators of the modes, or in terms of OPEs.

The commutation relation between mode operators $W^{(s)}_{m}$ and
$W^{(t)}_{n}$ can be expressed in terms of the fields as
\begin{equation}
\Big[W^{(s)}_{m},W^{(t)}_{n} \Big] = \bigg[\frac{1}{2\pi i}\oint \intd z\;
z^{m+s-1} W^{(s)} (z),\frac{1}{2\pi i} \oint \intd w\; w^{n+t-1} W^{(t)}
(w)\bigg] \, .
\end{equation}
Introducing the radial ordering symbol $\mathcal{R}$ for operators,\footnote{Analogously to the time ordering operator it reorders the operators according to the absolute value of the complex coordinate of the insertion point; smaller values are ordered to the right.} we can rewrite the expression above as
\begin{equation}
\Big[W^{(s)}_{m},W^{(t)}_{n} \Big] = \frac{1}{(2\pi i)^{2}} 
\Bigg\{\mathop{\oint\oint}_{|z|>|w|}- \mathop{\oint\oint}_{|z|<|w|}
\Bigg\}  \intd z\,\intd w\; z^{m+s-1}\,w^{n+t-1}
\mathcal{R}\Big(  W^{(s)} (z) W^{(t)}(w)\Big)\, .
\end{equation}
Assuming analyticity of the radially ordered operator product, we can
deform the contour of the $z$-integral and find
\begin{equation}\label{quantum:mode_commutator}
\Big[W^{(s)}_{m},W^{(t)}_{n} \Big] = \frac{1}{(2\pi i)^{2}}\oint_{0}\intd w
\oint_{w} \intd z\; z^{m+s-1}\,w^{n+t-1}
\mathcal{R}\Big(  W^{(s)} (z) W^{(t)}(w)\Big) \, .
\end{equation}
The commutator is therefore determined by the poles in the OPE of
$W^{(s)}$ and $W^{(t)}$.

Let us look at an example. The OPE of a (normalised) spin-$1$ current $J$ with
itself reads (we shall omit the radial ordering symbol $\mathcal{R}$
from now on),
\begin{equation}
J (z) J (w) = \frac{1}{(z-w)^{2}} + \text{regular} \, .
\end{equation}
For the modes $J_{m}$ in the mode decomposition 
$J(z)=\sum_{n\in\mathbb{Z}}J_{n}\,z^{-n-1}$ we find the commutator
\begin{align}
\big[J_{m},J_{n} \big] &=  \frac{1}{(2\pi i)^{2}}\oint_{0}\intd w
\oint_{w} \intd z\; z^{m}\,w^{n} \frac{1}{(z-w)^{2}}\nonumber\\
&= \frac{1}{2\pi i}\oint_{0}\intd w\; w^{n}\,m\, w^{m-1}\nonumber\\
&= m\,\delta_{m,-n} \, .
\end{align}
\begin{exercise}\hspace{-2pt}\hyperlink{sol:Virasoro}{*}\label{ex:Virasoro} 
Starting from the OPE of the energy-momentum tensor with
itself,
\begin{equation}\label{quantum:TTOPE}
T (z) T (w) = \frac{c}{2}\frac{1}{(z-w)^{4}} + \frac{2}{(z-w)^{2}}T
(w) + \frac{1}{z-w}\partial_{w}T (w) + \text{regular}\, ,
\end{equation}
show that its modes $L_{m}$ in the mode
decomposition $T(z)=\sum_{m\in\mathbb{Z}}L_{m}z^{-m-2}$ satisfy the
Virasoro algebra (see~\eqref{quantum:Virasoro}).
\end{exercise}

Another useful concept in conformal field theories is the
operator-state correspondence, which states that there is a one-to-one map
between local fields and states in the Hilbert space $\mathcal{H}$ (in radial
quantisation), on which the operator-valued fields act. The idea is
that in radial quantisation, a state at infinite Euclidean past
(corresponding to taking $z\to 0$ on the plane) is specified by the
insertion of a local operator at $z=0$.

In particular, we can look at the subspace of states which are
annihilated by the zero-mode $\bar{L}_{0}$ of the antiholomorphic
component $\overline{T} (\bar{z})$ of the energy-momentum tensor,
\begin{equation}
\mathcal{H}^{\text{hol}} = \{v\in\mathcal{H},\ \bar{L}_{0}v=0 \} \, .
\end{equation} 
The vectors in $\mathcal{H}^{\text{hol}}$ are in one-to-one
correspondence with the holomorphic currents. In
$\mathcal{H}^{\text{hol}}$ there is one vector $\Omega$ that is called
the vacuum. It is annihilated by all modes $W^{(s)}_{m}$ with $m>-s$,
\begin{equation}
W^{(s)}_{m} \Omega =0 \quad \text{for}\ m> -s \, .
\end{equation}
We denote the holomorphic field corresponding to a vector
$\phi\in\mathcal{H}^{\text{hol}}$ by $V(\phi;z)$. It has the property that
\begin{equation}
\lim_{z\to 0} V (\phi;z) \Omega = \phi \, ,
\end{equation}
so the insertion of the local field $V(\phi;z)$ at $z=0$ generates the
state $\phi$. 
\begin{exercise}\hspace{-2pt}\hyperlink{sol:operator_state}{*}\label{ex:operator_state}
Show that $W^{(s)} (z)= V \big(W^{(s)}_{-s}\Omega ;z\big)$.
\end{exercise}
The operator-state correspondence allows us to rewrite
the OPE of two fields. Consider the action of a field $V(\phi;z-w)$ on
a state $\psi$ (where we assume that both are eigenstates of $L_{0}$
with eigenvalues $h_{\phi}$ and $h_{\psi}$, respectively). It can be
expanded in powers of $z-w$,
\begin{equation}
V (\phi;z-w) \,\psi = \sum_{n\geq 0} (z-w)^{-h_{\phi}-h_{\psi}+n}
\,\chi_{n} \, ,
\end{equation}
where the vectors $\chi_{n}$ are eigenvectors of $L_{0}$ satisfying $L_{0}\,\chi_{n}=n\,\chi_{n}$. This
decomposition on the space of states has a corresponding decomposition
on the space of fields, which is just the OPE,
\begin{equation}\label{quantum:duality}
V (\phi ; z) V (\psi ;w) = V \big(V (\phi ;z-w)\psi;w\big) 
= \sum_{n\geq 0} (z-w)^{-h_{\phi}-h_{\psi}+n} \,V(\chi_{n} ;w)\, .
\end{equation}
We have seen above that the OPE determines the commutation relations
of the modes. With the help of the last equation, we can use the
commutation relations to determine OPEs. Let us look at the example of
the OPE of the energy-momentum tensor $T(z)=V(L_{-2}\Omega ;z)$ with itself. When we apply the
relation~\eqref{quantum:duality}, we obtain
\begin{align}
T (z) T (w) &= V\big(T (z-w)L_{-2}\Omega ;w \big)\nonumber\\
&= \sum_{n\geq 0} (z-w)^{n-4} V\big(L_{-n+2}L_{-2}\Omega ;w \big)\nonumber\\
&= \frac{1}{(z-w)^{4}}\frac{c}{2} + \frac{2}{(z-w)^{2}}T (w) + 
\frac{1}{(z-w)} \partial_{w}T (w) \nonumber\\
&\quad \ + V (L_{-2}L_{-2}\Omega ;w) + \dots 
\end{align}
by using the commutation relations of the modes $L_{m}$.
In the last step we used that $L_{-1}L_{-2}\Omega = L_{-3}\Omega$ is the state corresponding to $\partial T$ which can be checked straightforwardly.

The regular
term in the OPE is called the normal ordered product of the fields
whose OPE is considered. In this case we find that the normal ordered
product of $T$ with itself -- which we denote by $\boldsymbol{(}TT\boldsymbol{)}$ -- is given by
\begin{equation}
\boldsymbol{(}TT\boldsymbol{)}(w) = V (L_{-2}L_{-2}\Omega ;w) \, .
\end{equation}
\begin{exercise}\hspace{-2pt}\hyperlink{sol:NOP_of_currents}{*}
\label{ex:NOP_of_currents}
Show that the normal ordered product $\boldsymbol{(}W^{(s)}W^{(t)}\boldsymbol{)}(z)$ of two
holomorphic currents of spin $s$ and $t$, respectively, is given by 
$V(W^{(s)}_{-s}W^{(t)}_{-t}\Omega;z)$.
\end{exercise}

We have seen how we can determine the states corresponding to
the fields that appear in the OPE. Similar techniques can be used to obtain the fields themselves. By a straightforward
computation,\footnote{See, e.g., \cite[section 2.7]{Blumenhagen:2009zz} or \cite[section 6.5]{DiFrancesco:1997nk}.} we find that
\begin{equation}\label{quantum:TTmodes}
\boldsymbol{(}TT\boldsymbol{)}(z) = \sum_{n\in\mathbb{Z}} \bigg(\sum_{p\geq 2} L_{-p}L_{n+p} +
\sum_{p\leq 1} L_{n+p}L_{-p} \bigg) z^{-n-4} \, .
\end{equation}
Notice that the normal ordering that we defined using the OPE does not
coincide with the normal ordering of modes by their mode numbers.

As we said in the beginning, all fields can be generated from the set
of quasi-primary fields by taking derivatives. Quasi-primary fields
are specified by their behaviour under conformal transformations. This
translates into a specific structure of their OPE with the
energy-momentum tensor, namely that there is no third-order pole. 
For a quasi-primary field $\phi$ of weight $h$, the OPE with $T$ then
reads
\begin{equation}\label{quantum:OPE_T_quasiprimary}
T (z)\phi (w) = \text{higher-order poles} + \frac{0}{(z-w)^{3}} +
\frac{h\phi (w)}{(z-w)^{2}} + \frac{\partial_{w}\phi(w)}{z-w} +
\text{regular}\, .
\end{equation}
The
states corresponding to quasi-primary fields are also called
quasi-primary, they are characterised by the condition
\begin{equation}
\phi  \ \text{quasi-primary} \quad \Longleftrightarrow  \quad L_{1}\phi(0)\Omega  = 0 \, .
\end{equation}
The energy-momentum tensor is quasi-primary. The normal-ordered
product $\boldsymbol{(}TT\boldsymbol{)}$ is not, but 
\begin{equation}\label{quantum:quasiprimary_Tsquared}
\mathcal{T} = \boldsymbol{(}TT\boldsymbol{)} -\frac{3}{10}\partial^{2}T
\end{equation}
is, as we shall see now. The state corresponding to $\mathcal{T}$ is
\begin{equation}
\mathcal{T}(0)\Omega = \bigg( L_{-2}L_{-2}-\frac{3}{5} L_{-4}
\bigg)\Omega \, .
\end{equation}
Applying $L_{1}$ leads to
\begin{align}
L_{1}\mathcal{T} (0)\Omega &= L_{1} \bigg( L_{-2}L_{-2}-\frac{3}{5} L_{-4}
\bigg)\Omega = \bigg(3L_{-3} -\frac{3}{5}5L_{-3} \bigg)\Omega  = 0\, ,
\end{align}
so indeed $\mathcal{T}$ is quasi-primary.  
\begin{exercise}\hspace{-2pt}\hyperlink{sol:quasiprimary_current}{*}
\label{ex:quasiprimary_current}
Let $W^{(s)}(z)=\sum_{n\in\mathbb{Z}}W^{(s)}_{n}z^{-n-s}$ be a
holomorphic current of spin $s$. Show that $W^{(s)}$ is quasi-primary if
and only if
\begin{equation}\label{quantum:commutator_L_Ws}
\big[L_{m},W^{(s)}_{n}\big] = \big((s-1)m-n \big)W^{(s)}_{m+n}\quad
\text{for}\ m\in\{-1,0,1 \} \, .
\end{equation}
\end{exercise}

Fields $W^{(s)}$ whose modes satisfy~\eqref{quantum:commutator_L_Ws} for all $m$ are called \textsl{primary fields}.
We now have all the tools at our disposal to study the structure of
the quantum $\mathcal{W}$-algebras in more detail.

\subsection{The quantum $\cW_{3}$-algebra}

Let us consider the quantum $\mathcal{W}_{3}$-algebra generated by
modes $L_{m}$ and $W_n=W^{(3)}_{n}$. This is a prototypical example of a nonlinear $\cW$-algebra, and the first one that was considered~\cite{Zamolodchikov:1985wn}. As we discussed before, we follow here the idea to take the classical $\mathcal{W}_{3}$-algebra, and replace
the classical modes by operators and the Poisson bracket by the
commutator. This works fine as long as the Poisson bracket relations
are linear, the first obstacle is met when we consider the Poisson
bracket between two spin-3 modes, which contains a term quadratic in
the $L_{m}$'s. 

When going from the Poisson algebra to the
quantum algebra, we make the following ansatz:
\begin{subequations}\label{quantum:allW3brackets}
\begin{align}
[L_{m},L_{n}] &= (m-n)L_{m+n} + \frac{c}{12}m(m^2 -1)\delta_{m,-n}\,,\\
[L_{m},W_{n}] &= (2m-n)W_{m+n}\,,\\
[W_{m},W_{n}] &= \frac{1}{12}
\Big((m-n) (2m^{2}+2n^{2}-mn-8) L_{m+n}
+\frac{96}{c}\alpha (m-n) \Lambda_{m+n} \nonumber\\
&\qquad \quad + \frac{c}{12}m (m^{2}-1)
(m^{2}-4)\delta_{m,-n} \Big) \, .
\label{quantum:W3bracket}
\end{align}
\end{subequations}
This ansatz is motivated by the following considerations:
\begin{itemize}
\item The bracket for two $L_{m}$'s is not modified, so we find the Virasoro algebra as a subalgebra of the $\mathcal{W}_{3}$-algebra. The Jacobi identity for three $L_{m}$'s is automatically satisfied.
\item The linear commutator of $L_{m}$ with the modes $W_{n}$ is also unchanged; in this way also the Jacobi identity involving two $L_{m}$'s and one $W_{n}$ is directly satisfied. The modes $W_{n}$ can be seen to obey the expected commutation relations with the Virasoro modes $L_{m}$ corresponding to modes of a primary field of spin 3.
\item In the ansatz for the bracket of two $W$-modes we assume that the linear and central parts are unchanged,\footnote{There is no general reason why these terms should be unmodified, but in the case at hand it turns out to be the right guess.} and we put all modifications into a possible redefinition of the quadratic piece $\Lambda$. For convenience, we also put a free coefficient $\alpha$ in this term. 
\end{itemize}

We have to check that the proposed commutation relations are
consistent, i.e.\ that they satisfy the Jacobi identity. As already discussed, the Jacobi identity
involving three $L_{m}$'s, and the identity involving two $L_{m}$'s
and one $W_{n}$ are automatically satisfied.
Let us then consider
the Jacobi identity for one $L_{m}$ and two $W$-modes. Using the
fixed commutation relation between $L_{m}$ and $W_{n}$ we arrive
at the condition
\begin{equation}\label{quantum:JacobiLWW}
[L_{m},[W_{p},W_{q}]] = (2m-q) [W_{p},W_{q+m}]
+ (2m-p) [W_{p+m},W_{q}] \, .
\end{equation}
Both sides of the equation contain central, linear and quadratic
terms. From the Poisson bracket 
\begin{equation}\label{quantum:bracket_L_Lambda_2}
i\{\mathcal{L}_{m},\Lambda_{n} \} = (3m-n)\Lambda_{m+n} +
\frac{c}{6}m(m^{2}-1)\mathcal{L}_{m+n} 
\end{equation}
(see~\eqref{quantum:bracket_L_Lambda}) we see that for
$m\in\{-1,0,1\}$ there is no mixing between central, linear and
quadratic terms in the classical bracket. Because we did not modify the
linear and central terms when replacing the Poisson brackets by the commutators~\eqref{quantum:allW3brackets}, in the quantum algebra these terms will automatically cancel in~\eqref{quantum:JacobiLWW}, and we find that
\begin{equation}
[L_{m},\Lambda_{n}] = (3m-n)\Lambda_{m+n} \quad \text{for}\
m\in\{-1,0,1\}\, .
\end{equation}
We conclude that $\Lambda_{n}$ are the modes of a quasi-primary field
$\Lambda$ of spin $4$ (see~\eqref{quantum:commutator_L_Ws}). Up to
normalisation there is only one such field,\footnote{The space of states at spin $4$ is spanned by $L_{-4}\Omega$, $L_{-2}L_{-2}\Omega$ and $W_{-4}\Omega$. It is straightforward to check that there is only a one-dimensional subspace of quasi-primary states.} namely
$\mathcal{T}=\boldsymbol{(}TT\boldsymbol{)}-\frac{3}{10}\partial^{2}T$
which we introduced in the last subsection
(see~\eqref{quantum:quasiprimary_Tsquared}). So we set $\Lambda
=\mathcal{T}$.
\begin{exercise}\hspace{-2pt}\hyperlink{sol:modes_of_Lambda}{*}
\label{ex:modes_of_Lambda}
Work out the modes $\Lambda_{m}$ in terms of the Virasoro
modes $L_{n}$, and determine the function $f(m)$
in~\eqref{quantum:Lambda_modes}. Note that it might be useful to treat
the two cases of $m$ being even or odd separately.
\end{exercise}

The OPE of $T$ and $\Lambda$ is then
\begin{equation}
T (z) \Lambda (w) = \sum_{n\leq 4} (z-w)^{-n-2}\, V \big(L_{n}
(L_{-2}L_{-2}-\tfrac{3}{5}L_{-4})\Omega ;w\big) \, ,
\end{equation}
and a quick computation shows that
\begin{equation}
T (z) \Lambda (w) = \frac{1}{(z-w)^{4}} \bigg(c+\frac{22}{5} \bigg) T
(w) + \frac{4}{(z-w)^{2}}\Lambda (w) + \frac{1}{z-w} \partial\Lambda
(w) + \text{regular} \, .
\end{equation}
\begin{exercise}
Verify this.
\end{exercise}
The OPE determines the commutation relation of the modes, and by a
straightforward computation we find
\begin{equation}\label{quantum:commutator_L_Lambda}
[L_{m},\Lambda_{n}] = (3m-n)\Lambda_{m+n} + \frac{c+\frac{22}{5}}{6} m
(m^{2}-1)L_{m+n} \, .
\end{equation}
We observe that the quantum commutator differs from the corresponding
Poisson bracket (see~\eqref{quantum:bracket_L_Lambda_2}) by a shift
$c\to c+\frac{22}{5}$.

Let us go back to the Jacobi identity~\eqref{quantum:JacobiLWW} of one $L_{m}$ and two
$W_{n}$. For $|m|>1$, the different terms start to mix. The
terms in $\Lambda$ automatically cancel because the $\Lambda$-term on
the right-hand side of~\eqref{quantum:commutator_L_Lambda} is the same
as for the Poisson bracket. Similarly, the central terms
cancel. On the
left-hand side of~\eqref{quantum:JacobiLWW}, the commutator of $L_{m}$ with $\Lambda_{p+q}$ introduces a linear term which is -- compared to the Poisson
bracket -- modified by a factor $\alpha
\frac{c+\frac{22}{5}}{c}$. All other linear terms are
unmodified. Because the Jacobi identity holds for the Poisson bracket,
the factor has to be $1$, and therefore
\begin{equation}
\alpha = \frac{c}{c+\frac{22}{5}} \, .
\end{equation}
We conclude that with this choice of $\alpha$ and the identification
of $\Lambda$ with $\mathcal{T}$, the Jacobi identity is satisfied for
three $L_{m}$'s, two $L_{m}$'s and one $W_{n}$, and one
$L_{m}$ and two $W_{n}$. It remains to check the
Jacobi identity of three $W_{m}$'s. As we have already fixed all
ambiguities in our ansatz for the commutation relations, it is a
highly non-trivial requirement that also this last Jacobi identity holds. Checking the Jacobi identity is tedious. A more elegant way is to verify the associativity of the operator algebra by showing crossing symmetry of the four-point functions, which holds in the case of the $\cW_3$-algebra~\cite{Zamolodchikov:1985wn}.

\subsection{$\cW_{N}$ and the quantum Miura transform}\label{sec:quantumMiura}

It is in general a highly non-trivial question whether a given classical $\mathcal{W}$-algebra possesses a quantum deformation. The classical $\mathcal{W}_{N}$-algebras that we obtain from $\mathfrak{sl}(N)$ via a Drinfeld--Sokolov reduction of the corresponding affine Lie algebra (as we discussed in section~\ref{sec:symmetries}) can all be turned into quantum $\mathcal{W}_{N}$-algebras. How can one see this? Writing all Poisson brackets, deforming them to commutators and checking Jacobi identity does not appear to be a practicable approach. Various alternative approaches to this problem are discussed, e.g., in \cite{Bouwknegt:1992wg}. Here, we leverage upon the realisation of the classical $\mathcal{W}_N$-algebra in terms of free fields by the classical Miura transformation that we discussed in section~\ref{sec:beyondsl3} (see~\eqref{sym:classicalMiura}).
There is indeed an analogous construction, the quantum Miura transformation, by which the quantum $\mathcal{W}_N$-algebra is realised in terms of free fields.

We start with $N-1$ spin-1 currents $K_{a}$, $a=1,\dots ,N-1$, with the standard OPE
\begin{equation}
K_{a} (z) K_{b} (w) = \frac{\delta_{ab}}{(z-w)^{2}} + \text{regular} \, .
\end{equation}
We combine them into a vector $K= (K_{1},\dots ,K_{N-1})$. Let $\epsilon_{k}$ ($k=1,\dots ,N$) be $N$ vectors in $\mathbb{R}^{N-1}$ satisfying 
\begin{equation}
\epsilon_{i}\cdot \epsilon_{j} = \delta_{ij}-\frac{1}{N}
\end{equation}
with respect to the canonical scalar product (they can be thought of as the weight vectors of the vector representation of $\mathfrak{sl} (N)$). It follows that $\sum_{i}\epsilon_{i}=0$. With the help of these vectors we can form $N$ spin-1 currents 
\begin{equation}
J_{i} (z) = \epsilon_{i}\cdot K (z)
\end{equation}
as linear combinations of the components of $K$. They satisfy the OPE\footnote{This OPE should be compared to the Poisson bracket in~\eqref{sym:freePoisson}, the currents $J_k$ being the quantum analogues of the classical fields $v_k$.}
\begin{equation}
    J_j (z) J_k (w) = \frac{\delta_{jk}-\frac{1}{N}}{(z-w)^2} + \text{regular}\, .
\end{equation}
Similarly to the classical construction in~\eqref{sym:classicalMiura}, we now implicitly define fields $U_{s}$ of spin $s$ by the quantum Miura transform~\cite{Lukyanov:1990tf},
\begin{equation}
    \boldsymbol{\bigg(} \big(\alpha_0 \partial - J_1 \big)\cdots \big(\alpha_0 \partial - J_N \big) \boldsymbol{\bigg)} =  (\alpha_0 \partial)^N - \sum_{s=2}^{N} U_s (\alpha_0 \partial)^{N-s}\, .
\end{equation}
Here, the product on the left-hand side is the normal ordered product, and $\alpha_{0}$ is a parameter. By expanding the left-hand side in powers of the derivative, we find explicit expressions for the fields $U_{s}$ in terms of derivatives and normal ordered products of the $J_{i}$. Notice that there is no term of order $\partial^{N-1}$ because $\sum J_{i}=0$.

The fields $U_{s}$ defined in this way are the quantum analogues of the fields $u_s$ that we used in section~\ref{sec:beyondsl3}. They form a closed operator algebra. This is proven explicitly in~\cite{Lukyanov:1990tf}. Inside the free field algebra generated by the $J_{i}$, the quantum $\mathcal{W}_{N}$-algebra is realised as a subalgebra of fields which commute with a certain set of ``screening charges''. In this way it can be understood in a larger framework of a ``quantum Drinfeld--Sokolov reduction'' (see, e.g., \cite{Bouwknegt:1992wg}).

Consider as an example the field $U_{2}$ which arises from the coefficient of $(\alpha_{0}\partial)^{N-2}$. There are two contributions on the left-hand side, one coming from products of two $J_{i}$ and one from a derivative hitting a $J_{i}$,
\begin{equation}
U_{2} (z) = - \sum_{i<j} \boldsymbol{\big(} J_{i}J_{j} \boldsymbol{\big)} (z) +\alpha_{0} \sum_{j} (j-1)\partial J_{j} (z)\, .
\end{equation}
One can then check that it satisfies the standard OPE of an energy-momentum tensor,
\begin{equation}
U_{2} (z) U_{2} (w) = \frac{c/2}{(z-w)^{4}} + \frac{2}{(z-w)^{2}}U_{2} (w) + \frac{1}{z-w}\partial U_{2} (w) + \text{regular}
\end{equation}
with central charge
\begin{equation}
    c = (N-1) \big(1-N(N+1)\alpha_0^2\big) \, .
\end{equation}
We write $T (z)=U_{2} (z)$. For $N=2$ the construction stops here, and the algebra generated by the modes of $T$ is the Virasoro algebra.

In the following we sketch some results on the $\mathcal{W}_{N}$-algebras obtained by this construction without presenting details of the computation.

Consider the case $N>2$. The field at spin $3$, $U_{3} (z)$, turns out to not be quasi-primary. Its quasi-primary projection is
\begin{equation}
    W^{(3)} := U_3 -\alpha_0 \frac{N-2}{2} \partial T \, .
\end{equation}
One can then check that $W^{(3)}$ not only is quasi-primary, but primary. For $N=3$, there are no further fields, and we have reproduced the quantum $\mathcal{W}_{3}$-algebra. For $N>3$ the construction continues. Again, the field $U_{4}$ is not quasi-primary, and we denote its quasi-primary projection as
\begin{equation}
    \tilde{U}_4 = U_4  -\alpha_0 \frac{N-3}{2} \partial U_3 +\alpha_0^2 \frac{(N-2)(N-3)}{10} \partial^2 T \, .
\end{equation}
This field is not primary, but there is (up to normalisation) exactly one primary field of spin $4$ which is given by
\begin{equation}
    W^{(4)} = \tilde{U}_4 + \frac{(N-2)(N-3)}{2N} \frac{5-N(5N+7)\alpha_0^2}{17+5N-5N(N^2-1)\alpha_0^2}\Lambda \, .
\end{equation}
For any $N>3$, the identification of the primary fields $W^{(3)}$ and $W^{(4)}$ is unique up to normalisation. In our convention the normalisation is given by
\begin{equation}
W^{(s)} (z) W^{(s)} (w) = \frac{n_{s}\,c}{(z-w)^{2s}} + \text{subleading poles}
\end{equation}
with\footnote{The formula for $n_3$ has been checked explicitly, the expression for $n_4$ is an extrapolation from concrete computations at low $N$ and checked until $N=10$ with the help of the Mathematica package \texttt{OPEdefs} \cite{Thielemans:1991uw}.}
\begin{subequations}
\begin{align}\label{quantum:n3}
    n_3 &= \frac{(N-2)(4-N(N+2)\alpha_0^2)}{6N} \, , \\
    \label{quantum:n4}
 n_4 &= n_3 \frac{3(N+1)(N-3)(1-N(N-1)\alpha_0^2)(9-N(N+3)\alpha_0^2)}{2N(17+5N-5N(N^2-1)\alpha_0^2)}\, .
\end{align}
\end{subequations}
To better understand the structure of the $\mathcal{W}$-algebra, we consider the OPE of $W^{(3)}$ and $W^{(4)}$. Its leading term is
\begin{equation}
W^{(3)} (z) W^{(4)} (w) = \frac{c_{34}^{3}}{(z-w)^{4}} W^{(3)} (w) +  \text{subleading poles}
\end{equation}
with coefficient (checked until $N=10$ using again \texttt{OPEdefs} \cite{Thielemans:1991uw})
\begin{equation}\label{quantum:c343}
    c_{34}^3 = \frac{6 (N+1)(N-3)(1-N(N-1)\alpha_0^2)(9-N(N+3)\alpha_0^2)}{N(17+5N-5N(N^2-1)\alpha_0^2) } = 4 \frac{n_4}{n_3}\, .
\end{equation}
A normalisation independent quantity that characterises the $\mathcal{W}_{N}$-algebra is the ratio
\begin{equation}\label{quantum:WNcharacteristic}
    C (N,c)=\frac{\big( c_{34}^3 \big)^2}{n_4} = \frac{144(c+2)}{22+5c}\frac{(N-3)(2(N-1)(4N+3)+c(N+3))}{(N-2)((N-1)(3N+2)+c(N+2))}\, .
\end{equation}
As one can see, one can obtain very explicit results on the structure constants of the $\mathcal{W}_{N}$-algebra using the free field construction. In the next section we discuss how it can be used to also study $\mathcal{W}_{\infty}[\lambda]$.

\subsection{The $\cW_{\infty}[\l]$ quantum algebra and its triality} \label{sec:triality}

Until now, we have looked at the quantum version of the
$\mathcal{W}_{N}$-algebra. Eventually we shall be interested in the
quantum version of $\mathcal{W}_{\infty} [\lambda]$, where we have
higher-spin currents $W^{(s)}$ for every spin $s\geq 2$ (with
$W^{(2)}=T$). It can be constructed by observing that the structure constants (like~\eqref{quantum:n3}, \eqref{quantum:n4} or~\eqref{quantum:c343}) that appear in the OPEs of the $\mathcal{W}_{N}$-algebra are rational functions of $N$. These rational functions can be continued to real parameters, and we can simply replace $N$ by $\lambda$. In this way one automatically obtains a closed operator algebra that can be seen as quantum version of the classical $\mathcal{W}_{\infty} [\lambda]$ algebra.\footnote{This can be compared with the classical construction that we discussed in the previous chapter: the Poisson brackets obtained from the classical Miura transform with $N$ continued to $\lambda$ results in the same classical $\mathcal{W}_{\infty} [\lambda]$ algebra as the Drinfeld--Sokolov reduction based on $\mathfrak{hs}[\lambda]$.}

It has been observed in~\cite{Gaberdiel:2012ku}, further elaborated upon in \cite{Prochazka:2014gqa} and then proven in~\cite{Linshaw:2017tvv} that for fixed central charge $c$ there only exists a one-parameter family of quantum $\mathcal{W}$-algebras that is generated by one higher-spin current $W^{(s)}$ for each spin $s\geq 2$. It is uniquely characterised by the normalisation independent ratio $C=( c_{34}^{3})^{2}/n_{4}$ that we computed in the last section for the $\mathcal{W}_{N}$-algebras (see~\eqref{quantum:WNcharacteristic}). When we replace $N$ by $\lambda$, we find the characteristic parameter for the quantum $\mathcal{W}_{\infty} [\lambda]$-algebra:
\begin{equation}\label{quantum:CfromMiura}
C (\lambda ,c) =  \frac{144(c+2)}{22+5c}\frac{(\lambda -3)(2(\lambda -1)(4\lambda +3)+c(\lambda +3))}{(\lambda -2)((\lambda -1)(3\lambda +2)+c(\lambda +2))}\, .
\end{equation}
$\mathcal{W}_{\infty}$-algebras with the same parameter $C$ (and the same central charge) are isomorphic. As the above expression for $C$ involves a cubic rational function in $\lambda$, one observes that there are generically three different values of $\lambda$ that lead to the same value of $C$ (where the central charge $c$ is fixed), 
\begin{equation}
C (\lambda_{1},c) = C (\lambda_{2},c) = C (\lambda_{3},c) \, .
\end{equation}
One can show that for given $C$ and $c$ the solutions $\lambda_{i}$
satisfy \cite[section 2.2.3]{Prochazka:2014gqa} 
\begin{subequations}
\begin{align}
0&=\lambda_{1}\lambda_{2}+\lambda_{2}\lambda_{3}+\lambda_{3}\lambda_{1} \, , \\[5pt]
c&=(\lambda_{1}-1) (\lambda_{2}-1) (\lambda_{3}-1) \, , \\
C&=\frac{144 (c+2)}{22+5c}\frac{(\lambda_{1}-3)
(\lambda_{2}-3) (\lambda_{3}-3)}{(\lambda_{1}-2) (\lambda_{2}-2)
(\lambda_{3}-2)} \, .
\end{align}
\end{subequations}
This three-fold symmetry of $\mathcal{W}_{\infty}[\lambda]$ is called
triality~\cite{Gaberdiel:2012ku}. 

It is interesting to analyse the fate of triality in the classical limit. For $c\to \infty$, one obtains
\begin{equation}\label{quantum:Cclassical}
    C_{\mathrm{classical}}(\lambda)=\lim_{c\to \infty} C(\lambda,c)= \frac{144}{5}\frac{\lambda^{2}-9}{\lambda^{2}-4}\, .
\end{equation}
One can check that this coincides with the direct computation of $C_\mathrm{classical}$ using the explicit expressions for the Poisson brackets that can be found in~\cite[Appendix C]{Campoleoni:2011hg}. The classical remnant of triality is the symmetry $\lambda \mapsto -\lambda$, which is a symmetry of $\mathfrak{hs}[\lambda]$.

The above derivation of triality is not the original one, and it is instructive to sketch the arguments in~\cite{Gaberdiel:2012ku} that led to its discovery because one gets a better idea of the general structure of the algebra (see also \cite{Vollenweider:2015, Kelm:2016pad}). For a moment, we forget the results from the quantum Miura transformation, and more generally ask what is the family of quantum $\mathcal{W}$-algebras corresponding to the classcial $\mathcal{W}_\infty$-algebras.

When we consider a $\mathcal{W}$-algebra generated by $T$ and primary fields $W^{(s)}$ ($s=3,4,\dots$), the commutator of the modes of the spin-3 current will get an extra term compared to the bracket~\eqref{quantum:W3bracket} in the $\mathcal{W}_3$-algebra,
\begin{equation}
[W^{(3)}_{m},W^{(3)}_{n}] = 2 (m-n) W^{(4)}_{m+n} + \dots \, ,
\end{equation}
involving the spin-4 current. This is a linear term that we directly
get from the Poisson brackets of the classical $\mathcal{W}$-algebra. With this commutator, the normalisation
of $W^{(3)}$ and $W^{(4)}$ are fixed (if also the term involving the Virasoro modes is fixed to its standard normalisation).  If one then looks at $[W^{(3)},W^{(4)}]$, there is a piece proportional to $W^{(3)}$, and
its coefficient $\tilde{C}$ is a true characteristic of the quantum algebra,
\begin{align}
[W^{(3)}_{n},W^{(4)}_{n}] = &\dots W^{(5)}_{m+n} + \dots
\Lambda^{(5)}_{m+n}+\dots \theta^{(6)}_{m+n}\nonumber\\
&-\tilde{C} (n^{3}-5m^{3}-3mn^{2}+5m^{2}n-9n+17m) W^{(3)}_{m+n} \, .
\end{align}
From computing Jacobi identities for higher fields, one finds that all
higher structure constants are determined by this coefficient
$\tilde{C}$ as was observed in~\cite{Gaberdiel:2012ku} and proven in~\cite{Linshaw:2017tvv}. This statement does not rely on the fact that one started from a known classical algebra, but it is valid for any algebra that one
builds out of higher-spin currents $W^{(s)}$, one for each spin $s\geq 2$. The coefficient $\tilde{C}$ is related to the coefficient $C=(c_{34}^3)^2/n_4$ by
\begin{equation}\label{quantum:CandCtilde}
    C=\frac{2016}{5}\tilde{C}\, .
\end{equation}
We know the $\lambda$-dependence of the coefficient $\tilde{C}$ for the classical
Poisson algebra (which follows from~\eqref{quantum:Cclassical}, see also~\cite{Gaberdiel:2012ku}),
\begin{equation}
\tilde{C}_{\text{classical}} (\lambda) = \frac{1}{14}
\frac{\lambda^{2}-9}{\lambda^{2}-4} \, .
\end{equation}
What is the corresponding relation in the quantum algebra? The quantum Miura transformation gives one answer, but in general the question is not well-defined. By definition, it is clear what we mean
by $\tilde{C}$ in the quantum algebra, but it is a priori not clear what we mean by
$\lambda$ in the quantum algebra. We have learned that we can characterise our quantum
$\mathcal{W}_{\infty}$ algebra by the central charge $c$ and the
coefficient $\tilde{C}$, so where does $\lambda$ come into play? We expect
that corresponding to the classical family of $\mathcal{W}_{\infty}[\lambda]$,
we have a family of quantum $\mathcal{W}_{\infty}[\lambda]$-algebras
which are characterised by the coefficient $\tilde{C}(\lambda,c)$ in dependence of
$\lambda$ and the central charge $c$. We require that the classical
algebra arises from the quantum algebra in the limit $c\to \infty$, so
we demand that
\begin{equation}
\lim_{c\to\infty}\tilde{C}(\lambda,c) = \tilde{C}_{\text{classical}} (\lambda) =  \frac{1}{14}
\frac{\lambda^{2}-9}{\lambda^{2}-4} \, .
\end{equation}
This alone does of course not fix $\tilde{C}(\lambda,c)$. On the other hand, we
know that if $\lambda=N\in\mathbb{N}$, the classical algebra can be
truncated to an algebra generated by $N-1$ currents. We can demand
that this is also true for the quantum family
$\mathcal{W}_{\infty}[\lambda]$, namely that they reduce to the
quantum $\mathcal{W}_{N}$-algebra for $\lambda=N$. Although the structure
constants of the quantum $\mathcal{W}_{N}$-algebra are not completely
known, one can fix $\tilde{C}(N,c)$ indirectly by requiring that one finds known representations of the quantum $\mathcal{W}_N$-algebra \cite{Gaberdiel:2012ku}. This then fixes
$\tilde{C}(N,c)$, but it still does not fix the function
$\tilde{C}(\lambda,c)$ completely. On the other hand, one finds a unique
solution if one requires $C(\lambda,c)$ to be a rational function of
$\lambda$, which is then given by
\begin{equation}\label{quantum:C_of_lambda}
\tilde{C} (\lambda ,c) =
\frac{1}{14}\frac{c+2}{c+\frac{22}{5}}\frac{(\lambda -3)(2(\lambda -1)(4\lambda +3)+c(\lambda +3))}{(\lambda -2)((\lambda -1)(3\lambda +2)+c(\lambda +2))}\, .
\end{equation}
One should be aware that the outlined procedure contains some
arbitrariness, and the solution for $\tilde{C}(\lambda,c)$ given
in~\eqref{quantum:C_of_lambda} depends on the choices we made. For
instance, we could have required instead that we
recover $\mathcal{W}_{N}$ for $\lambda=-N$ ($\mathfrak{hs}[\lambda]$ has the symmetry $\lambda
\mapsto -\lambda$), and we would have obtained a different solution
(indeed, $\tilde{C}(\lambda,c)$ in~\eqref{quantum:C_of_lambda} is not
invariant under $\lambda\mapsto -\lambda$). 

The solution in~\eqref{quantum:C_of_lambda} coincides with the result~\eqref{quantum:CfromMiura} obtained from the quantum Miura transformation (taking into account the relation~\eqref{quantum:CandCtilde} between $C$ and $\tilde{C}$). This should not come as a surprise because the quantum Miura transformation by construction yields the $\mathcal{W}_N$-algebra for $\lambda=N$ and the coefficients that appear are rational functions of $\lambda$.

We focused here on the structure of the $\cW_\infty[\lambda]$ algebra, but we wish to stress that its representation theory has been studied, e.g., in \cite{Prochazka:2014gqa} and this led to identify an interesting link with the affine $\widehat{\mathfrak{gl}}_1$ Yangian \cite{Tsymbaliuk:2014,Prochazka:2015deb, Gaberdiel:2017dbk}.

\subsection{Conformal field theories with $\cW$-symmetry}

We now look for candidates for a CFT dual of higher-spin gauge
theories in three dimensions. There are several families of CFTs known with
$\mathcal{W}_{N}$- or $\mathcal{W}_{\infty}$-symmetry:
\begin{itemize}
\item $\mathcal{W}_{N}$-minimal models \cite{Bais:1987zk} labelled by a positive integer $k$ (the
level): for $0<c<N-1$ there are only finitely many unitary irreducible
representations of $\mathcal{W}_{N}$ for a given $c$. In fact within this range there is
only a discrete set of values for $c$, for which there exist unitary
representations at all:
\begin{align}
c_{N,k} & = (N-1)\left(1-\frac{N(N+1)}{(N+k)(N+k+1)}\right)\nonumber\\
&= 2k \left(1-\frac{(k+1)(k+\frac{3N+1}{2})}{(k+N)(k+N+1)}\right) .
\end{align}
For each such value, one can build a CFT whose spectrum is given by
the corresponding unitary representations -- these are the
$\mathcal{W}_{N}$-minimal models.
\item conformal $\mathfrak{sl}(N)$ Toda theories \cite{Fateev:2007ab}: this is a continuous family of
CFTs with central charges $c\geq (N-1)(1+4N(N+1))$ with a continuous
spectrum of primary fields.
\item $m$ complex free bosons: the singlet sector of the chiral algebra gives rise to a 
$\mathcal{W}_{\infty}[\lambda =1]$-algebra with $c=2m$ \cite{Bakas:1990ry,Gaberdiel:2013jpa}.
\item $m$ complex free fermions: their symmetry algebra contains a 
$\mathcal{W}_{\infty}[\lambda =0]$-algebra with $c=m-1$ as a subalgebra \cite{Gaberdiel:2013jpa}. It can be obtained by a $U(1)$ coset construction from the free fermion theory.
\end{itemize}
Let us focus on the $\mathcal{W}_{N}$-minimal models. They can be
realised by a procedure known as coset construction \cite{Goddard:1986ee}. One starts with the
affine algebra $\widehat{\mathfrak{sl}(N)}$ with generators $J_{m}^{a}$ ($m\in
\mathbb{Z}$, $a=1,\dots ,\dim \mathfrak{sl}(N)$) and a central element $\hat{k}$,
obeying the commutation relations
\begin{equation}\label{quantum:affineLie}
[J^{a}_{m},J^{b}_{n}] = f^{ab}{}_{c}\, J^{c}_{m+n} + \hat{k}\,\gamma^{ab} \,m\,
\delta_{m,-n} \, .
\end{equation}
Here, $f^{ab}{}_{c}$ are the structure constants of $\mathfrak{sl}(N)$. We now
build the algebra $U_{k}(\mathfrak{sl}(N))$ which is obtained from the universal
enveloping algebra of the affine $\widehat{\mathfrak{sl}(N)}$ by dividing out the ideal
generated by $\hat{k}-k\cdot \mathbf{1}$ for some integer level $k$,
\begin{equation}
U_{k}(\mathfrak{sl}(N)) = \mathcal{U} (\widehat{\mathfrak{sl}(N)})/\langle \hat{k}-k\cdot
\mathbf{1}\rangle \, .
\end{equation}
When we consider the tensor product of such algebras at level $k$
and level $1$, it contains the diagonally embedded algebra
$U_{k+1}(\mathfrak{sl}(N))$,
\begin{equation}
U_{k+1} (\mathfrak{sl} (N)) \subset U_{k} (\mathfrak{sl} (N)) \otimes U_{1} (\mathfrak{sl} (N)) \, .
\end{equation}
We then define the coset algebra to be the commutant of $U_{k+1}
(\mathfrak{sl} (N))$ inside $U_{k} (\mathfrak{sl} (N)) \otimes U_{1} (\mathfrak{sl} (N))$, i.e.\ the
algebra of all elements that commute with all elements of $U_{k+1} (\mathfrak{sl} (N))$. The
coset algebra is often denoted by (the slightly misleading
quotient notation)
\begin{equation}
\frac{\mathfrak{sl}(N)_{k}\times \mathfrak{sl}(N)_{1}}{\mathfrak{sl}(N)_{k+1}} := \left\{a \in U_{k}
(\mathfrak{sl} (N))\otimes U_{1}(\mathfrak{sl} (N)): \ \forall b\in U_{k+1}
(\mathfrak{sl} (N))\ \  ab=ba \right\} \, .
\end{equation}
This realises the $\mathcal{W}_{N}$-algebra at central charge $c_{N,k}$.
For example, the spin-2 current (the energy-momentum tensor) $T$ is
constructed as 
\begin{equation}
T = T^{\mathfrak{sl}(N),k} + T^{\mathfrak{sl}(N),1} - T^{\mathfrak{sl}(N),k+1} \, ,
\end{equation}
where 
\begin{equation}
T^{\mathfrak{sl}(N),k} = \frac{1}{2(k+N)} \kappa_{ab}\boldsymbol{(}J^{a}J^{b}\boldsymbol{)}
\end{equation}
is the Sugawara energy-momentum tensor constructed in $U_{k}(\mathfrak{sl}(N))$.
\begin{exercise}
Denote by $J^{(k),a}_{m}$ and $J^{(1),a}_{m}$ the generators of $U_{k}(\mathfrak{sl}(N))$
and $U_{1}(\mathfrak{sl}(N))$, respectively. The generators of the embedded
$U_{k+1}(\mathfrak{sl}(N))$ are then given by
\begin{equation}
J^{(k+1),a}_{m} = J^{(k),a}_{m}\otimes \mathbf{1} + \mathbf{1}\otimes
J^{(1),a}_{m} \, .
\end{equation}
The modes $L^{(k)}_{m}$ of the Sugawara energy-momentum tensor
$T^{\mathfrak{sl}(N),k}$ satisfy
\begin{equation}
[L^{(k)}_{m},J^{(k),a}_{n}] = -nJ^{(k),a}_{m+n} \, .
\end{equation}
The analogous relation holds for level $1$, and also for the modes
$L^{(k+1)}_{m}$ built out of the embedded generators $J^{(k+1),a}_{n}$.\\
Show that the modes
\begin{equation}
L_{m}= L^{(k)}_{m}\otimes \mathbf{1} + \mathbf{1}\otimes L^{(1)}_{m} -L^{(k+1)}_{m}
\end{equation}
commute with all generators $J^{(k+1),a}_{n}$ and therefore are part
of the coset algebra.
\end{exercise}

The power of the coset construction lies in the possibility to
construct representations of $\mathcal{W}$-algebras out of
representations of affine Lie algebras. Let $\mathcal{H}^{N,k}_{R}$
and $\mathcal{H}^{N,1}_{R'}$ be the representation spaces of the unitary
irreducible representations $R$ and $R'$ of
$U_{k}(\mathfrak{sl}(N))$ and $U_{1}(\mathfrak{sl}(N))$, respectively. We then consider the
decomposition of the tensor product with respect to the subalgebra
$U_{k+1}(\mathfrak{sl}(N))$ and obtain
\begin{equation}
\mathcal{H}^{N,k}_{R}\otimes\mathcal{H}^{N,1}_{R'} =
\bigoplus_{S}\mathcal{H}_{(R,R';S)}\otimes \mathcal{H}^{N,k+1}_{S} \, .
\end{equation} 
The spaces $\mathcal{H}_{(R,R';S)}$ then form a representation space for
the coset algebra. In this way, one obtains unitary representations of
the $\mathcal{W}_{N,k}$ algebra. In the cases that are of interest here, they are irreducible, but this is not true in general.

\begin{figure}
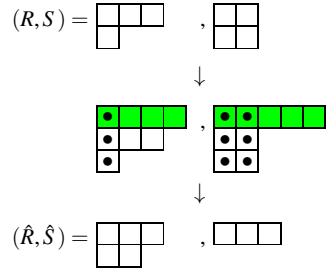

    \centering
    \begin{tabular}{llcl}
        $(R,S) =$ &\ydiagram{3,1} &\ ,\  & \ydiagram{2,2}\\[4mm]
        && $\downarrow$ & \\[2mm]
        &\begin{ytableau}*(green)\bullet & *(green) & *(green) & *(green)\\ \bullet &  & \\ \bullet\end{ytableau} &\ ,\ & \begin{ytableau}*(green)\bullet & *(green) \bullet& *(green) & *(green) & *(green) \\ \bullet &\bullet \\ \bullet & \bullet\end{ytableau}\\[6mm]
        && $\downarrow$ & \\[2mm]
        $(\hat{R},\hat{S})=$&\ydiagram{3,2} &\ ,\ & \ydiagram{3}
    \end{tabular}
    \label{fig:exampleYoung}
    \caption{An example for $N=3$ and $k=4$: starting from Young diagrams $R$ and $S$ we add a row of $k=4$ and $k+1=5$ boxes (in green), respectively. Then we delete the columns of length $N=3$ (marked with a bullet).}
\end{figure}

The unitary dominant highest-weight representations of $U_{k}(\mathfrak{sl}(N))$
are labelled by Young diagrams $R$ with maximally $N-1$ rows and
maximally $k$ columns. For given representations $R$ and $R'$, not all
representations $S$ can occur in the decomposition above. The
selection rule can be formulated as the requirement that the total
number of boxes in $R$ and $R'$ equals the number of boxes in $S$ up
to multiples of $N$. So if we denote the number of boxes of a Young
diagram $R$ by $B(R)$, we demand that
\begin{equation}
B(R)+B (R') = B (S) \ \text{mod}\ N\, .
\end{equation}
For level $1$, there are only $N$ different Young diagrams allowed
(the empty diagram or diagrams consisting of only one column with length
$1$ to $N-1$),
\ytableausetup{boxsize=.95em}
\begin{equation}
0\, ,\ \square\, ,\ \begin{array}{l}
\square  \\[-5.5pt] 
\square
\end{array}\, , \ \dots ,\ \left. \begin{array}{l}
\square\\[-5.5pt]
\square\\[-5.5pt]
\mspace{4mu}\vdots \\[-1.5pt]
\square
\end{array}\right\} (N-1)\,.
\end{equation}
By the above selection rule, for given $R$ and $S$,
there is a unique representation $R'$ such that $S$ occurs in the
decomposition of the product of $R$ and $R'$. The coset
representations that appear in the decomposition can therefore be
labelled by pairs $(R,S)$. This labelling is, however, not unique:
there can be different pairs $(R_{1},S_{1})$ and $(R_{2},S_{2})$ that
label the same representation
\begin{equation}
\mathcal{H}_{(R_{1},S_{1})} \cong \mathcal{H}_{(R_{2},S_{2})} \, .
\end{equation}
To get all pairs that label the same representation as $(R,S)$, we can
perform the following algorithm: we add a row of length $k$ on the top
of the Young diagram $R$, and a row of length $k+1$ on top of
$S$. Then we reduce the new diagrams by killing all columns of length
$N$, and we arrive at a pair $(\hat{R},\hat{S})$ that labels the same
coset representation (see figure~\ref{fig:exampleYoung}). The orbit of pairs under this operation has
length $N$, so there will be $N$ pairs which label the same coset representation.
The conformal weight $h_{R,S}$ of the ground states of the representation
$\mathcal{H}_{(R,S)}$ is given in terms of the row lengths $r_{i}$ and
$s_{i}$ of the Young diagrams $R$ and $S$, respectively. Namely,
introduce $t_{i}:= (k+N+1)r_{i}- (k+N)s_{i}$, then
\begin{equation}\label{quantum:confweight}
h_{(R,S)} = \frac{1}{2(k+N)(k+N+1)} \left(\sum_{i=1}^{N-1} t_{i}
(t_{i}+2N-2i) +\sum_{i,j=1}^{N-1}t_{i} (t_{j}+2N-2j)\right) \, .
\end{equation}
The $\mathcal{W}_{N}$-minimal models are CFTs whose Hilbert space
is built precisely from those coset representations
\begin{equation}
\mathcal{H} = \bigoplus_{(R,S)}' \mathcal{H}_{(R,S)}\otimes
\mathcal{H}_{(R,S)}\, ,
\end{equation}
where the prime on the sum means that we only sum over inequivalent
pairs. Notice that we have two copies of the $\mathcal{W}$-algebra
present (one in the holomorphic and one in the antiholomorphic
sector), therefore we find tensor products of representations in the
total Hilbert space.

After this brief excursion into minimal models, we have the
prerequisites to discuss the proposal for a CFT dual of higher-spin
gauge theories in three dimensions. For more details on minimal models
see~\cite{Fateev:1987zh,Bais:1987zk,Gaberdiel:2012uj}, or more generally on coset models see, e.g., \cite{DiFrancesco:1997nk}. 

\subsection{Minimal-model holography}\label{sec:holography}

We now have all ingredients to look 
for CFTs that are dual to higher-spin gauge theories on
AdS$_{3}$. We do not have a quantum description of the higher-spin
theories, so we only have control over the theories for small
gravitational coupling, $G/\adsR \ll 1$. For the dual CFT this means to
consider large central charges, 
\begin{equation}
c=\frac{3\adsR}{2G}\gg 1 \, .
\end{equation}
When we want to propose a duality, we need to have a family of CFTs
that has a large $c$ limit.

Consider the $\mathcal{W}_{N}$-minimal models. Their central charge is
bounded by $\min (N-1 ,2k)$, so to allow for an arbitrarily high central
charge, we have to consider a limit where both $N$ and $k$ go to
infinity. On the other hand, it seems that in this way the
$\mathcal{W}$-algebra structure never stabilises, and it is unclear a
priori how one could obtain a $\mathcal{W}_{\infty}[\lambda]$-algebra
with a finite value of $\lambda$. Here, triality comes to the
rescue. For central charge $c=c_{N,k}$, and $\lambda_{1}=N$, the other
two values of $\lambda$ leading to the same $\mathcal{W}$-algebra
structure are
\begin{equation}
\lambda_{2}= \frac{N}{N+k}\quad ,\quad \lambda_{3} = -\frac{N}{N+k+1}\, .
\end{equation}
Therefore if we keep the ratio 
\begin{equation}
\Lambda=\frac{N}{N+k}
\end{equation}
fixed while sending $N$ and $k$ to infinity, the $\mathcal{W}$-algebra
structure approaches (in the appropriate sense) $\mathcal{W}_{\infty}[\Lambda]$.
This is very encouraging because it means that we can indeed find a
family of CFTs that have a large $c$ limit in which the symmetry
algebra is given by a $\mathcal{W}_{\infty}$-algebra \cite{Gaberdiel:2010pz}. This limit of
$\cW_N$-minimal models has been dubbed 't~Hooft limit, in analogy with the limit of Yang-Mills theories typically entering the boundary side of the AdS/CFT correspondence.

Besides the algebra, we can also analyse the spectrum. The simplest
primary state in the $\mathcal{W}_{N}$-minimal models (except for the
vacuum) is the state where
\begin{equation}
(R,S) = (\square , 0)\, .
\end{equation}
Here, $0$ stands for the empty Young diagram. By
considering~\eqref{quantum:confweight} we see that the corresponding
conformal weight is
\begin{equation}
h_{(\square,0)} = \frac{N-1}{2N} \left(1+\frac{N+1}{N+k} \right)
\xrightarrow{N,k\to\infty} \frac{1}{2} (1+\Lambda) \, .
\end{equation}
As the primary fields in the minimal models have identical conformal weights in the left- and right-moving sector, we find an operator with scaling dimension
\begin{equation}
\Delta =h+\bar{h} = 1+\Lambda
\end{equation}
in the spectrum.

According to the AdS/CFT correspondence (see, e.g., \cite{Ammon:2015wua}) this operator should
correspond to a scalar bulk field with a definite scaling behaviour at
the conformal boundary of AdS, which in turn restricts its mass to be
\begin{equation} \label{scalar-mass}
M^{2}=\frac{\Delta (\Delta -2) }{\adsR^2}= \frac{\Lambda^{2}-1}{\adsR^2}\, .
\end{equation}
Hence, in addition to the higher-spin gauge fields, which are responsible for
the $\mathcal{W}$-symmetry of the CFT, we also expect a massive scalar
field in the bulk with a specific mass.

Indeed it turns out that the $\mathfrak{hs}[\lambda]$ higher-spin gauge theories
can be coupled to a massive scalar for
$M^{2}=\frac{\Lambda^{2}-1}{\adsR^2}$. To describe those theories we have to go beyond
the Chern--Simons formulation -- this will be the subject of the
following section.

Before doing so, let us consider further states in the CFT
spectrum. Another natural example to consider is the state
\begin{equation}
(R,S) = (\overline{\square},0) \, ,
\end{equation}
where $\overline{\square}$ denotes the Young diagram conjugate to
$\square$,
\begin{equation}
\overline{\square} = \left. \begin{array}{l}
\square\\[-5.5pt]
\square\\[-5.5pt]
\mspace{4mu}\vdots \\[-1.5pt]
\square
\end{array}\right\} (N-1)\, .
\end{equation}
Whereas $\square$ corresponds to the fundamental representation of
$\mathfrak{sl}(N)$, $\overline{\square}$ corresponds to the antifundamental
representation.\footnote{In general, the Young diagram of the
representation conjugate to a representation with Young diagram $R$
(with row lengths $r_{i}$) has row lengths $\bar{r}_{i}=r_{1}-r_{N-i+1}$.}
The state $(\overline{\square},0)$ has the same conformal weight as
$(\square,0)$, so we expect to find two scalars of the same mass (and with
the same scaling behaviour at the boundary) in
the bulk theory (or one complex scalar). 
Indeed, if one works out the contribution of all representations of the form $(R,0)$ to the CFT partition function, it precisely equals the contribution of one complex scalar to the one-loop partition function on thermal AdS~\cite{Gaberdiel:2011zw}. This again is very encouraging!

Similarly, we can also consider the representation 
\begin{equation}
(R,S) = (0,\square) \, .
\end{equation}
Here the conformal weight is
\begin{equation}
h_{(0,\square)} = \frac{N-1}{2N}\left(1-\frac{N+1}{N+k+1} \right) 
\xrightarrow{N,k\to\infty} \frac{1}{2} (1-\Lambda) \, .
\end{equation}
On the bulk side, this again looks like the contribution of a scalar
of the same mass $M^{2}=\frac{\Lambda^{2}-1}{\adsR^2}$ but with a different scaling
behaviour at the boundary (corresponding to a different quantisation). However, as we discuss below, this contribution should not be interpreted as a scalar.

On the other hand, we can for example look at the representation
\begin{equation}
(R,S) = (\square,\square)\, .
\end{equation}
Here we find the conformal weight
\begin{equation}
h_{(\square,\square)} = \frac{N^{2}-1}{2N(N+k)(N+k+1)}
\xrightarrow{N,k\to\infty}
\frac{\Lambda^{2}}{2N} \, ,
\end{equation}
which goes to zero in the limit. We find a similar behaviour for
representations $(R,R)$, so it seems that we find infinitely many
states whose conformal weight approaches zero. From the bulk side, this
would correspond to a huge number of ultra-light fields, which are 
hard to interpret within the AdS/CFT correspondence.

So in spite of the encouraging matches that we observed, the spectrum is not easy to interpret in the holographic setting. One could think of different explanations:
\begin{itemize}
\item The difficulties with the interpretation indicate that there might not be a consistent quantisation of the classical higher-spin gauge theory, and that instead they could be resolved by adding to the spectrum additional perturbative fields. One proposal going in this direction involves a higher-spin theory on AdS$_3\times$S$^1$ \cite{Chang:2013izp,Jevicki:2013kma}. Another option might be that the only solution is to embed the higher-spin states in string theory in a tensionless limit (see the discussion in section~\ref{sec:extra}).
\item The 't~Hooft limit has to be modified. Indeed, it seems that the
unwanted part of the spectrum decouples in the limit \cite{Gaberdiel:2010pz}. Taking limits of CFTs can have subtle effects on the spectrum \cite{Roggenkamp:2003qp}, and one could try to define a limit in which the unwanted states are not present.
\item The 't~Hooft limit is not truly a semi-classical limit. Although $c$ goes to infinity, the spectrum changes when we let $N$ and $k$
go to infinity, so it is a priori not clear whether this is really the
appropriate limit.
\end{itemize}
In view of the last option, one might look for other ways of taking a
semi-classical limit of the minimal models. Indeed, there is a
proposal \cite{Castro:2011iw, Gaberdiel:2012ku, Perlmutter:2012ds} which is more a semi-classical continuation than a
semi-classical limit: we consider the $\mathcal{W}_{N}$-minimal
models, and in all quantities that we can compute, the level $k$ is
analytically continued such that it approaches $k\to -N-1$.  In this
limit the central charge behaves as
\begin{equation}
c_{N,k} \xrightarrow{k\to -N-1} \frac{N (N^{2}-1)}{N+k+1}\, ,
\end{equation}
so it goes to infinity. The conformal weights of some of the states
behave as
\begin{itemize}
\item $(\square,0)$: 
\begin{equation}
h_{(\square,0)} = \frac{N-1}{2N}\left(1+\frac{N+1}{N+k} \right)
\longrightarrow \frac{1}{2} (1-N)
\end{equation}
corresponding to a scalar of mass $M^{2}=\frac{N^{2}-1}{\adsR^2}$.
\item $(0,\square)$:
\begin{equation}\label{weight-charge}
h_{(0,\square)} = \frac{N-1}{2N}\left(1-\frac{N+1}{N+k+1} \right)
\longrightarrow -\frac{c}{2N^{2}}\, .
\end{equation}
The conformal weight grows with the central charge, so one would
expect those states to correspond to non-perturbative states on the
bulk side.
\item $(\square,\square)$:
\begin{equation}
h_{(\square,\square)} = \frac{N^{2}-1}{2N (N+k) (N+k+1)}
\longrightarrow -\frac{c}{2N^{2}} \, ,
\end{equation}
so again those correspond to non-perturbative states.
\end{itemize}
In this semi-classical limit (or continuation) there is no apparent
problem: we find the contributions of one complex scalar
(corresponding to $(\square,0)$ and $(\overline{\square},0)$) and
non-perturbative contributions that should correspond to classical
solutions of the higher-spin gauge theory. The problem of finding
infinitely many light states has disappeared. 

There is indeed a class of classical solutions of the higher-spin
gauge theories, called conical
defect solutions, that explains the full spectrum \cite{Castro:2011iw}.
These solutions can be labelled
by a Young diagram $S$, and in the holographic picture they
correspond to the states $(0,S)$. This can be confirmed by comparing
the higher-spin charges carried by the solutions to the higher-spin
charges of the CFT states \cite{Castro:2011iw, Campoleoni:2013iha}. In this interpretation, the states $(R,S)$ correspond
to excitations of the scalar on the classical solutions \cite{Perlmutter:2012ds}.

The picture that emerges in this semi-classical limit nicely matches
the spectra on the two sides. On the other hand, it is not obvious how to make rigorous sense of the continuation as a limit of CFTs. This is furthermore complicated by the fact that negative conformal weights appear, as it is manifest in \eqref{weight-charge}, so that  the corresponding CFTs cannot be unitary. This non-unitary but more controllable version of AdS/CFT has been further studied in \cite{Campoleoni:2017xyl} (see also \cite{Raeymaekers:2014kea, Raeymaekers:2020gtz}), while the very different behaviour of the $(R,0)$ and $(0,S)$ states in this limit has been used to argue that they should be treated differently in the 't Hooft limit too. In particular, this led to refine the original conjecture of \cite{Gaberdiel:2010pz} postulating that also in the 't Hooft limit the $(R,0)$ states should correspond to a perturbative scalar, while the $(0,S)$ states should correspond to non-perturbative bulk states \cite{Gaberdiel:2012ku, Perlmutter:2012ds}.

There are a number of further checks that have been performed on the
proposed dualities. In particular, one can compare correlators on the
two sides of the correspondence. For example, the 3-point correlators $\langle O\bar{O}J^{(s)}\rangle$ for the scalar primary $O$ (corresponding to $(\square,0)$), its complex conjugate $\bar{O}$ and a higher-spin current $J^{(s)}$ match on the two sides \cite{Ammon:2011ua}. Another check shows that the 3-point functions of perturbative $(R,0)$ states factorise in the limit in a way that is consistent with the interpretation that they correspond to multiparticle states built from the elementary scalar field $O$ \cite{Chang:2011vka}. Further checks and details of the proposed duality are discussed in the review~\cite{Gaberdiel:2012uj}.

The higher-spin AdS$_3$/CFT$_2$ correspondence has been generalised in various ways, for example to even spin minimal-model holography (where only even spins appear in the spectrum) \cite{Ahn:2011pv,Gaberdiel:2011nt}, supersymmetric minimal-model holography \cite{Creutzig:2012ar} or matrix-extended minimal-model holography \cite{Gaberdiel:2013vva,Creutzig:2013tja}. The latter also appears to describe a subsector of a correspondence of string theory on AdS$_3\times$S$^3\times$T$^4$ to a symmetric orbifold theory on the boundary \cite{Gaberdiel:2014cha}, as we shall briefly discuss in section~\ref{sec:extra}.

\section{Coupling to matter}\label{sec:scalar}

In this section, we discuss the coupling of a propagating scalar field to higher-spin gauge fields. To this end, we start in section~\ref{sec:unfoldedscalar} by reformulating the equations of motion of a free scalar field on AdS$_3$ in a so-called \textsl{unfolded} form, that allows to couple it to a higher-spin background. 
We then present the oscillator formulation of $\mathfrak{hs}[\lambda]$, that provides a useful tool to handle the unfolded equations of motion. The latter is also used in the approach by Prokushkin--Vasiliev \cite{Prokushkin:1998bq} towards a non-linear theory of a complex scalar coupled to higher-spin fields that we introduce in section~\ref{sec:nonlineartheory}. Our presentation partly follows the one in~\cite{Kessel:2016}.

\subsection{Free scalar field in the unfolded formulation}\label{sec:unfoldedscalar}

The goal is to describe a scalar field that is coupled to gravity and to higher-spin fields. For this purpose, we need a description of the scalar that is compatible with the first-order frame-like formulation of higher-spin gravity that has been discussed in the previous sections. A first-order action describing the coupling of a scalar to gravity can be obtained by introducing a single auxiliary field; see, e.g., \cite[chapter I.5]{castellani1991supergravity}.
On the other hand, it is unclear how this formulation could be modified so as to allow one to couple higher-spin fields too.
An alternative first-order formulation that appears better adapted to discuss such coupling is provided by the unfolded formulation, which we introduce below.\footnote{Actions describing the coupling of higher-spin Chern--Simons theories to (topological) matter have also been introduced by using a $2$-form and a $0$-form to model the matter sector \cite{Fujisawa:2013lua, Fujisawa:2013ima}.}

\subsubsection{Matter fields}
\label{ssec:matterUndeformed}

To motivate the unfolded re-formulation of the Klein--Gordon equation, we consider for simplicity a flat background with a constant vielbein $\bar{h}^a = \delta_\mu{}^a\, dx^\mu$. We can expand the differential of a scalar field in this basis of one-forms,
\begin{equation}\label{scalar:unfoldingIdentityComponent}
    \intd \Phi =\bar{h}_a\, C^a \, ,
\end{equation}
with coefficient functions $C^a=\bar{h}^{\mu a}\,\partial_\mu \phi$ that encode the derivatives of the scalar field. Taking another differential we obtain
\begin{equation}
    0=\intd^2 \Phi = -\bar{h}_a\wedge \intd C^a\,.
\end{equation}
The solution of this equation can be parameterised by a symmetric tensor $\tilde{C}^{ab}$,
\begin{equation}
    \intd C^a=\bar{h}_b \,\tilde{C}^{ab} \,.
\end{equation}
The coefficients are given by second derivatives of the scalar field,
\begin{equation}
    \tilde{C}^{ab}=\bar{h}^{\mu a}\bar{h}^{\nu b}\,\partial_\mu \partial_\nu \Phi\, .
\end{equation}
Imposing the Klein--Gordon equation $\Box \Phi=M^2\Phi$ constrains the trace part, and we can write
\begin{equation}\label{scalar:analogofKG}
    \intd C^a=\bar{h}_b\, C^{ab} + \frac{1}{3}\, M^2 \,\bar{h}^a\, \Phi\, 
\end{equation}
with an undetermined symmetric traceless part $C^{ab}$. The structure continues: one can apply a differential to~\eqref{scalar:analogofKG} which leads to a condition on $\intd C^{ab}$, and its solution is parameterised by a symmetric traceless tensor $C^{abc}$, and so on. At the end, one arrives at a system of first-order equations
\begin{equation}\label{scalar:UnfoldedScalarFlat}
    \intd C^{a_1\cdots a_n} = \bar{h}_b\, C^{a_1\cdots a_n b}+\frac{n}{3}\, M^2 \,\bar{h}^{(a_1}\,C^{a_2\cdots a_n)}\, .
\end{equation}

\begin{exercise}\hspace{-2pt}\hyperlink{sol:consistencyscalarflat}{*}
\label{ex:consistencyscalarflat}
    Show the consistency of~\eqref{scalar:UnfoldedScalarFlat} by checking that the differential $\intd$ annihilates the right-hand side.
\end{exercise}

In this way, one has encoded the Klein--Gordon equation in a first-order system of \emph{unfolded form}. In general, equations of motion are said to be in unfolded form if they possess the structure
\begin{equation}
\intd \Phi^i = F^i(\Phi) \,,
\end{equation}
where $\Phi^i$ denotes various fields of possibly different form degrees (see, e.g., \cite{Vasiliev:1995dn, Bekaert:2004qos, Didenko:2014dwa, Rahman:2015pzl} for reviews). Consistency ($\intd^2=0$) requires
\begin{equation}
    F^j\wedge \frac{\partial F^i}{\partial \Phi^j}=0\,.
\end{equation}
Note that in particular the equations of motion for the gauge fields \eqref{curv_gen} are in unfolded form. Obtaining a formulation of the scalar on the same footing appears to be a good strategy if we want to couple the scalar to a gauge field background.

What we have presented for a flat background can also be repeated for a scalar on AdS$_3$ with a background vielbein $\bvb^a$ and spin connection $\bsc^a$. Following the same strategy as above and replacing the differential $\intd$ by the Lorentz covariant differential $\nabla$ that was introduced in~\eqref{eq:lorentzDer}, we arrive at a first order system of the form
\begin{equation}\label{scalar:UnfoldedKGAdS}
    \nabla C^{a_1\cdots a_n} = \bvb_b\, C^{a_1\cdots a_n b}+\left(\frac{n}{3}\, M^2-\frac{n(n-1)}{\adsR^2}\right) \bvb^{(a_1}\,C^{a_2\cdots a_n)}\, .
\end{equation}
The difference to~\eqref{scalar:UnfoldedScalarFlat} comes from the fact that the square of the covariant differential $\nabla$ is not zero, but given by the AdS curvature (see~\eqref{eq:lorentzsquared}). 

\begin{exercise}\hspace{-2pt}\hyperlink{sol:consistencyunfoldedKGAdS}{*}
\label{ex:consistencyunfoldedKGAdS}
    Check consistency of~\eqref{scalar:UnfoldedKGAdS} by applying $\nabla$ on both sides and using~\eqref{eq:lorentzsquared}.
\end{exercise}

Eventually we want to couple the scalar to higher-spin fields. As we presented in section~\ref{sec:CSformulation}, higher-spin fields can be encoded in Chern--Simons connections $A$ and $\tilde{A}$ with values in a Lie algebra with generators $J_{a_1\cdots a_{s-1}}$ which are symmetric and traceless in the indices $a_n$. An unfolded scalar consists of a system of symmetric traceless tensors, and they can be contracted with these generators to obtain an algebra valued object that could be used to describe a coupling to higher-spin fields. Concretely, we need a Lie algebra where all values of the spin occur once. A natural framework is the algebra $\mathfrak{hs}[\lambda]$, that we introduced in section~\ref{sec:CSformulation} considering traceless symmetrised products of the $\mathfrak{sl}(2,\mathbb{R})$ generators $J_a$ in the universal enveloping algebra. The scalar field itself has no indices and should be contracted with the identity, therefore we have to consider the associative algebra behind $\mathfrak{hs}[\lambda]$, that includes the unit and that we denoted as $B[\lambda]$ in \eqref{CS:Blambda}. 
To realise two copies of this algebra (which is needed to realise the full higher-spin algebra) we introduce a variable $\phi$ which squares to $1$ and commutes with the generators $J_a$,
\begin{equation}
    \phi^2=1\,,\quad \phi J_a = J_a \phi\,.
\end{equation}
In this way one obtains the algebra $B[\lambda]\oplus B[\lambda]$, where projections on the first and second component are given by
\begin{equation}
\Pi_\pm = \frac{1}{2} (1\pm \phi)\, .
\end{equation}
The $\mathfrak{sl}(2,\mathbb{R})\oplus \mathfrak{sl}(2,\mathbb{R})$ Lie subalgebra is spanned by $J_a^\pm=\Pi_\pm J_a$, and the generators of the AdS$_3$ isometry algebra are then realised as 
\begin{equation} \label{ads-algebra-phi}
    L_a = J_a \, ,\qquad 
    P_a = \frac{G}{\adsR}\phi J_a\,.
\end{equation}
The scalar field and its undetermined derivatives can be combined into an algebra valued 0-form (we use a calligraphic letter for the components $\pC^{a_1\cdots a_n}$ as they can differ from the coefficients $C^{a_1 \cdots a_n}$ in normalisation)
\begin{equation}
    \pC(x) = \sum_{n=0}^\infty \pC^{a_1 \cdots a_n}(\phi|x) J_{\{a_1}\dots J_{a_n\}}\,,
\end{equation}
where the component without indices that comes with the unit element is the scalar field itself, $\pC^\varnothing=\Phi$, and we recall that braces denote a symmetrisation and a traceless projection as, e.g., in \eqref{eq:matrices}.
The Lorentz covariant derivative that we introduced in~\eqref{eq:lorentzDer} can be generalised to any differential form $F$, and can be formulated in terms of algebra valued objects as 
\begin{equation} \label{nablaF}
\nabla F = \d F + \bsc \wedge F - (-1)^{|F|} F \wedge  \bsc \,,
\end{equation}
where $\bsc= \bsc^b L_b$ is the AdS background spin connection.
Using the commutator \eqref{eq:hscommutatorsl2}, one can check that the action of the Lorentz covariant derivative on the zero-form defined above is given by
\begin{align} \label{nablaC}
\nabla \pC = \sum_{n=0}^\infty  (\underbrace{\d \pC^{a_1\cdots a_n} + n \,\epsilon^{bc(a_1}\,\bsc_b \,\pC^{a_2\cdots a_n)}_c}_{\nabla \pC^{a_1\cdots a_n}})J_{\{a_1}\dots J_{a_n\}}\,.
\end{align}
To encode the unfolded Klein--Gordon equation~\eqref{scalar:UnfoldedKGAdS} in an equation involving objects valued in this algebra, a natural ansatz where the right-hand side is linear in $\pC$ and in the vielbein\footnote{Notice that the Lie-algebra valued vielbein and spin-connection we use in this section take values in the whole AdS$_3$ isometry algebra, so that they are closer to those we introduced in \eqref{eq:EH3} rather than to those we used in \eqref{eq:sl2EHaction} and in section~\ref{sec:CSformulation}. The same conventions will be used for their higher-spin generalisations, that will thus take values in the $\mathfrak{hs}[\lambda] \oplus \mathfrak{hs}[\lambda]$ algebra. \label{new-vielbein}} $\bvb=\bvb^a J_a\phi$ is
\begin{equation}\label{scalar:firstansatzUnfolded}
    \nabla \pC = \frac{\alpha_1}{\adsR} \bvb \,\pC +\frac{\alpha_2}{\adsR} \pC\, \bvb \,,
\end{equation}
with some constants $\alpha_1$, $\alpha_2$. 
This ansatz needs to satisfy consistency when applying the covariant differential $\nabla$ again. Acting on the left-hand side of~\eqref{scalar:firstansatzUnfolded} leads to
\begin{align}
    \nabla^2 \pC &= \sum_{n=0}^\infty \nabla^2 \pC^{a_1\cdots a_n}J_{\{a_1}\dots J_{a_n\}}\nonumber\\
    &= \sum_{n=0}^\infty \frac{n}{\adsR^2}\bvb^b \wedge \bvb^{(a_1}\,\pC^{a_2\cdots a_n)}{}_b\, J_{\{a_1}\dots J_{a_n\}}\nonumber\\
    &= \frac{1}{\adsR^2}[\pC,\bvb\wedge \bvb]\,. \label{scalar:nablaSquaredonC}
\end{align}
\begin{exercise}
    Check~\eqref{scalar:nablaSquaredonC}.
\end{exercise}
Acting with $\nabla$ on the right-hand side of~\eqref{scalar:firstansatzUnfolded} gives
\begin{align}
    \nabla\Big(\frac{\alpha_1}{\adsR} \bvb \pC +\frac{\alpha_2}{\adsR} \pC \bvb\Big) &= -\frac{\alpha_1}{\adsR} \,\bvb\wedge\nabla\pC + \frac{\alpha_2}{\adsR} \,\nabla\pC\wedge\bvb\nonumber\\
    &= -\frac{\alpha_1^2}{\adsR^2}\, \bvb\wedge\bvb \,\pC + \frac{\alpha_2^2}{\adsR^2}\, \pC \,\bvb\wedge\bvb\, .\label{scalar:RHSconsistency}
\end{align}
Comparing~\eqref{scalar:nablaSquaredonC} and~\eqref{scalar:RHSconsistency}, we conclude that 
\begin{equation}
    \alpha_1^2 = \alpha_2^2 = 1\,.
\end{equation}
Hence, the right-hand side of~\eqref{scalar:firstansatzUnfolded} is given by the commutator or by the anticommutator. The commutator would lead to a system of uncoupled equations for the components $\pC^{a_1\cdots a_n}$ because the commutator of $\bvb$ with $J_{\{a_1}\dots J_{a_{s-1}\}}$ produces again a generator that corresponds to the same spin $s$. Such an equation does not describe a physical scalar field, but describes the so-called twisted sector of the Prokushkin--Vasiliev model \cite{Prokushkin:1998bq} as we shall discuss in section~\ref{sec:twistedfields}. For the anticommutator, we are left with the possibilities $\alpha_1=\alpha_2=\pm 1$. As we are in principle free to swap the sign of the vielbein, we can set $\alpha_1=\alpha_2=-1$, and we arrive at the equation
\begin{equation}\label{scalar:ansatzScalarUnfolded}
    \nabla \pC =  -\frac{1}{\adsR}\{\bvb ,\pC \} \,.
\end{equation}

The anticommutator of $J_a$ with another basis element is given by
\begin{multline}
    \big\{J_a , J_{\{b_1}\dots J_{b_n\}}\big\}
    = 2\, J_{\{a}J_{b_1}\dots J_{b_n\}} \\ + \frac{2n}{2n+1}\left(C_2 - \frac{n^2-1}{4}\right)\left(\eta_{a(b_1}J_{\{b_2}\dots J_{b_n)\}} - \frac{n-1}{2n-1}\eta_{(b_1 b_2}\,J_{\{b_3}\dots J_{b_n)}J_{a\}}\right) .
\end{multline}
With this information, we can now extract the components of~\eqref{scalar:ansatzScalarUnfolded}. The identity component gives 
\begin{equation}
    \nabla \Phi =  -\frac{2}{3\adsR}\,C_2 \,\phi \,\pC^a \,\bvb_a\,,
\end{equation}
leading to
\begin{equation}\label{scalar:CaAsDerivative}
    \pC^a = -\frac{3\adsR}{2 C_2}\, \phi \,\bvb^{\mu a} \partial_\mu \Phi \,.
\end{equation}
The $J_a$-component of~\eqref{scalar:ansatzScalarUnfolded} is given by
\begin{equation}
    \nabla_{\!\mu} \pC^a = -\frac{2}{\adsR}\, \bvb_\mu{}^a \Phi \,\phi - \frac{4}{5\adsR}\left(C_2 -\frac{3}{4}\right) \bvb_{\mu c}\, \pC^{ca}\phi\,.
\end{equation}
When one contracts this with the inverse background vielbein $\bvb^\mu{}_a$, one obtains
\begin{equation}
    \bvb^\mu{}_a \nabla_{\!\mu} \pC^a = -\frac{6}{\adsR}\, \phi\, \Phi\, .
\end{equation}
Inserting the expression~\eqref{scalar:CaAsDerivative} for $\pC^a$, we finally arrive at a Klein--Gordon equation
\begin{equation}\label{scalar:KGequationfromunfolded}
    \Box \Phi = \bar{g}^{\mu \nu}\bar{\nabla}_{\!\mu}\partial_\nu \Phi = \underbrace{\frac{4}{\adsR^2}\,C_2 }_{M^2} \Phi\, ,
\end{equation}
where $\bar{\nabla}$ denotes the Levi--Civita covariant derivative of the AdS background. The mass squared is given by
\begin{equation}\label{unfolded-scalar-mass}
    M^2 = \frac{4}{\adsR^2}\,C_2 = \frac{\lambda^2-1}{\adsR^2}\, .
\end{equation}
The other components of~\eqref{scalar:ansatzScalarUnfolded} do not lead to any further conditions, and~\eqref{scalar:ansatzScalarUnfolded} provides an unfolded formulation of the scalar on AdS as before.
Because $\Phi$ can depend on $\phi$, we get two scalar fields $\Pi_\pm\Phi$ of the same mass. The mass depends on the parameter $\lambda$ of the algebra $\mathfrak{hs}[\lambda]$. Only for this value of the mass it is possible to write the unfolded equation for the scalar in a way that is compatible with the algebra structure and to couple the scalar also to a non-trivial higher-spin background. Let us also stress that the value for the mass we obtained in \eqref{unfolded-scalar-mass} precisely fits the prediction we made in \eqref{scalar-mass} only looking at the spectrum of minimal models in the 't Hooft limit. This had to be expected since the value of the mass is fixed by symmetry considerations both on the CFT side and in the previous bulk computation, but this is a first encouraging evidence of the consistency of the holographic conjecture that we reviewed in section~\ref{sec:holography}.

The generalisation to an arbitrary higher-spin background is straightforward. We replace the vielbein and the spin connection by their higher-spin generalisations $e$ and $\omega$ and obtain
\begin{equation}\label{scalar:scalarinarbitrarybg}
    \intd\pC + [\omega , \pC] = -\frac{1}{\adsR}\{e,\pC\}\, .
\end{equation}
This is consistent as long as $\omega$ and $e$ satisfy the equations of motion~\eqref{curv_gen} (see exercise~\ref{ex:consistencyscalarinHSbackground} below).
Because of the $\phi$-dependence, we can split this in two equations. Each component of $\pC$ we split as
\begin{equation}
    \pC^{a_1\cdots a_n}(\phi\vert x) = C^{a_1\cdots a_n} (x) \Pi_+ + \tilde{C}^{a_1\cdots a_n}(x) \Pi_-\, ,
\end{equation}
and we obtain two $B[\lambda]$-valued fields $C$ and $\tilde{C}$, such that $\pC= C \Pi_+ + \tilde{C}\Pi_-$. As in section~\ref{sec:CSformulation}, we introduce gauge fields
\begin{subequations}
\begin{align}
    A &= \sum_{n=1}^\infty A^{a_1\cdots a_n} J_{\{a_1}\dots J_{a_n\}} = \sum_{n=1}^\infty \Big( \omega^{a_1\cdots a_n} + \frac{1}{\adsR}e^{a_1\cdots a_n}\Big)J_{\{a_1}\dots J_{a_n\}}\, ,\\
    \tilde{A} &= \sum_{n=1}^\infty \tilde{A}^{a_1\cdots a_n} J_{\{a_1}\dots J_{a_n\}} = \sum_{n=1}^\infty \Big( \omega^{a_1\cdots a_n} - \frac{1}{\adsR}e^{a_1\cdots a_n}\Big)J_{\{a_1}\dots J_{a_n\}}\, ,
\end{align}
\end{subequations}
which take values in $\mathfrak{hs}[\lambda]\subset B[\lambda]$. Higher-spin vielbeins and spin connections taking values in the whole higher-spin algebra can then be defined as $\omega= \frac{1}{2}(A+\tilde{A})$ and $e=\phi\frac{\adsR}{2}(A-\tilde{A})$ (see also footnote \ref{new-vielbein}). Finally, we obtain from~\eqref{scalar:scalarinarbitrarybg} two equations (each being an equation in $B[\lambda]$)
\begin{subequations}\label{scalar:unfoldedinHSbackground}
\begin{align}
    \intd C &= -A \,C + C \,\tilde{A} \, ,\label{scalar:unfoldedinHSbackgroundA}\\
    \intd\tilde{C} &= -\tilde{A} \,\tilde{C} + \tilde{C} \,A \, .\label{scalar:unfoldedinHSbackgroundB}
\end{align}
\end{subequations}
For backgrounds that include non-trivial higher-spin gauge fields these equations lead to two equivalent generalisations of the Klein--Gordon equation involving higher derivatives, see~\cite{Ammon:2011ua}.

\begin{exercise}\hspace{-2pt}\hyperlink{sol:consistencyscalarinHSbackground}{*}
\label{ex:consistencyscalarinHSbackground}
Check the consistency of~\eqref{scalar:unfoldedinHSbackground} for flat connections $A$ and $\tilde{A}$.
\end{exercise}

\subsubsection{Oscillator realisation of $\mathfrak{hs}[\lambda]$}\label{sec:oscillators}

In this section, we introduce an oscillator realisation of the higher-spin algebra $\mathfrak{hs}[\lambda]$. This is a useful tool to handle infinite-dimensional algebras (see, e.g., \cite{Vasiliev:1989re, Joung:2014qya, Basile:2016goq} and references therein for some applications of oscillator techniques in this and related contexts),
and we shall rely on it in section~\ref{sec:nonlineartheory} to set up a full non-linear interacting theory of a scalar coupled to higher-spin gauge fields along the lines of \cite{Prokushkin:1998bq}.

We begin by introducing an abstract algebra with generating objects $y_\ga$ with $\ga=0,1$ and a multiplication denoted by $\star$, and demand the commutation relations
\begin{equation}
\label{eq:yundefCommuationRel}
\left[y_\ga, y_\gb \right]_\star = 2i \, \epsilon_{\ga \gb} \,.
\end{equation}
Indices can be lowered and raised by using the antisymmetric epsilon tensor $\epsilon_{\ga \gb}$ and its inverse $\epsilon^{\ga \gb}$ with $\epsilon_{01}=\epsilon^{01}=1$, 
\begin{equation}
\label{eq:yproperties}
y^\ga := \epsilon^{\ga \gb} y_\gb \,, \qquad
y_\ga = y^\gb \epsilon_{\gb \ga} \,, 
\end{equation}
where we have used $\epsilon^{\ga \gamma} \epsilon_{\beta \gamma} = \delta^\ga{}_\gb$.

These oscillators allow us to realise $\mathfrak{\mathfrak{sl}}(2,\mathbb{R})$, which is isomorphic to $\mathfrak{sp}(2,\mathbb{R})$.
Using the commutator \eqref{eq:yundefCommuationRel}, one can indeed 
show that the generators
\begin{equation} \label{eq:AdSGenerators}
L_{\ga \gb} = -\frac{i}{2}\, y_{(\ga} \star y_{\gb)} 
\end{equation}
satisfy the $\mathfrak{sp}(2,\mathbb{R})$ commutation relations
\begin{equation}
\label{commutationrelationsLL}
[L_{\alpha\beta},L_{\alpha'\beta'}]_\star = \epsilon_{\alpha\alpha'}L_{\beta\beta'} + \epsilon_{\beta \alpha'}L_{\alpha \beta'} + \epsilon_{\alpha \beta'}L_{\beta \alpha'} + \epsilon_{\beta\beta'}L_{\alpha\alpha'}\,.
\end{equation}

\begin{exercise}\hspace{-2pt}\hyperlink{sol:isomorphism}{*}
\label{ex:isomorphism}
Using the commutation relations \eqref{commutationrelationsLL}, verify the isomorphism $\mathfrak{sl}(2,\mathbb{R}) \simeq \mathfrak{sp}(2,\mathbb{R})$ by checking that the generators $L_{-1}=\frac{1}{2} L_{11}$, $L_{+1}=\frac{1}{2} L_{00}$ and $L_0=\frac{1}{2} L_{10}=\frac{1}{2} L_{01}$ obey the $\mathfrak{sl}(2,\mathbb{R})$ commutation relations $[L_m,L_n]=(m-n)L_{m+n}$. 
\end{exercise}

This specific realisation of $\mathfrak{sl}(2,\mathbb{R})$ can be used to obtain a realisation of a specific higher-spin algebra. Indeed, symmetrised polynomials in the $y$-oscillators of even degree form an associative algebra under star multiplication. This associative algebra is isomorphic to $B[\lambda]$ for $\lambda=\frac{1}{2}$, as can be seen by comparing with the quadratic Casimir of the oscillator realisation of $\mathfrak{sp}(2,\mathbb{R})$,
\begin{equation}
\label{eq:quadCasimirValue}
C_2 := -\frac{1}{8}\, L^{\ga \gb} \star L_{\ga \gb} = -\frac{1}{16}\, \{ L^{\ga \gb}, L_{\ga \gb}  \}_\star = -\frac{3}{16} \,,
\end{equation}
which indeed corresponds to $\lambda=\frac12 $ in \eqref{CS:Blambda}.

\begin{exercise}\label{ex:C2valueundeformed}
    Verify~\eqref{eq:quadCasimirValue} using~\eqref{eq:yundefCommuationRel}.
\end{exercise}

The higher-spin algebra is obtained from the associative algebra $B[\lambda]$ by
\begin{equation}
\label{eq:higherspinalgebraFromAssociative}
B[\lambda] = \mathbb{C} \oplus \mathfrak{hs}[\lambda] \,,
\end{equation}
where, with respect to \eqref{fromB-to-hs}, we stressed that we can also complexify the algebra. Therefore, even monomials in the oscillators of degree at least 2 form a basis of a Lie algebra (with respect to the star commutator) that is isomorphic to $\mathfrak{hs}[\lambda]$ for $\lambda=\frac12$. 

The star product that we introduced abstractly can also be realised as a Moyal--Weyl product. To this end, we can think of the oscillators as ordinary commuting variables,
\begin{equation}
\label{eq:ycommuatativity}
y^\ga y^\gb = y^\gb y^\ga \,,
\end{equation}
where we distinguish this product from the previous star product by omitting the product symbol.

For functions $f,g$ of the oscillators $y_\ga$, one then defines the star product by 
\begin{equation}
\label{eq:starproducty}
(f \star g )(y)=\frac{1}{(2\pi)^2}\int \intd^2u\, \intd^2v\, \, f(y+u) \, g(y+v) \exp{(iv u)}\,,
\end{equation}
where we used the notation $  v \,u = - u \, v := v^\ga u_\ga$. Let us stress that the contraction of two identical commuting oscillators vanishes,
\begin{equation}
\label{eq:ossciSquared}
y \, y = u \, u=v \, v = 0 \,.
\end{equation}
Furthermore, we shall use the notation
\begin{equation}
\partial^y_\ga = \frac{\partial}{\partial y^\ga} \,, \qquad \partial^\ga_y := \epsilon^{\ga \gb} \partial^y_\gb \,,
\end{equation}
from which we deduce the following relations
\begin{subequations}
\begin{align}
\partial^y_\ga y^\gb &= -\partial^\gb_y y_\ga = \delta_\ga^\gb\,, \label{eq:counterIntuitiveDeriv}\\
\partial^\ga_y y^\gb &= \epsilon^{\ga \gb} \,, \\
\partial^y_\ga y_\gb &= \epsilon_{\ga \gb} \,.
\end{align}
\end{subequations}
As a word of warning we remark that relation \eqref{eq:counterIntuitiveDeriv} implies that $\partial^\ga_y =  - \frac{\partial}{\partial y_\ga}$.

A differential version of the star product (as long as the functions are analytic) can also be obtained from its definition \eqref{eq:starproducty}:
\begin{align}
(f \star g)(y) = f(y) e^{-i \overset{\leftarrow}{\partial}_y \overset{\rightarrow}{\partial}_y } g(y) \,. \label{eq:differentialFormOfStarProduct}
\end{align}
\begin{exercise}\hspace{-2pt}\hyperlink{sol:differentialStarProduct}{*}
\label{ex:differentialStarProduct} 
Use Taylor's theorem, $f(y+u)=e^{u \,\partial^{y}}f(y)$ and the identity $\delta^{(2)}(u)= \frac{1}{(2 \pi)^2} \int \, \intd^2 v \, e^{i v u}$ to obtain the differential version of the star product.
\end{exercise}

If one of the factors is just an oscillator variable, we obtain the following expressions for the star product (which follow straightforwardly from~\eqref{eq:differentialFormOfStarProduct}):
\begin{subequations}
\label{eq:basicStarProd}
\begin{align}
y_\ga \star f(y) &= (y_\ga + i \partial^y_\ga) f(y) \,, \\
f(y) \star y_\ga &= (y_\ga - i \partial^y_\ga) f(y) \,.
\end{align}
\end{subequations}
From these we obtain the relations
\begin{subequations}
\label{eq:starCommAndAnticomm}
\begin{align}
\left[y_\ga, f(y) \right]_\star &= 2i \, \partial^y_\ga f(y) \,,\label{eq:starCommutator}\\
\frac12 \left\{ y_\ga , f(y) \right\}_\star &= y_\ga f(y)\, . \label{eq:starAntiCommutator}
\end{align}
\end{subequations}
We observe from~\eqref{eq:starAntiCommutator} that the symmetrised star-product of $y_\ga$\nobreakdash-oscillators is simply the ordinary product, $y_{(\ga} \star y_{\gb)} = y_\ga \,y_\gb$. On the other hand, when we insert $f(y)=y_\gb$ into~\eqref{eq:starCommutator}, we recover the star commutation relation~\eqref{eq:yundefCommuationRel} of the oscillators.

\begin{exercise}\hspace{-2pt}\hyperlink{sol:starproductRel}{*}\label{ex:starproductRel} 
Check equations \eqref{eq:basicStarProd} and \eqref{eq:starCommAndAnticomm}.
\end{exercise}

\begin{exercise}\hspace{-2pt}\hyperlink{sol:starCommAndAntiComm}{*}
\label{ex:starCommAndAntiComm}
Using \eqref{eq:basicStarProd} and~\eqref{eq:starCommAndAnticomm} check the identities
\begin{subequations}
\begin{align}
[L_{\ga \gb}, f(y)]_\star &= 2 \, y_{(\ga} \partial^y{}_{\gb)} \, f(y) \,, \label{eq:LComm}\\
\{ L_{\ga \gb}, f(y) \}_\star &= -i \left( y_\ga y_\gb - \partial^y_\ga \partial^y_\gb \right) \, f(y) \label{eq:LAntiComm}\,.
\end{align}
\end{subequations}
Use the latter identity to verify once again the result~\eqref{eq:quadCasimirValue} for the eigenvalue of the Casimir operator.
\end{exercise} 

With the construction above, we could realise $B[\lambda]$ for the specific value of $\lambda=\frac{1}{2}$.
To get an oscillator realisation for other values of $\lambda$, we introduce \emph{deformed oscillators} $\hat{y}_\ga$ with $\alpha=0,1$ obeying the commutation relations  
\begin{equation}
\label{eq:deformedCommuationRelationsyy}
[ \hat{y}_\ga, \hat{y}_\gb ]_\star = 2i \, \epsilon_{\ga \gb} ( 1 + \nu k) \,,
\end{equation}
where $\nu \in \mathbb{R}$ is some real parameter, while the symbol $k$ denotes an object called \emph{outer Kleinian} obeying
\begin{equation}
\label{eq:outerKleinian}
k \,\hat{y}_\ga = - \hat{y}_\ga \,k \,, \qquad k^2=1 \,.
\end{equation}
As we shall see shortly, the parameter $\nu$ is related to the parameter $\lambda$ that appears in $B[\lambda]$. Note that for $\nu=0$, eq.~\eqref{eq:deformedCommuationRelationsyy} reduces to the undeformed commutation relations \eqref{eq:yundefCommuationRel} discussed previously.
One can then show that
\begin{equation}
L_{\ga \gb}=-\frac{i}{2}\, \hat{y}_{(\ga} \star \hat{y}_{\gb)}
\end{equation}
fulfils the $\mathfrak{sp}(2,\mathbb{R})$ commutation relations \eqref{commutationrelationsLL}. This can be seen by observing that
\begin{align}
[L_{\ga \gb},\hat{y}_\gamma]_\star &= -\frac{i}{2} \hat{y}_{(\ga} \star [\hat{y}_{\gb)},\hat{y}_\gamma]_\star + \frac{i}{2} [\hat{y}_\gamma,\hat{y}_{(\beta}]_\star \star \hat{y}_{\ga)} \notag \\ &= - \epsilon_{\gamma (\beta} \,\hat{y}_{\ga)} ( 1 + \nu k) - \epsilon_{\gamma ( \beta} \,\hat{y}_{\ga)} (1 - \nu k) \notag \\[5pt]
&= -2 \epsilon_{\gamma ( \ga} \,\hat{y}_{\gb)} \label{eq:commuatatorLwithy}\,.
\end{align}
Note that we have used $\hat{y}_\ga \,k = - k\, \hat{y}_\ga$ to obtain the second line which is crucial for the final expression to be $\nu$ independent. By using \eqref{eq:commuatatorLwithy} twice, it immediately follows that $L_{\ga \gb}$ obeys the $\mathfrak{sp}(2,\mathbb{R})$ commutation relations \eqref{commutationrelationsLL}.
For the quadratic Casimir, we obtain 
\begin{equation}
\label{eq:CasimirDeformed}
C_2 = -\frac{1}{8}\, L^{\ga \gb} \star L_{\ga \gb} = - \frac{1}{16} \left(3 + 2 \nu k - \nu^2\right) .
\end{equation} 

\begin{exercise}\hspace{-2pt}\hyperlink{sol:deformedComm}{*}
\label{ex:deformedComm} 
Verify this statement by using the commutation relations \eqref{eq:deformedCommuationRelationsyy} to show that
\begin{equation}
\hat{y}^\ga \star \hat{y}_\ga = -2i (1 + \nu k)\,, \qquad L_{\ga \gb} = -\frac{i}{2}\, \hat{y}_{\ga} \star \hat{y}_{\gb} - \frac{1}{2}\, \epsilon_{\ga \gb} (1 + \nu k) \,,
\end{equation}
and use these relations to calculate the value of $C_2$.
\end{exercise} 

The quadratic Casimir~\eqref{eq:CasimirDeformed} is not a pure number, but contains the outer Kleinian $k$. By using the projectors 
\begin{equation}
\label{eq:kProjector}
P_\pm = \frac12 (1 \pm k)
\end{equation}
to restrict to a definite parity of $k=\pm1$, we can then realise the associative higher-spin algebra $B[\lambda]$ (see~\eqref{CS:Blambda}) for arbitrary values of $\lambda$ 
with the identification
\begin{equation}
\lambda= \frac12 (\nu \pm 1) \quad\text{for} \ k=\mp 1\, . 
\end{equation}
A basis for this realisation is given by even and symmetric monomials in the deformed oscillators $\hat{y}_\ga$ projected on a sector of definite $k$-parity, i.e. 
\begin{equation}
L_{\a_1\cdots \a_{2n}} := P_\pm \, \hat{y}_{(\ga_1} \star \dots \star \hat{y}_{\ga_{2n})} \,.
\end{equation}
By turning the associative algebra into a Lie algebra and using the decomposition \eqref{eq:higherspinalgebraFromAssociative}, a realisation for $\mathfrak{hs}[\lambda]$ for arbitrary $\lambda$ is thus obtained.

\subsubsection{Higher-spin and matter fields using oscillators}\label{sec:eqs-oscillators}

We now wish to rewrite the unfolded equations for the higher-spin and matter fields introduced in section~\ref{ssec:matterUndeformed} using the oscillator realisation of $\mathfrak{hs}[\lambda]$, in order to facilitate the introduction of interactions in the following section.
We start by realising the AdS$_3$ background in this setup. The AdS$_3$ isometry algebra is $\mathfrak{so}(2,2)$. 
This algebra is isomorphic to two copies of $\mathfrak{\mathfrak{sl}}(2,\mathbb{R})\cong \mathfrak{sp}(2,\mathbb{R})$, that we realise as before by introducing an additional variable $\phi$ with the properties
\begin{align}
\phi^2 = 1 \, , &&  \hat{y}_\ga \,\phi = \phi \,\hat{y}_\ga \, , && \phi\, k=k\,\phi \,.
\end{align}
As in section~\ref{ssec:matterUndeformed}, the two commuting copies of $\mathfrak{sp}(2,\mathbb{R})$ (and thus the AdS$_3$ isometry algebra $\mathfrak{so}(2,2)$) are obtained with the help of the projectors $\Pi_\pm=\frac12 (1 \pm \phi)$ by introducing the generators
$\Pi_+ L_{\ga \gb}$ and $\Pi_- L_{\ga \gb}$.

Using this realisation of $\mathfrak{so}(2,2)$, we can describe the AdS$_3$ background by defining the connections 
\begin{subequations}\label{AAbar}
\begin{align}
A^{AdS} &= -\frac12 \Big(\bsc+\frac{1}{\adsR}\bvb\Big)^{\ga \gb} \, L_{\ga \gb} \,, \\
\tilde{A}^{AdS} &= -\frac12 \Big(\bsc-\frac{1}{\adsR}\bvb\Big)^{\ga \gb} \, L_{\ga \gb} \,,
\end{align}
\end{subequations}
where $\bvb$ and $\bsc$ describe the AdS$_3$ vielbein and spin connection, respectively. We normalise the vielbein $\bvb$ such that 
\begin{subequations}
\begin{align}
\bvb_{\sm \, \ga \gb} \, \bvb_\sn{}^{\ga \gb} &= -\frac{1}{2}\, g_{\sm \sn} \,, \label{eq:vielbeinContrSpinorInd} \\
\bvb^\sm{}_{\ga_1 \ga_2} \, \bvb_\sm{}^{\gb_1 \gb_2} &= -\frac{1}{2}\, \delta^{(\gb_1}{}_{\!\ga_1}\delta^{\gb_2)}{}_{\!\ga_2} \label{eq:vielbeiContrSpaceInd} \,.
\end{align}
\end{subequations}
Here, first letters of the Greek alphabet always describe spinorial indices, $\ga,\gb,\dots \in \{1,2\}$, whereas letters $\mu,\nu,\dots $ denote spacetime indices.
Note that these identities allow us to convert the spinorial indices of a completely symmetric tensor $f^{\ga_1 \cdots \ga_{2s}}$ to spacetime indices:
\begin{equation}
f^{\alpha_1\cdots\alpha_{2s}}\,\bvb^{\sm_1}{}_{\alpha_1\alpha_2}\cdots \bvb^{\sm_s}{}_{\alpha_{2s-1}\alpha_{2s}} = \big(-\thalf\big)^s f^{\sm_1\cdots \sm_s}\,.
\end{equation}
The tensor $f^{\sm_1 \cdots \sm_s}$ is completely symmetric and traceless.
\begin{exercise}\hspace{-2pt}\hyperlink{sol:tracelessness}{*}
\label{ex:tracelessness}
Prove that $f^{\sm_1 \cdots \sm_s}$ is traceless by using \eqref{eq:vielbeiContrSpaceInd}.
\end{exercise}
The AdS$_3$ connections $A^{AdS}$ and $\tilde{A}^{AdS}$ satisfy the zero-curvature condition~\eqref{CSeom}.
As in section~\ref{se:3dGrav}, it is convenient to combine $A^{AdS}$ and $\tilde{A}^{AdS}$ into a single connection
\begin{equation}
\label{eq:AdSConnection}
\mathcal{A}^{AdS}= \Pi_+ A^{AdS} + \Pi_-\tilde{A}^{AdS}= -\frac12 \bsc^{\ga \gb} L_{\ga \gb} - \frac{1}{2G} \bvb^{\ga \gb} P_{\ga \gb} = \bsc + \frac{1}{\adsR}\bvb \,,
\end{equation}
where we defined $P_{\ga \gb}:= \phi\frac{G}{\adsR} L_{\ga \gb}$ consistently with \eqref{ads-algebra-phi}. $\cA^{AdS}$ is also a solution of the Chern--Simons equation of motion~\eqref{CSeom},
\begin{equation}
\label{eq:backgroundEOM}
\d \mathcal{A}^{AdS} + \mathcal{A}^{AdS} \wedge \star \mathcal{A}^{AdS} = 0 \,.
\end{equation}

As in section~\ref{sec:CSformulation}, we describe higher-spin fields by considering connections $A$ and $\tilde{A}$ to be valued in $\mathfrak{hs}(\lambda)$, which now we realise using the oscillators,
\begin{equation}
\mathcal{A} =  \Pi_+ A + \Pi_-\tilde{A} = \omega + \frac{1}{\adsR}e \,.
\end{equation}
Here $\omega$ and $e$ are even polynomials in the $\hat{y}_\ga$-oscillators, i.e.
\begin{subequations}
\label{eq:hsvielbeinspinconnection}
\begin{align}
\omega &= \frac12 \sum_{s=0}^\infty \frac{i}{(2s)!}\, \omega_{\ga_1 \cdots \ga_{2s}}(x) \, P_\pm\,\hat{y}^{(\ga_1}\star \dots\star \hat{y}^{\ga_{2s)}} \,, \\
e &= \frac12 \sum_{s=0}^\infty \frac{i}{(2s)!}\, e_{\ga_1 \cdots \ga_{2s}}(x) \frac{\phi}{\adsR}\, P_\pm\,\hat{y}^{(\ga_1}\star \dots\star \hat{y}^{\ga_{2s})} \,.
\end{align}
\end{subequations}
In complete analogy to our discussion in section~\ref{sec:frame-like}, the components $e_{\ga_1 \cdots \ga_{2s}}(x)$ and $\omega_{\ga_1 \cdots \ga_{2s}}(x)$ are the higher-spin vielbeins and spin connections, respectively. They correspond to tensors which are fully symmetric and traceless.

On an AdS$_3$ background $\mathcal{A}^{AdS}=\bsc + \frac{1}{\adsR}\bvb $, the zero-curvature conditions~\eqref{eq:frameLikeFreeEOM} translate to the equation of motion 
\begin{equation}
\label{eq:eomGaugePhys}
D^{AdS} \mathcal{A} := \d \mathcal{A} + \mathcal{A}^{AdS} \wedge \star \mathcal{A} + \mathcal{A} \wedge \star \mathcal{A}^{AdS}= 0 \,.
\end{equation}
Here, $D^{AdS}$ denotes the AdS$_3$ covariant derivative that includes the background spin connection and the vielbein. On a differential form $F$ of degree $|F|$, it acts as
\begin{align}
\label{eq:AdSCovariantDerivative}
D^{AdS} F &:= \d F + \mathcal{A}^{AdS} \wedge \star F - (-1)^{|F|} \, F \wedge \star \mathcal{A}^{AdS} \,, 
\end{align}
and it is related to the Lorentz covariant differential as
\begin{equation}
D^{AdS} F = \nabla F + \frac{1}{\adsR}\bvb \wedge \star F - \frac{1}{\adsR}(-1)^{|F|} F \wedge \star \bvb \,.
\end{equation}
The equation of motion~\eqref{eq:eomGaugePhys} is invariant under the gauge transformation
\begin{equation}
\label{eq:WGaugeVariation}
\delta \mathcal{A} = D^{AdS} \xi\,,
\end{equation}
with some arbitrary zero-form $\xi$ generically depending on $\hat{y}^\ga$, $\phi$ and spacetime coordinates $x^\mu$. The gauge invariance follows immediately from the fact that the covariant derivative $D^{AdS}$ is nilpotent
\begin{align}\label{scalar:nilpotencyofDAdS}
D^{AdS} D^{AdS} F = 0 \,.
\end{align}

\begin{exercise}\hspace{-2pt}\hyperlink{sol:nilpotency}{*}\label{ex:nilpotency} 
Check the nilpotency~\eqref{scalar:nilpotencyofDAdS}.
\end{exercise}

After having discussed the higher-spin fields, we now reformulate the unfolded scalar in terms of the oscillators. It is described by a zero-form
\begin{equation}
\label{eq:pCComponents}
\pC(y,\phi|x) = \sum_{s=0}^\infty \frac{1}{s!} \pC_{\ga_1 \cdots \ga_{s}}(\phi,k|x) \, \hat{y}^{(\ga_1}\star  \dots \star \hat{y}^{\ga_{s})} \,,
\end{equation}
where we restrict the sum to even $s$.

Consistently with \eqref{nablaC}, the action of the Lorentz covariant derivative on this zero-form is given by
\begin{align}
\nabla \pC = \sum_{s=0}^\infty \frac{1}{s!} (\underbrace{\d \pC_{\ga_1 \cdots \ga_{s}} + s \,\bsc^{\beta}{}_{\ga_1} \pC_{\beta \ga_2 \cdots \ga_{s}}}_{\nabla \pC_{\ga_1 \cdots \ga_{s}}})\hat{y}^{(\ga_1}\star \dots\star \hat{y}^{\ga_{s})} \,,
\end{align}
where we defined the Lorentz covariant derivative acting on the fully symmetric components $\pC_{\ga_1 \cdots \ga_{s}}$. 
The Klein--Gordon equation on AdS$_3$ in unfolded form \eqref{scalar:ansatzScalarUnfolded} in the oscillator formalism then reads
\begin{equation}
\label{eq:eomZeroForm}
\nabla \pC = -\{\bvb, \pC \}_\star \, .
\end{equation}

We reformulated the whole free theory in the oscillator language. In the remaining part of this subsection, we shall revisit the proof that \eqref{eq:eomZeroForm} is equivalent to the Klein--Gordon equation by using oscillator techniques, so as to familiarise with them. Our path towards interactions restarts in section~\ref{sec:twistedfields}, where another class of fields that enter the interacting theory is discussed.

To analyse the content of eq.~\eqref{eq:eomZeroForm}, it is convenient to first express the anticommutator of $L_{\ga_1 \ga_2}$ with an oscillator that is given by
\begin{multline}
\label{eq:deformedLAnticommutator}
\{ L_{\ga_1 \ga_2} , \hat{y}_{(\gb_1} \star \dots \star \hat{y}_{\gb_s)}\}_\star =- i \, \hat{y}_{(\ga_1} \star y_{\ga_2} \star \dots \star \hat{y}_{\gb_s)} \\  - i s \frac{\nu^2 - 2k \nu - (s^2 -1)}{s+1} \epsilon_{\ga_1 (\gb_1} \epsilon_{|\ga_2| \gb_2 } \hat{y}_{\beta_3} \star \dots \star \hat{y}_{\beta_s)}
\end{multline}
for even s.
\begin{exercise}\hspace{-2pt}\hyperlink{sol:deformedSym}{*}
\label{prob:deformedSym}  
Verify this statement by first using the commutation relation \eqref{eq:deformedCommuationRelationsyy} to show that
\begin{multline}
\label{eq:firstsymIdent}
\hat{y}_{\ga} \star  \hat{y}_{(\gb_1} \star  \dots \star \hat{y}_{\gb_s)} =  \hat{y}_{(\ga} \star \hat{y}_{\gb_1} \star \dots \star \hat{y}_{\gb_s)}  \\ + i \left( s + \frac{\frac12(1-(-1)^s)+s}{s+1} \nu k \right) \epsilon_{\ga (\gb_1} \hat{y}_{\gb_2} \star \dots \star \hat{y}_{\gb_s)}
\end{multline}
and analogously 
\begin{multline}
\hat{y}_{(\gb_1} \star  \dots \star \hat{y}_{\gb_s)} \star \hat{y}_{\ga} =  \hat{y}_{(\ga} \star \hat{y}_{\gb_1} \star \dots \star \hat{y}_{\gb_s)}  \\ - i \left( s + \frac{\frac12(1-(-1)^s)+(-1)^{s+1}s}{s+1} \nu k \right) \epsilon_{\ga (\gb_1} \hat{y}_{\gb_2} \star \dots \star \hat{y}_{\gb_s)} \,. \label{eq:secIdent}
\end{multline}
By using both results twice deduce \eqref{eq:deformedLAnticommutator}. Recall that \eqref{eq:deformedLAnticommutator} holds for even $s$ only. 
\end{exercise}

We now evaluate~\eqref{eq:eomZeroForm} by replacing $\bvb= -\frac{\phi}{2 \adsR}\bvb^{\ga \gb}L_{\ga\gb}$ and using~\eqref{eq:deformedLAnticommutator}. In components, \eqref{eq:eomZeroForm} is then given by
\begin{equation}
\nabla \pC_{\ga_1 \dots \ga_s} = \frac{\phi}{2i \adsR} \left( s (s-1) \bvb_{(\ga_1 \ga_2} \pC_{\ga_3 \cdots \ga_s)}-\left( 1 - \frac{\nu(\nu-2k)}{(s+1)(s+3)}\right) \bvb^{\gb_1 \gb_2} \pC_{\gb_1 \gb_2 \ga_1 \cdots \ga_s} \right)\,.
\end{equation}
To analyse this equation further, we contract its spacetime index with a vielbein $\bvb^\sm_{\gamma_1 \gamma_2}$ which yields
\begin{multline}
\bvb^\sm_{\gamma_1 \gamma_2} \nabla_\sm \pC_{\ga_1 \cdots \ga_s} = \frac{\phi}{2i \adsR} \bigg( s (s-1) \, \underbrace{\bvb^\sm_{\gamma_1 \gamma_2} \bvb_{\sm \, (\ga_1 \ga_2}}_{\sim \; \epsilon_{\gamma_1 (\alpha_1} \epsilon_{\alpha_2 |\gamma_2|}} \pC_{\ga_3 \cdots \ga_s)} \\
-\frac43 \, \adsR^2 \, \underbrace{\frac3{4\adsR^2} \left( 1 - \frac{\nu(\nu-2k)}{(s+1)(s+3)}\right)}_{=:M_s^2} \, \underbrace{\bvb^\sm_{\gamma_1 \gamma_2} \bvb_\sm^{\beta_1 \beta_2}}_{=-\frac12 \delta^{(\gb_1}_{\gamma_1}\delta^{\gb_2)}_{\gamma_2}} \pC_{\beta_1 \beta_2 \ga_1 \cdots \ga_s} \bigg)  \,.
\label{eq:contrVB}
\end{multline}
To analyse this equation further, we first consider its symmetrisation over all $\gamma_i$ and $\ga_i$ indices. The first summand on the right hand side does not contribute, and we obtain
\begin{equation}
\label{eq:pCderivComponents}
\pC_{\ga_1 \dots \ga_s}=\frac{3i}{M_{s-2}^2} \, \frac{\phi}{\adsR} \, \bvb^\sm{}_{(\ga_1 \ga_2} \nabla_\sm \pC_{\ga_3 \cdots \ga_s)} \,,
\end{equation}
where we also used $\phi^2=1$. From this expression we see that by recursion, all higher tensor components $\pC_{\ga_1 \cdots \ga_s}$  can be expressed as derivatives of the lowest component $\Phi=\pC_{\varnothing}=\pC|_{y=0}$. To read off the remaining information contained in~\eqref{eq:contrVB} we now consider its projection to the part that is antisymmetric under $\gamma_1 \leftrightarrow\ga_1$ and $\gamma_2 \leftrightarrow \ga_2$. To this end, we contract it with $\epsilon^{\gamma_1 \ga_1} \epsilon^{\gamma_2 \ga_2}$, which annihilates the second term on the right-hand side. With the help of the identity~\eqref{eq:vielbeinContrSpinorInd} we obtain
\begin{equation}
\label{eq:antisymmZeroForm}
\epsilon^{\gamma_1 \ga_1} \epsilon^{\gamma_2 \ga_2} \, \bvb^\sm_{\gamma_1 \gamma_2} \nabla_\sm \pC_{\ga_1 \cdots \ga_s}=-\frac{3}{4i} \, \frac{\phi}{\adsR} \, s (s-1) \, \pC_{\ga_3 \cdots \ga_s} \,.
\end{equation}
Evaluating this equation for $s=2$ implies
\begin{equation}
-\frac{3}{2i} \frac{\phi}{\adsR} \, \Phi =\bvb_\sm^{\ga \gb} \nabla^\sm \pC_{\ga \gb} = \frac{3i}{M_0^2} \, \frac{\phi}{\adsR} \; \bvb_\sm^{\ga \gb} \nabla^\sm \; \bvb^\sn_{\ga \gb} \nabla_\sn \Phi \,,  
\end{equation}
where we have used \eqref{eq:pCderivComponents} in the last step.
Using $\nabla \bvb = 0$, which follows from the $\phi$-component of the AdS$_3$ flatness condition~\eqref{eq:backgroundEOM}, we obtain the Klein--Gordon equation
\begin{equation}
\label{eq:KGequation}
g^{\sm \sn} \bar{\nabla}_\sm \bar{\nabla}_\sn \Phi = M^2 \Phi \,,
\end{equation}
with 
\begin{equation}
M^2=M_0^2=-\frac3{4 \adsR^2}\left( 1 - \frac{\nu(\nu-2k)}{3}\right)=\frac{4}{\adsR^2}C_2\, ,
\end{equation} 
which reproduces our previous result~\eqref{scalar:KGequationfromunfolded}.
We have thus seen how to reformulate free higher-spin and matter fields in terms of the oscillators.

\subsubsection{Twisted fields}\label{sec:twistedfields}

Before we move to the nonlinear equations, let us compare the equations of motion for the higher-spin gauge fields encoded in $\mathcal{A}$ and the scalar field encoded in $\pC$ on an AdS background. The equation of motion $D^{AdS} \mathcal{A}=0$ for the one-form $\mathcal{A}$ can be written as
\begin{align}
    D^{AdS} \mathcal{A}&= \nabla \mathcal{A} + \bvb \wedge \star \mathcal{A} + \mathcal{A} \wedge \star \bvb \nonumber\\
    &= \nabla \mathcal{A} + [\bvb_\sm , \mathcal{A}_\sn]_\star \, \d x^\sm \wedge \d x^\sn \, .
\end{align}
We observe that --~as we are used to in gauge theories~-- the Lie bracket of the gauge algebra occurs; we say that the one-form $\mathcal{A}$ transforms in the adjoint representation of the higher-spin algebra. On the other hand, the equation of motion~\eqref{eq:eomZeroForm} for $\pC$ involves the anticommutator; for this reason the zero-form $\pC$ is said to transform in the \emph{twisted adjoint representation}. Recall that to obtain the Klein--Gordon equation, the occurrence of the anticommutator in \eqref{eq:eomZeroForm} is essential as it relates different components of the zero-form $\pC$ contrary to the commutator.\footnote{This can be seen by comparing the star product identities for the anticommutator \eqref{eq:deformedLAnticommutator} with the analogous expression \eqref{eq:commuatatorLwithy} for the commutator.}  

The equation of motion for $\pC$ cannot directly be written with the help of the covariant derivative $D^{AdS}$ because $D^{AdS} \pC$ involves the commutator rather than the anticommutator. There is an elegant way to again bring it into the form $D^{AdS} (\,\cdot\,)=0$. 
To this end, let us introduce an additional variable $\psi$ obeying
\begin{align}
\psi^2 = 1 \,, && \psi\, \phi = - \phi\, \psi \,, && \psi \,\hat{y}_\ga = \hat{y}_\ga \,\psi \,.
\end{align}
We can then rewrite the equation of motion \eqref{eq:eomZeroForm} as
\begin{equation}
\label{eq:eomZeroFormPhys}
D^{AdS} \left( \pC \,\psi \right) = \nabla (\pC \,\psi) - \frac12 \bvb^{\ga \gb} \left[ \phi\, L_{\ga \gb}\, , \pC \,\psi \right]_\star = (\nabla \pC + \left\{ \bvb, \pC \right\}_\star) \psi = 0 \,,
\end{equation}
where we have used the fact that the Lorentz covariant derivative $\nabla$ is independent of $\phi$ and $\left[ \phi f(\hat{y}), g(\hat{y},\phi) \psi \right]_\star=\left\{ \phi f(\hat{y}),g(\hat{y},\phi) \right\}_\star \psi$, which follows immediately from $\psi \phi=-\phi \psi$. 

After having introduced $\psi$, at the formal level it is natural to also include ``twisted fields'' in the description and to consider the combinations 
\begin{subequations}
\begin{align}
\boldsymbol{A}(\hat{y},\phi,\psi|x) &= \mathcal{A}(\hat{y},\phi|x) + \tilde{\mathcal{A}}(\hat{y},\phi|x) \, \psi \,, \\
\flC(\hat{y},\phi,\psi|x) &= \pC(\hat{y},\phi|x) \,\psi + \sC(\hat{y},\phi|x) \,,
\end{align} 
\end{subequations}
where the fields are functions of $\psi$. Note that the twisted field $\tilde{\mathcal{A}}$ for the one-form is given by the $\psi$-dependent part whereas the zero-form has the opposite decomposition, and we shall see in the next section that this extension of the field content has been used by Prokushkin and Vasiliev to introduce interactions.

The twisted zero-form $\sC$ obeys the equation of motion
\begin{equation}
\label{eq:twistedZeroForm}
D^{AdS} \sC = 0 \,,
\end{equation}
corresponding to the adjoint representation.
This equation of motion does not mix different components of $\sC$ due to the fact that it only contains commutators of $L_{\ga \gb}$.\footnote{It corresponds to the other solution of the consistency condition for the ansatz~\eqref{scalar:firstansatzUnfolded} of an unfolded equation for the scalar.} Therefore, it is not equivalent to a Klein--Gordon equation. Note also that by \eqref{eq:WGaugeVariation} this is precisely the form of a covariantly constant gauge parameter $\xi$. In a slight abuse of terminology, the components of the twisted zero-form $\sC$ are therefore sometimes referred to as Killing tensors. 

On the other hand, the twisted one-form $\tilde{\mathcal{A}}$ is defined to satisfy the equation
\begin{equation}
D^{AdS} (\tilde{\mathcal{A}} \,\psi) = 0 \,,
\end{equation}
and it corresponds to the twisted adjoint representation.
Note that this relation does not decompose into independent equations for each component of $\tilde{\mathcal{A}}$ and therefore it does not describe a multiplet of higher-spin fields. 

The equations of motion for twisted and untwisted fields can be summarised as
\begin{subequations}
\label{eq:freeEoMVasiliev}
\begin{align}
D^{AdS} \boldsymbol{A} &= 0 \,, \\
D^{AdS} \flC &= 0 \,, \label{eq:freeEoMVasilievB}
\end{align}
\end{subequations}
which are invariant under
\begin{subequations}
\begin{align}
\delta \boldsymbol{A} &= D^{AdS} \xi \,,\\
\delta \flC &= 0 \,,
\end{align}
\end{subequations}
where $\xi$ is an arbitrary zero-form which can now also depend on $\psi$ in addition to $\phi$, $\hat{y}^\ga$ and $x^m$. The non-linear equations of motion that we shall discuss in the following section will reproduce~\eqref{eq:freeEoMVasiliev} upon linearisation.


\subsection{Non-Linear Theory}
\label{sec:nonlineartheory}
In this section, we introduce Vasiliev's approach to a non-linear theory of higher-spin gauge fields coupled to matter. It is based on unfolded equations for master fields which encode the equations of motion of the physical fields. Their linearisation leads to the free equations of motion for higher-spin and scalar fields discussed in section~\ref{sec:eqs-oscillators}, together with the free equations of motion for the twisted fields that we introduced in section~\ref{sec:twistedfields}. For the remainder of this section, we set the AdS radius to one ($\adsR=1$) to simplify notation.

\subsubsection{Master fields and Vasiliev Equations}

To formulate Prokushkin--Vasiliev theory \cite{Prokushkin:1998bq, Prokushkin:1998vn}, we first introduce another type of commuting oscillators $z_\ga$ in addition to the $y_\ga$ and define a star product for functions of both these oscillators as
\begin{align}
\label{eq:starproductZ}
(f\star g)(y,z)=\frac{1}{(2\pi)^2}\int \intd^2u\, \intd^2v\, \, f(&y+u,z+u)\,g(y+v,z-v) \exp{(iv^\ga u_\ga)}\,. 
\end{align}
This definition reduces to the previous star product \eqref{eq:starproducty} for functions that only depend on $y$. This looks like a restriction to the undeformed oscillators, but, as we shall see, the deformed algebra arises upon the expansion around a non-trivial background value for the master field describing the scalar field.
The new variables $z_\ga$ satisfy
\begin{align}
    [z_\ga,z_\gb]_\star &= - 2 i \,\epsilon_{\ga\gb}\,, & [z_\ga,y_\gb]_\star &=0\,, \label{scalar:zstarrelations}
\end{align}
and -- similarly to~\eqref{eq:starCommutator} -- we have 
\begin{subequations}
\begin{align}
[y_\alpha,f(y,z)]_\star &= 2i\,\partial^y_\alpha f(y,z)\,, \\[5pt]
[z_\alpha,f(y,z)]_\star &= -2i\,\partial^z_\alpha f(y,z)\,. \label{eq:zalphacommutatorzderivative}
\end{align}
\end{subequations}
We now introduce master fields $\MW$, $\MB$ and $\MS_\alpha$, which depend on $y_\ga$- and $z_\ga$-oscillators, as well as on the variables $\phi$, $\psi$ and the idempotent outer Kleinian $k$ that will allow us to also realise the undeformed algebra later. 
$k$ anticommutes with the oscillators $y_\ga$ and $z_\ga$, but commutes with $\phi$ and $\psi$,
\begin{align}
    k\,y_\ga &= -y_\ga k\,, & k\,z_\ga &= -z_\ga\, k \,,& k\,\phi & =\phi\, k\,, & k\,\psi &= \psi\,k\,, & k^2 &=1\,.
\end{align}
The $z$-independent part of $\MW$ and $\MB$ are interpreted as gauge field $\boldsymbol{A}$ and scalar field $\flC$,
\begin{subequations}
\begin{align}
\MW(y,z,\phi,\psi,k|x) \big|_{z=0}&= -\boldsymbol{A}(y,\phi,\psi,k|x)\,, \\
\MB(y,z,\phi,\psi,k|x) \big|_{z=0}&= \flC(y,\phi,\psi,k|x)  \,.
\end{align}
\end{subequations}
The $z$-dependent part describes yet another 
auxiliary sector besides that of twisted fields already included in $\boldsymbol{A}$ and $\flC$. To formulate the nonlinear equations for the master fields, we need to introduce another idempotent variable $\rho$, that anticommutes with $k$ and commutes with the rest,
\begin{align}
    \rho^2=1 \,, && \rho k = - k \rho \,, && \rho \phi = \phi \rho\,,&& \rho\psi=\psi\rho\,&& \rho y_\ga = y_\ga \rho \,,&&\rho z_\ga =z_\ga\rho \,.
\end{align}
The Prokushkin--Vasiliev equations then read\footnote{We have chosen to make the $\rho$-dependence manifest to have $\rho$-independent master fields.}
\begin{subequations}
\label{eq:threedVasilievmasterequations}
\begin{align}
\intd \MW&=\MW \wedge\star \MW\,,  \label{eq:threedVasilievdW}\\
\intd \MB&=[\MW, \MB]_\star\,,\label{eq:threedVasilievdB}\\
\rho \,\intd \MS_\ga&= [\MW,\,\rho\MS_{\ga}]_\star\,,\label{eq:threedVasilievdS}\\
0&=[\MB  , \rho\MS_\ga]_\star \, ,\label{eq:threedVasilievBS} \\
[\rho\MS_\ga,\,\rho\MS_\gb]_\star &= - 2i \epsilon_{\ga \gb} (1 +  \MB \star k\varkappa) \label{eq:threedVasilievSS}\,,
\end{align}
\end{subequations}
where we defined the (inner) Kleinian $\varkappa := e^{i\,yz}$. It satisfies
\begin{equation}
\label{eq:innerKleinianConditions}
\varkappa \star \varkappa = 1 \,, \qquad
\varkappa \star f(y,z) \star \varkappa = f(-y,-z) \,.
\end{equation}
The equations~\eqref{eq:threedVasilievmasterequations} are invariant with respect to the gauge transformations
\begin{subequations}
\label{eq:mastergaugevariations}
\begin{align}
\delta \MW &=\intd \Mxi - [\MW,\Mxi]_\star\,,\label{eq:mastergaugevariationW}\\
\delta \MB &=[\Mxi,\MB]_\star\,,\\
\rho\,\delta \MS_\ga &= [\Mxi,\,\rho \MS_\ga]_\star \label{eq:Sgaugevariation}\,,
\end{align}
\end{subequations}
where the gauge parameter $\Mxi = \Mxi(y,z,\phi,\psi,k|x)$ depends on the same variables as the master fields. We call the first two Prokushkin--Vasiliev equations \eqref{eq:threedVasilievdW}-\eqref{eq:threedVasilievdB} \emph{dynamical equations} because they are similar to the unfolded equations for the gauge and scalar fields, now generalised to the corresponding master fields.\footnote{The first equation has the form of a zero-curvature condition, but the dependence of the master fields on the oscillators $z_\ga$ leads to a non-trivial interaction between higher-spin and scalar fields.} The other three equations \eqref{eq:threedVasilievdS}-\eqref{eq:threedVasilievSS} will be referred to as \emph{non-dynamical equations}.

We are interested here in bosonic higher-spin and matter fields, and this requires to apply a bosonic projection on the master fields,\footnote{Without the projection, the master fields also encode fermionic fields, and the theory describes a higher-spin generalisation of $\mathcal{N}=2$ supergravity \cite{Prokushkin:1998bq}.}
\begin{align}
\label{eq:bosonicProjection}
\varkappa \star \MW \star \varkappa = \MW \,, && \varkappa \star \MB \star \varkappa = \MB \,, && \varkappa \star \MS_\ga \star \varkappa = - \MS_\ga \,.
\end{align}
As one can see from~\eqref{eq:innerKleinianConditions}, this constrains $\MW$ and $\MB$ to even functions in $y_\ga$, $z_\ga$, and $\MS_\ga$ to an odd function.
As in~\cite{Prokushkin:1998bq}, we introduce a hermitian conjugation for the variables as
\begin{align}
\label{eq:conjugation}
(y_\alpha)^\dagger = y_\alpha\,,&& (z_\alpha)^\dagger =  -z_\alpha\,,&& \phi^\dagger = \phi\,,&& \psi^\dagger = \psi \,,&& k^\dagger=k\,,&& \rho^\dagger = \rho\,,
\end{align}
and require the fields to satisfy the (anti-)hermiticity conditions
\begin{align}
\label{eq:hermiticityconditions}
&\MW^\dagger = -\MW\,, &\MS_\alpha^\dagger = -\MS_\alpha\,, &&\MB^\dagger = \MB\,.
\end{align}
It is straightforward to verify that equations~\eqref{eq:threedVasilievmasterequations} are compatible with imposing these conditions. The gauge transformations~\eqref{eq:mastergaugevariations} that are consistent with the \mbox{(anti-)}hermiticity conditions have gauge parameters that satisfy
\begin{equation}\label{bosonicprojgauge}
    \varkappa \star \Mxi \star \varkappa = \Mxi \,, \qquad (\Mxi)^\dagger = - \Mxi \, . 
\end{equation}
As the master field $\MB$ contains the unfolded scalar as $\flC = \sC + \pC\psi$, the condition $\MB^\dagger = \MB$ implies
\begin{align} \label{conjugation}
(\Phi_\pm)^\dagger = \Phi_\mp\,,
\end{align}
where $\Phi_\pm = \Pi_\pm \pC(y=0)$. The theory therefore contains one complex scalar field. On the other hand, the condition $\MW^\dagger = -\MW$ leads to a reality condition for the higher-spin fields. A further truncation of the theory to describe a single real scalar field has been also introduced in \cite{Prokushkin:1998bq, Bonezzi:2015igv}.

\subsubsection{Higher-spin background solutions}
\label{sec:adsvaccumvasiliev}

In this section we consider solutions of the Vasiliev equations~\eqref{eq:threedVasilievmasterequations} with a vanishing scalar, $\pC=0$. We make the ansatz of a constant value for the twisted scalar field by setting
\begin{equation}
    \MB^{(0)} = \nu\, 
\end{equation}
with a constant parameter $\nu$ (which will turn out to be the parameter $\nu$ of the deformed oscillators). This directly solves two of Vasiliev's equations, namely~\eqref{eq:threedVasilievdB} and~\eqref{eq:threedVasilievBS}. The remaining equations are
\begin{subequations}
    \begin{align}
        \intd \MW&=\MW \wedge\star \MW\,, \label{scalar:eqforW} \\
\rho\,\intd \MS_\ga&= [\MW,\,\rho \MS_{\ga}]_\star\,,\label{scalar:backgroundeqforW}\\
[\rho\MS_\ga,\,\rho\MS_\gb]_\star &= - 2i \epsilon_{\ga \gb} (1 +  \nu k\varkappa) \,.\label{scalar:backgroundeqforS}
    \end{align}
\end{subequations}
For $\nu=0$, the last equation is solved by $\MS_\ga=z_\ga$ (this follows directly from~\eqref{scalar:zstarrelations}). A solution for $\nu\not=0$ is given by 
\begin{equation}\label{scalar:backgroundvalueS}
    \MS_\ga=\hat{z}_\ga := z_\ga + \nu \,w_\ga\,k\,,
\end{equation}
where
\begin{equation}
    w_\ga = (z_\ga +y_\ga)\int_0^1 \intd t\,t\, e^{it\,yz}\,.
\end{equation}

\begin{exercise}
    Check that~\eqref{scalar:backgroundvalueS} solves~\eqref{scalar:backgroundeqforS}.
\end{exercise}

One can also realise the deformed oscillators $\hat{y}_\ga$ in a similar way by setting
\begin{equation}
    \hat{y}_\ga = y_\ga + \nu \,w_\ga \star k\varkappa\,.
\end{equation}
Then
\begin{subequations}
\begin{align}
    [\rho\,\hat{z}_\ga,\,\rho\,\hat{z}_\gb]_\star &= 2i \epsilon_{\ga \gb} (1 +  \nu k\varkappa)\,,\\
    [\hat{y}_\ga,\hat{y}_\gb]_\star &= 2i \epsilon_{\ga \gb} (1 +  \nu k)\,,\\
    [\hat{y}_\ga,\,\rho\,\hat{z}_\gb]_\star &= 0\,.
\end{align}
\end{subequations}
For the background value~\eqref{scalar:backgroundvalueS} for $\MS_\ga$, the equation~\eqref{scalar:backgroundeqforW} becomes
\begin{equation}
    [\MW,\,\rho\,\hat{z}_\ga]_\star =0\, .
\end{equation}
To solve this equation, we choose $\MW$ to be a combination of (star) products of $\hat{y}_\ga$, $\phi$, and $k$. The remaining equation~\eqref{scalar:eqforW} is then solved by any $\MW=-\cA(\hat{y}_\ga,\phi,k)$ which solves the flatness condition
\begin{equation}
    \intd\cA + \cA\wedge \star \cA = 0\, ,
\end{equation}
which is precisely the equation of motion for a Chern--Simons field $\cA$. When we project to a definite $k$-parity, for example by requiring $P_+\cA=\cA$, we can identify $\cA$ as a higher-spin gauge field for the algebra $\mathfrak{hs}[\lambda]\oplus \mathfrak{hs}[\lambda]$ with $\lambda=\frac{1}{2}(\nu-1)$. We conclude that every solution of Chern--Simons theory for this algebra gives rise to a solution of Vasiliev equations with constant (twisted) scalar $\MB=\nu$.

\subsubsection{Linear Perturbations}
\label{sec:linearVasiliev}
We shall now consider linear perturbations around the higher-spin background of the last section. To this end, we expand
\begin{subequations}
\begin{align}
\MS_\ga &= \hat{z}_\ga + \epsilon \MS_\ga^{(1)} + \epsilon^2 \MS_\ga^{(2)}+ \dots\,,\\
\MW &= -\cA + \epsilon \MW^{(1)}+ \epsilon^2 \MW^{(2)} + \dots \,,\\
\MB &= \nu + \epsilon \MB^{(1)} + \epsilon^2 \MB^{(2)}+ \dots \,,
\end{align}
\end{subequations}
with a formal expansion parameter $\epsilon$. Inserting this into the Prokushkin--Vasiliev equations~\eqref{eq:threedVasilievmasterequations}, we find at linear order in $\epsilon$: 
\begin{subequations}
\begin{align}
      0&=\intd\MW^{(1)} + \cA\wedge\star \MW^{(1)} + \MW^{(1)}\wedge\star \cA \,,\\
      0&=\intd\MB^{(1)} + [\cA,\MB^{(1)}]_\star  \,,\label{scalar:linearVasilievB}\\
      [\MW^{(1)},\,\rho\hat{z}_\ga]_\star &= \rho\,\intd\MS_\ga^{(1)} + [\cA,\,\rho\MS_\ga^{(1)}]_\star  \,,\\
      [\MB^{(1)},\,\rho\hat{z}_\ga]_\star &=0\,, \label{scalar:linearVasilievD}\\
      [\rho \hat{z}_\ga,\,\rho \MS^{(1)\ga}]_\star &= -2i\,\MB^{(1)}\star k\varkappa\,.
\end{align}
\end{subequations}
As discussed in~\cite{Prokushkin:1998bq}, the general solution of the equation~\eqref{scalar:linearVasilievD} for the scalar master field is any function of $\hat{y}_\ga$, $\phi$, $\psi$, $k$, and the spacetime coordinates,
\begin{equation}
    \MB^{(1)} = \flC^{(1)}(\hat{y},\phi,\psi,k|x) \, .
\end{equation}
From~\eqref{scalar:linearVasilievB} we then recover the free equation of motion~\eqref{eq:freeEoMVasilievB} for the unfolded scalar $\flC^{(1)}$ in a higher-spin background,
\begin{equation}
    D\flC^{(1)} = 0 \, .
\end{equation}

\subsubsection{Subtleties of the twisted sector}

As we have shown above, the linearised Vasiliev equations reproduce the expected equations for the unfolded scalar. In the analysis of the remaining equations at the linear level, we shall find that the twisted part of the gauge field is sourced by the scalar field. A priori this would be problematic since one would wish to decouple the unphysical twisted sector. It was shown in~\cite{Vasiliev:1992ix}, however, that one can remove this source term by a field redefinition as we review in the following.

To simplify the discussion, from now on we set $\nu=0$.
The variable $k$ then is not necessary any more: we can redefine $\MB\to \MB k$ and take all fields otherwise to be independent of $k$. Furthermore, the variable $\rho$ drops out of the equations, and we have $\hat{z}_\ga=z_\ga$. For bosonic fields (satisfying \eqref{bosonicprojgauge}), the linear equations then simplify to
\begin{subequations}
\begin{align}
      D\MW^{(1)} &=0   \,,\label{scalar:linearVasilievnu0A}\\
      D\MB^{(1)}&=0  \,,\label{scalar:linearVasilievnu0B}\\
      2i\partial_\ga^z \MW^{(1)} &= D\MS_\ga^{(1)}   \,,\label{scalar:linearVasilievnu0C}\\
      \partial_\ga^z \MB^{(1)} &=0\,, \label{scalar:linearVasilievnu0D}\\
       \partial_\ga^z  \MS^{(1)\ga} &= \MB^{(1)}\star \varkappa\,, \label{scalar:linearVasilievnu0E}
\end{align}
\end{subequations}
where we also used~\eqref{eq:zalphacommutatorzderivative} to replace the commutator with $z_\ga$ by a derivative.

The non-dynamical equations \eqref{scalar:linearVasilievnu0C}-\eqref{scalar:linearVasilievnu0E} are first-order differential equations with respect to $z$, and they can be solved using the following general results\footnote{The second equation~\eqref{homotopyuncontracted} only has a solution if the following compatibility condition is satisfied, 
\[
\partial^\ga_z g_\ga(y,z)=0 \,,
\]
which holds in the case of~\eqref{scalar:linearVasilievnu0C} to which we are going to apply this solution.}
\begin{subequations}
\label{eq:homotopyIntegrals}
\begin{align}
\partial^z_\ga f^\ga(y,z)&=g(y,z) &  \Longrightarrow &&  f_\ga (y,z)&= \partial^z_\ga \epsilon(y,z) + z_\ga \,\homo{1}{g(y,z)} \, , \label{homotopycontracted}\\
\partial^z_\ga f(y,z)&=g_\ga(y,z) & \Longrightarrow &&  f(y,z)&=\epsilon(y) + z^\ga \homo{0}{g_\ga(y,z)} \label{homotopyuncontracted}\,.
\end{align}
\end{subequations}
Here, $\homo{n}{\bullet}$ stands for \emph{homotopy integrals} defined as
\begin{equation}
\label{eq:homotopyintegralsdefinition}
\homo{n}{f}(z):=\int_0^1  \intd t\,t^n\, f(tz)\,,
\end{equation}
and $\epsilon (y,z)$ and $\epsilon(y)$ are arbitrary functions of the indicated arguments.

Therefore, the solutions for the non-dynamical equations are given by (where we suppress the dependence on $\phi$, $\psi$ and $x$)
\begin{subequations}
\label{eq:zdependenceorder1}
\begin{align}
\MB^{(1)}&=\flC^{(1)}(y) \,, \label{eq:zdependenceBorder1} \\
\MS^{(1)}_\ga&= \partial^z_\ga \epsilon^{(1)}(y,z) + z_\ga \,\homo{1}{\flC^{(1)} \star \varkappa}\,, \label{eq:zdependenceAorder1}\\
\MW^{(1)}&= -\boldsymbol{A}^{(1)}(y) -\frac{i}{2} z^\ga \homo{0}{D \MS^{(1)}_{\ga}}\,.\label{eq:zdependenceWorder1}
\end{align}
\end{subequations}
Note that these equations fully determine the $z_\ga$-dependence of the master fields. This statement also holds at higher orders in perturbation theory. 

We now impose the following gauge condition on the master field $\MS_\ga$,
\begin{equation}
\label{eq:SchwingerFockGaugeCondition}
z^\ga \MS_\ga = 0 \,,
\end{equation}
which is usually referred to as \emph{Schwinger--Fock gauge}.\footnote{In some parts of the literature, the Schwinger--Fock gauge is also known as the Vasiliev gauge.} 
The Schwinger--Fock gauge implies that the homogeneous solution $\partial^z_\ga \epsilon^{(1)}(y,z)$ in \eqref{eq:zdependenceAorder1} vanishes. This is because $z^\ga \partial^z_\ga$ is the $z_\ga$-number operator which implies
that in the Schwinger--Fock gauge $\epsilon^{(1)}(y,z)=\epsilon^{(1)}(y)$ and therefore 
\begin{align}
\label{eq:homogenousSFGVanishing}
\partial^z_\ga \epsilon^{(1)}(y,z)=0\,.
\end{align}
To determine the residual gauge transformations, we also expand the gauge parameter starting from linear order,
\begin{equation}
    \xi = \epsilon\, \xi^{(1)}+\dots 
\end{equation}
The first order transformations are then
\begin{subequations}
\begin{align}
\delta \MW^{(1)} &= D \xi^{(1)}  \,, \\
\delta \MB^{(1)} &=0 \,, \\
\delta \MS_{\ga}^{(1)} &= 2i\,\partial^z_\ga \xi^{(1)} \,, \label{eq:transformationA}
\end{align}
\end{subequations}
where we used the restrictions~\eqref{bosonicprojgauge} on $\Mxi$ from the bosonic projection. When we compare with the gauge-fixing condition~\eqref{eq:homogenousSFGVanishing} we find
\begin{align}
z^\ga \delta \MS^{(1)}_\ga = 2i\,z^\ga \partial_\ga^z \xi^{(1)}(y,z) \overset{!}{=} 0 && \Longrightarrow && \xi^{(1)}(y,z) = \xi^{(1)}(y)\,.
\end{align}
It follows that the residual gauge freedom preserving the Schwinger--Fock gauge is given at first order by $z$-independent gauge parameters $\xi^{(1)}(y)$.

Plugging the solutions \eqref{eq:zdependenceorder1} of the non-dynamical equations in the dynamical equations \eqref{scalar:linearVasilievnu0A} and \eqref{scalar:linearVasilievnu0B}, we obtain in Schwinger--Fock gauge
\begin{subequations}
\begin{align}
D \left(\pC^{(1)}(y)\psi+ \sC^{(1)}(y) \right)&=0 \,,\label{eq:linearFirst}\\
D \left(\mathcal{A}^{(1)}(y) + \tilde{\mathcal{A}}^{(1)}(y) \psi \right)&= -\frac{i}{2}D \left( z^\ga \Gamma_0 \Big\langle D  \big(  z_\ga \homo{1}{\flC^{(1)} \star \varkappa} \big)\Big\rangle  \right) \,. \label{eq:firstOrderDynWEq} 
\end{align}
\end{subequations}
Notice that the fields still depend on $\phi$ and the spacetime variable $x$.
The left-hand side of \eqref{eq:firstOrderDynWEq} is $z$-independent, and therefore also its right-hand side has to share this property. This implies that we can evaluate it for $z=0$ as all $z$-dependent terms have to cancel out anyway. Note however that the star product \eqref{eq:starproductZ} does not commute with setting the $z_\ga$-oscillators to zero. Therefore, we need to first evaluate all star products and only afterwards set all $z$-dependent factors to zero. 

To simplify the computations, we now restrict the background to AdS, $\cA=\cA^{AdS}$.
After some algebra, one obtains
\begin{multline}
\label{eq:sourceterm}
D^{AdS} \!\!\left( z^\ga \Gamma_0 \Big\langle D^{AdS}  \big(  z_\ga \homo{1}{\flC^{(1)} \star \varkappa} \big) \Big\rangle \right) \bigg|_{z=0} \\ = \tfrac{1}{8} \, E^{\ga\gb} \, (y_\ga+i\partial^u_\ga) (y_\gb+i\partial^u_\gb) \, \pC^{(1)}(u) \psi\big|_{u=0} \,,
\end{multline}
where we have defined 
\begin{equation}
\label{eq:twoFromEDefTmp}
E^{\ga \gb}=\bvb^{\ga}{}_\gamma \wedge \bvb^{\gb \gamma} \,.
\end{equation}
It was shown by Vasiliev \cite{Vasiliev:1992ix} that one can remove the source term \eqref{eq:sourceterm} by a field redefinition of the twisted one-form $\tilde{\cA}^{(1)} \to \tilde{\cA}^{(1)} + M$ with
\begin{equation}
\label{eq:fieldRedefVas}
M = -\frac{i}{8}\phi\,\bvb^{\ga\gb} \, \int^1_0 \intd t \,(t^2 - 1) \, (y_\ga + it^{-1} \partial_\ga^{y} )(y_\gb + it^{-1} \partial_\gb^{y} ) \,\pC^{(1)} (ty)\,.
\end{equation}
After performing this field redefinition, we obtain the equations of motion 
\begin{subequations}
\begin{align}
D^{AdS}\!\! \left(\pC^{(1)}(y)\psi+ \sC^{(1)}(y) \right)&=0 \,,\\
D^{AdS}\!\! \left(\cA^{(1)}(y) + \tilde{\cA}^{(1)}(y) \psi \right)&= 0 \,, 
\end{align}
\end{subequations}
which indeed coincide with the free unfolded equations \eqref{eq:freeEoMVasiliev} introduced earlier. Therefore, the Prokushkin--Vasiliev equations provide us with a non-linear theory of higher--spin gauge fields coupled to a complex scalar field (as well as additional twisted fields). Note that we can choose vanishing solutions for the twisted fields, i.e. $\sC^{(1)}=\tilde{\cA}^{(1)}=0$, as their equations of motion do not contain any source terms (after we have performed the field redefinition \eqref{eq:fieldRedefVas}).

\begin{exercise}\hspace{-2pt}\hyperlink{sol:backreaction}{*}
\label{prob:backreaction}
Check that \eqref{eq:sourceterm} holds. Consider the case $\sC^{(1)}=0$ and neglect all terms involving the background spin connection for simplicity.
\end{exercise}

\begin{exercise}\hspace{-2pt}\hyperlink{sol:removalBackreaction}{*}\label{prob:removalBackreaction}
Show that the field redefinition \eqref{eq:fieldRedefVas} indeed removes the source term \eqref{eq:sourceterm}.
\end{exercise}

\subsubsection{Comments on higher orders}
\label{sec:higherOrderVasiliev}

In the last section, we have seen how to perturbatively extract equations of motion from Vasiliev equations~\eqref{eq:threedVasilievmasterequations}. At first order, we have verified in this way that the linearised equations of motion describe a free scalar field and free higher-spin fields (together with a twisted sector that can however be decoupled at this order). In principle, one can now go on and compute the nonlinear contributions to the equations of motion. Field redefinitions as the one we discussed for the twisted sector at linear order become necessary at higher orders also in the untwisted sector.

This becomes first apparent at second order when one analyses the backreaction of the matter fields to the higher-spin fields. 
The naive computation results in a non-local backreaction, and a non-local field redefinition is needed to bring the backreaction to the expected local form \cite{Kessel:2015kna}. On the other hand, one cannot allow for arbitrary non-local field redefinitions because one could otherwise remove all interactions \cite{Prokushkin:1998bq}. The necessity of a particular non-local field redefinition can be interpreted by saying that the identification of the physical field inside the master fields has to be modified (for example \cite{Korybut:2022kdx} by using a shifted version of the homotopy integrals~\eqref{eq:homotopyintegralsdefinition}). It is an open question what the correct prescription is at all orders. These issues might be related to the non-localities that appear in four dimensions, for a discussion of this topic see, e.g., \cite{Boulanger:2015ova, Skvortsov:2015lja, Sleight:2017pcz, Ponomarev:2017qab, Didenko:2018fgx}.

Let us also stress that these issues are relevant to determine the status of the Prokushkin--Vasiliev model as a candidate bulk dual within minimal model holo\-graphy. Indeed, it is unclear what the twisted sector could correspond to on the CFT side. As we mentioned, accepting certain non-local field redefinitions, the twisted sector can be decoupled up to second order, but it is not known if a similar scenario applies also to all orders. To conclude, we mention that, while Vasiliev's equations in three and four dimensions have been originally defined using oscillator techniques that are peculiar to these dimensions, interacting equations of motion have been later defined in a language that applies to arbitrary spacetime dimensions \cite{Vasiliev:2003ev}. A priori, these equations of motion do not involve a twisted sector, so that their three-dimensional instance could provide an alternative proposal for a coupling of matter to higher spin fields.

\section{Summary and further developments}\label{sec:extra}

These lecture notes focus on three-dimensional models describing the interactions of relativistic massless higher-spin fields and their couplings to scalar matter, mainly on an anti-de Sitter background, as well as on their holographic description in terms of $\cW_N$-minimal models. 
To this end, an introduction to quantum $\cW$-symmetries and to their realisation in conformal field theory is also provided. The selected material has been chosen so as to provide the bases to appreciate the key ideas underlying minimal-model holography \cite{Gaberdiel:2010pz}, which we review in section~\ref{sec:holography}. 

In section~\ref{sec:CS}, we first review the Chern--Simons formulation of three-dimensional gravity, and we then show how it can be naturally extended to describe higher spins by enlarging the gauge algebra. We focus on extensions involving $\mathfrak{sl}(N,\mathbb{R}) \oplus \mathfrak{sl}(N,\mathbb{R})$ gauge algebras, which describe fields of spin $2,3,\ldots,N$ on AdS$_3$, and on their $N \to \infty$ limit, involving two copies of the infinite-dimensional $\mathfrak{hs}[\lambda]$ algebra, whose construction we review in section~\ref{sec:CSformulation}.

In section~\ref{sec:symmetries}, we identify the asymptotic symmetries of the 
previous field theories, starting from the example of gravity and then moving to the higher-spin case, showing that they are given by non-linear $\cW$-algebras. We study asymptotic symmetries using  Hamiltonian techniques, and we stress the link with the mathematical literature on the Drinfeld--Sokolov reduction. In particular, in section~\ref{sec:beyondsl3} we compute the asymptotic symmetries of $\mathfrak{sl}(N,\mathbb{R})$ Chern--Simons theories in a basis in which their algebra admits at most quadratic non-linearities and in which explicit expressions for the structure constants can be derived. We achieve this goal by encoding the relevant conditions into suitable pseudodifferential operators, following the standard treatment in the mathematical literature. We then briefly discuss how these techniques can be adapted to deal with the infinite-dimensional $\mathfrak{hs}[\lambda]$ case.

In section~\ref{sec:quantum}, we deal with the quantum version of the $\cW$-algebras introduced in the previous section, discussing the normal-ordering subtleties brought by the non-linearities in the commutators. We first present the example of the $\cW_3$-algebra, and we then discuss generic $\cW_N$-algebras using their realisation in terms of the Miura transform, which is the quantum counterpart of the classical construction of section~\ref{sec:beyondsl3}. This allows us to eventually consider the $N \to \infty$ limit, resulting in the $\cW_\infty[\lambda]$ family of infinite-dimensional non-linear algebras, labelled by the parameter $\lambda$ and by the central charge. We discuss the emergence of a remarkable triality symmetry relating different values of these parameters, and we introduce conformal field theories admitting the previous symmetry algebras. We begin by defining $\cW_N$-minimal models, whose global symmetries are given by $\cW_N$-algebras,  and we then define their large-$N$ limit that enters minimal-model holography, highlighting the crucial role played by triality. We argue that the resulting CFT should admit a holographic description in terms of a higher-spin theory coupled to scalar matter, and we discuss the criticalities of this proposal.

The Chern--Simons theories that we consider in section~\ref{sec:CS} only describe massless fields of spin $s \geq 2$. Coupling them to matter introduces additional difficulties that we discuss in section \ref{sec:scalar}. We then present the Prokushkin--Vasiliev model \cite{Prokushkin:1998bq}, that provides an explicit example of such a coupling. It does so using an approach, dubbed unfolded formalism, in which the matter equations of motion are reformulated in a first order form and using an infinite number of auxiliary fields, encoding the derivatives of the Klein--Gordon field. We first review the peculiarities of this formulation of the dynamics for a free Klein--Gordon field on AdS$_3$, and then we show how to describe its propagation on a higher-spin background. We conclude by reviewing how one can include scalar backreaction along the lines of \cite{Prokushkin:1998bq}. The latter reference relies on an oscillator realisation of the $\mathfrak{hs}[\lambda]$ algebras (introduced in section~\ref{sec:CSformulation}) that we also detail along the way.

Most of the material reviewed here already found various applications that go well beyond the study of minimal-model holography. For this reason, we close these lecture notes with a quick overview of selected further developments related to higher spins in three dimensions.

\paragraph{\textbf{Higher-spin black holes}} 
Shortly after the observation that higher-spin gauge theories in three spacetime dimensions have extended asymptotic symmetries \cite{Henneaux:2010xg,Campoleoni:2010zq}, 
higher-spin generalisations of the BTZ black hole \cite{Banados:1992wn, Banados:1992gq} have been constructed \cite{Gutperle:2011kf}.
Since higher-spin gauge transformations can map a given metric into another one with a different causal structure (see, e.g., \cite{Ammon:2011nk, Castro:2011fm}), the proposed definition of higher-spin black holes relies on their thermal properties rather than on the presence of an event horizon. In practice, these solutions are first defined in Euclidean spacetime, assuming the same torus topology as that of a BTZ black hole and demanding regularity of the Chern--Simons connection in its interior, and then continued to Lorentzian signature; see, e.g., \cite{Ammon:2012wc, Perez:2014pya, Castro:2016tlm} for a review. This somehow indirect way of defining higher-spin black holes is well adapted to look for holographic counterparts of these classical solutions \cite{Kraus:2011ds, Gaberdiel:2013jca, Tan:2016bhs}, but introduces various subtleties. For instance, the actual dependence of the black-hole entropy on higher-spin charges have been debated in the literature \cite{Banados:2012ue, Perez:2012cf, Campoleoni:2012hp, Perez:2013xi, deBoer:2013gz}, before leading to a finer characterisation of these solutions \cite{Bunster:2014mua, deBoer:2014fra}. Possible generalisations of the usual metric-based definition of a Lorentzian causal structure have also been discussed in \cite{Kraus:2012uf, Castro:2016ehj}.

\paragraph{\textbf{Semiclassical methods to compute observables in CFTs with $\cW$-symmetry}} 
Independently of the details of a precise holographic duality, minimal-model holography suggests the option to compute several, possibly non-local, observables in conformal field theories with extended $\cW$-symmetries by means of semiclassical computations in higher-spin Chern--Simons theories. This option has been explored, e.g., in \cite{Ammon:2013hba, deBoer:2013vca, Perlmutter:2013paa, deBoer:2014sna, Perlmutter:2016pkf, Narayan:2019ove, Alday:2020qkm, Zhao:2022wnp} often relying on the following logic: once a holographic prescription to compute a CFT observable by means of a semiclassical computation in gravity is known, in three dimensions one can first reformulate it in a Chern--Simons language and then naturally extend it to higher-spin theories.

\paragraph{\textbf{Higher spins and strings}}
As mentioned in the introduction, it is expected that string theory in the tensionless limit develops higher-spin symmetries; see, e.g., \cite{Sagnotti:2011jdy, Rahman:2015pzl} for a review. This has been made precise for strings on $\mathrm{AdS}_3\times S^3\times T^4$, where it was shown that the tensionless limit has a CFT dual given by a symmetric orbifold of $T^4$, which indeed shows a large symmetry algebra of global higher-spin charges that should signal the presence of higher-spin gauge fields in the bulk \cite{Gaberdiel:2014cha, Gaberdiel:2015mra, Eberhardt:2018ouy}. The latter non-linear algebra, whose wedge algebra has been dubbed as higher-spin square, contains a $\cN = 4$ generalisation of the algebra $\cW_\infty[\lambda]$ that we discussed in these notes, but it is much bigger.
A higher-spin theory displaying this symmetry was built from scratch in \cite{Raeymaekers:2019dkc}. 

\paragraph{\textbf{Three-dimensional higher-spin theories in flat space}}
In the very first paper on the subject \cite{Blencowe:1988gj}, it was already noticed that the Chern--Simons formulation of higher-spin gauge theories does not require a cosmological constant. In the last years, three-dimensional gravity proved to be an interesting setup where to test various ideas on flat space holography and this naturally led to explore higher-spin gauge theories on three-dimensional Minkowski space; see, e.g., \cite{Prohazka:2017lqb} for a review. Their asymptotic symmetries have been characterised in \cite{Afshar:2013vka, Gonzalez:2013oaa, Ammon:2017vwt} and the representation theory of the resulting algebras has been studied in \cite{Ammon:2020fxs, Campoleoni:2016vsh}. 
Coupling these models to matter is instead much subtler than in AdS; see, e.g., \cite{Ammon:2017vwt}. The main reason that allows to easily construct higher-spin theories in three-dimensional Minkowski space is that the higher-spin algebras discussed in section~\ref{sec:CSformulation} admit contractions that can be neatly interpreted as an appropriate gauge algebra for a Chern--Simons theory in Minkowski space. These steps are much subtler in four dimensions due to the presence of additional generators; on the other hand, the structure of three-dimensional flat-space higher-spin algebras recently suggested a path to introduce similar contractions in four dimensions too \cite{Campoleoni:2021blr}.

\paragraph{\textbf{Metric-like formulation}}
In these notes we have reviewed Chern--Simons higher-spin gauge theories, which are based on a first-order frame-like formulation of the dynamics. In principle, such theories can also be formulated as second-order theories in terms of metric-like Fronsdal fields, adding interaction vertices to the free theories reviewed in section~\ref{sec:FronsdalFormulation}. In \cite{Campoleoni:2012hp, Fredenhagen:2014oua, Campoleoni:2014tfa} it was shown how one can perturbatively obtain such field theories from the Chern--Simons formulation in a weak field expansion, but similarly to gravity they involve interaction vertices of arbitrarily high order in the fields (see also \cite{Fujisawa:2012dk, Fujisawa:2013lua} for an alternative proposal to rewrite Chern--Simons actions in terms of metric-like fields involving additional auxiliary fields). Alternatively, one can classify perturbatively all possible interactions between higher-spin gauge fields directly within the metric-like formulation \cite{Mkrtchyan:2017ixk, Kessel:2018ugi}. This approach confirms --~in agreement with the Chern--Simons formulation~-- that cubic vertices determine all interactions of higher-spin gauge fields, in the sense that no further independent coupling constants can enter the interacting action \cite{Fredenhagen:2018guf, Grigoriev:2020lzu}. Although this perturbative approach is less efficient in describing massless fields compared to the Chern--Simons formulation, let us stress that it provides an alternative and conceptually straightforward setup where to analyse and classify possible matter couplings, with potential applications in the quest for alternative holographic models with respect to the Prokushkin--Vasiliev model. 

\paragraph{\bf{Other higher-spin theories in three dimensions}}
We focused on massless bosonic fields and their couplings to scalar matter, but other classes of higher-spin theories have been studied too in three dimensions. Massless fermions can be described employing Chern--Simons theories with gauge superalgebras, in the same spirit of the Chern--Simons formulation of three-dimensional supergravity; see, e.g., \cite{Achucarro:1986uwr, Henneaux:1999ib, Henneaux:2012ny, Hanaki:2012yf, Tan:2012xi, Candu:2014yva}. Supersymmetric extensions of the Prokushkin--Vasiliev bosonic model reviewed here have also been introduced \cite{Prokushkin:1998bq}, as well as supersymmetric extensions of minimal-model holography \cite{Creutzig:2011fe, Candu:2012jq, Beccaria:2013wqa}. On constant-curvature backgrounds one can also introduce non-propagating three-dimensional generalisations of partially-massless theories, see e.g.\ \cite{Grigoriev:2019xmp, Grigoriev:2020lzu}, while on a Minkowski background one can consider similar theories with exotic indecomposable spectra \cite{Boulanger:2023tvt}. All these classes of non-propagating fields admit interactions that can always be rewritten in a Chern--Simons form \cite{Grigoriev:2020lzu}. 
Massive higher-spin fields, instead, do propagate local degrees of freedom in three dimensions too \cite{Binegar:1981gv}; this does not allow for a compact description in terms of Chern--Simons actions, but their interaction vertices have been studied perturbatively, e.g., in \cite{Zinoviev:2021cmi, Zinoviev:2023vvz}. Topologically-massive higher-spin theories have been also considered, e.g., in \cite{Damour:1987vm, Chen:2011vp, Bagchi:2011vr, Bergshoeff:2011pm, Chen:2011yx, Boulanger:2014vya, Dalmazi:2021dgp} and massive higher-spin supermultiplets have been considered, e.g., in \cite{Bergshoeff:2009tb, Buchbinder:2015mta, Kuzenko:2016qwo}. Fractional and continuous spin representations and their field theory realisations have been explored too \cite{Boulanger:2013naa, Boulanger:2015uha, Schuster:2014xja}.

\paragraph{\bf{Non-relativistic three-dimensional higher-spin theories}}
We already mentioned that the higher-spin algebras of section~\ref{sec:CSformulation} allow for contractions that can be used to define higher-spin theories in Minkowski space via the Chern--Simons formulation. Other contractions are also possible, and they have been shown to lead to Chern--Simons theories that can be interpreted as non-relativistic field theories \cite{Gary:2012ms, Afshar:2012nk, Gary:2014mca, Bergshoeff:2016soe, Chernyavsky:2019hyp, Caroca:2022byi, Concha:2022muu}, with potential applications in the study of two-dimensional condensed matter systems. Higher-rank tensorial fields in $2+1$ dimensions have also been employed in effective theories describing various features of the fractional quantum Hall effect; see, e.g., \cite{Golkar:2016thq, Cappelli:2015ocj, Liu:2018swy}.


\begin{acknowledgement}
We thank G.~Lucena G\'omez and P.~Kessel for collaboration on this project at an initial stage. We also thank M.~Henneaux, P.~Kessel, O.~Kr\"uger, K.~Mkrtchyan, S.~Pfenninger, T.~Prochazka, J.~Raeymaekers and S.~Theisen for collaboration on the topics reviewed in these lecture notes as well as G.~Barnich, N.~Boulanger, A.~Delfante, J.~Fischer, D.~Francia, M.~Gaberdiel, H.~Gonz\'alez, R.~Lomartire, T.~Nutma, S.~Pekar, M.~Riegler, B.~Oblak, A.~Sagnotti, E.~Schnabel, E.~Skvortsov, and M.~Taronna for discussions. AC is a research associate of the Fund for Scientific Research -- FNRS. His work was partially supported by FNRS through the grants No.\ F.4503.20, T.0022.19 and T.0047.24. We acknowledge the Galileo Galilei Institute for Theoretical Physics in Florence for the opportunity to present a part of the material collected in these lecture notes.
\end{acknowledgement}

\appendix
\section{Frame-like formulation in arbitrary dimension}\label{app:arbitraryD}

In the main text, we have presented the frame-like formulation of higher-spin fields in three spacetime dimensions with the action~\eqref{eq:frameAction}. In this appendix, we contrast it with the frame-like formulation in generic spacetime dimensions~\cite{Vasiliev:1980as,Lopatin:1987hz} (see also \cite{Ponomarev:2022vjb} for a review).
 
In arbitrary spacetime dimensions, a frame-like free action can be obtained by introducing a vielbein-like field $e^{a_1 \cdots a_{s-1}}$ and a spin-connection-like field $\omega^{b,a_1 \cdots a_{s-1}}$. Both fields are traceless in all the Lorentz indices, fully symmetric in $a_i$ and the spin connection obeys the irreducibility condition $\omega^{(a_1,a_2 \cdots a_s)}=0$, which will play a key role in the proof of most of the identities discussed in the following. When $D=3$, this auxiliary field can be dualised as $\omega^{a(s-1)} \sim \omega^{c, b (a_1 \cdots a_{s-2}}\epsilon^{a_{s-1})}{}_{bc}  $ to recover the field we introduced in \eqref{eq:def-fields}. 

The frame-like action is given by\footnote{For some specific values of $D$, as for instance $D=3$, additional terms may be written. Notice also that here we presented the free action as, e.g., in \cite{Zinoviev:2008ze}. 
An equivalent presentation of this action, used in \cite{Vasiliev:1980as, Ponomarev:2022vjb}, can be obtained by rewriting the terms quadratic in the spin connection thanks to the identity
\begin{align*}
    0 &= (D+1)\,\bvb^{[d}\ww \omega_{d,}{}^{a|q_{s-2}|}\ww \omega^{b,c}{}_{q_{s-2}}\ww \underbrace{\bvb^{p_1}\ww\cdots\ww\bvb^{p_{D-3}]} \epsilon_{p_1 \cdots p_{D-3}abc}}_{K_{abc}}\\
    &= (D-2)\,\bvb^d \ww \omega_{d,}{}^{aq_{s-2}}\ww \omega^{b,c}{}_{q_{s-2}}\ww K_{abc}\\
    &\phantom{=} + \bvb^a \ww \left( \omega_{d,}{}^{bq_{s-2}}\ww \omega^{d,c}{}_{q_{s-2}}+\frac{1}{s-1}\, \omega^{b,}{}_{q_{s-1}}\ww\omega^{c,q_{s-1}}\right)\ww K_{abc}\,,
\end{align*}
where the first line vanishes because of the antisymmetrisation over $D+1$ indices and the two terms in the third line are those that enter \eqref{action_anyD}.}
\begin{multline} \label{action_anyD}
S  = \frac{1}{16\pi G}
\int K_{abc} \ww \bigg( e^{a}{}_{q_{s-2}} \ww \nabla \omega^{b,\,c q_{s-2}} - \frac{1}{2(D-2)(s-1)}\, \bvb^a \ww \omega^{b,}{}_{q_{s-1}} \ww \omega^{c,\,q_{s-1}} \\
 - \frac{1}{2(D-2)}\, \bvb^a \ww \omega_{d,\,q_{s-2}}{}^b \ww \omega^{d,\,cq_{s-2}} + \frac{s(D+s-4)}{2(D-2)\adsR^2}\, \bvb^a \ww e^b{}_{q_{s-2}} \ww e^{cq_{s-2}} \bigg) \, ,
\end{multline}
where
\begin{equation}
K_{abc} := \e_{p_1 \cdots p_{D-3}abc}\, \bvb^{p_1} \ww \cdots \ww \bvb^{p_{D-3}} \, ,
\end{equation}
and where an index with a subscript denotes a group of symmetrised indices. For instance, $e^{a_{s-1}} := e^{a_1\cdots a_{s-1}}$. Moreover, repeated covariant or contravariant indices denote a symmetrisation. For instance, $A^a B^a := \frac{1}{2} \left( A^{a_1} B^{a_2} + A^{a_2} B^{a_1} \right)$.
This action is invariant under the gauge transformations
\begin{subequations}
\begin{align}
\delta e^{a_{s-1}} & = \nabla\x^{\,a_{s-1}} + \bvb_b \L^{b,\,a_{s-1}} \, , \label{varE_anyD} \\[5pt]
\delta \omega^{b,\,a_{s-1}} & = \nabla \L^{b,\,a_{s-1}} + \bvb_c \theta^{\,bc,\,a_{s-1}} + \frac{(s-1)(D+s-4)}{(D-2)\adsR^2} \left( \bvb^b \x^{\,a_{s-1}} - \bvb^a \x^{\,ba_{s-2}} \right) \nn \\ &\phantom{=} - \frac{(s-1)(s-2)}{(D-2)\adsR^2} \left( \h^{ab} \bvb_c \x^{\,ca_{s-2}} - \h^{aa} \bvb_c \x^{\,ba_{s-3}} \right) ,\label{varO_anyD}
\end{align}
\end{subequations}
where we fixed conventionally the relative factor in the first variation. The terms in $\eta^{ab}$, here and in the following gauge variations, just implement a traceless projection, in agreement with the properties of the fields that are varied. 

For $s=2$, the action \eqref{action_anyD} reduces to the linearisation of the Einstein-Hilbert action. For arbitrary values of $s$, the previous action contains all possible terms that are quadratic in the higher-spin fields, while the relative coefficients are fixed by demanding the presence of a gauge symmetry of the form $\delta e^{a_{s-1}} = \nabla\x^{\,a_{s-1}} +\ldots$ and $\delta \omega^{b,\,a_{s-1}} = \nabla \L^{b,\,a_{s-1}} + \ldots$ These gauge transformations involve parameters with the same tensorial structure as the fields and generalise the symmetry under local translations and local Lorentz transformations of linearised gravity. This requirement fixes uniquely the coefficients in the action, but notice that the result displays an additional symmetry generated by a parameter $\theta^{bc, a_1 \cdots a_{s-1}}$ which is fully traceless and satisfies the irreducibility condition $\theta^{b(a_1, a_2 \cdots a_{s})} = 0$.

In the following, we shall show that the equations of motion implied by the action \eqref{action_anyD} are equivalent to the Fronsdal equations on AdS of \eqref{eq:fronsdalEqAds}. To this end, it will be convenient to rewrite them in terms of (linearised) higher-spin curvatures, that are gauge-invariant two-forms linear in the fields. They read 
\begin{subequations}
\begin{align}
\cT^{a_{s-1}} & = \nabla e^{a_{s-1}} + \bvb_b \ww \omega^{b,\,a_{s-1}} \, , \\[5pt]
\cR^{\,b,\,a_{s-1}} & = \nabla \L^{b,\,a_{s-1}} + \frac{(s-1)(D+s-4)}{(D-2)\adsR^2} \left( \bvb^b \ww e^{a_{s-1}} - \bvb^a \ww e^{ba_{s-2}} \right) \nn \\
& - \frac{(s-1)(s-2)}{(D-2)\adsR^2} \left( \h^{ab} \bvb_c \ww e^{ca_{s-2}} - \h^{aa} \bvb_c \ww e^{bca_{s-3}} \right) + \bvb_c \ww \Omega^{\,bc,\,a_{s-1}} \, ,
\end{align}
\end{subequations}
where in the following we shall often denote $\cT^{a_{s-1}}$ as the torsion two-form.
Notice that asking for gauge invariance with respect to $\L^{b,a_1 \cdots a_{s-1}}$ requires to introduce a further connection that transforms as
\begin{equation}\label{delta-omega}
\begin{split}
& \delta \O^{\,bb,\,a_{s-1}} = \nabla \theta^{\,bb,\,a_{s-1}} - \frac{2(s-2)(D+s-3)}{D\,\adsR^2} \left( \bvb^b \L^{b,\,a_{s-1}} - \frac{s-1}{s-2}\, \bvb^a \L^{b,\,ba_{s-2}} \right) \\
& + \frac{2(s-1)(s-2)}{D(D-2)\adsR^2}\, \bigg( \frac{D+s-3}{s-1}\, \h^{bb} \bvb_c \L^{c,\,a_{s-1}} - \frac{D+2s-6}{s-2}\, \h^{ab} \bvb_c \L^{c,\,ba_{s-2}} \\
& + \frac{(s-3)(D-2)}{s-2}\, \h^{ab} \bvb_c \L^{b,\,ca_{s-2}} - (D-2)\, \h^{aa} \bvb_c \L^{b,\,bca_{s-3}} + \h^{aa} \bvb_c \L^{c,\,bba_{s-3}} \bigg) + \cdots \, .
\end{split}
\end{equation}
The procedure iterates: building a gauge-invariant curvature for $\Omega^{b_2, a_{s-1}}$ would require to introduce yet another connection that would naturally transform as $\delta \Omega^{b_3,a_{s-1}} = \nabla \theta^{b_3,a_{s-1}} + \ldots$ and the new gauge parameter will appear in the terms that we omitted in \eqref{delta-omega}. Proceeding in this way one would introduce the additional auxiliary fields $\O^{b_t,a_{s-1}}$ with $2 \leq t \leq s-1$ that enter Vasiliev's equations in four and higher dimensions (see, e.g., \cite{Fradkin:1986ka, Lopatin:1987hz}), while this step will not be relevant in the ensuing discussion.

We can now move to study the equations of motion. The variation of the action with respect to the spin connection gives
\begin{equation}
\delta_\omega S =\frac{1}{16\pi G} \int K_{abc} \ww \left( \cT^a{}_{q_{s-2}} \ww \delta \omega^{b,\,cq_{s-2}} + \cK^{abc} \right) ,
\end{equation}
where the three-form $\cK^{abc}$ reads 
\begin{equation}\label{rest_T}
\begin{split}
& \cK^{abc} = \bvb_f \ww \omega^{f,\,a}{}_{q_{s-2}} \ww \delta \omega^{b,\,cq_{s-2}} \\
& + \frac{1}{D-2} \left( \frac{1}{(s-1)}\, \bvb^a \ww \omega^{b,}{}_{q_{s-1}} \ww \delta \omega^{c,\,q_{s-1}} + \bvb^a \ww \omega_{d,\,q_{s-2}}{}^b \ww \delta \omega^{d,\,cq_{s-2}} \right) .
\end{split}
\end{equation}
Its contribution to the variation vanishes thanks to
\begin{equation}\label{trick}
K_{abc} \ww \cK^{abc} = - 3!(D-3)!\, \delta^{mnr}_{abc}\, \cK_{mnr;}{}^{abc} \,\bvb\,\intd^Dx = 0 \,,
\end{equation}
where $\bvb$ denotes the determinant of the background vielbein, $\delta^{mnr}_{abc} = \delta^m{}_{[a} \delta^n{}_{b} \delta^r{}_{c]}$, and we introduced the form components as $\cK^{abc} = \bvb^m \ww \bvb^n \ww \bvb^r \cK_{mnr;}{}^{abc}$. The last identity can be verified by substituting the explicit expression \eqref{rest_T}.

The equation of motion for the spin connection therefore reads
\begin{equation}
K_{abq} \ww \cT_{q_{s-2}}{}^{a} = 0
\end{equation} 
since this expression already has the same irreducibility properties under permutations of its indices as the spin connection.
Taking the wedge product with another background vielbein and introducing the components of the two-form $\cT$ as in \eqref{trick} leads to
\begin{equation}
\cT_{aq;\,bq_{s-2}} = 0 \, .
\end{equation} 
In spite of the symmetrisation over the indices $q$, this expression suffices to conclude that the full torsion tensor vanishes, since its components can be rewritten as
\begin{equation}
\cT_{ab;\,q_{s-1}} = \sum_{k=1}^{s-1} \cT_{a(b;\,q_1\cdots \widehat{q_k}\cdots  q_{s-1})q_k} - (s-2)\cT_{aq;\,bq_{s-2}}=0\, .
\end{equation} 

The variation of the action with respect to the vielbein reads
\begin{equation}
\begin{split}
\delta_e S & =\frac{1}{16\pi G}   \int K_{abq} \ww \left( \nabla \omega^{a,b}{}_{q_{s-2}} + \frac{s(D+s-4)}{(D-2)\adsR^2}\, \bvb^a \ww e^b{}_{q_{s-2}} \right)\ww \delta e^{q_{s-1}} \\
& = \frac{1}{16\pi G} \int K_{abq} \ww \left( \cR^{a,\,b}{}_{q_{s-2}} - \bvb_c \ww \O^{\,ac,\,b}{}_{q_{s-2}} \right) \ww \delta e^{q_{s-1}}\,.
\end{split}
\end{equation} 
When we rewrite this in components, we obtain
\begin{equation}
\delta_e S =\frac{1}{16\pi G} \int \intd^Dx\,\bvb \, \delta_{abq}^{mnr} \left( \cR_{mn;}{}^{a,\,b}{}_{q_{s-2}} - \delta_{[m}{}^c \O_{n];\,c}{}^{a,\,b}{}_{q_{s-2}} \right) \delta e_{r;}{}^{q_{s-1}} \, .
\end{equation} 
Notice however that the functional derivative with respect to the hook component\footnote{By hook component we mean the projection of the vielbein on its irreducible component satisfying $e^{q;\,q_{s-1}} = 0$.} of $e^{r;\,q_s-1}$ vanishes due to the invariance under local Lorentz transformations of the action, which implies that the latter actually does not depend on this component. Therefore, it is enough to consider the variation of the action under variations $\delta e^{q;\,q_{s-1}}$. The symmetrisation eliminates the terms in $\O$, so that one can rewrite the equations of motion in terms of $\cR$ even if $\O$ does not enter the action \eqref{action_anyD}. This results in the equations of motion
\begin{equation}\label{Rrelation}
\h_{qq}\, \cR_{ab;}{}^{a,b}{}_{q_{s-2}} -  \cR_{aq;}{}^{a,}{}_{q_{s-1}} + \cR_{aq;\,q,}{}^{a}{}_{q_{s-2}} = 0\, .
\end{equation} 
One can also check explicitly that $\Omega$ drops out of this particular projection of the curvature tensor, which displays the same symmetries under permutation of its indices as the field we varied. Taking a trace of \eqref{Rrelation} and substituting back, we finally arrive at the equations of motion
\begin{subequations}
\begin{align}
\cT_{mn;\,q_{s-1}} & = 0 \, , \label{torsion_anyD} \\
\cR_{aq;}{}^{a,}{}_{q_{s-1}} & = 0\, . \label{curvature_anyD}
\end{align}
\end{subequations}

The torsion equation \eqref{torsion_anyD} allows one to express the following part of the spin connection in terms of the vielbein:
\begin{equation}\label{solving_torsion}
\omega_{[m;\,n],q_{s-1}} = \nabla_{\![m} e_{n];q_{s-1}} \, .
\end{equation}
This is actually the only part of the spin connection entering the equation of motion \eqref{curvature_anyD},\footnote{This is the case because $\omega_{[m;\,n],q_{s-1}}$ is actually the only part of the spin connection entering the action \eqref{action_anyD}. Indeed, $\omega_{m;\,n,\,q_{s-1}}$ can be decomposed into its symmetric and antisymmetric components in the indices $m$ and $n$, and the former can be further decomposed as
\[
\omega_{m;\,m,\,q_{s-1}} = 2\,\omega_{[q;\,m],mq_{s-2}} + \frac{s-2}{s-1} \left( \omega_{m;\,m,\,q_{s-1}} + \frac{s-1}{2}\, \omega_{q;\,q,\,q_{s-3}mm} \right) .
\]
The first term can be expressed in terms of the vielbein using again \eqref{solving_torsion}, while the combination between parentheses corresponds to the $\{s-1,2\}$ Young projection of $\omega_{m;\,n,\,q_{s-1}}$. As such, it can be gauged away using the Stueckelberg symmetry with parameter $\theta$ in \eqref{varO_anyD}, thus implying that this combination actually does not even enter the action \eqref{action_anyD}.}  
that can be rewritten as
\begin{equation}
\begin{split}
0 = \cR_{aq;}{}^{a,}{}_{q_{s-1}} = & \ \nabla^a \omega_{[q;\,a],q_{s-1}} + (s-1)\,\nabla_{\!q} \omega_{[a;\,q],\,q_{s-2}}{}^a \\
& + \frac{s-1}{2\adsR^2} \left( (D+s-4)\, \phi_{q_s} + \frac{s}{2}\, \h_{qq}\, \phi_{q_{s-2}a}{}^a \right) ,
\end{split}
\end{equation}
where we also introduced the symmetric component of the vielbein,
\begin{equation}
\phi_{q_{s}} = e_{q;\,q_{s-1}} \, .
\end{equation}
Substituting the solution \eqref{solving_torsion} to the torsion constraint and using the commutator
\begin{equation}
\nabla_{\![\m} \nabla_{\!\n]} v^a = \frac{1}{\adsR^2} \, \bvb_{[\m}{}^b \bvb_{\n]}{}^a v_b\,,
\end{equation}
one eventually recovers Fronsdal's equation \eqref{eq:fronsdalEqAds} for the field $\phi_{q_{s}}$.

\section{More on the classical Miura transformation}\label{sec:Miura}

We start by analysing the Drinfeld--Sokolov reduction of $\mathfrak{gl}(N)$ in the single-row gauge (where $u_1$ can be different from zero). The transformations are generated by the charges
\begin{align}
    Q(\lambda) &= -\frac{k}{2\pi} \int \intd\theta\ \tr \lambda u\nonumber\\
    &= -\frac{k}{2\pi} \int \intd\theta\, \frac{6}{N(N^2-1)} \mathrm{tr}_{N\times N} \lambda u\nonumber\\
    &= -\frac{k}{2\pi}\frac{6}{N(N^2-1)}\int \intd\theta\, (\lambda_{11}u_1 + \dots +\lambda_{N1}u_N)\nonumber\\
    &=-\frac{k}{2\pi}\frac{6}{N(N^2-1)}\int \intd\theta\ \res (L\,\lambda^1 )\,.
\end{align}
Here, we expressed the normalised trace $\tr$ (that satisfies $\tr L_0^2=\frac{1}{2}$) in terms of the matrix trace, and in the last step we wrote the integrand in terms of the residue (see~\eqref{sym:defofres}) of the product of the differential operators $L$ (containing the fields $u_j$, see~\eqref{sym:defofL}) and $\lambda^1$ (containing the elements of the first column of the matrix $\lambda$, see~\eqref{sym:defoflambda1}).

We can infer the Poisson bracket by requiring that the charges generate the transformation~\eqref{sym:transformationofL} of the fields $u_j$ (encoded in $L$),
\begin{subequations}
\begin{align}
    \delta_\lambda L &= \{ Q(\lambda),L\}\\
    \Longrightarrow \ 
    L(\lambda^1 L)_+ -(L\lambda^1)_+ L &= -\frac{k}{2\pi}\frac{6}{N(N^2-1)}\left\{\int \intd\theta'\ \res (L\,\lambda^1 ) ,L \right\} .
\end{align}
\end{subequations}
From here we can derive the Poisson bracket of linear functionals of the field $u_j$:
\begin{align}
    &\left\{ \int \intd\theta' \ \res L(\theta')\lambda^1(\theta') , \int \intd\theta\  \res L(\theta)\mu^1 (\theta)\right\} \nonumber\\
    &\qquad = \int \intd\theta \ \res \left(\left\{ \int \intd\theta' \res L(\theta')\lambda^1(\theta'), L(\theta)\right\}\mu^1(\theta)\right)\nonumber\\
    &\qquad = - \frac{2\pi N(N^2-1)}{6k} \int \intd\theta \ \res \left( L (\lambda^1 L)_+ \mu^1 - (L\lambda^1)_+ L \mu^1\right)(\theta)\nonumber\\
    &\qquad = \frac{2\pi N(N^2-1)}{6k} \int \intd\theta \ \res \left( (L\lambda^1)_+ L \mu^1 - (\lambda^1 L)_+ \mu^1 L\right)(\theta)\,. 
\end{align}
In the last step we used that the integral over a residue is cyclic,
\begin{equation}
    \int \intd\theta\ \res XY = \int \intd\theta\ \res YX\, .
\end{equation}
The Poisson bracket can be straightforwardly generalised to arbitrary functionals $F$ and $G$ of the fields,
\begin{equation}\label{Miura:Poissonbracket}
    \{ F,G\} = \frac{2\pi N(N^2-1)}{6k} \int \intd\theta \ \res \left( \left( L \frac{\delta F}{\delta L}\right)_+ L\frac{\delta G}{\delta L} - \left( \frac{\delta F}{\delta L}L\right)_+ \frac{\delta G}{\delta L}L  \right) . 
\end{equation}
Here, the functional derivative by the $u$'s is encoded in the pseudo-differential operator setting as
\begin{equation}
    \frac{\delta F}{\partial L} = \sum_{k=1}^N \partial^{-N+k-1} \frac{\delta F}{\delta u_k}\, .
\end{equation}
Then for example
\begin{equation}
    \frac{\delta}{\delta L} \int \intd\theta\ \res L \lambda^1 = \sum_{k=1}^N \partial^{-N+k-1} \frac{\delta}{\delta u_k}\sum_{i=1}^N u_i \lambda_{i1} = \lambda^1\, .
\end{equation}
We now use this formulation to show that the Poisson bracket is induced by the Poisson bracket of free fields via the Miura transformation. To this end we employ the following theorem stating that for a factorised differential operator $L=L_1 L_2$ we have
\begin{multline}\label{Miura:KWthm}
    \int \intd\theta\ \res \left( \left( L \frac{\delta F}{\delta L}\right)_+ L\frac{\delta G}{\delta L} - \left( \frac{\delta f}{\delta L}L\right)_+ \frac{\delta G}{\delta L}L \right) \\
    = \sum_{i=1}^2 \int \intd\theta\ \res \left( \left( L_i \frac{\delta F}{\delta L_i}\right)_+ L_i\frac{\delta G}{\delta L_i} - \left( \frac{\delta F}{\delta L_i}L_i\right)_+ \frac{\delta G}{\delta L_i}L_i \right) .
\end{multline}
For a proof we follow~\cite{Dickey:1997wia}. We denote the order of $L_i$ by $N_i$, then $N=N_1+N_2$ is the order of $L$.
First, we note that
\begin{equation}\label{Miura:deltaF}
    \delta F =  \int \intd\theta\ \res \frac{\delta F}{\delta L}\delta L = \int \intd\theta\ \res \left( L_2 \frac{\delta F}{\delta L} \delta L_1 + \frac{\delta F}{\delta L} L_1 \delta L_2 \right) ,
\end{equation}
and hence
\begin{equation}\label{Miura:deltaF1and2}
    \frac{\delta F}{\delta L_1} = \left(L_2 \frac{\delta F}{\delta L} \right)_{-} + \dots \quad \text{and} \quad \frac{\delta F}{\delta L_2} = \left(\frac{\delta F}{\delta L} L_1\right)_{-}+ \dots \, ,
\end{equation}
where $\dots$ refers to terms of the form $\partial^{-k} a_k$ with $k>N_i$ which are undetermined by~\eqref{Miura:deltaF}, but which are irrelevant for the following computations.  
Then
\begin{align}
    &\int \intd\theta\ \res \left( \left( L_1 \frac{\delta F}{\delta L_1}\right)_+ L_1\frac{\delta G}{\delta L_1} - \left( \frac{\delta F}{\delta L_1}L_1\right)_+ \frac{\delta G}{\delta L_1}L_1 \right)\nonumber \\
    &= -\int \intd\theta\ \res \left( \left( L_1 L_2\frac{\delta F}{\delta L}\right)_- L_1\left(L_2\frac{\delta G}{\delta L}\right)_- - \left( L_2\frac{\delta F}{\delta L}L_1\right)_- \left( L_2\frac{\delta G}{\delta L}\right)_-L_1 \right)\nonumber\\
    &= -\int \intd\theta\ \res \bigg( \left( L_1 L_2\frac{\delta F}{\delta L}\right)_- L_1 L_2\frac{\delta G}{\delta L} - \left( L_2\frac{\delta F}{\delta L}L_1\right)_- L_2\frac{\delta G}{\delta L} L_1 \nonumber \\
    &\qquad \qquad -\left( L_1 L_2\frac{\delta F}{\delta L}\right)_- L_1\left(L_2\frac{\delta G}{\delta L}\right)_+ + \left( L_2\frac{\delta F}{\delta L}L_1\right)_- \left( L_2\frac{\delta G}{\delta L}\right)_+ L_1 \bigg)\nonumber\\
    &= -\int \intd\theta\ \res \bigg( \left( L_1 L_2\frac{\delta F}{\delta L}\right)_- L_1 L_2\frac{\delta G}{\delta L} - \left( L_2\frac{\delta F}{\delta L}L_1\right)_- L_2\frac{\delta G}{\delta L} L_1 \bigg) \, ,
\end{align}
where in the last step one can omit in the second row the subscript $-$ such that the terms cancel. A similar computation gives
\begin{align}
    &\int \intd\theta\ \res \left( \left( L_2 \frac{\delta F}{\delta L_2}\right)_+ L_2\frac{\delta G}{\delta L_2} - \left( \frac{\delta F}{\delta L_2}L_2\right)_+ \frac{\delta G}{\delta L_2}L_2 \right)\nonumber \\
    &= -\int \intd\theta\ \res \bigg( \left(  L_2\frac{\delta F}{\delta L}L_1\right)_-  L_2\frac{\delta G}{\delta L}L_1 - \left( \frac{\delta F}{\delta L}L_1 L_2 \right)_- \frac{\delta G}{\delta L} L_1 L_2\bigg) \,.
\end{align}
Using these results to evaluate the right-hand side of~\eqref{Miura:KWthm} we straightforwardly obtain the left-hand side.

Using the above theorem iteratively on the Miura transformation $L=L_1 \dots L_N$ with $L_i=\partial +v_i$ (see \eqref{sym:classicalMiura}) we see that the Poisson bracket for the fields $u_k$ is induced by a Poisson bracket\footnote{This is the content of the Kupershmidt--Wilson theorem ~\cite{Dickey:1997wia}.} for the fields $v_i$ which in terms of modes reads
\begin{align}
    \{ v_{i,m} , v_{i,n} \} &= 
    \left\{ \frac{1}{2\pi}\int \intd\theta \ v_i (\theta)e^{im\theta} , \frac{1}{2\pi}\int \intd\theta \ v_i (\theta)e^{in\theta} \right\}\nonumber\\
    &=
    \frac{N(N^2-1)}{12\pi k} \int \intd\theta \ \res \Big( \left( L_i \partial^{-1} e^{im\theta}\right)_+ L_i \partial^{-1} e^{in\theta} \nonumber\\
    &\qquad \qquad \qquad \qquad \qquad - \left( \partial^{-1} e^{im\theta}L_i\right)_+ \partial^{-1}e^{in\theta}L_i  \Big) \nonumber\\
    &= \frac{N(N^2-1)}{12\pi k} \int \intd\theta \  ( in) e^{i(m+n)\theta} \nonumber\\
    &= -\frac{N(N^2-1)}{6 k} im\,\delta_{m,-n}\, .
\end{align}
These are the Poisson brackets of a free field.

We now discuss the Drinfeld--Sokolov reduction based on $\mathfrak{sl}(N)$. When we constrain the variations of $L$ such that $u_1=0$, there is some ambiguity in the definition of the functional derivative. Demanding that  
\begin{equation}
    \delta F = \int \intd\theta\, \res \frac{\delta F}{\delta L} \delta L
\end{equation}
for the constrained variations, we see that we have the freedom to redefine $\frac{\delta F}{\delta L}\to \frac{\delta F}{\delta L} + \partial^{-N}a(\theta)$ for an arbitrary function $a$. We know that the Poisson bracket is given by
\begin{equation}
    -\frac{k}{2\pi}\frac{6}{N(N^2-1)}\left\{\int \intd\theta'\ \res (L\,\lambda^1 ) ,L \right\} = L(\lambda^1 L)_+ -(L\lambda^1)_+ L
\end{equation}
where $\lambda^1$ satisfies $\res [L,\lambda^1]=0$ (see~\eqref{sym:slNreduction}). This can be generalised to a Poisson bracket of arbitrary functionals $F$ and $G$ as in~\eqref{Miura:Poissonbracket}, but where we now use the ambiguity of the functional derivative to require
\begin{equation}\label{Miura:ambiguityFD}
    \res \left[L,\frac{\delta F}{\delta L}\right] = \res \left[L,\frac{\delta G}{\delta L}\right] = 0 \, .
\end{equation}
The Poisson bracket can be reproduced by free fields. For a factorised operator $L=L_1 L_2$, the computation above leading to the result~\eqref{Miura:KWthm} carries over as long as we still use the relation~\eqref{Miura:deltaF1and2}  for the functional derivative with respect to $L_1$ and $L_2$ in terms of $\frac{\delta F}{\delta L}$. We then have
\begin{equation}
   \sum_i \res \left[L_i,\frac{\delta F}{\delta L_i}\right] = \res \left[L_1,\left(L_2 \frac{\delta F}{\delta L}\right)_-\right] + \res \left[ L_2, \left(\frac{\delta F}{\delta L}L_1\right)_-\right] ,
\end{equation}
where the subscript $-$ can be omitted. Writing out the commutators, two terms cancel, and we arrive at
\begin{equation}
   \sum_i \res \left[L_i,\frac{\delta F}{\delta L_i}\right] = \res \left[L,\frac{\delta F}{\delta L}\right] .
\end{equation}
Fixing the ambiguity of the functional derivative by requiring~\eqref{Miura:ambiguityFD} is then equivalent to 
\begin{equation}
    \sum_i \res \left[L_i,\frac{\delta F}{\delta L_i}\right] = 0\, .
\end{equation}
For $L=(\partial + v_1)\dots(\partial +v_N)$, the requirement on the functional derivative becomes
\begin{equation}
    \sum_{i=1}^N \frac{\delta F}{\delta v_i} = 0\, .
\end{equation}
When we apply this to the modes of the fields $v_i$, we find 
\begin{equation}
    \frac{\delta}{\delta v_k(\theta)} \int \intd\theta' \, v_i(\theta')e^{im\theta'} = \left(\delta_{ik}-\frac{1}{N}\right)e^{im\theta}\, .
\end{equation}
The Poisson brackets then become
\begin{equation}
    \{ v_{i,m} , v_{j,n} \} = -\frac{N(N^2-1)}{6 k}\left(\delta_{ij}-\frac{1}{N}\right) im\,\delta_{m,-n}\, ,
\end{equation}
which is the mode version of the result stated in~\eqref{sym:freePoisson}.

\section{Constrained Hamiltonian systems}\label{sec:hamiltonian}

In this appendix, we briefly summarise some important concepts that are useful to discuss Hamiltonian systems with constraints. For simplicity, we restrict the discussion to finite-dimensional systems, but similar considerations apply to field theories in Hamiltonian form. See~\cite{Gitman1990, Wipf:1993xg, Henneaux:1994lbw} for ampler introductions to this material.

We consider a dynamical system with Hamiltonian $H$ on a phase space with canonical coordinates $q^1,\dots,q^n$, $p_1,\dots,p_n$ and the Poisson bracket 
\begin{equation}
    \{q^i,p_j\} = \delta^i{}_j \,. 
\end{equation}
We then assume that the system is confined to a constraint surface $M$ defined by the vanishing of some phase space functions $\phi_i(q,p)$ ($i=1,\dots,N$), which we denote as constraints. We demand a certain regularity for these functions such that locally around a point in $M$, we can take the functions $\phi_i$ as the first $N$ coordinates of a local coordinate system on phase space. An equality for phase space functions that only holds on the constraint surface $M$ is denoted by the symbol $\approx$, so in particular we have
\begin{equation}
    \phi_i \approx 0\, .
\end{equation}
We further assume that the constraints are consistent with the time evolution, i.e., we demand that $\dot{\phi}_i \approx 0$.
Notice that on the constraint surface, the Hamiltonian can be modified by adding functions that vanish on the constraint surface,\footnote{If $H$ is derived from a Lagrangian theory in which some of the relations between coordinates and momenta are not invertible, then it is necessary to add the corresponding constraints to $H_T$; see, e.g., \cite{Wipf:1993xg}.}
\begin{equation}
    H_T = H +\sum u^i \phi_i\, ,
\end{equation}
for some functions $u^i$. Consistency of the constraints with time evolution is then formulated as
\begin{equation} \label{constr-evolution}
    \dot{\phi}_i = \{ \phi_i , H_T \}\approx \{\phi_i,H\}+\sum_j u^j \{ \phi_i,\phi_j\} \approx 0\,,
\end{equation}
so that adding the terms in $u^i$ to the total Hamiltonian 
can avoid the introduction of further constraints.
When the matrix 
\begin{equation}
    C_{ij}=\{\phi_i,\phi_j\}
\end{equation}
is not degenerate on the constraint surface ($\det (C_{ij})\not\approx 0$), the Lagrange multipliers $u^i$ are determined by consistency. Otherwise, there is some freedom to redefine them, which signals the appearance of a gauge symmetry.

A constraint $\phi$ that has vanishing Poisson brackets with all constraints on the constraint surface,
\begin{equation}
    \{\phi,\phi_i\}\approx 0\quad \text{for all}\ i\,,
\end{equation}
is called a \emph{first-class constraint}, otherwise it is called \emph{second-class}. 
If a first-class constraint $\phi$ is present, then, as we emphasised above, the matrix $C_{ij}$ is degenerate and the total Hamiltonian $H_T$ is not uniquely defined, leaving the ambiguity of adding a term $u\,\phi$ which is not fixed by \eqref{constr-evolution}. An infinitesimal time evolution of a phase space function then has the ambiguity
\begin{equation}\label{delta_f}
    \delta f = u\,\delta t\,\{f,\phi\}\, . 
\end{equation}
The different results of time evolution have to be considered as equivalent, and we should interpret~\eqref{delta_f} as a gauge transformation, with $u$ interpreted as the gauge parameter per unit of time. 
A priori, this freedom can be fixed by introducing additional constraints to be interpreted as gauge-fixing conditions. A full gauge fixing is then leading to an extended set of constraints that are all of second class.

If only second-class constraints are present (i.e.~if the matrix $C_{ij}$ is not degenerate on the constraint surface), there are no gauge symmetries, and we can take $M$ as a constrained phase space with a new Poisson bracket (called the \emph{Dirac bracket}) defined by
\begin{equation}\label{DefDiracbracket}
    \{f,g\}_* = \{f,g\}-\sum_{ij} \{f,\phi_i\}C^{ij}\{\phi_j,g\}\,,
\end{equation}
where $C^{ij}$ denotes the coefficients of the inverse of the matrix $(C_{ij})$. The Dirac bracket satisfies the Jacobi identity and
\begin{align}
    \{f,\phi_i\}_* &\approx 0\,, & \dot{f} &\approx \{f,H \}_*\,.
\end{align}
Therefore, the dynamical system can be treated as a system without constraints defined only on the constraint surface.

\section{Solutions to selected exercises}\label{sec:solutions}

\solu{ex:torsion}{
\hypertarget{sol:torsion}{Considering} the relation $\e^{abc} \cT_c = 0$ and writing explicitly the spacetime indices one obtains
\begin{equation}
\omega_{[\m}{}^a e_{\n]}{}^b - \omega_{[\m}{}^b e_{\n]}{}^a = \e^{abc} \pr_{[\m} e_{\n]c} \, .
\end{equation}
Contracting with the inverse vielbein $e^\n{}_{\!b}$ one then gets
\begin{equation}\label{eqsol:torsion}
\omega_\m{}^a + \omega_b{}^b e_\m{}^a = 2 \,\e^{abc} e^\n{}_{\!b}\, \pr_{[\m} e_{\n]c} \, .
\end{equation}
The trace of the spin connection $\omega_b{}^b := e^\n{}_{\!b} \omega_\n{}^b$ can be computed by contracting \eqref{eqsol:torsion} with $e^\m{}_{\!a}$. Substituting the result in the same equation one obtains \eqref{eq:spinsolved}.
}

\solu{ex:diffeos}{
\hypertarget{sol:diffeos}{Plugging} $\xi^a = e_\m{}^a v^\m$ and $\Lambda^a = \omega_\m{}^a v^\m$ into the variation $\delta e_\m{}^a$ (see~\eqref{eq:localsymmEH}), we obtain after some algebra
\begin{equation}
     \delta e_\m{}^a = \, v^\n \partial_\n e_\m{}^a + e_\n{}^a \partial_\m v^\n + 2v^\n \big( \partial_{[\m}e_{\n]}{}^a + \epsilon^a{}_{bc} \omega_{[\m}{}^b e_{\n]}{}^c \big)\, .
\end{equation}
The last term in parentheses contains the components of the torsion $\cT^a$ (see~\eqref{eq:torsion-spin2}) and vanishes on-shell.}

\solu{ex:sl2EHaction}{
\hypertarget{sol:sl2EHaction}{Plugging} the definitions $A=(\omega^a + \frac1{\adsR} e^a)J_a$ and $\tilde{A}=(\omega^a - \frac1{\adsR} e^a)J_a$ in \eqref{eq:differenceCS}, 
and using partial integration to show that $\int \tr (\intd e \wedge \omega ) = \int \tr ( e \wedge \intd \omega )$, we obtain
\begin{align}
\frac{\adsR}{16 \pi G} \int \tr \left( \frac4\adsR e \ww \intd \omega  + \frac{4}\adsR e \ww \omega \ww \omega + \frac{4}{3 \adsR^3} e \ww e \ww e \right) ,
\end{align}
where we have also used the cyclicity of the trace, that is $\tr( e\ww \omega \ww \omega ) = \tr ( \omega \ww e \ww \omega ) = \tr (\omega \ww \omega \ww e)$ and similarly for other terms. Recalling $R=\intd \omega + \omega \wedge \omega$, we see that this is precisely \eqref{eq:sl2EHaction}.

Using the definition of the bilinear form \eqref{eq:bilinearFormSl2}, one can then rewrite the action \eqref{eq:sl2EHaction} in terms of $e^a$ and $\omega^a$ obtaining the Einstein--Hilbert action \eqref{eq:EH2}. The latter expression is valid for arbitrary values of the cosmological constant, so that the same is true for  \eqref{eq:sl2EHaction}.
}

\solu{ex:linearizationHS}{
\hypertarget{sol:linearizationHS}{The} zero-curvature condition~\eqref{eq:curvatureConstraint2} can be rewritten (using the definition~\eqref{curvature_3D} of $\cR^{a_1 \cdots a_{s-1}}$) as 
\begin{equation}
    d \omega^{a_1 \cdots a_{s-1}} + (s-1) \epsilon_{bc_1}{}^{(a_1}\delta_{c_2}^{a_2} \cdots \delta_{c_{s-1}}^{a_{s-1})}\, \Big(  \bsc^{b} \ww \omega^{c_1 \cdots c_{s-1}} + \frac{1}{\adsR^2} \,  \bvb^{b} \ww e^{c_1 \cdots c_{s-1}} \Big)= 0\,.
\end{equation}
When we compare this to~\eqref{eom_after_linearisation}, we can directly read off the commutation relations~\eqref{eq:hscommutatorsl2}. The analogous inspection of~\eqref{eq:curvatureConstraint1} leads to the same result.

To determine the trace, one can insert $e=\bvb_a J^a + e_{a_1 \cdots a_{s-1}}J^{a_1 \cdots a_{s-1}}$ and $\omega=\bsc_a J^a + \omega_{a_1 \cdots a_{s-1}}J^{a_1 \cdots a_{s-1}}$ in the action~\eqref{eq:sl2EHaction} and expand to quadratic order in the higher-spin fields. Comparison to the higher-spin action~\eqref{eq:frameAction} then leads to the results~\eqref{killing_J} for the trace $\tr$.
}

\solu{ex:LWcommrel}{%
\hypertarget{sol:LWcommrel}{The} relation~\eqref{LWcommrel} for $m=-1$,
\begin{equation}
[L_{-1},W^s_n] = -(s-1+n)W^s_{n-1} \, ,
\end{equation} 
directly follows from the definition~\eqref{Wsrecursion}. For $m=0$, one can prove inductively that 
\begin{equation}
[L_0,W^s_n] = -n \,W^s_n \, .
\end{equation}
For this purpose one starts from
\begin{equation}
[L_0, W^s_{s-1}]=[L_0,L_1^{s-1}]=-(s-1)L_1^{s-1}=-(s-1)W^s_{s-1}\, ,
\end{equation}
and in the induction step one uses Jacobi identity:
\begin{align}
    [L_0,W^s_m] &= -\frac{1}{m+s}[L_0,[L_{-1},W^s_{m+1}]]\nonumber\\
    &=-\frac{1}{m+s}\big( [L_{-1},[L_0,W^s_{m+1}]] + [[L_0,L_{-1}],W^s_{m+1}]  \big)\nonumber\\
    &=-\frac{1}{m+s}\big( -(m+1)[L_{-1},W^s_{m+1}] + [L_{-1},W^s_{m+1}] \big)\nonumber\\
    &= -m\, W^s_m\, .
\end{align}
The relation~\eqref{LWcommrel} for $m=1$,
\begin{equation}
[L_1,W^s_n] = (s-1-n) W^s_{n+1}\, ,
\end{equation} 
can be proven analogously. It trivially holds for $n=s-1$, and for $n=s-2$ one obtains 
\begin{equation}
[L_1,W^s_{s-2}]= -\frac{1}{2s-2}[L_1,[L_{-1},L_1^{s-1}]
= -\frac{1}{2s-2}[[L_1,L_{-1}],L_1^{s-1}] = 
W^s_{s-1}\, .
\end{equation} 
Similarly one can prove the relation for lower $n$ by induction.

Again by employing Jacobi identity one concludes that
\begin{equation}
[L_0,[W^{s}_m,W^{t}_n]] = [[L_0,W^{s}_m],W^{t}_n] + [W^{s}_m,[L_0,W^{t}_n]] = -(m+n)[W^{s}_m,W^{t}_n]\, ,
\end{equation}
hence $[W^{s}_m,W^{t}_n]$ has to be a linear combination of terms with mode number $m+n$.
}

\solu{ex:WWcommrel}{
\hypertarget{sol:WWcommrel}{Thanks} to the results of exercise~\ref{ex:LWcommrel}, the commutator with the highest mode number, $[ W^3_2 , W^3_1 ]$, can only be proportional to $W^4_3$,
\begin{equation}\label{WWhighestmodenumber} 
[ W^3_2 , W^3_1 ] = \alpha W^4_3 \, .
\end{equation} 
Indeed, it cannot be proportional to any $W^s_3$ with $s > 4$ since the Jacobi identities imply that it is annihilated by the adjoint action of $L_1$ and this property is satisfied only by $W^4_3$.

In this particular example, this first result already suffices to rule out the presence of $W^3_{m+n}$ in the expansion of the commutator in a basis of the $\mathfrak{hs}[\lambda]$ algebra. This is so because there are at most $10$ independent commutators $[ W^3_m , W^3_n ]$ and we must already have $7$ independent $W^4_{m+n}$ generators in the expansion. The remaining $3$ independent generators can thus only be proportional to the 3 independent components of $L_{m+n}$.

The same outcome could be obtained by considering that $[ W^3_2 , W^3_0 ]$ is the only commutator with ``total mode number'' $2$. As a result, it must be proportional to $W^4_2$, thus excluding a contribution in $W^3_2$. At the next step, on the other hand, one encounters two commutators with mode number $1$, i.e.\ $[ W^3_2 , W^3_{-1} ]$ and $[ W^3_1 , W^3_0 ]$, so that they can both be proportional to a linear combination of $W^4_1$ and $L_1$. Notice that this argument can be generalised to fix the structure of commutators involving generators with the same ``total angular momentum'' quantum number as follows:
\begin{equation}
[ W^s_m , W^s_n ] = \sum_{k=1}^{s-1} c_k[m,n]\, W^{2(s-k)}_{m+n} \, .
\end{equation}
We conclude that for $s=3$ we have the Lie bracket
\begin{equation}
[ W^3_m , W^3_n ] = c_0 [m,n] \,W^4_{m+n} + c_1 [m,n] \, L_{m+n}\, .
\end{equation} 
Both sides of the equation have a definite behaviour under the adjoint action with $L_p$, and this completely fixes the dependence on $m$ and $n$ of each $c_i[m,n]$ up to a constant factor. Indeed, acting on the commutator~\eqref{WWhighestmodenumber} with $L_{-1}$ leads (upon using the Jacobi identity on the left-hand side) to
\begin{equation}
(-3) [ W^3_2 , W^3_0 ] = \alpha \,(-6) \,W^4_2 \quad  \Longrightarrow \quad  [ W^3_2 , W^3_0 ] = 2\alpha\, W^4_2\, , 
\end{equation} 
which determines the coefficient $c_0[2,0]$ in terms of $c_0[2,1]=\alpha$. Acting once more with $L_{-1}$ we obtain
\begin{equation}
(-4) [ W^3_1 , W^3_0 ] + (-2) [ W^3_2 , W^3_{-1} ] = 2\alpha\,(-5)\, W^4_1 \, .
\end{equation}
Additionally we can look for a linear combination of the terms on the right-hand side that vanishes when we apply $L_1$ and therefore has to be proportional to $L_1$,
\begin{equation}
(-3) [ W^3_1 , W^3_0 ] + [ W^3_2 , W^3_{-1} ] = \frac{5}{2}\beta\, L_1 \, ,
\end{equation} 
where the factor $\frac{5}{2}$ has been chosen to get the final result in a convenient form. The previous two equations determine $[W^3_1,W^3_0]$ and $[W^3_2,W^3_{-1}]$, and one finds
\begin{equation}
    [W^3_1,W^3_0]=\alpha\,W^4_1 - \frac{1}{2}\beta\,L_1 \,,\qquad
    [W^3_2,W^3_{-1}]=3\alpha\,W^4_1 + \beta\,L_1\, .
\end{equation}
Analogously, all other commutators $[W^3_m,W^3_n]$ are determined in this way leaving only two constants $\alpha$ and $\beta$,
\begin{subequations}
\begin{align}
    [W^3_1,W^3_{-1}]&=2\alpha\,W^4_0 -\frac{\beta}{2}\,L_0 \,,\\
    [W^3_2,W^3_{-2}]&=4\alpha\,W^4_0 + 4\beta\,L_0 \,,\\
    [W^3_1,W^3_{-2}]&=3\alpha\,W^4_{-1}+\beta\,L_{-1} \,,\\
    [W^3_0,W^3_{-1}]&=\alpha\,W^4_{-1} - \frac{\beta}{2}\,L_{-1} \,,\\
    [W^3_0,W^3_{-2}]&=2\alpha\,W^4_{-2} \,,\\
    [W^3_{-1},W^3_{-2}]&=\alpha\,W^4_{-3}\, .
\end{align}
\end{subequations}
One can then explicitly check that the formula~\eqref{W3W3commutator} precisely summarises these commutators.
}

\solu{ex:W3mexplicit}{%
\hypertarget{sol:W3mexplicit}{Starting} from $W^3_2=(L_1)^2$ one can use the recursion relation~\eqref{Wsrecursion} to obtain
\begin{subequations}
\label{explicitexpressionsforsl3gen}
\begin{alignat}{2}
    W^3_1 &= -\frac{1}{4}[L_{-1},W^3_2] &&= \frac{1}{2}\big(L_0 L_1 + L_1 L_0\big) \,,\\
    W^3_0 &= -\frac{1}{3}[L_{-1},W^3_1] &&= \frac{1}{6}\big(L_{-1}L_1+L_1 L_{-1} +4 (L_0)^2 \big) \,,\\
    W^3_{-1} &= -\frac{1}{2}[L_{-1},W^3_0] &&= \frac{1}{2}\big(L_{-1}L_0 + L_0 L_{-1}\big) \,,\\[5pt]
    W^3_{-2} &= -[L_{-1},W^3_{-1}] &&= (L_{-1})^2\, .
\end{alignat}
\end{subequations}
In a similar way we can determine the first few basis elements $W^4_m$ starting from $W^4_3=(L_1)^3$:
\begin{subequations}
\begin{alignat}{2}
    W^4_2 &= -\frac{1}{6}[L_{-1},W^4_3]&&=L_{(0}L_1 L_{1)} \,,\\
    W^4_1 &= -\frac{1}{5}[L_{-1},W^4_2]&&= \frac{1}{5} L_{(-1}L_1 L_{1)} + \frac{4}{5} L_{(0} L_0 L_{1)}\, .
\end{alignat}
\end{subequations}
Notice that
\begin{subequations}
\begin{align}
    L_{(-1}L_1 L_{1)} &= \frac{1}{2}\big(L_{-1}L_1^2 + L_1^2L_{-1}\big)-\frac{1}{3}L_1 \,,\\
    L_{(0} L_0 L_{1)} &= \frac{1}{2}\big(L_0^2L_1 +L_1L_0^2\big)-\frac{1}{6}L_1\nonumber \\
    &=C_2 L_1 +\frac{1}{2}\big(L_{-1}L_1^2 + L_1^2 L_{-1}\big)-\frac{2}{3}L_1\,,
\end{align}
\end{subequations}
which can be checked by elementary computations. Using the explicit value for the Casimir we can write $W^4_1$ as
\begin{equation}
    W^4_1 = \frac{1}{2}\big(L_{-1}L_1^2+L_1^2L_{-1}\big)+\frac{\lambda^2-4}{5}L_1 \, .
\end{equation}
From the explicit expressions for $W^3_m$ we compute
\begin{equation}
    [W^3_2,W^3_1] = \frac{1}{2}[L_1^2,L_0L_1+L_1L_0]=2\,L_1^3 = 2\,W^4_3\, ,
\end{equation}
and by comparison to the general form~\eqref{W3W3commutator} of the commutator we conclude that $\alpha[\lambda]=2$. A straightforward computation shows that
\begin{align}
    [W^3_2,W^3_{-1}] &= \frac{1}{2}[L_1^2,L_{-1}L_0+L_0L_{-1}]\nonumber\\
    &=3\big(L_1^2 L_{-1} + L_{-1}L_1^2\big) + (\lambda^2-4)L_1\, .
\end{align}
On the other hand, using $\alpha[\lambda]=2$ in the general commutator~\eqref{W3W3commutator} we obtain
\begin{equation}
    [W^3_2,W^3_{-1}] = 6 \,W^4_1 +\beta[\lambda] \,L_1\, ,
\end{equation}
and by comparison with the previous expression we can read off
\begin{equation}
    \beta[\lambda] = \frac{4-\lambda^2}{5}\, .
\end{equation}
}

\solu{ex:PoissonbracketofGs}{
\hypertarget{sol:PoissonbracketofGs}{From} the definition~\eqref{sym:PBforfunctionals} for the Poisson bracket we obtain
\begin{align}
    \{G(\Lambda),G(\Gamma)\} &= \frac{2\pi}{k}\int_{D_2}\intd \rho \intd \theta\ \tr \left( \frac{\delta G(\Lambda)}{\delta A_\rho} \frac{\delta G(\Gamma)}{\delta A_\theta} - (\Lambda \leftrightarrow \Gamma) \right)\nonumber \\
    &= \frac{k}{2\pi} \int_{D_2}\intd \rho \intd \theta\ \tr \big( (\partial_\theta \Lambda + [A_\theta,\Lambda])(-\partial_\rho \Gamma - [A_\rho,\Gamma]) - (\Lambda \leftrightarrow \Gamma) \big)\,,
\end{align}
where we used the variation~\eqref{varG} of $G$. Rearranging the terms and using partial integration in $\theta$, we find
\begin{align}
    \{G(\Lambda),G(\Gamma)\} &= \frac{k}{2\pi} \int_{D_2}\intd \rho \intd \theta\ \tr \big(\partial_\rho (\Lambda \partial_\theta \Gamma) - \partial_\rho(A_\theta [\Lambda,\Gamma]) +[\Lambda,\Gamma]F_{\rho\theta}\big)\nonumber\\
    &= G([\Lambda,\Gamma]) + \frac{k}{2\pi}\int_{S^1}\intd\theta \ \tr (\Lambda \partial_\theta \Gamma)\,,
\end{align}
reproducing~\eqref{sym:algebraofG}.
}

\solu{ex:DScondition}{%
\hypertarget{sol:DScondition}{From}~\eqref{gauge-fixed-A_2} we read off the relation between $a_\theta$ and $A_\theta$, which then determines the $\rho$-dependence. Considering the expansion of $a_\theta$ in terms of the basis elements $L_m$ and $W_m^{s}$, we have to understand the $\rho$-dependence of $b^{-1}(\rho)W^{s}_m b(\rho)$. From the commutation relations~\eqref{LWcommrel_2} we see that
\begin{equation}
    L_0 \,W^s_m = W^s_m (L_0 -m)
\end{equation}
which means that whenever we move a mode $W^s_m$ through $L_0$ from the right, $L_0$ is shifted by $-m$. Hence
\begin{equation}
    e^{-\rho L_0}\, W^s_m \,e^{\rho L_0} = e^{m\rho}\,W^s_m\,.
\end{equation}
Any mode $W^s_m$ with a positive mode number $m$ then corresponds to a term that diverges for $\rho \to \infty$. Imposing the AdS condition~\eqref{sym:AdSconditions} then leads to the expansion~\eqref{DScond_1} of $a_\theta$. A similar argument leads to the expansion~\eqref{DScond_2} of $\tilde{a}_\theta$.
}

\solu{ex:checkPoissonL}{%
\hypertarget{sol:checkPoissonL}{We} can rewrite~\eqref{deltaL} as
\begin{align}
    \delta \cL (\theta')&= \int \intd \theta \,\delta(\theta-\theta')\Big(\epsilon(\theta) \, \cL'(\theta) + 2 \,\epsilon'(\theta) \,\cL (\theta)+\frac{1}{2}\,\epsilon'''(\theta) \Big)\nonumber\\
    &= \int \intd \theta \,\epsilon (\theta)\Big( -\delta(\theta-\theta')\cL'(\theta) -2\,\delta'(\theta-\theta')\,\cL(\theta) -\frac{1}{2} \delta'''(\theta-\theta')\Big)
    \, .
\end{align}
When we compare this with~\eqref{deltaLcanonical} where we insert the charge~\eqref{integratedcharge},
\begin{align}
    \delta \cL(\theta') = \frac{k}{2\pi} \int \intd\theta \,\epsilon(\theta) \{\cL(\theta),\cL(\theta')\}\,,
\end{align}
we read off the Poisson bracket~\eqref{PoissonL}.
}

\solu{ex:deltaaSL3}{
\hypertarget{sol:deltaaSL3}{We} start from the highest-weight form~\eqref{sym:sl3hwgofa} of $a$ and a general transformation parameter $\lambda$ given in~\eqref{sym:sl3gaugeparameter}. Using the commutation relations~\eqref{sl3commrel} of $\mathfrak{sl}(3,\mathbb{R})$, one can compute
\begin{align}
    \delta_\lambda a &= \lambda' + [a,\lambda]\nonumber\\ 
    &= (\epsilon'+\epsilon^0)L_1\nonumber\\ 
    &\quad +((\epsilon^0)'+2\epsilon^{-1}-2 \cL \epsilon+ 4 \cW \chi) L_0\nonumber\\ 
    &\quad +((\epsilon^{-1})'-\cL \epsilon^0 + \cW \chi^1)L_{-1}\nonumber\\ 
    &\quad +(\chi' + \chi^1) W_2\nonumber\\ 
    &\quad +((\chi^1)'+2\chi^0-4\cL \chi)W_1\nonumber\\ 
    &\quad +((\chi^0)'+3\chi^{-1}-3\cL \chi^1)W_0\nonumber\\
    &\quad + ((\chi^{-1})'+4\chi^{-2}-2\cL \chi^0-4\cW \epsilon)W_{-1}\nonumber\\ 
    &\quad + ((\chi^{-2})'-2\cL \chi^{-1}-2\cW \epsilon^0)W_{-2}\, .\label{sl3deltaaexplicit}
\end{align}
Requiring the vanishing of all coefficients except for those of $L_{-1}$ and $W_{-2}$, one obtains a system of differential equations that allows us to express all $\epsilon^m$ and $\chi^n$ in terms of (derivatives of) $\epsilon$ and $\chi$:
\begin{subequations}
\begin{align}
    \epsilon^0 &= -\epsilon' \,,\\ 
    \epsilon^{-1}&= \frac{1}{2} \epsilon'' + \cL \epsilon - 2 \cW \chi \,,\\ 
    \chi^1 &= -\chi' \,,\\ 
    \chi^0 &= \frac{1}{2}\chi''+2\cL \chi \,,\\
    \chi^{-1}&= -\frac{1}{6}\chi^{(3)}-\frac{5}{3}\cL\chi'-\frac{2}{3}\cL'\chi \,,\\ 
    \chi^{-2}&= \frac{1}{24}\chi^{(4)} + \frac{2}{3}\cL \chi''+\frac{7}{12}\cL'\chi' +\frac{1}{6}\cL'' \chi+\cL^2\chi+\cW \epsilon\, .
\end{align}
\end{subequations}
Inserting these expressions into~\eqref{sl3deltaaexplicit} one can read off the transformation~\eqref{sym:sl3transformofcW} of $\cW$.
}

\solu{ex:deltaw}{
\hypertarget{sol:deltaw}{The} transformation of $\cW$ is given in~\eqref{sym:sl3transformofcW}.
Similarly to~\eqref{sym:expansionofell} we expand
\begin{equation}\label{expansionofw}
    \cW(\theta) = \frac{1}{k}\sum_{n\in \mathbb{Z}} \cW_n e^{-in\theta}\, .
\end{equation}
With $\chi=0$ and $\epsilon=\epsilon_{(0)}e^{im\theta}$ we obtain
\begin{equation}
    \frac{1}{k}\sum_n \delta \cW_n e^{-in\theta} = \epsilon_{(0)} \sum_n (-in)e^{-i(n-m)\theta} \cW_n + 3 \epsilon_{(0)} \sum_n (im)e^{-i(n-m)\theta}\cW_n\,.
\end{equation}
From here we can deduce the transformation $\delta \cW_n$ as in~\eqref{sym:deltaw}.
}

\solu{ex:transfofw}{%
\hypertarget{sol:transfofw}{The} transformation of $\cW$ with $\epsilon=0$ and $\chi=\chi_{(0)}e^{im\theta}$ can be obtained from~\eqref{sym:sl3transformofcW}. Expanding $\cW$ and $\cL$ in Fourier modes as in~\eqref{expansionofw} and~\eqref{sym:expansionofell}, we find
\begin{align}
\sum_n \delta \cW_n e^{-in\theta} & = \frac{k}{24}(im)^5 \chi_{(0)}e^{im\theta} + \frac{5}{6}(im)^3\chi_{(0)} \sum_n \cL_n e^{i(m-n)\theta}\nonumber\\ 
&\quad +\frac{15}{12}(im)^2 \chi_{(0)}\sum_n(-in)\cL_n e^{i(m-n)\theta}\nonumber\\ 
&\quad +\frac{9}{12} (im)\chi_{(0)}\sum_n (-in)^2\cL_n e^{i(m-n)\theta}\nonumber\\ 
&\quad +\frac{1}{6}\chi_{(0)} \sum_n (-in)^3\cL_n e^{i(m-n)\theta}\nonumber\\ 
&\quad +\frac{1}{k}\frac{8}{3}(im)\chi_{(0)}\sum_{n,p}\cL_n \cL_{-p} e^{i(m-n+p)\theta}\nonumber\\ 
&\quad +\frac{1}{k}\frac{8}{3}\chi_{(0)} \sum_{n,p}\cL_n\cL_{-p}e^{i(m-n+p)\theta}\,.
\end{align}
By changing the summation variable as $n\mapsto n+m$ in the first four lines and $n\mapsto n+m+p$ in the last two lines we can deduce the transformation of $\cW_n$, 
\begin{align}
    \delta \cW_n & = i\frac{k}{24} m^5\chi_{(0)}\delta_{m+n,0} \nonumber\\ 
    & \phantom{=}\, +\frac{i}{12} \chi_{(0)}\big( -10m^3+15m^2(n+m)-9m(n+m)^2+2(n+m)^3\big)\cL_{m+n}\nonumber\\ 
    & \phantom{=}\, +\frac{i}{k}\frac{8}{3}\sum_p (m+p)\cL_{n+m+p} \cL_{-p}\, .
\end{align}
The sum in the last term can be rewritten as 
\begin{align}
    \sum_p (m+p)\cL_{n+m+p} \cL_{-p} &=  \frac{1}{2}\left(\sum_p (m+p)\cL_{n+m+p} \cL_{-p}+ \sum_p (-n-p)\cL_{n+m+p} \cL_{-p} \right)\nonumber\\ 
    &= (m-n)\sum_p \cL_{n+m+p} \cL_{-p} \, ,
\end{align}
where in the first step in one of the sums we have changed the summation label $p\mapsto -m-n-p$.
After rearrainging some terms we arrive at the result for $\delta \cW_n$ given in~\eqref{sym:transfofw}.
}

\solu{ex:WWbracket}{%
\hypertarget{sol:WWbracket}{The} transformation of $a$ is generated by the charge $Q(\lambda)$ through the Dirac bracket,
\begin{equation}
    \delta_\lambda a = \{ Q(\lambda),a\}\, .
\end{equation}
Generalising the discussion that led to~\eqref{integratedcharge}, the charge $Q(\lambda)$ for a gauge parameter $\lambda$ of the form~\eqref{sym:sl3gaugeparameter} is given as 
\begin{equation}
    Q(\lambda) = \frac{k}{2\pi} \int \intd\theta\, \big(\epsilon(\theta)\cL(\theta)-\chi(\theta) \cW (\theta)\big)
\end{equation}
(note that $\tr( W_2 W_{-2})=1$ (see~\eqref{sl3trace}) compared to $\tr (L_1 L_{-1})=-1$). If $\epsilon=0$ and $\chi=\chi_{(0)}e^{im\theta}$, we find 
\begin{align}
Q(\lambda)
&= -\frac{k}{2\pi} \int \intd\theta\, \chi_{(0)} e^{im\theta} \frac{1}{k}\sum_n \cW_n e^{-in\theta}\nonumber\\ 
&= -\chi_{(0)} \cW_m\, .
\end{align}
We then get 
\begin{align}
i\{ \cW_m,\cW_n\} &= -\frac{i}{\chi_{(0)}}\delta \cW_n\nonumber\\ 
& = \frac{1}{12} \bigg(\frac{k}{2}m^{5}\,\delta_{m+n,0} - (m-n) (2m^{2}+2n^{2}-mn)\cL_{m+n} \nonumber\\
&\qquad \qquad \quad + \frac{16}{k} (m-n)\sum_{q\in \mathbb{Z}}\cL_{m+n+q}\cL_{-q} \bigg)\, ,\label{WWbracketintermediateresult}
\end{align}
where we used the result~\eqref{sym:transfofw} for $\delta \cW_n$.
Finally, we want to express this in terms of the modes $\hat{\mathcal{L}}_{m}=-\cL_{m}+\frac{k}{4}\,\delta_{m,0}$ and the quadratic field $\Lambda$ given in~\eqref{sym:DefofLambda},
\begin{align}
    \Lambda_p & = \sum_q \hat{\mathcal{L}}_{p+q} \hat{\mathcal{L}}_{-q}\nonumber\\ 
    &= \sum_q \cL_{p+q}\cL_{-q} + \frac{k}{2} \hat{\mathcal{L}}_p -\left(\frac{k}{4} \right)^2\delta_{p,0}\, .
\end{align}
Using this result in~\eqref{WWbracketintermediateresult} and replacing $k=\frac{c}{6}$, we obtain the Dirac bracket given in~\eqref{sym:WWbracket}.
}

\solu{ex:sl3ugauge}{%
\hypertarget{sol:sl3ugauge}{The} transformation of $a$ with a general matrix-valued parameter $\lambda$ takes the form
\begin{equation}\label{sl3ugaugedeltaa}
    \delta_\lambda a = \begin{pmatrix}
        \lambda'_{11}-\sqrt{2}\lambda_{12} + u_{11} &  \lambda'_{12} - \sqrt{2}\lambda_{13}-u_2\lambda_{11} + u_{12} &  \lambda'_{13} -u_3 \lambda_{11} + u_{13}\\
        \lambda'_{21} + \sqrt{2}(\lambda_{11}-\lambda_{22}) & \lambda'_{22}+\sqrt{2}(\lambda_{12}-\lambda_{23})-u_2 \lambda_{21} & \lambda'_{23}+\sqrt{2}\lambda_{13}-u_3 \lambda_{21}\\ 
        \lambda'_{31}+\sqrt{2}(\lambda_{21}-\lambda_{32}) & \lambda'_{32}+\sqrt{2}(\lambda_{22}-\lambda_{33})-u_2 \lambda_{31} & \lambda'_{33} + \sqrt{2}\lambda_{23}-u_3\lambda_{31}
    \end{pmatrix},\,
\end{equation}
where
\begin{equation}
    u_{1j} = u_2 \lambda_{2j}+u_3\lambda_{3j}\, .
\end{equation}
Requiring that the second and third row vanish, we can express the coefficients $\lambda_{ij}$ in terms of the entries $\lambda_{j1}$ of the first column as we discuss now. First notice that the 3-1-entry and the 2-1-entry of~\eqref{sl3ugaugedeltaa} allows us to conclude 
\begin{subequations}
\begin{align}
    \lambda_{32} &= \frac{1}{\sqrt{2}}\lambda'_{31} + \lambda_{21} \,,\\ 
    \lambda_{22} &= \frac{1}{\sqrt{2}} \lambda'_{21} + \lambda_{11} \, .
\end{align}
\end{subequations}
Now we can use the 3-2-entry to determine 
\begin{equation}
    \lambda_{33} = \frac{1}{2}\lambda''_{31}+\sqrt{2} \lambda'_{21} + \lambda_{11} - \frac{1}{\sqrt{2}} u_2 \lambda_{31} \, .
\end{equation}
Using the entries 3-3, 2-3 and 2-2 of~\eqref{sl3ugaugedeltaa} we arrive at
\begin{subequations}
\begin{align}
    \lambda_{23} &= -\frac{1}{2\sqrt{2}} \lambda^{(3)}_{31}-\lambda''_{21} -\frac{1}{\sqrt{2}}\lambda'_{11}+\frac{1}{2}\partial(u_2 \lambda_{31}) + \frac{1}{\sqrt{2}} u_3 \lambda_{31}\,,\\ 
    \lambda_{13}&=\frac{1}{4}\lambda^{(4)}_{31}+\frac{1}{\sqrt{2}}\lambda^{(3)}_{21} + \frac{1}{2}\lambda''_{11}-\frac{1}{2\sqrt{2}}\partial^2 (u_2 \lambda_{31})-\frac{1}{2}\partial(u_3 \lambda_{31})+\frac{1}{\sqrt{2}}u_3 \lambda_{21}\,,\\ 
    \label{lambda12}
    \lambda_{12} &= -\frac{1}{2\sqrt{2}}\lambda^{(3)}_{31}-\frac{3}{2}\lambda''_{21}-\sqrt{2}\lambda'_{11}+\frac{1}{2}\partial(u_2\lambda_{31})+\frac{1}{\sqrt{2}}(u_2\lambda_{21}+u_3\lambda_{31})\,.
\end{align}
\end{subequations}
It remains to implement the condition that the 1-1-entry of~\eqref{sl3ugaugedeltaa} vanishes. Using the expression~\eqref{lambda12} of $\lambda_{12}$, we get a condition on $\lambda_{11}$,
\begin{equation}
    \lambda'_{11}= -\frac{1}{6}\lambda^{(3)}_{31}-\frac{1}{\sqrt{2}}
    \lambda''_{21}+\frac{1}{3\sqrt{2}} \partial (u_2\lambda_{31})\,.
\end{equation}
We can then read off the transformation of $u_2$ and $u_3$ from the 1-2- and 1-3-entry of~\eqref{sl3ugaugedeltaa}, and we arrive at~\eqref{deltalambdau1} and~\eqref{deltalambdau2}.%
} 

\solu{ex:pdoconstraint}{%
\hypertarget{sol:pdoconstraint}{From} the expression~\eqref{sym:slNsinglerowdeltaa} for $\delta_\lambda a_{ij}$ we find
\begin{equation}
    \sum_{j=1}^{N}\delta_\lambda a_{ij}\partial^{N-j} =\sum_{j=1}^{N} \big(\lambda_{ij}'- \lambda_{i-1\,j} + \lambda_{i\,j+1} - u_{j}\lambda_{i1}\big)\partial^{N-j}\, .
\end{equation}
We rewrite
\begin{equation}
    \lambda_{ij}'\partial^{N-j} = \partial \circ \big(\lambda_{ij}\partial^{N-j}\big) - \lambda_{ij}\partial^{N-j+1}\, ,
\end{equation}
and obtain (using the definition~\eqref{sym:deflambdai} of $\lambda_i$)
\begin{align}
    \sum_{j=1}^{N}\delta_\lambda a_{ij}\partial^{N-j} &= \partial \circ \lambda_i -\lambda_{i-1} -\sum_{j=1}^{N} \big(\lambda_{ij}\partial^{N-j+1} - \lambda_{i\,j+1}\partial^{N-j} + \lambda_{i1}u_{j}\partial^{N-j}\big)\nonumber\\ 
    &= \partial \circ \lambda_i -\lambda_{i-1}-\lambda_{i1}\partial^N -\lambda_{i1}\sum_{j=1}^{N} u_{j}\partial^{N-j}\, ,
\end{align}
from which the result~\eqref{sym:pdoconstraint} follows.
}

\solu{ex:solutionofrecursion}{%
\hypertarget{sol:solutionofrecursion}{We} insert the proposed solution~\eqref{sym:solutionofrecursion} into the right-hand side of the recursion relation~\eqref{sym:recursionrelation} (setting $i$ to $N-i$) and obtain
\begin{align}
\partial \lambda_{N-i}-\lambda_{N-i\,1}L &= \partial^{i+1}\big( \lambda^1 L\big)_+ -\partial \big(\partial^i \lambda^1\big)_+ L -\lambda_{N-i\,1}L \nonumber\\
&=\partial^{i+1}\big( \lambda^1 L\big)_+ -\partial\!\!\! \sum_{j=N+1-i}^N \!\!\!\partial^{-N-1+i+j} \lambda_{j1}\, L -\lambda_{N-i\,1}L \nonumber\\
&= \partial^{i+1}\big( \lambda^1 L\big)_+ -\sum_{j=N-i}^N \partial^{-N+i+j} \lambda_{j1}\, L \nonumber\\
&= \partial^{i+1}\big( \lambda^1 L\big)_+ -\big(\partial^{i+1}\lambda^1 \big)_+ L \, ,
\end{align}
which coincides with $\lambda_{N-i-1}$ given by the proposed solution~\eqref{sym:solutionofrecursion}.
}

\solu{ex:transformationofL}{%
\hypertarget{sol:transformationofL}{We} insert the solution~\eqref{sym:solutionofrecursion} for $\lambda_i$ into the expression~\eqref{sym:transformationofLunsimplified} for the transformation of $L$ and obtain
\begin{align}
    \delta_\lambda L &= \partial \big(\partial^{N-1} \big(\lambda^1 L\big)_+ - \big(\partial^{N-1}\lambda^1\big)_+L\big)-\lambda_{11} L\nonumber \\
    &\quad +\sum_{k=2}^N u_k \big(\partial^{N-k}\big(\lambda^1 L\big)_+-\big(\partial^{N-k}\lambda^1\big)_+L\big)\nonumber\\
    &= L\big(\lambda^1 L\big)_+ -\big(\partial^N \lambda^1\big)_+L -\sum_{k=2}^N u_k \big(\partial^{N-k}\lambda^1\big)_+L\nonumber\\
    &= L\big(\lambda^1 L\big)_+ - \big(L\lambda^1\big)_+L\, ,
\end{align}
which gives the desired result~\eqref{sym:transformationofL}.
}

\solu{ex:bracket_L_Lambda}{%
\hypertarget{sol:bracket_L_Lambda}{\begin{align}
i\{\hat{\mathcal{L}}_{m} ,\Lambda_{n} \} &= 
\sum_{q\in \mathbb{Z}}i\{\hat{\mathcal{L}}_{m} , \hat{\mathcal{L}}_{n+q}\hat{\mathcal{L}}_{-q} \}\nonumber\\
&= \sum_{q\in \mathbb{Z}}
\Big( (m-n-q)\hat{\mathcal{L}}_{m+n+q}+\delta_{m,-n-q}\frac{c}{12}m
(m^{2}-1)\Big) \hat{\mathcal{L}}_{-q} \nonumber \\
&\quad +
\sum_{q\in \mathbb{Z}}
\hat{\mathcal{L}}_{n+q}\Big( (m+q)\hat{\mathcal{L}}_{m-q} +\delta_{m,q}\frac{c}{12}m (m^{2}-1)\Big)\nonumber \\
&= \sum_{q\in \mathbb{Z}} (m-n-q)\hat{\mathcal{L}}_{m+n+q}\hat{\mathcal{L}}_{-q} +
\sum_{q\in \mathbb{Z}}(2m+q)\hat{\mathcal{L}}_{n+m+q}\hat{\mathcal{L}}_{-q} \nonumber\\
&\quad + \frac{c}{6}m (m^{2}-1)\hat{\mathcal{L}}_{m+n}\nonumber\\
&=(3m-n) \Lambda_{m+n} + \frac{c}{6}m (m^{2}-1)\hat{\mathcal{L}}_{m+n}\, ,
\end{align}}
where in the third equation we shifted the summation variable in the
second sum, $q\to q+m$.%
}

\solu{ex:Virasoro}{%
\hypertarget{sol:Virasoro}{We} rewrite the commutator of the modes $L_{m}$ and $L_{n}$ of $T(z)$
as a contour integral (see~\eqref{quantum:mode_commutator}),
\begin{align}
\big[L_{m},L_{n}\big] &= 
\frac{1}{(2\pi i)^{2}} \oint_{0}\intd w
\oint_{w} \intd z\; z^{m+1}\,w^{n+1} \mathcal{R}\Big(  T (z) T (w)\Big) \nonumber\\
&= \frac{1}{(2\pi i)^{2}} \oint_{0}\intd w
\oint_{w} \intd z\;  z^{m+1}\,w^{n+1} \bigg( \frac{c/2}{(z-w)^{4}} + \frac{2T(w)}{(z-w)^{2}}
+ \frac{\partial_{w}T (w)}{z-w} \bigg) \, ,
\end{align}
where we inserted the singular part of the OPE of $T$ with
itself~\eqref{quantum:TTOPE}. Now we firstly evaluate the residue
integral in $z$, succeeded by the residue integral in $w$:
\begin{align}
\big[L_{m},L_{n}\big] &= \frac{1}{2\pi i} \oint_{0}\intd w\;
w^{n+1} 
\bigg(\frac{c}{2}\cdot \frac{1}{6} (m+1)m (m-1)w^{m-2} \nonumber\\
&\mspace{160mu}+ 2T(w) (m+1)w^{m}+ \partial_{w}T (w)w^{m+1} \bigg)\nonumber\\
&= \frac{c}{12} m (m^{2}-1) \delta_{m+n,0} + 2 (m+1)L_{m+n} \nonumber\\
&\quad  + \frac{1}{2\pi i} \oint_{0}\intd w\; \sum_{p\in\mathbb{Z}} L_{p}
(-p-2) w^{m+n-p-1}\nonumber\\
&= \frac{c}{12} m (m^{2}-1) \delta_{m+n,0} + (m-n)L_{m+n} \, .
\end{align}
}

\solu{ex:operator_state}{%
\hypertarget{sol:operator_state}{We} want to find the state corresponding to the current $W^{(s)}(z)$. For
that purpose we act with $W^{(s)} (z)$ on the vacuum and take $z\to
0$. Notice that the modes $W^{(s)}_{m}$ annihilate the vacuum for
$m>-s$, so that we find
\begin{equation}
W^{(s)} (z)\Omega = \sum_{m \leq -s} z^{-m-s} \, W^{(s)}_{m}\Omega\ 
\xrightarrow{z\to 0} \ W^{(s)}_{-s}\Omega \, .
\end{equation}
The state corresponding to $W^{(s)}$ is therefore $W^{(s)}_{-s}\Omega$.
}

\solu{ex:NOP_of_currents}{
\hypertarget{sol:NOP_of_currents}{The} OPE of two currents of spin $s$ and $t$, respectively, is given by
\begin{equation}
W^{(s)} (z) W^{(t)} (w) = V \big(W^{(s)} (z-w)W^{(t)}_{-t}\Omega ;w \big)
= \sum_{n\leq t} (z-w)^{-n-s}\,
V\big(W^{(s)}_{n}W^{(t)}_{-t}\Omega ;w\big)\, .
\end{equation}
The normal ordered product $\boldsymbol{(}W^{(s)}W^{(t)}\boldsymbol{)}$ of
the fields occurs in the term where the exponent of $(z-w)$ is zero,
therefore
\begin{equation}
\boldsymbol{(}W^{(s)}W^{(t)}\boldsymbol{)} (w) =  V\big(W^{(s)}_{-s}W^{(t)}_{-t}\Omega ;w\big)\, .
\end{equation}
}

\solu{ex:quasiprimary_current}{
\hypertarget{sol:quasiprimary_current}{If} the commutator of $L_{m}$ and $W_{n}^{(s)}$ is given
by~\eqref{quantum:commutator_L_Ws}, then we find
\begin{equation}
L_{1}W_{-s}^{(s)}\Omega = \big[ L_{1},W_{-s}^{(s)}\big]\Omega =
(2s-1) W_{-s+1}^{(s)}\Omega = 0\, ,
\end{equation}
hence $W^{(s)}$ is quasi-primary.

On the other hand, if $W^{(s)}$ is quasi-primary, it satisfies the
OPE~\eqref{quantum:OPE_T_quasiprimary} (with weight $h=s$). The OPE
determines the commutation relation of the modes
by~\eqref{quantum:mode_commutator}, so we find
\begin{align}
\Big[L_{m}, W^{(s)}_{n} \Big] &= \frac{1}{(2\pi i)^{2}}\oint_{0}\intd w
\oint_{w} \intd z\; z^{m+1}\,w^{n+s-1} T (z) W^{(s)}(w)\nonumber\\
&=  \frac{1}{(2\pi i)^{2}}\oint_{0}\intd w
\oint_{w} \intd z\; z^{m+1}\,w^{n+s-1}\nonumber\\
&\quad \times \bigg(\text{higher-order poles} + \frac{0}{(z-w)^{3}} +
\frac{sW^{(s)}(w)}{(z-w)^{2}} + \frac{\partial_{w}W^{(s)}(w)}{z-w} \bigg)\, .
\end{align}
For $-1\leq m\leq 1$, the term $z^{m+1}$ is at most quadratic, so the
higher-order poles in $(z-w)$ (fourth and higher) do not contribute to
the residue integral. Specialising on this case ($m\in\{-1,0,1\}$) we get
\begin{align}
\Big[L_{m}, W^{(s)}_{n} \Big] &= \frac{1}{(2\pi i)^{2}}\oint_{0}\intd w
\oint_{w} \intd z\; z^{m+1}\,w^{n+s-1} \bigg(\frac{sW^{(s)} (w)}{(z-w)^{2}} + \frac{\partial_{w}W^{(s)}(w)}{z-w} \bigg)\nonumber\\
&= \frac{1}{2\pi i} \oint_{0}\intd w\; w^{n+s-1} \bigg(s(m+1)w^{m}W^{(s)} (w) +w^{m+1}\partial_{w}W^{(s)} (w)\bigg)\nonumber\\
&= \sum_{p\in\mathbb{Z}} 
\frac{1}{2\pi i} \oint_{0}\intd w\;   w^{m+n-p-1} \big(s (m+1) + (-p-s) \big)W_{p}^{(s)}\nonumber\\
&= \big((s-1)m-n \big) W^{(s)}_{m+n} \, .
\end{align}
}

\solu{ex:modes_of_Lambda}{
\hypertarget{sol:modes_of_Lambda}{The} modes of $\partial^{2}T$ are 
\begin{equation}
(\partial^{2}T)_{m} = (m+2) (m+3)L_{m}\, ,
\end{equation}
the modes of $\boldsymbol{(}TT\boldsymbol{)}$ are
(see~\eqref{quantum:TTmodes})
\begin{equation}
\boldsymbol{(}TT\boldsymbol{)}_{m} = \sum_{p\geq 2}L_{-p}L_{m+p}
+\sum_{p\geq 1}L_{m+p}L_{-p} \, ,
\end{equation}
so that the modes of $\Lambda$ are
\begin{equation}
\Lambda_{m} = \sum_{p\geq 2}L_{-p}L_{m+p}
+\sum_{p\geq 1}L_{m+p}L_{-p} -\frac{3}{10} (m+2) (m+3) L_{m}\, .
\end{equation}
We now want to rewrite this such that the quadratic terms are mode
normal ordered (meaning that the $L_{m}$ with greater mode number
appears on the right). Let us consider the first sum. The operators
are mode normal ordered for $2p\geq -m$, therefore all terms that
appear are mode normal ordered for $m\geq -4$. (Similarly the terms in
the second sum are always mode normal ordered for $m\leq -2$.)

Assume now that $m<-4$. Then the second sum is already mode normal
ordered, and for the first sum we can write
\begin{equation}
\sum_{p\geq 2}L_{-p}L_{m+p} = \sum_{p\geq 2}:L_{-p}L_{m+p}: +
\sum_{p=2}^{N_{m}}[L_{-p},L_{m+p}] \, ,
\end{equation}
where 
\begin{equation}
N_{m} = \left\{\begin{array}{ll}
-\frac{m+2}{2} & \text{for}\ m\ \text{even}\\
-\frac{m+1}{2} & \text{for}\ m\ \text{odd.}
\end{array} \right.
\end{equation}
Using the commutation relations (notice that by assumption $m\not= 0$)
we find
\begin{align}
\sum_{p\geq 2}L_{-p}L_{m+p} &= \sum_{p\geq 2}:L_{-p}L_{m+p}: +
\sum_{p=2}^{N_{m}}\big(-m-2p \big)L_{m}\nonumber\\
&= \sum_{p\geq 2}:L_{-p}L_{m+p}: +\big( -m (N_{m}-1) -N_{m} (N_{m}+1)+2 \big)L_{m}\nonumber\\
&= \sum_{p\geq 2}:L_{-p}L_{m+p}: + g(m)L_{m} \, ,
\end{align}
with 
\begin{equation}
g (m) = \left\{\begin{array}{ll}
\frac{1}{4} (m+2) (m+4) & \text{for}\ m\ \text{even}\\
\frac{1}{4} (m+3)^{2} & \text{for}\ m\ \text{odd.}
\end{array} \right.
\end{equation}
Similarly one can show that for all $m\in\mathbb{Z}$ we have
\begin{equation}
\boldsymbol{(}TT\boldsymbol{)}_{m} = \sum_{p\geq 2}:L_{-p}L_{m+p}:
+\sum_{p\geq 1}:L_{m+p}L_{-p}: + g (m)L_{m} \, ,
\end{equation}
and therefore 
\begin{equation}
\Lambda_{m} =  \sum_{p\geq 2}:L_{-p}L_{m+p}:
+\sum_{p\geq 1}:L_{m+p}L_{-p}: + f (m)L_{m}
\end{equation}
with 
\begin{equation}
f (m) = \left\{\begin{array}{ll}
-\frac{1}{20} (m^{2}-4) & \text{for}\ m\ \text{even}\\
-\frac{1}{20} (m^{2}-9) & \text{for}\ m\ \text{odd.}
\end{array} \right.
\end{equation}
}

\solu{ex:consistencyscalarflat}
{
\hypertarget{sol:consistencyscalarflat}{When} we apply the differential $\intd$ to~\eqref{scalar:UnfoldedScalarFlat}, we arrive at the condition for consistency,
\begin{equation}
    \bar{h}_b\wedge \intd C^{a_1\dots a_n b}+\frac{n}{3}M^2\,\bar{h}^{(a_1}\wedge \intd C^{a_2\dots a_n)} = 0\,.
\end{equation}
Inserting~\eqref{scalar:UnfoldedScalarFlat}, we indeed find
\begin{align}
    \bar{h}_b\wedge \Big(\bar{h}_c\,C^{a_1\dots a_n bc} + \frac{n+1}{3}M^2 \,\bar{h}^{(a_1}C^{a_2\dots a_nb)}\Big)\nonumber\\
    +\frac{n}{3}M^2 \bar{h}^{(a_1}\wedge \Big(\bar{h}_c\,C^{a_2\dots a_n)c}+\frac{n-1}{3}M^2\,\bar{h}^{a_2}C^{a_3\dots a_n)}\Big)&=0\,,
\end{align}
where one uses that $\bar{h}_b \wedge \bar{h}_c$ is antisymmetric in $b$ and $c$.
}

\solu{ex:consistencyunfoldedKGAdS}
{
\hypertarget{sol:consistencyunfoldedKGAdS}{We} apply the Lorentz covariant differential $\nabla$ to~\eqref{scalar:UnfoldedKGAdS}, and we obtain after using~\eqref{eq:lorentzsquared}
\begin{equation}\label{consistencyfirststep}
    \frac{n}{\adsR^2}\bvb_b\wedge \bvb^{(a_1}C^{a_2\dots a_n)b} = -\bvb_b \wedge \nabla C^{a_1\dots a_n b}-\Big(\frac{n}{3}M^2-\frac{n(n-1)}{\adsR^2}\Big)\bvb^{(a_1}\wedge \nabla C^{a_2\dots a_n)}\,.
\end{equation}
On the right-hand side, we insert the expression~\eqref{scalar:UnfoldedKGAdS} for $\nabla C^{\dots}$, and check that both sides agree.
We only have to check this for the terms proportional to $\frac{1}{\adsR^2}$, the rest of the calculation is similar to the consistency of~\eqref{scalar:UnfoldedScalarFlat} (see exercise~\ref{ex:consistencyscalarflat}). The $\frac{1}{\adsR^2}$-terms on the right-hand side of~\eqref{consistencyfirststep} are
\begin{multline}
     \frac{n(n+1)}{\adsR^2} \bvb_b \wedge\bvb^{(a_1}C^{a_2\dots a_n b)} + \frac{n(n-1)}{\adsR^2} \bvb^{(a_1}\wedge \bvb_{b}C^{a_2\dots a_n) b}\\
    = \Big(\frac{n^2}{\adsR^2}-\frac{n(n-1)}{\adsR^2}\Big) \bvb_b \wedge \bvb^{(a_1}C^{a_2\dots a_n)b}\,,
\end{multline}
which indeed agrees with the left-hand side of~\eqref{consistencyfirststep}.
}

\solu{ex:consistencyscalarinHSbackground}
{
\hypertarget{sol:consistencyscalarinHSbackground}{Applying} the differential to the right-hand side of~\eqref{scalar:unfoldedinHSbackgroundA} and inserting back~\eqref{scalar:unfoldedinHSbackgroundA} one obtains
\begin{align}
    \intd(-AC + C\tilde{A}) &= -\intd A\,C+A\wedge \intd C + \intd C\wedge \tilde{A} + C\,\intd\tilde{A} \nonumber\\
    &= -\intd A\,C+A\wedge (-A\,C+C\,\tilde{A}) + (-A\,C+C\,\tilde{A})\wedge \tilde{A} + C\,\intd\tilde{A}\nonumber\\
    &= -(\intd A+A\wedge A)\,C+C\,(\intd\tilde{A}+\tilde{A}\wedge \tilde{A})\,,
\end{align}
which indeed vanishes when $A$ and $\tilde{A}$ are solutions of the Chern--Simons equation of motion~\eqref{CSeom}. The argument is similar for~\eqref{scalar:unfoldedinHSbackgroundB}.
}

\solu{ex:isomorphism}
{
	\hypertarget{sol:isomorphism}{Using} \eqref{commutationrelationsLL} and the fact that the epsilon tensor $\epsilon_{\ga \gb}$ is antisymmetric and normalized such that $\epsilon_{01}=1$, we obtain
 \begin{subequations}
	\begin{align}
	\left[ L_{11} , L_{00} \right]_\star &= 4\, \epsilon_{10} L_{10} = -4\, L_{10} \,, \\
	\left[ L_{01} , L_{00} \right]_\star &= 2\, \epsilon_{10} L_{00} = -2\, L_{00}  \,, \\
	\left[ L_{01} , L_{11} \right]_\star &= 2\, \epsilon_{01} L_{11} = 2\, L_{11} \,. 
	\end{align}
\end{subequations}
    Therefore by identifying $L_0=\frac{1}{2} L_{10}$, $L_{+1}=\frac{1}{2} L_{00}$ and $L_{-1}=\frac{1}{2} L_{11}$ we recover the $\mathfrak{sl}(2,\mathbb{R})$ commutation relations.
}

\solu{ex:differentialStarProduct}
{
\hypertarget{sol:differentialStarProduct}{The} differential version of the star product can be obtained by straightforward computation:
\begin{align}
(f\star g)(y) &= \frac{1}{(2\pi)^2} \int \intd^2u \, \intd^2v\, \,e^{iv u} \, e^{u \,\partial^{y'}}f(y') \, e^{v \,\partial^{y''}} g(y'') \big|_{y'=y''=y} \notag \\
&= \frac{1}{(2\pi)^2} \int \intd^2u \, \intd^2v\, \, e^{iv (u - i \partial^{y''})} \, e^{u \partial^{y'}}f(y')\, g(y'') \big|_{y'=y''=y} \notag \\
&= f(y) \,e^{-i \overset{\leftarrow}{\partial}_y \overset{\rightarrow}{\partial}_y }\, g(y) \,, 
\end{align}
where we have used Taylor's theorem, $f(y+u)=e^{u \,\partial^{y}}f(y)$, in the first line and the identity $\delta^{(2)}(u)= \frac{1}{(2 \pi)^2} \int \, \intd^2 v \, e^{i v u}$ in the last step. The arrow on the derivative indicates on which function the derivative acts (in the above formula $\overset{\leftarrow}{\partial}$ acts on $f$) without introducing additional signs.
}

\solu{ex:starproductRel}
{
\hypertarget{sol:starproductRel}{The} most convenient way to check \eqref{eq:basicStarProd} is using \eqref{eq:differentialFormOfStarProduct} from which we obtain
\begin{align}
y_\ga \star f(y) &= y_\ga \big(1-i \overset{\leftarrow}{\partial}{}^\sigma_y \overset{\rightarrow}{\partial}{}_\sigma^y\big) f(y) \notag \\ &= \big(y_\ga - i (\partial^\sigma_y y_\ga) \partial_\sigma^y\big) f(y) \notag \\ &= (y_\ga + i \partial_\ga^y) f(y)  \,,
\end{align}	
where we have used $\partial^\sigma_y y_\ga = \epsilon^{\sigma \delta} \partial^y_\delta y_\ga = \epsilon^{\sigma \delta} \epsilon_{\delta \ga}=-\delta^\sigma_\ga$.
In complete analogy one derives $f(y) \star y_\ga = (y_\ga - i \partial_\ga^y) f(y)$. From these results \eqref{eq:starCommAndAnticomm} follows immediately, 
\begin{equation}
y_\ga \star f(y) - f(y) \star y_\ga = 2i  \, \partial^y_\ga  f(y) \,,
\end{equation}
and analogously for the anticommutator.
}

\solu{ex:starCommAndAntiComm}
{
	\hypertarget{sol:starCommAndAntiComm}{To} prove these identities, it is important to remember that $L_{\ga \gb}=-\frac{i}{2}y_{(\ga} \star y_{\gb)}$ (see~\eqref{eq:AdSGenerators}). Using \eqref{eq:starCommAndAnticomm} we then obtain
	\begin{align}
	\left[ L_{\ga \gb}, f(y) \right]_\star &= -\frac{i}{2} \left[ y_{(\ga} \star y_{\gb)}, f(y)\right]_\star \notag \\ &=  -\frac{i}{2} \left( y_{(\ga} \star \left[ y_{\gb)}, f(y)\right]_\star +  \left[ y_{(\ga}, f(y)\right]_\star \star y_{\gb)}   \right) \notag \\
	&= \{ y_{(\ga }, \partial^y_{\gb)} f(y) \}_\star \notag \\
	&= 2 \, y_{(\ga} \,\partial_{\gb)}^y f(y) \,,
	\end{align}
    which confirms~\eqref{eq:LComm}. To prove the result \eqref{eq:LAntiComm} most efficiently we employ~\eqref{eq:basicStarProd} and compute
    \begin{align}
    \{ L_{\ga \gb}, f(y) \}_\star &= -\frac{i}{2} \left\{ y_{(\ga} \star y_{\gb)}, f(y)  \right\}_\star \notag \\ &= - \frac{i}{2} \left( \left( y + i\partial^y \right)_{(\ga} \left( y + i\partial^y \right)_{\gb)} f(y) + \left( y - i\partial^y \right)_{(\ga} \left( y - i\partial^y \right)_{\gb)} f(y) \right) \notag \\ &= -i \left( y_{\ga} y_{\gb} - \partial^y_\ga \partial^y_\gb \right) f(y) \,.
    \end{align}
    In the last step we have used $\partial^y_{(\ga} y_{\gb)}=0$. Applying the result for the anticommutator we can straightforwardly check \eqref{eq:quadCasimirValue}:
    \begin{align}
    -\frac14 \{L_{\ga \gb},L^{\ga \gb}\}_\star=  \frac14 i (y_\ga y_\gb - \partial^y_\ga \partial^y_\gb ) (- \frac{i}{2}) y^\ga y^\gb = -\frac34 \,.  
    \end{align}
    To obtain the last equation we have used the fact that $y_\ga y^\ga=0$ and $\partial^y_\ga \partial^y_\gb y^\ga y^\gb = \delta^\ga_\gb \partial_\ga^y y^\gb + 2 \partial^y_\ga y^\ga=6$.
}

\solu{ex:deformedComm}
{
	\hypertarget{sol:deformedComm}{First} we check the intermediate result given in the exercise
	\begin{align}
	\hat{y}^\ga \star \hat{y}_\ga = \frac12 \epsilon^{\ga \gb} \left[\hat{y}^\ga, \hat{y}^\gb\right]_\star = i \epsilon^{\ga \gb} \epsilon_{\ga \gb} ( 1 + \nu k) = -2 i (1 + \nu k) \,.
	\end{align}
	Similarly one derives
	\begin{align}
	L_{\ga \gb}=-\frac{i}{4} \left( \hat{y}_\ga \star \hat{y}_\gb + \hat{y}_\gb \star \hat{y}_\ga \right) =-\frac{i}{2} \hat{y}_{\ga} \star \hat{y}_{\gb} - \frac12 \epsilon_{\ga \gb} (1 + \nu k) \,.
	\end{align}
    These results can be used to check the value of the quadratic Casimir
    \begin{align}
    C_2 =& - \frac{1}{32} \big(-i \hat{y}_\ga \star \hat{y}_\gb - \epsilon_{\ga \gb} (1 + \nu k) \big) \big( -i \hat{y}^\ga \star \hat{y}^\gb - \epsilon^{\ga \gb} (1 + \nu k) \big) \nonumber \\
    =& -\frac{1}{32} \big( - \hat{y}^\ga \star \hat{y}^\gb \star \hat{y}_\gb \star \hat{y}_\ga  + 2i \, \hat{y}^\ga \star \hat{y}^\gb \, (1 + \nu k) - 2 (1+\nu k)^2  \big) \nonumber \\
    =& -\frac{1}{16} (3+2\nu k - \nu^2) \,,
    \end{align}
    where we have used $\hat{y}_\ga k = - k \hat{y}_\ga$ and $k^2=1$ to obtain the last equation.
}

\solu{ex:tracelessness}
{
	\hypertarget{sol:tracelessness}{Combining} \eqref{eq:vielbeiContrSpaceInd} and the fact that $f^{\sm_1 \cdots \sm_s}$ is completely symmetric it follows that
	\begin{align}
	f_{\sn}{}^{\sn \sm_3 \cdots \sm_s} \sim f^{\ga_1 \cdots \ga_{2s}} \,\epsilon_{\ga_1 \ga_3} \epsilon_{\ga_2 \ga_4} \,\bvb^{\sm_3}_{\ga_5 \ga_6} \cdots \bvb^{\sm_s}_{\ga_{2s-1} \ga_{2s}} = 0 \,.
	\end{align}
}

\solu{ex:nilpotency}
{
\hypertarget{sol:nilpotency}{Explicit} computation gives
\begin{align}
D^{AdS} D^{AdS} F &=\begin{aligned}[t]&(\intd \,\cdot\, - \cA^{AdS} \wedge \star \,\cdot\, + (-1)^{|F|+1} \,\cdot\, \wedge \star \cA^{AdS})\nonumber\\
&(\intd F - \cA^{AdS} \wedge \star F + (-1)^{|F|} F \wedge \star \cA^{AdS})\end{aligned} \\ 
&= \begin{aligned}[t]&F \wedge \star ( \intd \cA^{AdS} - \cA^{AdS} \wedge \star \cA^{AdS}) \\
&- (\intd\cA^{AdS} - \cA^{AdS} \wedge \star \cA^{AdS}) \wedge \star F  \\
& + (\cA^{AdS}-\cA^{AdS}) \wedge \star \intd F  \\  
&  + \intd F \wedge \star \left( (-1)^{|F|} \cA^{AdS} + (-1)^{|F|+1} \cA^{AdS} \right)  \\
& + (-1)^{|F|} \left( \cA^{AdS} \wedge \star F \wedge \star \cA^{AdS} - \cA^{AdS} \wedge \star F \wedge \star \cA^{AdS} \right) \,.\end{aligned}
\end{align}
The first two terms vanish upon imposing the equations of motion for $\cA^{AdS}$ while the other terms are identically zero.
}

\solu{prob:deformedSym}{
\hypertarget{sol:deformedSym}{In} order to simplify notation, we define
\begin{align}
\hat{y}_{\ga(s)} = \hat{y}_{(\ga_1} \star \dots \star \hat{y}_{\ga_s)} \,.
\end{align}	
We first show \eqref{eq:firstsymIdent}. We start with the fully symmetrised expression
\begin{equation}
    \hat{y}_{(\ga} \star \hat{y}_{\gb(s))} = \frac1{s+1} \left( \hat{y}_\ga \star \hat{y}_{\gb(s)} + \sum_{j=1}^s \hat{y}_{(\gb_1} \star \dots \hat{y}_{\gb_j} \star \hat{y}_{|\ga|} \star \hat{y}_{\gb_{j+1}} \star \dots \star \hat{y}_{\gb_s)}  \right) \, ,
\end{equation}
and in all terms on the right-hand side, we bring $\hat{y}_\ga$ to the left by using the commutation relations \eqref{eq:deformedCommuationRelationsyy}. We arrive at
\begin{align}
    \hat{y}_{(\ga} \star \hat{y}_{\gb(s))} =  \hat{y}_\ga \star \hat{y}_{\gb(s)} - \sum_{j=1}^s \sum_{m=0}^{j-1} 2i\,(1+\nu k (-1)^{j-1-m}) \epsilon_{\ga (\gb_1} \hat{y}_{\gb_2} \star \dots \star \hat{y}_{\gb_s)} \,,
\end{align}
where we also used that $k$ anticommutes with the $\hat{y}_\gb$.
After evaluating the sum, we obtain
\begin{align}
\hat{y}_{(\ga} \star \hat{y}_{\gb(s))}  = \hat{y}_\ga \star \hat{y}_{\gb(s)} - i \left( s + \frac{\frac12(1-(-1)^s)+s}{s+1} \nu k \right) \epsilon_{\ga (\gb_1} \hat{y}_{\gb_2} \star \dots \star \hat{y}_{\gb_s)} \,,
\end{align}
which confirms \eqref{eq:firstsymIdent}. The identity \eqref{eq:secIdent} can be derived along very similar lines.

Using this formula, it follows that
\begin{equation}
\hat{y}_{\ga_1}\star\hat{y}_{\ga_2}\star\hat{y}_{\gb(s)} = \hat{y}_{\ga_1}\star \left( \hat{y}_{(\ga_2} \star \hat{y}_{\gb(s))} + i \left( s + \tfrac{\frac12(1-(-)^s)+s}{s+1} \nu k \right) \epsilon_{\ga_2 (\gb_1} \hat{y}_{\gb_2} \star \dots \star \hat{y}_{\gb_s)} \right) .
\end{equation}
Applying the same formula again, one obtains that 
\begin{align}
&\hat{y}_{\ga_1}\star\hat{y}_{\ga_2}\star\hat{y}_{\gb(s)}  \nonumber\\
& \qquad  = \hat{y}_{(\ga_1}\star\hat{y}_{\ga_2}\star\hat{y}_{\gb(s))} + i \left( s + 1 + \tfrac{\frac12(1-(-)^{s+1})+s+1}{s+2} \nu k \right) \epsilon_{\ga_1(\ga_2} \hat{y}_{\gb(s))} \nonumber\\ & 
\qquad \qquad + i \left( s - \tfrac{\frac12(1-(-)^s)+s}{s+1} \nu k \right) \bigg\{ \tfrac{s+1}{s} \, \epsilon_{\ga_2(\ga_1} \hat{y}_{\gb(s))} -\tfrac{1}{s}\epsilon_{\ga_2 \ga_1} \hat{y}_{\gb(s)}\nonumber\\
& \qquad \qquad \qquad + i \left( s-1 + \tfrac{\frac12(1-(-)^{s-1})+s-1}{s} \nu k \right) \epsilon_{\ga_1(\gb_1}\epsilon_{|\ga_2| \gb_2} \hat{y}_{\gb_3} \star \dots \star \hat{y}_{\gb_s)} \bigg\} \,.
\end{align}
Following completely analogous steps, one can show that $\hat{y}_{\gb(s)}\star\hat{y}_{\ga_1}\star\hat{y}_{\ga_2}$ is equal to
\begin{align}
&\hat{y}_{\gb(s)}\star\hat{y}_{\ga_1}\star\hat{y}_{\ga_2}\nonumber \\
&\qquad = \hat{y}_{(\gb(s)}\star\hat{y}_{\ga_1}\star\hat{y}_{\ga_2)} - i \left( s + 1 + \tfrac{\frac12(1-(-)^{s+1})+(-)^{s+2}(s+1)}{s+2} \nu k \right) \epsilon_{\ga_2(\ga_1} \hat{y}_{\gb(s))} \nonumber\\ 
& \qquad\qquad - i \left( s + \tfrac{\frac12(1-(-)^s)+(-)^{s+1} s}{s+1} \nu k \right) \bigg\{ \tfrac{s+1}{s} \, \epsilon_{\ga_1(\ga_2} \hat{y}_{\gb(s))} -\tfrac{1}{s}\epsilon_{\ga_1 \ga_2}  \hat{y}_{\gb(s)}\nonumber\\
& \qquad \qquad \qquad - i \left( s-1 + \tfrac{\frac12(1-(-)^{s-1})+(-)^s(s-1)}{s} \nu k \right) \epsilon_{\ga_1(\gb_1}\epsilon_{|\ga_2| \gb_2} \hat{y}_{\gb_3} \star \dots \star \hat{y}_{\gb_s)} \bigg\} \,.
\end{align}
Using these two results (and simplifying them for even $s$), it is straightforward to check that \eqref{eq:deformedLAnticommutator} indeed holds when $s$ is even.
}

\solu{prob:backreaction}{
\hypertarget{sol:backreaction}{Two} identities are convenient to use
\begin{subequations}\label{staridentities}
\begin{align}
\left[ f(y) \star  z_\ga g(y,z) \right]_{z=0} &= \left[ i \partial^y_\ga f(y) \star g(y,z) \, \right]_{z=0} \,, \\[5pt]
z^\ga \left( f(y) \star z_\ga g(y,z) \right) &= i z^\ga \left( \partial^y_\ga f(y) \star g(y,z) \right)\,.
\end{align}
\end{subequations}
Both identities can be easily checked using the integral representation of the star product. For example, the first identity can be derived as follows
\begin{align}
&\frac{1}{(2\pi)^2} \int \intd^2u \, \intd^2v \, e^{i vu} \, f(y+u) \, (z-v)_\ga \, g(y+v,z-v) \big|_{z=0} \nonumber\\
=& \frac{1}{(2\pi)^2} \int \intd^2u \, \intd^2v \, e^{i vu} \,  f(y+u) \, (-v_\ga) \, g(y+v,-v) \nonumber\\ 
=& \frac{1}{(2\pi)^2} \int \intd^2u \, \intd^2v \, (-i \partial^u_\ga) e^{i vu} \,  f(y+u) \, g(y+v,-v) \nonumber\\
=& \frac{1}{(2\pi)^2} \int \intd^2u \, \intd^2v \, e^{i vu} \,  (i \partial^u_\ga) f(y+u) \, g(y+v,-v) \nonumber\\
=& \frac{1}{(2\pi)^2} \int \intd^2u \, \intd^2v \, e^{i vu} \,  (i \partial^y_\ga f(y+u) ) \, g(y+v,-v) \nonumber\\
=& \left( i \partial^y_\ga f(y) \star g(y,z) \, \right)_{z=0}  \,.
\end{align}
The other identity follows along similar lines. Now, consider the equation of motion for the linear perturbations in the one-form,
\begin{align}
D^{AdS} \boldsymbol{A}^{(1)} &=  -\frac{i}{2} D^{AdS} \, z^\ga \homo{0}{D^{AdS}  \MS^{(1)}_\ga} \big|_{z=0}\nonumber\\
&= -\frac{i}{2} \left\{ \cA^{AdS} , z^\ga \Gamma_0 \Big\langle \big[ \cA^{AdS}, z_\ga t^{(1)} \big]_\star\Big\rangle  \right\}_\star \big|_{z=0} \,,
\end{align}
where we have used the fact that $[\intd , z_\ga] = 0$ and $\cA^{AdS}$ denotes the AdS background. Furthermore, we have defined $\MS^{(1)}_\ga = z_\ga s^{(1)}$ where 
\begin{equation}
    s^{(1)} = \homo{1}{\flC^{(1)}(y)\star\varkappa} = \Gamma_1\Big\langle \big(\pC^{(1)}(y)\psi + \sC^{(1)}(y) \big)\star\varkappa\Big\rangle\,.
\end{equation}
We shall now apply the identities~\eqref{staridentities} to the equation of motion,
\begin{align}
D^{AdS} \boldsymbol{A}^{(1)} &= \frac{i}{2}\left\{ \partial^\ga_y \cA^{AdS} \; , \; \Gamma_0 \Big\langle \big[ \partial^y_\ga \cA^{AdS} \;, \; s^{(1)} \big]_\star\Big\rangle \right\}_\star \bigg|_{z=0} \,.
\end{align}
The derivative of $\cA^{AdS}$ is given by
\begin{equation}
    \partial^\ga_y \bar{\cA} = - \frac{i}{2} \left(\bvb^{\ga\gb} \phi + \bsc^{\ga\gb} \right)y_\gb\, ,
\end{equation}
and hence we find
\begin{equation}
    D^{AdS} \boldsymbol{A}^{(1)}  = -\frac{i}{8} \left\{ \left(\bvb^{\ga\gb} \phi + \bsc^{\ga\gb} \right)y_\gb ,\Gamma_0 \Big\langle \big[\left(\bvb_\ga{}^{\gc} \phi + \bsc_\ga{}^\gc \right)y_\gc, s^{(1)} \big]_\star\Big\rangle \right\}_\star \bigg|_{z=0} \,.
\end{equation}
We shall only focus on the terms proportional to $\bvb \wedge \bvb$ and $\pC^{(1)}$ (the other terms vanish which can be shown using the same methods that we use in the following). Let us denote the projection to this part by $|_{\bvb\bvb\pC}$, and we get
\begin{equation}
    D^{AdS} \boldsymbol{A}^{(1)} \big|_{\bvb\bvb\pC} = -\frac{i}{8} \bvb^{\ga\gb} \wedge \bvb_\ga{}^{\gc} \left\{ y_\gb ,\Gamma_0 \Big\langle \big\{y_\gc, \homo{1}{\pC^{(1)}\psi\star\varkappa} \big\}_\star\Big\rangle \right\}_\star \bigg|_{z=0} \,.
\end{equation}
Next we use
\begin{equation}
    \homo{1}{\pC^{(1)}(y)\star\varkappa} = \homo{1}{\pC^{(1)}(-z) e^{i\,yz}} =
    \int_0^1 \intd\tau \,\tau \,e^{i\tau\,yz}\,\pC^{(1)}(-\tau z) 
\end{equation}
and the identity
\begin{equation}
    \{y_\ga, f(y,z) \}_\star = 2 (y-i\partial^z)_\ga f(y,z)\, 
\end{equation}
that can be straightforwardly verified using the integral representation of the star product. We obtain (using the two-form $E$ defined in~\eqref{eq:twoFromEDefTmp})
\begin{align}
    D^{AdS} \boldsymbol{A}^{(1)} \big|_{\bvb\bvb\pC} &=
    \frac{i}{2}E^{\ga \gb} \, \int\limits_0^1\!\!\!\int\limits_0^1 \intd t \, \intd \tau \, (y-i\partial^z)_\ga \big(y-i \tfrac{1}{t} \partial^z\big)_\gb \, \tau \, \pC^{(1)}(-t \tau z) \, e^{i t \tau yz} \psi \big|_{z=0} \nonumber\\
    &= \frac{i}{2}E^{\ga \gb} \, \int\limits_0^1\!\!\!\int\limits_0^1 \intd t \, \intd \tau \, ((1-\tau t) y-i\partial^z)_\ga \big((1-\tau) y-i \tfrac{1}{t} \partial^z\big)_\gb \, \tau \, \pC^{(1)}(-t \tau z) \psi \big|_{z=0} \nonumber\\
    &= \frac{i}{2}E^{\ga \gb} \, \int\limits_0^1\!\!\!\int\limits_0^1 \intd t \, \intd \tau \, ((1-\tau t) y+i t\tau \partial^{u})_\ga ((1-\tau) y+i \tau \partial^{u})_\gb \, \tau \, \pC^{(1)}(u) \psi \big|_{u=0} \,.
\end{align}
It remains to evaluate the integrals:
\begin{subequations}
\begin{align}
& yy:\; \; \int_0^1 \int_0^1 \intd t \intd\tau \; (1-t \tau) (1-\tau) \tau = \frac{1}{8}\,,
\\& \partial \partial: \; \; \int_0^1 \int_0^1  \intd t \intd\tau \; t \tau^3 = \frac{1}{8}\,,
\\& y \partial: \; \;\int_0^1 \int_0^1 \intd t \intd\tau \; (1-t \tau) \tau^2 = \frac{5}{24}\,,
\\& \partial y: \; \;\int_0^1 \int_0^1 \intd t \intd\tau \; (1-\tau) t\tau^2 = \frac{1}{24}\,.
\end{align}
\end{subequations}
The $y \partial$ and $\partial y$ contributions can be combined because $E^{\ga\gb}$ is symmetric. We therefore obtain the final result
\begin{equation}
D^{AdS} \boldsymbol{A}^{(1)} \big|_{\bvb\bvb\pC} = \frac{i}{16} E^{\ga \gb} (y_\ga + i \partial^u_\ga)(y_\gb + i \partial^u_\gb) \pC^{(1)}(u) \psi  |_{u=0} \,.
\end{equation}
Using the same methods, one can show that all contributions involving the twisted zero-form $\sC^{(1)}$ cancel out.
}

\solu{prob:removalBackreaction}{
\hypertarget{sol:removalBackreaction}{Our} aim is to show that for $M$ defined in~\eqref{eq:fieldRedefVas} we have
\begin{equation}
    D^{AdS}\big( M\,\psi \big) = -\frac{i}{16} E^{\ga\gb}  (y_\ga + i \partial^u_\ga)(y_\gb + i \partial^u_\gb) \pC^{(1)}(u) \psi  |_{u=0} \,.
\end{equation}
We introduce a $0$-form 
\begin{equation}
    F_{\ga\gb} =  \int^1_0 \intd t \,(t^2 - 1) \, (y_\ga + it^{-1} \partial_\ga^{y} )(y_\gb + it^{-1} \partial_\gb^{y} ) \,\pC^{(1)} (ty) \, ,
\end{equation}
such that $M=-\frac{i}{8}\phi \,\bvb^{\ga\gb}F_{\ga\gb}$. Then
\begin{equation}
   D^{AdS}\big( M\,\psi \big) = -\frac{i}{8}\phi \bigg( \intd\bvb^{\ga\gb} F_{\ga\gb} - \bvb^{\ga\gb} \wedge \big( \intd F_{\ga\gb} - [\bsc ,F_{\ga\gb}]_* -\{ \bvb ,F_{\ga\gb} \}_* \big) \bigg) \psi \, .
\end{equation}
We compute
\begin{equation}
    [\bsc ,F_{\ga\gb}]_* = \bsc^{\gamma \gd}y_{(\gamma}\partial_{\gd)} F_{\ga\gb} \quad ,\quad  \{ \bvb ,F_{\ga\gb} \}_* = -\frac{i}{2} \bvb^{\gamma\gd} \big(y_\gamma y_\gd - \partial_\gamma^y \partial_\gd^y\big)F_{\ga\gb}\,\phi \, .
\end{equation}
To evaluate $\intd F_{\ga\gb}$ we use the equation of motion for $\pC^{(1)}$ which implies
\begin{equation}
    \intd\pC^{(1)} (ty) = -\frac{i}{2} \bvb^{\gamma\gd}\big( t^2 \,y_\gamma y_\gd - t^{-2}\,\partial_\gamma^y \partial_\gd^y \big) \pC^{(1)} (ty) \phi + \bsc^{\gamma\gd} \,y_{(\gamma}\partial_{\gd)}^y\,\pC^{(1)} (ty) \, .
\end{equation}
The last ingredient we need is 
\begin{equation}
    \intd\bvb^{\ga\gb} = -2\,\bsc^{\gamma(\ga} \wedge \bvb_\gamma{}^{\gb)}
\end{equation}
which can be inferred from $d\cA^{AdS} =\cA^{AdS}\wedge *\cA^{AdS}$. Combining everything we observe that in $B$ all contributions containing $\bsc$ drop out, and we arrive at
\begin{align}
    D^{AdS}\big( M\,\psi \big)=& \frac{1}{16} \bvb^{\ga\gb} \wedge \bvb^{\gamma\gd} \int_0^1 \intd t \, (t^2-1)  \nonumber \\
    &\qquad \Bigg\{ \bigg(yy+\frac{2i}{t}y\partial -\frac{1}{t^{2}}\partial\partial \bigg)_{(\ga\gb)} 
    \bigg(t^2 y y - \frac{1}{t^{2}} \partial \partial\bigg)_{\gamma \gd} \nonumber\\
    & \qquad\qquad  - \big( y y - \partial \partial\big)_{\gamma \gd} \bigg(yy+\frac{2i}{t}y\partial -\frac{1}{t^{2}}\partial\partial \bigg)_{(\ga\gb)} \Bigg\} \pC^{(1)}(ty) \psi \, .
\end{align}
Here we introduced the notation
\begin{align}
    \bigg(yy+\frac{2i}{t}y\partial -\frac{1}{t^{2}}\partial\partial \bigg)_{(\ga\gb)} &= y_\ga y_\gb +\frac{i}{t}\big( y_\ga\partial^y_\gb +y_\gb\partial^y_\ga \big) -\frac{1}{t^2}\partial^y_\ga\partial^y_\gb \nonumber\\
    &= \bigg( y_{(\ga}+\frac{i}{t}\partial^y_{(\ga}\bigg)
    \bigg( y_{\gb)}+\frac{i}{t}\partial^y_{\gb)}\bigg) \, .
\end{align}
The two-form built from the vielbein can be written as
\begin{equation}
    \bvb^{\ga\gb}\wedge \bvb^{\gamma\gd} = E^{(\gamma |(\ga} \epsilon^{\gb)|\gd)}\,.
\end{equation}
When the epsilon tensor hits two adjacent $y$'s or two adjacent derivatives, there is no contribution, otherwise it produces the $y$-counting operator
\begin{equation}
    y\cdot \partial^y = \epsilon^{\gb\gd}\, y_\gd \partial^y_\gb \, .
\end{equation}
We now commute all $y$'s to the left of the counting operator, and all derivatives to the right.
After some algebra, and defining $u=ty$ such that $\partial^y=t\,\partial^u$ we arrive at
\begin{align}
    D^{AdS}\big( M\,\psi \big) & = \frac{1}{16} E^{\ga\gamma} \int_0^1 \intd t \, (t^2-1)  \nonumber \\
    &\qquad \bigg\{ 
    i\,y_\ga y_\gamma \bigg(4t +(t^2-1) \frac{1}{t} y\cdot \partial^y\bigg)  
    -y_\ga \bigg(4t +(t^2-1) \frac{1}{t} y\cdot \partial^y\bigg) \partial^u_\gamma\nonumber\\
    &\qquad\ -y_\gamma \bigg(4t +(t^2-1) \frac{1}{t} y\cdot \partial^y\bigg) \partial^u_\ga
    -i\,\bigg( 4t +(t^2-1) \frac{1}{t} y\cdot \partial^y \bigg) \partial^u_\ga \partial^u_\gamma 
     \bigg\} \pC^{(1)}(ty) \psi \, .
\end{align}
Now we use $(y\cdot \partial^y)f(ty) =t\frac{\intd}{\intd t}f(ty)$, and we finally obtain
\begin{align}
    D^{AdS}\big( M\,\psi \big) & = \frac{1}{16} E^{\ga\gamma}  \int_0^1 \intd t \, (t^2-1)  \nonumber \\
    &\qquad  
    \bigg( (t^2-1)\frac{\intd}{\intd t} +4t\bigg) \big(  iy_\ga y_\gamma -y_\ga \partial^u_\gamma -y_\gamma \partial^u_\ga -i\partial^u_\ga \partial^u_\gamma \big) 
    \pC^{(1)}(ty) \psi \nonumber\\
    & = \frac{1}{16} E^{\ga\gamma}  \int_0^1 \intd t \, \frac{\intd}{\intd t}\Big( 
    (t^2-1)^2 \big(  iy_\ga y_\gamma -y_\ga \partial^u_\gamma -y_\gamma \partial^u_\ga -i\partial^u_\ga \partial^u_\gamma \big) 
    \pC^{(1)}(ty) \psi \Big)\nonumber\\
    & = -\frac{i}{16} E^{\ga\gamma} (y_\ga +i\partial_\ga^u)(y_\gamma +i\partial_\gamma^u)\pC^{(1)}(u)\Big|_{u=0} \psi\, .
\end{align}}



\bibliographystyle{JHEP}
\bibliography{references}

\end{document}

%% file: commands.tex
\usepackage[T1]{fontenc}

\usepackage[colorlinks=true,linkcolor=black,citecolor=black,urlcolor=blue,filecolor=black]{hyperref}

\usepackage{mathptmx}       
\SetSymbolFont{operators}{bold}{OT1}{txr}{bx}{n}
\DeclareSymbolFont{letters}{OML}{txmi}{m}{it}
\SetSymbolFont{letters}{bold}{OML}{txmi}{bx}{it}
\usepackage{helvet}         
\usepackage{courier}        
\usepackage{makeidx}         
\usepackage{graphicx}        
\usepackage{multicol}        
\usepackage[bottom]{footmisc}

\usepackage[nosort]{cite}
\usepackage{amsmath}
\usepackage{amsfonts}
\usepackage{amssymb}
\usepackage{empheq}
\usepackage{ytableau}

\usepackage{refcount}
\newcommand{\test}[1]{\ifnum\thesolution=\getrefnumber{#1}
{\bf \color{green} Correct number!}
\else
{\bf \color{red} Warning: Corresponding exercise number is~\ref{#1}!!}
\fi}
\newcommand{\solu}[2]{\setcounter{solution}{\getrefnumber{#1}-1}\begin{solution}

#2
\end{solution}}
\usepackage[colorinlistoftodos,draft]{todonotes} 

\renewcommand{\d}{\textrm{d}}

\newcommand{\MW}{\mathcal{W}}

\newcommand{\MB}{\mathcal{B}}
\newcommand{\MS}{\mathcal{S}}

\newcommand{\Mxi}{\mathcal{\xi}}
\newcommand{\ga}{\alpha}
\newcommand{\gb}{\beta}
\newcommand{\gc}{\gamma}
\newcommand{\gd}{\delta}

\newcommand{\thalf}{\tfrac{1}{2}}

\newcommand{\intd}{\text{d}}
\newcommand{\adsR}{\ell}

\newcommand{\flC}{{\mathbf{C}}} 
\newcommand{\pC}{\mathcal{C}} 
\newcommand{\sC}{{\tilde{\mathcal{C}}}} 
\newcommand{\homo}[2]{{\Gamma_{#1}\langle#2\rangle}}

\newcommand{\bvb}{\bar{e}} 
\newcommand{\bsc}{\bar{\omega}} 

\def\sn{\nu}
\def\sm{\mu}

\def\a{\alpha}

\def\g{\gamma}

\def\e{\epsilon}

\def\h{\eta}

\def\l{\lambda}
\def\L{\Lambda}
\def\m{\mu}
\def\n{\nu}
\def\x{\xi}

\def\p{\pi}

\def\r{\rho}

\def\O{\Omega}

\newcommand{\cA}{{\cal A}}

\newcommand{\cF}{{\cal F}}

\newcommand{\cH}{{\cal H}}

\newcommand{\cK}{{\cal K}}
\newcommand{\cL}{{\cal L}}

\newcommand{\cN}{{\cal N}}

\newcommand{\cP}{{\cal P}}

\newcommand{\cR}{{\cal R}}

\newcommand{\cT}{{\cal T}}

\newcommand{\cW}{{\cal W}}

\newcommand{\be}{\begin{equation}}
\newcommand{\ee}{\end{equation}}
\newcommand{\bea}{\begin{eqnarray}}
\newcommand{\eea}{\end{eqnarray}}

\newcommand{\nn}{\nonumber}

\newcommand{\tr}{\mathrm{tr}\,}
\newcommand{\res}{\mathrm{res}\,}

\newcommand{\ww}{\wedge}

\newcommand{\pr}{\partial}